\documentclass[12pt]{article}

\usepackage{amsmath}
\usepackage{amsfonts}
\usepackage{amssymb}
\usepackage[hiresbb]{graphicx}
\usepackage{caption}
\usepackage{subcaption}
\usepackage{float}
\usepackage{ascmac}
\usepackage{tikz}
\usepackage{url}
\usepackage{booktabs}
\usepackage[utf8]{inputenc}
\usepackage{listings}
\usepackage{xcolor}
\usepackage{here}
\usepackage[sectionbib,round]{natbib}
\definecolor{codegreen}{rgb}{0,0.6,0}
\definecolor{codegray}{rgb}{0.5,0.5,0.5}
\definecolor{codepurple}{rgb}{0.58,0,0.82}
\definecolor{backcolour}{rgb}{0.95,0.95,0.92}
\lstdefinestyle{mystyle}{
    backgroundcolor=\color{backcolour},
    commentstyle=\color{codegreen},
    keywordstyle=\color{magenta},
    numberstyle=\tiny\color{codegray},
    stringstyle=\color{codepurple},
    basicstyle=\footnotesize,
    breakatwhitespace=false,
    breaklines=true,
    captionpos=b,
    keepspaces=true,
    numbers=left,
    numbersep=5pt,
    showspaces=false,
    showstringspaces=false,
    showtabs=false,
    tabsize=2
}
\usepackage{authblk}
\newcommand{\be}{\begin{equation}}
\newcommand{\ee}{\end{equation}}
\newcommand{\beq}{\begin{eqnarray}}
\newcommand{\eeq}{\end{eqnarray}}

\title{\Large\bf Price Stability of Cryptocurrencies as a Medium of Exchange}
\author[1]{Tatsuru Kikuchi}
\author[2]{Toranosuke Onishi}
\author[1]{Kenichi Ueda}
\affil[1]{\small \it Faculty of Economics, The University of Tokyo}
\affil[2]{\small \it Tokio Marine \& Nichido Fire Insurance Co., Ltd.}
\date{\small (November, 2021)}
\begin{document}
\maketitle
\setcounter{tocdepth}{3}
\begin{abstract}
We present positive evidence of price stability of cryptocurrencies as a medium of exchange. For the sample years from 2016 to 2020, the prices of major cryptocurrencies are found to be stable, relative to major financial assets. Specifically, after filtering out the less-than-one-month cycles, we investigate the daily returns in US dollars of the major cryptocurrencies ({\it i.e.}, Bitcoin, Ethereum, and Ripple) as well as their comparators ({\it i.e.}, major legal tenders, the Euro and Japanese yen, and the major stock indexes, S\&P 500 and MSCI World Index). We examine the stability of the filtered daily returns using three different measures. First, the Pearson correlations increased in later years in our sample. Second, based on the dynamic time-warping method that allows lags and leads in relations, the similarities in the daily returns of cryptocurrencies with their comparators have been present even since 2016. Third, we check whether the cumulative sum of errors to predict cryptocurrency prices, assuming stable relations with comparators' daily returns, does not exceeds the bounds implied by the Black-Scholes model. This test, in other words, does not reject the efficient market hypothesis. 
\end{abstract}

\section{Introduction}
A key question on cryptocurrencies is whether they can be used as money, a medium of exchange. Many argue they cannot. It is said to be primarily because they are not backed by any valuable goods and because their price movements are too volatile to use as money. 

This paper presents positive evidence of price stability of cryptocurrencies as a medium of exchange. For this purpose, we examine price stability after filtering out the high frequency components. For example, the Euro or Japanese yen are known to be quite volatile against the US dollar but still they are used in the daily lives by people in the Euro area countries or in Japan because the high-frequency exchange rate volatility does not matter much for daily real-goods transactions, as long as their average values are stable against the US dollar.

Specifically, we apply the frequency-filtering technique developed by \citet{MW}, abbreviated as the MW filter hereafter, to the daily returns of major cryptocurrencies, legal tenders, and stock indexes. We essentially get rid of the less-than-one-month ({\it i.e.}, 20 business days) movements in their daily returns. Then, we compare the stability in the filtered daily returns of cryptocurrencies against those of legal tenders and stock indexes. For cryptocurrencies, we pick the three major ones, that is, Bitcoin (BTC), Ethereum (ETH), and Ripple (XRP). For major legal tenders to compare, we pick the Euro (EUR) and Japanese yen (JPY), and we pick the S\&P 500 (S\&P500) and MSCI World Index (MSCI) for the major stock indexes. All values are defined in terms of US dollars. The daily data are taken from the beginning of 2016 to the end of 2020.

To see the stability of these prices based on the MW-filtered daily returns, we use three different measures. The first one is the Pearson correlation as is also used by \citet{MW} in a different context. We find that the MW filter makes cryptocurrency returns closer to the other financial asset returns. Moreover, the correlations in daily returns between cryptocurrencies and other assets increased in later years in our sample. 

The second one is a measure based on the dynamic time warping (DTW) algorithm. In general, DTW is a method that calculates an optimal match between two time series data. For the Pearson correlation, we compare, for example, BTC returns with S\&P 500 returns in each same day. However, in DTW, the data sequences are {\it warped} non-linearly in the time dimension to see a similar pattern allowing for lags and leads. DTW is a widely applied algorithm in non-economic fields, such as speech pattern recognition, though some studies have already used DTW in economics, for example, \citet{DTW1}, \citet{DTW2}, and \citet{DTW3}. Based on the DTW-based similarity measure, we find that the similarity in daily returns of cryptocurrencies with other assets has been present even since 2016, the beginning of our sample, and not much has changed throughout our sample period. Note that the MW filtering also does the job by removing erratic ups and downs in the similarity over the years for some pairs of cryptocurrencies and other assets.

The third one is the cumulative sum (CUSUM) test, a test of the structural stability under the efficient market hypothesis. It is simply a two-sample Kolmogorov-Smirnov test in statistics. It was first developed based on the recursive residuals by \citet{Brown}, and the method was refined using the OLS residuals by \citet{PK}. Some discussions on the boundaries were made by \citet{Zeileis2}\footnote{R package was also developed by \citet{Zeileis}.}. The CUSUM test is essentially based on the Black-Scholes model with the assumption of stable relationships with other financial assets. In economics literature, for example, the stability of money-demand models has been examined by the CUSUM test (see \citet{Stock}, \citet{PCBK} and \citet{Elliott}). 

We find that cryptocurrency daily returns, based on the whole sample period, do pass the CUSUM test after the MW filtering, though they do not pass before the MW filtering. As for the other assets (EUR, JPY, S\&P500, and MSCI), all of them essentially pass the test either before or after the MW filtering. In other words, cryptocurrency price movements have been enjoying stable relations with other major assets over the whole sample period, from 2016 to 2020. We also conduct the CUSUM tests for later years only, that is, samples starting from 2017, 2018, or 2019. BTC starting 2018, and JPY starting 2018 do not pass the test before the MW filtering. However, after the MW filtering, only ETH starting 2018 fails the test. 

There have been a few studies that examined the price stability of cryptocurrencies, such as \citet{Martin}. A typical finding is a high price volatility of cryptocurrency compared with traditional financial assets. Some similarity between Bitcoin and gold has been discussed in \citet{Anne}. A more comprehensive study has been done in \citet{Yhlas}, who analyzed the underlying factors that influence prices of cryptocurrencies, stocks, and gold in both the short- and long-term. Also, \citet{KKK} show a somewhat stable relationship across prices of Bitcoin, gold, and S\&P 500 based on a GARCH model. Moreover, \citet{Shams} shows that cryptocurrency prices co-move a lot, perhaps based on SNS-based demand effects. Most of these existing studies, however, methodologically requires some sort of stationarity or parametric stability, for which \citet{MW} renounce for key economic and financial time series data and instead advocate to use nonparametric or semi-parametric models, such as the Pearson correlation after the MW filtering. Note that DTW is also nonparametric, and the CUSUM test is the test for a stable relationship itself, not taking it as given.

A few studies on cryptocurrency price movements have shown that the high frequency movements may be contaminated by inefficient market forces. This is also a key reason to get rid of such high frequency components of price movements. For example, \citet{AH} find that a technical cost issue ({\it i.e.}, hash rate) matters for determining the daily raw returns of Bitcoin. Moreover, \citet{LSW} show that pump-and-dump schemes have been prevalent in cryptocurrency markets. Such a scheme is an illegal act in the US stock market as it is backed by a small group of people who buy up a target cryptocurrency up to a target price level and then sell it, all under close coordination, to obtain huge profits.

A fact that a cryptocurrency is not backed by real assets cannot be used as a reason to dismiss a cryptocurrency's role as a medium of exchange. The theory-of-money literature has shown the existence of a fiat money that is not backed by any real assets. Indeed, money has a role when financial contracts are not completely enforceable due to costs (especially for small transactions), and such an environment is often considered as trading with strangers for whom contract enforcement is difficult (\citet{Bewley}, \citet{Townsend}, and \citet{KW}). Moreover, intrinsically valueless money has a value only when people think it is accepted by other people, who think it is accepted by yet other people, who think…(repeated infinitely). This means that money is in essence a \textit{bubble} and has to be \textit{common knowledge} (\citet{Chwe}). Given these formal theoretical definitions in economics, cryptocurrencies (or sea shells in old times) are legitimate candidates of money. What has remained is an empirical question as to whether an acceptance by the people is stable or not.

The rest of the paper is organized as follows. In Section \ref{sec3}, we give a detailed description of the data for the cryptocurrencies and other financial assets. In Section \ref{sec4}, we perform a standard spectrum analysis to see the frequency distributions. Section \ref{sec5} gives a brief explanation of the MW filter. In Section \ref{sec6}, we show the Pearson correlation before and after the MW filtering. Section \ref{sec7} explains the DTW method and results. In Section \ref{sec8}, we discuss the structural stability based on the CUSUM test. Section \ref{sec9} concludes.

\section{Data}\label{sec3}
We use daily time series data from January 1, 2016, to December 31, 2020, for three major cryptocurrencies, Bitcoin (BTC), Ethereum (ETH), and Ripple (XRP). Those data are obtained through the API provided by the Cryptocompare (\url{https://min-api.cryptocompare.com/}). We evaluate the stability of cryptocurrency values compared with the other financial data, such as S\&P500, gold, MSCI World Index (MSCI), Japanese Yen (JPY), and the Euro (EUR). While the historical data of gold prices and the MSCI World Index are obtained from Investing.com (\url{https://investing.com/}), the historical data of S\&P500 is obtained from Yahoo finance (\url{https://finance.yahoo.com/}), and the historical data of JPY and EUR are obtained from the API provided by Cryptocompare (\url{https://min-api.cryptocompare.com/}). 

The prices of the cryptocurrencies and other financial assets are all defined in US dollars. We let $P_t$ denote the price of a financial asset at time $t$ and $P_{t-1}$ the price of the asset at time $t-1$, and $r_t$ the (logarithmic) returns of the asset at $t$ by
\be
  r_t = \log \frac{P_t}{P_{t-1}} \times 365 \;.
\ee
The descriptive statistics of the daily returns of cryptoprices and the key financial assets for the whole sample period are described in Table \ref{tab:Table1}\footnote{The skewness measures symmetricity of the distribution. The kurtosis tells us whether the data is heavy-tailed or light-tailed, relative to a normal distribution. Both skewness and kurtosis of normal distribution are equal to $0$.}. Figure \ref{fig:Fig1} shows the 50-day moving averages of the daily returns of the cryptocurrencies and other financial assets.

\section{Short-Run and Long-Run Movements}\label{sec4}
The short-run and long-run price stability appear a bit differently in the return data. The daily movements of cryptocurrencies are in general very large, relative to the traditional financial assets. This means that when we make a Fourier transformation from the time domain to the frequency domain, it is expected to see a large amount of high-frequency component in their spectrum. 

A power spectrum analysis is a useful tool to see the frequency distribution of some financial assets data. Let us consider the daily return of asset data $r_t$ of length $T=n \, \Delta t = 1.258 \times 10^5$, where $n=1258$ is the number of daily data and $\Delta t = 100$ is taken as the sampling interval, and the frequency is defined by $f=1/T = 1/(n \, \Delta t) = 7.949 \times 10^{-6}$. Then, the discretized frequency and the discretized time are given by $f_k = k/(n \, \Delta t) ~ (k=1, \cdots , n)$ and $t_j = j \, \Delta t ~ (j=1, \cdots , n)$, respectively. With this setup, the discrete Fourier transformation (DFT) of the return of asset data $r_j$ is defined by
\be
 Q_k = \sum_{j=1}^{n} r_j \exp(- i \, 2 \pi j \, k/n) \;,
\ee
where $i = \sqrt{-1}$ is the imaginary unit. The discretization of time with the sampling period $\Delta t$ implies a limitation of frequency $f$ to the band $f \in [-\frac{1}{2 \, \Delta t},~ \frac{1}{2 \; \Delta t} ]$ as frequencies outside the range are folded inside by the finite sampling (see, for example, \citet{HM}). This boundary frequency is called the Nyquist frequency, $f_{\text{NQ}} = 1/(2 \, \Delta t)$. 

The power spectrum density $I_\text{PSD} (f_k)$ for the return of asset data $r_t$ is then calculated by
\be
 I_\text{PSD} (f_k) = \frac{2 \, \Delta t^2}{n} Q_k \, Q_k^* \;,
\ee
where $*$ stands for the complex conjugate. In practice, we use the Fast Fourier Transformation (FFT) to calculate the Fourier transformation of the return of asset data, and take the sampling period as $\Delta t = 100$, hence, the Nyquist frequency is $f_{\text{NQ}} = 0.005$. 

The power spectrum density for the returns of cryptocurrencies and key financial assets are shown in Figures \ref{fig:Spec_BTC}, \ref{fig:Spec_ETH}, \ref{fig:Spec_XRP}, \ref{fig:Spec_JPY}, \ref{fig:Spec_EUR}, \ref{fig:Spec_GOLD}, \ref{fig:Spec_SP500}, and \ref{fig:Spec_MSCI}. These figures typically show that returns of cryptocurrencies and key financial assets vary intensively with frequency. In the frequency region higher than $f_{\text{NQ}} = 0.005$, the noises seem dominant over the structurally important signals. On the other hand, the structural signals are quite clear in the frequency region lower than $f_{\text{NQ}} = 0.005$ ({\it i.e.}, $5 \times 10^{-3}$).

\section{Frequency Filtering}\label{sec5}
Now we know that the short-run movements of cryptocurrencies are somewhat strange, we focus on the rest by filtering out the short-run movements. A proper way of using a high-frequency filtering method for noisy non-stationary data has been proposed in \citet{MW}, who use an orthogonal projection of time series data $y_j (j=1,\cdots,n)$ onto the space spanned by the cosine function based on low-frequency periodic vectors. Their approach is an extension of the discrete cosine transformation (DCT) for the time series $y_j$,
\beq
  \hat{y}_1 &=& \sqrt{\frac{1}{n}} \sum_{j=1}^{n} y_j \; , \nonumber \\
  \hat{y}_k &=& \sqrt{\frac{2}{n}} \sum_{j=1}^{n} \cos \left[ (j-1)
  \left( k - \frac{1}{n} \right) \frac{\pi}{n} \right] y_j \; , ~ \text{for}~ k=2,\cdots,n \; . 
\eeq

The long-run projection of M\"{u}ller-Watson is made by using the following transformation, truncating the approximation with $q < n$,
\beq
  \hat{y}_1 &=& \sqrt{\frac{1}{n}} \sum_{j=1}^{q} y_j \; , \nonumber \\
  \hat{y}_k &=& \sqrt{\frac{2}{n}} \sum_{j=1}^{q} \cos \left[ (j-1) 
  \left(k - \frac{1}{2} \right) \frac{\pi }{n} \right] y_j \; , ~ \text{for}~ k=2,\cdots,q \;.  
\eeq

Here, up to the frequency that is represented by parameter $q$ are extracted. More specifically, $\hat{y}_q$ is the data that is filtered to extract only frequencies lower than $\hat{f}=1/\hat{T} = q/(2 n)$ or $\hat{T}=2 n/q$ period. We use the following condition to decide the value of parameter $q$.
\be
 q = \left[ \frac{n}{10} \right] + 3 \;,
\ee
where $\left[ \frac{n}{10} \right]$ is the integer part of $\frac{n}{10}$. For example, in the case of BTC, we set these parameters as $n = 1420 ~ ({\it i.e.}, q=145), ~ \hat{T} = 19,~ \hat{f}= 0.0526$. This roughly means that we take out short-run cycles occurring for less than $20$ business days, which is essentially one month in the calendar\footnote{In the literature, a similar method of high-frequency filtering has been proposed and named as the separating information maximum likelihood (SIML) method in \citet{KS}. We use an R script provided by Sato, for whom we are grateful.}.

In other words, the MW filtering smooths the higher frequency movements in more than one month (20 business days) as shown in the power spectrum density in Figures \ref{fig:Spec_BTC}, \ref{fig:Spec_ETH}, \ref{fig:Spec_XRP}, \ref{fig:Spec_JPY}, \ref{fig:Spec_EUR}, \ref{fig:Spec_GOLD}, \ref{fig:Spec_SP500}, and \ref{fig:Spec_MSCI}. Note that the MW filtering also somewhat smooths the low frequencies. 

\section{Similarity to the Major Financial Assets: The M\"{u}ller-Watson (2018) Approach}\label{sec6}
Recall that our objective in this study is to see the stability of prices or returns of cryptocurrencies relative to key financial assets. The time series similarity could be measured by the degree of comovements across two or more variables. How to evaluate comovements is discussed extensively in \citet{MW}. They renounce the cointegtation approach and instead propose to use the Pearson correlation after filtering out high-frequency movements in the data. 

Recall that the Pearson correlation is defined for two time series of data $x_t$ and $y_t$ as
\be
  P(x_t, y_t) = \frac{\text{cov}\left(x_t,y_t \right)}{\sigma_x ~ \sigma_y} 
  = \frac{\mathbb{E}\left[ \left(x_t - \mu_x \right)\left(y_t - \mu_y \right) \right]}{\sigma_x ~\sigma_y} \; . 
\ee
In order to show the importance of the M\"{u}ller-Watson filtering, we show the correlation matrices for both before and after filtering. The correlations before the M\"{u}ller-Watson filtering are shown in Tables \ref{fig:Fig1b}, \ref{fig:Fig2b}, \ref{fig:Fig3b}, \ref{fig:Fig4b}, and \ref{fig:Fig5b} for 2016, 2017, 2018, 2019, and 2020, respectively. The correlations after the M\"{u}ller-Watson filtering are shown in Tables \ref{fig:Fig1a}, \ref{fig:Fig2a}, \ref{fig:Fig3a}, \ref{fig:Fig4a}, and \ref{fig:Fig5a} for 2016, 2017, 2018, 2019, and 2020, respectively. 

The correlation matrices indicate the long-run stability of cryptocurrencies relative to the major financial assets. More specifically, we find three patterns. First, the correlations between the returns of cryptocurrencies and key financial assets becomes more significant after the M\"{u}ller-Watson filtering for each year. Second, the returns of cryptocurrencies and key financial assets become more and more correlated to each other in later years. Third, the correlations among cryptocurrencies are getting stronger over time.

\section{Dynamic Time Warping}\label{sec7}
Next, we investigate the similarities of cryptocurrency returns to key financial asset returns by looking at the Dynamic Time Warping (DTW). This is often used in pattern recognition, as developed by \citet{It} and \citet{SC} (for a review, see \citet{MM}), which allows leads and lags of data sequences over time when measuring the similarity.

A set of time series, which are the returns of assets in our case, $x = \{x_t \}=(x_1, \cdots, x_n)$ and $y = \{y_t \}=(y_1, \cdots, y_m)$, can be expressed in terms of the warping path, $\Lambda =(w_1, \cdots, w_K)$. The warping path is a contiguous set of matrix elements that define a mapping between $x_t$ and $y_t$. A typical element $w_{\ell}$ is represented by $(i,j)$, that selects $x_i$ from $x$ and $y_j$ from $y$.

Formally, the DTW distance between two given sequences, $x$ and $y$, can be calculated by
\be
  D(x, y) 
  = \min_{\Lambda^*} ~ \sum_{(i,j) \in \Lambda^*} d(x_i,  y_j) \;,
\ee
where $\Lambda^*$ is the warping path that minimizes the cumulative distance of all mapped point-pairs on the path. Although the element distance function $d(x_i,y_j)$ could take one of several forms, normally and here, it is given by the Euclidean distance, that is,
\be
  d(x_i, y_j) = \sqrt{\left( x_i - y_j \right)^2 } \;.
\ee

The best warping path $\Lambda^*$ is found using a dynamic programming approach to align two sequences. Going through all possible paths is combinatorially explosive, as pointed out by \citet{BC}, hence, the possible warping paths need to be restricted. When we apply the DTW method to two economic data series, we think it natural to constrain the possible warping paths by following three conditions.

The first constraint is the boundary condition. This requires the warping path to start and finish in diagonally opposite corner cells ({\it i.e.}, the starting date and the ending date of the sample) of the warping path matrix,
\be
w_1 = (1,1) ~\text{and} ~ w_K = (n,m) \;. 
\ee
The second constraint is the continuity condition. This constraint limits the path transitions to adjacent points in time,
\be
w_\ell - w_{\ell -1} \in \{ (1,0) \;, (0,1) \;, (1,1) \} \;.
\ee
The third constraint is the monotonicity condition. This constraint preserves the time-order of points.
\be
\text{If}~w_\ell = (i,j) ~\text{and} ~ w_{\ell+1} = (i^\prime,j^\prime) \;, 
~\text{then}~ i \leq i^\prime  ~\text{and} ~ j \leq j^\prime \;.
\ee

In practice, the DTW distance is calculate in a recursive way. First, we construct two subsequences $\tilde{x}_i = (x_1,\cdots,x_i)$ for $i =1,\cdots,n$ and $\tilde{y}_j = (y_1,\cdots,y_j)$ for $j =1,\cdots,m$ from an assumed optimal path sequence. Then, we define the cost function for each subsequence as
\be
 \gamma(i, j) = D(\tilde{x}_i, \tilde{y}_j) ~ \text{for}~i =1,\cdots,n ~\text{and}~j =1,\cdots,m \; .
\ee

The cost function $\gamma(i, j)$ specifies the total cost of an assumed optimal warping path starting from $w_1 = (1,1)$ and ending at $w_\ell = (i,j)$.

Second, we find the cost function $\gamma(i,j)$ in the recursive way. Namely, it is computed iteratively using a nested loop according to the following formula,
\be
  \hat{\gamma}(i, j) = d(x_i,  y_j) + \min \{\gamma(i-1, j), \, \gamma(i, j-1), \, \gamma(i-1, j-1) \} \; , 
\ee
and
\be
  \hat{\gamma}(i, j) = \gamma(i, j) \; ,
\ee
where $i =1,\cdots,n$ and $j = 1,\cdots,m$, with appropriate initial values\footnote{The python code is available upon request.}. Once the cost function is found, it is the DTW distance, $D(x,y) = \gamma(n,m)$, and the optimal warping path $\Lambda^*$ is identified. For example, the paired data ({\it i.e.}, the DTW optimal warping path $\Lambda^*$) between BTC and GOLD in 2020 is shown in Figure \ref{fig:FigBTC}. 

Note that the Pearson correlation always compares the same period data $(x_t, y_t)$. For the DTW distance, the periods to compare are related $(x_i, y_j)$, though $i$ and $j$ are not necessary the same time $t$. For example, if $x=\{x_t \}$ and $y=\{y_t \}$ follow the simple sine and cosine functions, respectively, in the time dimension, then they are orthogonal, and thus their Pearson correlation is zero. However, they look the same when one of their phases is shifted by $\pi$, thus, their DTW distance is $0$, which means their shapes are exactly the same.

It is useful to normalize the DTW distance relative to some base value. Let us use the maximum of the DTW distance between a pair of key financial assets ({\it i.e.}, JPY, EUR, GOLD, S\&P500, and MSCI), $D_{\text{max}}$, as the base value for all other DTW distances. That is, we normalize the DTW distance as follows,
\be
\bar{D}(x, y) = \frac{D(x, y)}{D_{\text{max}}} \;.
\ee

We show the normalized DTW distance $\bar{D}(x, y)$ before the M\"{u}ller-Watson filtering in Tables \ref{fig:Fig11}, \ref{fig:Fig12}, \ref{fig:Fig13}, \ref{fig:Fig14}, and \ref{fig:Fig15} for each sample year. The corresponding results after the M\"{u}ller-Watson filtering are shown in Tables \ref{fig:Fig11q}, \ref{fig:Fig12q}, \ref{fig:Fig13q}, \ref{fig:Fig14q}, and \ref{fig:Fig15q}. 

There are several salient features which appear in these Tables. First, the strengthening trend in similarities among returns of cryptocurrencies and key financial assets seem less clear than in the Pearson correlations (Tables \ref{fig:Fig1b} and \ref{fig:Fig2b}). Before the MW filtering, they, especially BTC, seem to behave similarly even in 2016, that is the initial period of our sample. These similarities dwindled in 2017 until 2019. In 2020, it recovered especially for BTC. However, after the MW filtering, similarities are less fluctuating and relatively stable, especially for BTC, in almost all the sample years. 

\begin{figure}[H]
  \centering
  \caption{50 days moving average returns (annualized, \%) of cryptocurrencies and key financial assets}
  \label{fig:Fig1}
  \begin{subfigure}[H]{0.8\textwidth}
  \centering
  \caption{50 days moving average returns (annualized, \%) of cryptocurrencies (BTC, ETH, and XRP)}
  \label{fig:Fig1a}
  \includegraphics[width=1.0\linewidth]{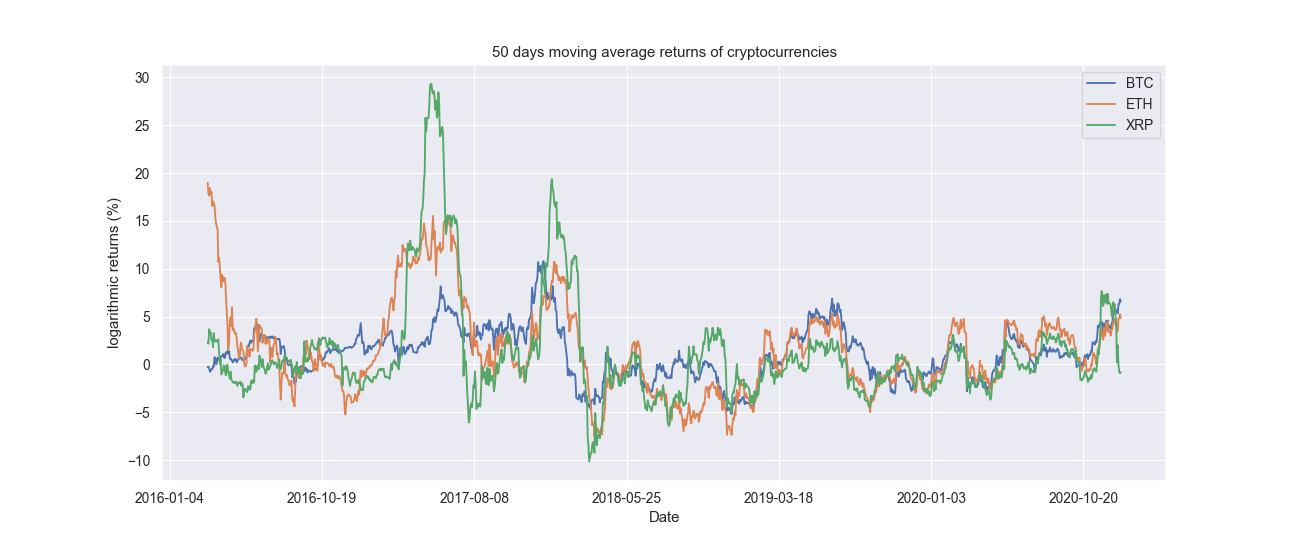}
  \end{subfigure}
  \hfill
  \begin{subfigure}[H]{0.8\textwidth}
  \centering
  \caption{50 days moving average returns (annualized, \%) of JPY, EUR, and GOLD}
  \label{fig:Fig1b}
  \includegraphics[width=1.0\linewidth]{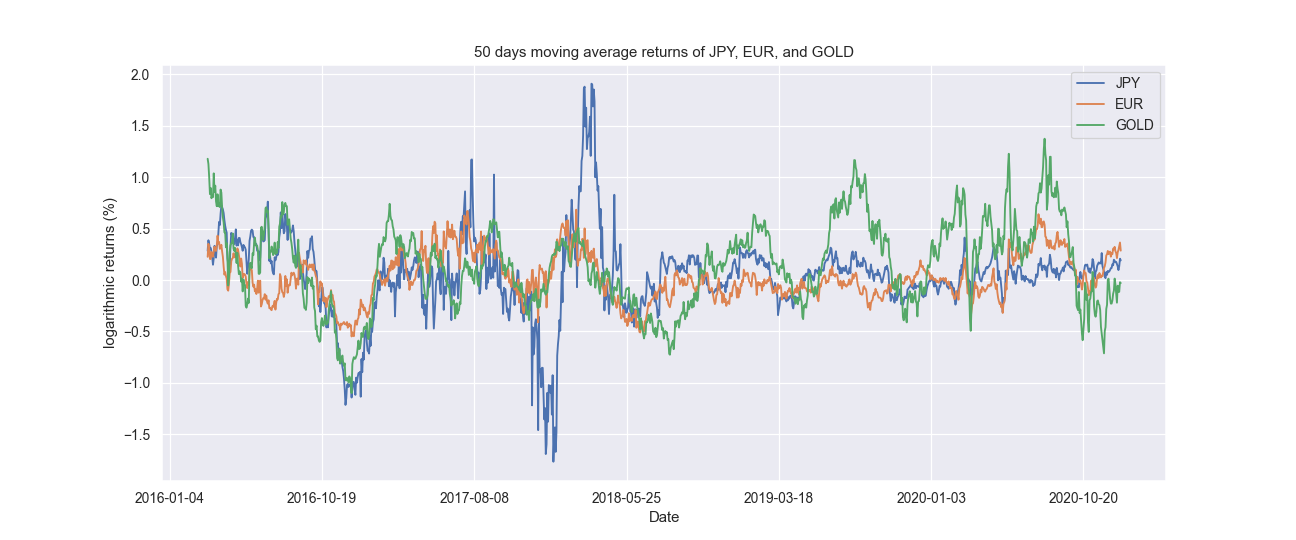}
  \end{subfigure}
  \hfill
  \begin{subfigure}[H]{0.8\textwidth}
  \centering
  \caption{50 days moving average returns (annualized, \%) of S\&P500 and MSCI}
  \label{fig:Fig1c}
  \includegraphics[width=1.0\linewidth]{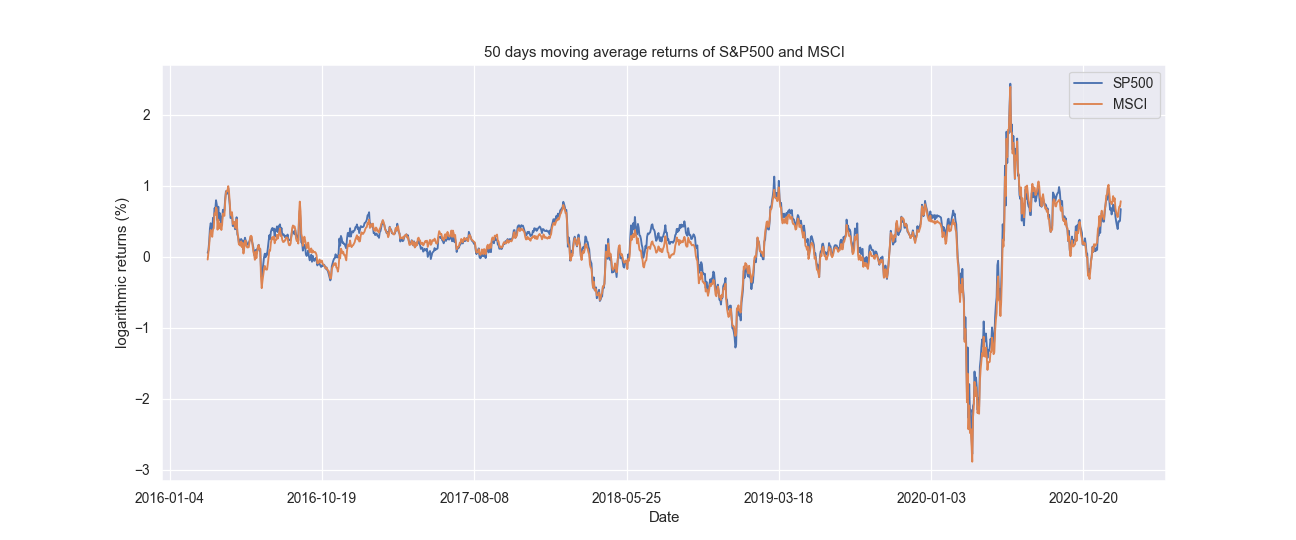}
  \end{subfigure}
\end{figure}

\begin{figure}[H]
  \centering
  \caption{The power spectrum density (PSD) of BTC}
  \label{fig:Spec_BTC}
  \begin{subfigure}[H]{1.0\textwidth}
  \centering
  \caption{The power spectrum density (PSD) of BTC (before the M\"{u}ller-Watson filtering)}
  \includegraphics[width=1.0\linewidth]{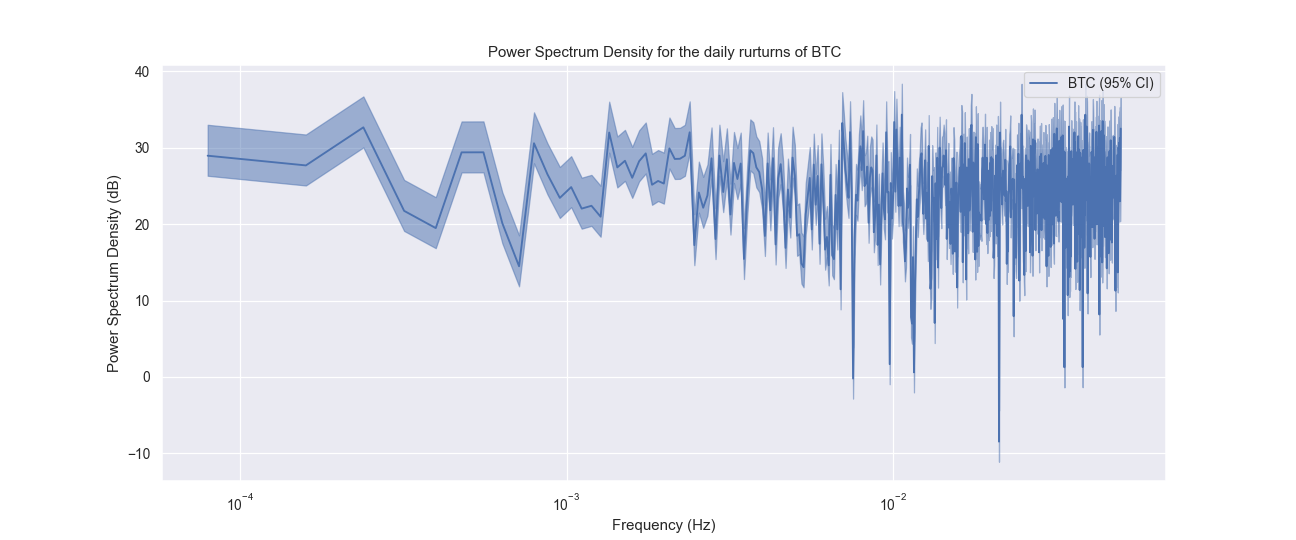}
  \end{subfigure}
  \hfill
  \begin{subfigure}[H]{1.0\textwidth}
  \centering
  \caption{The power spectrum density (PSD) of BTC (after the M\"{u}ller-Watson filtering)}
  \includegraphics[width=1.0\linewidth]{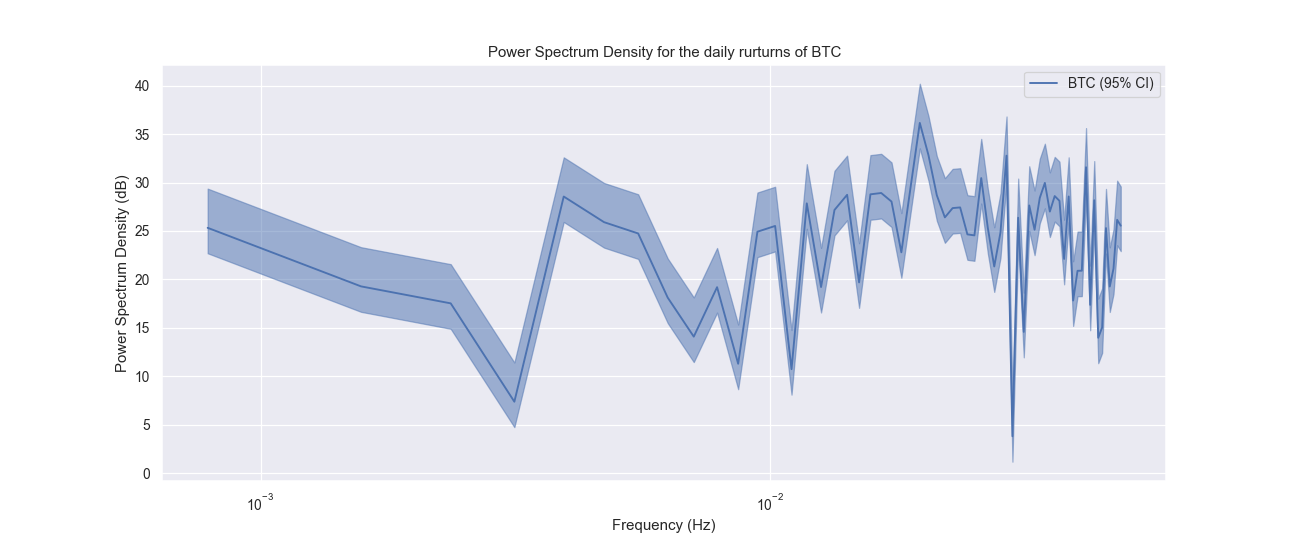}
  \end{subfigure}
\end{figure}

\begin{figure}[H]
  \centering
  \caption{The power spectrum density (PSD) of ETH}
  \label{fig:Spec_ETH}
  \begin{subfigure}[H]{1.0\textwidth}
  \centering
  \caption{The power spectrum density (PSD) of ETH (before the M\"{u}ller-Watson filtering)}
  \includegraphics[width=1.0\linewidth]{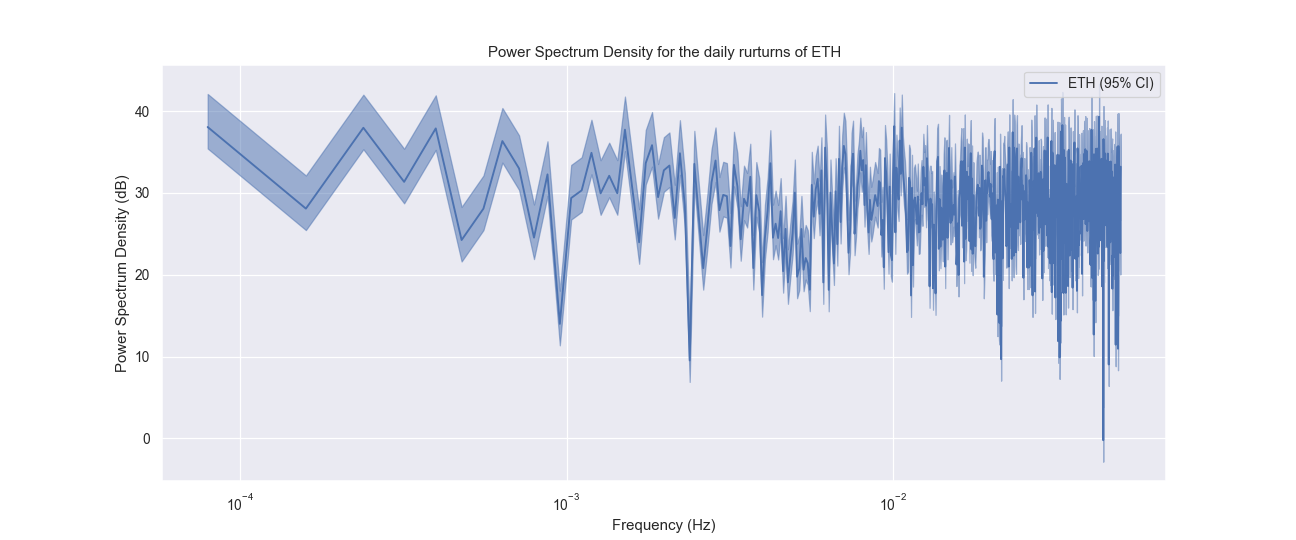}
  \end{subfigure}
  \hfill
  \begin{subfigure}[H]{1.0\textwidth}
  \centering
  \caption{The power spectrum density (PSD) of ETH (after the M\"{u}ller-Watson filtering)}
  \includegraphics[width=1.0\linewidth]{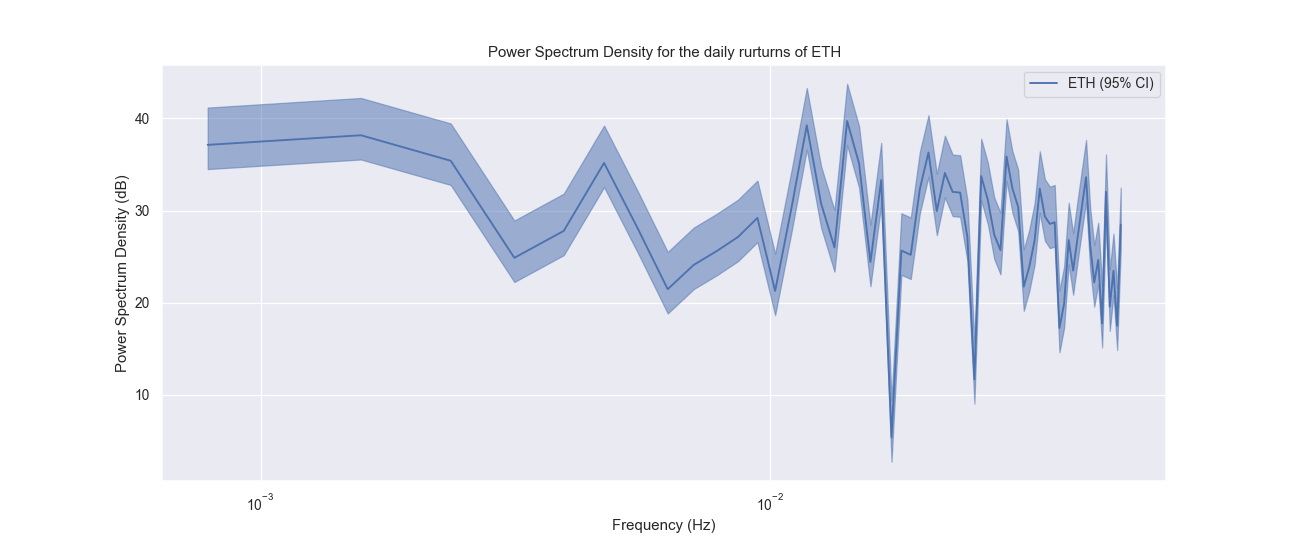}
  \end{subfigure}
\end{figure}

\begin{figure}[H]
  \centering
  \caption{The power spectrum density (PSD) of XRP}
  \label{fig:Spec_XRP}
  \begin{subfigure}[H]{1.0\textwidth}
  \centering
  \caption{The power spectrum density (PSD) of XRP (before the M\"{u}ller-Watson filtering)}
  \includegraphics[width=1.0\linewidth]{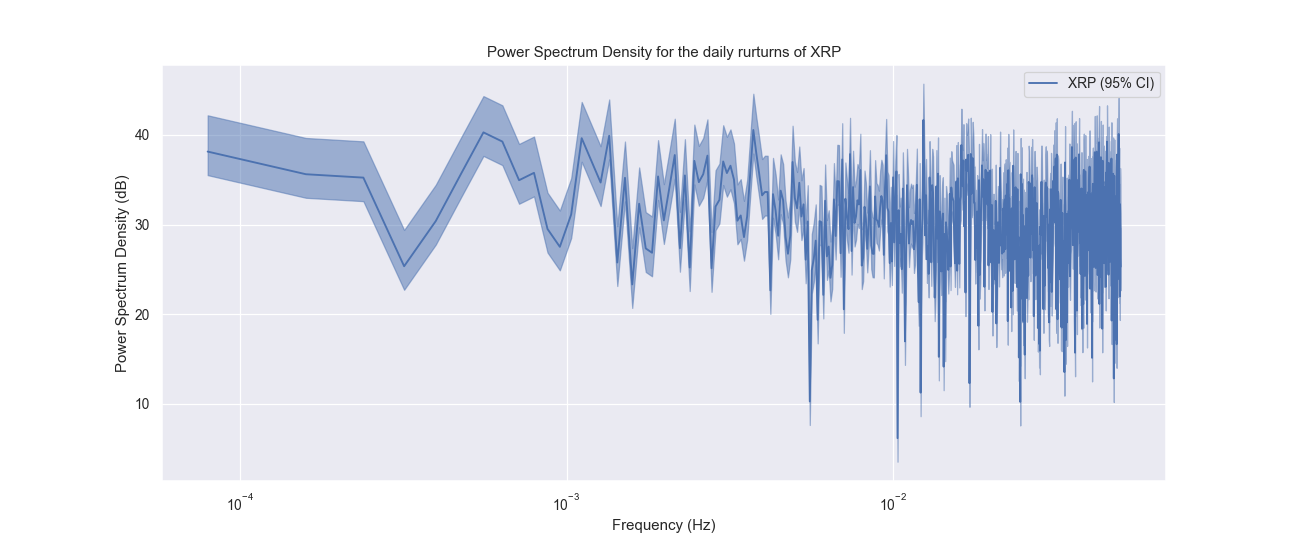}
  \end{subfigure}
  \hfill
  \begin{subfigure}[H]{1.0\textwidth}
  \centering
  \caption{The power spectrum density (PSD) of XRP (after the M\"{u}ller-Watson filtering)}
  \includegraphics[width=1.0\linewidth]{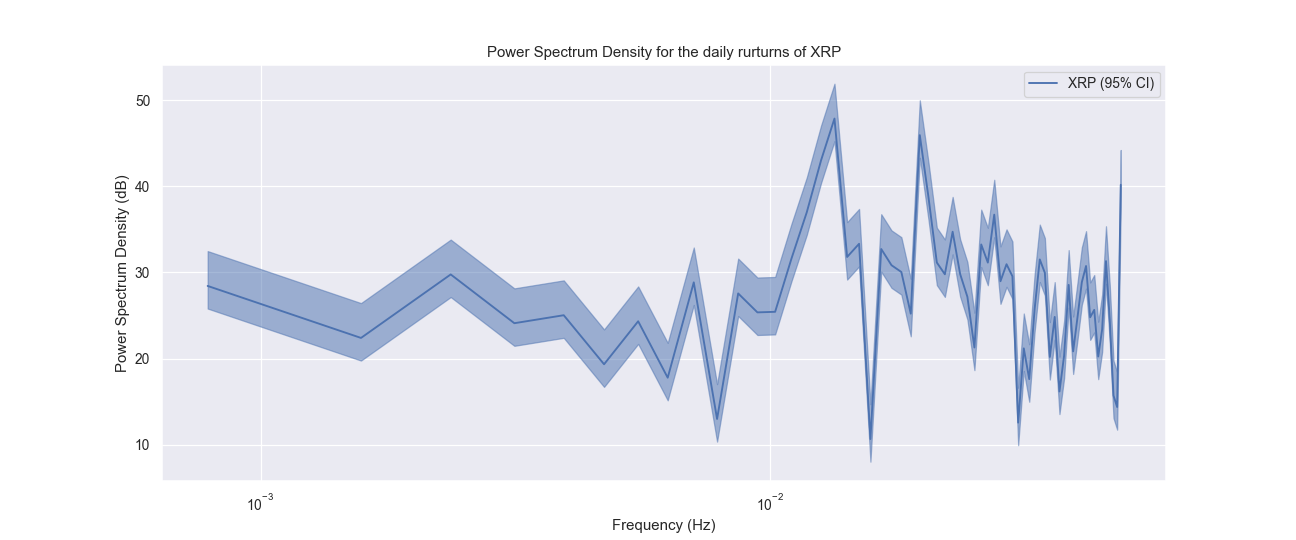}
  \end{subfigure}
\end{figure}

\begin{figure}[H]
  \centering
  \caption{The power spectrum density (PSD) of JPY}
  \label{fig:Spec_JPY}
  \begin{subfigure}[H]{1.0\textwidth}
  \centering
  \caption{The power spectrum density (PSD) of JPY (before the M\"{u}ller-Watson filtering)}
  \includegraphics[width=1.0\linewidth]{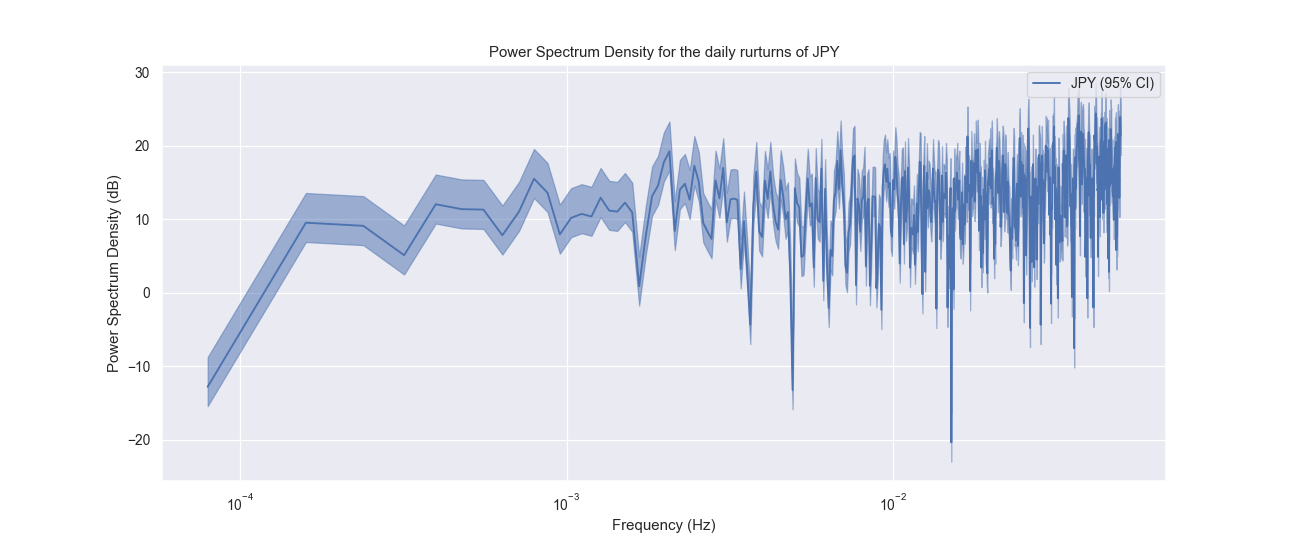}
  \end{subfigure}
  \hfill
  \begin{subfigure}[H]{1.0\textwidth}
  \centering
  \caption{The power spectrum density (PSD) of JPY (after the M\"{u}ller-Watson filtering)}
  \includegraphics[width=1.0\linewidth]{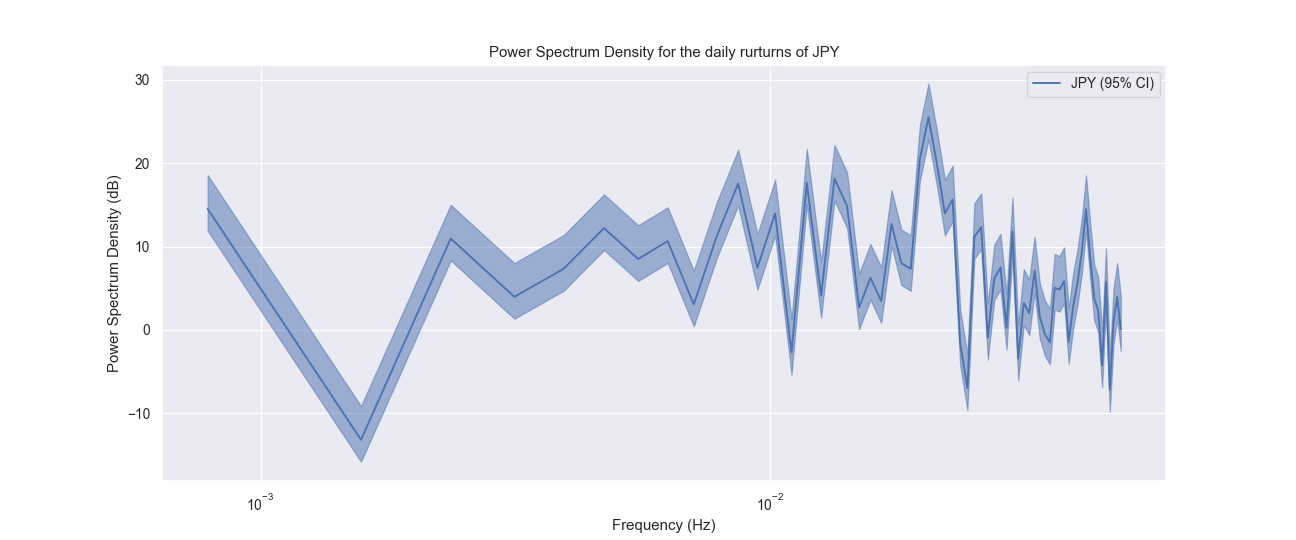}
  \end{subfigure}
\end{figure}

\begin{figure}[H]
  \centering
  \caption{The power spectrum density (PSD) of EUR}
  \label{fig:Spec_EUR}
  \begin{subfigure}[H]{1.0\textwidth}
  \centering
  \caption{The power spectrum density (PSD) of EUR (before the M\"{u}ller-Watson filtering)}
  \includegraphics[width=1.0\linewidth]{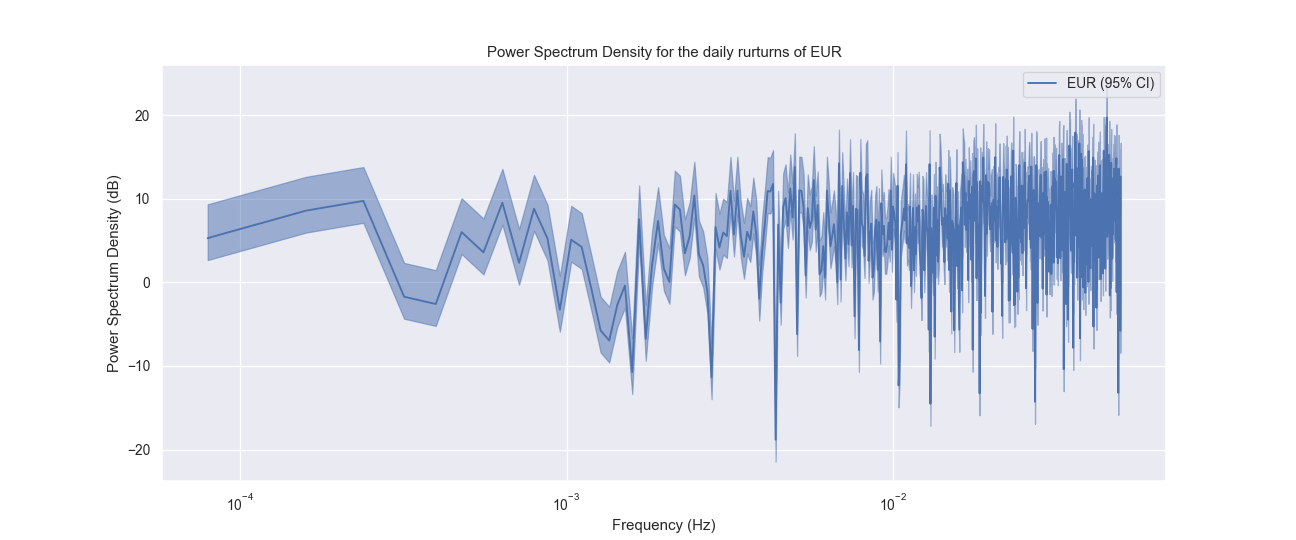}
  \end{subfigure}
  \hfill
  \begin{subfigure}[H]{1.0\textwidth}
  \centering
  \caption{The power spectrum density (PSD) of EUR (after the M\"{u}ller-Watson filtering)}
  \includegraphics[width=1.0\linewidth]{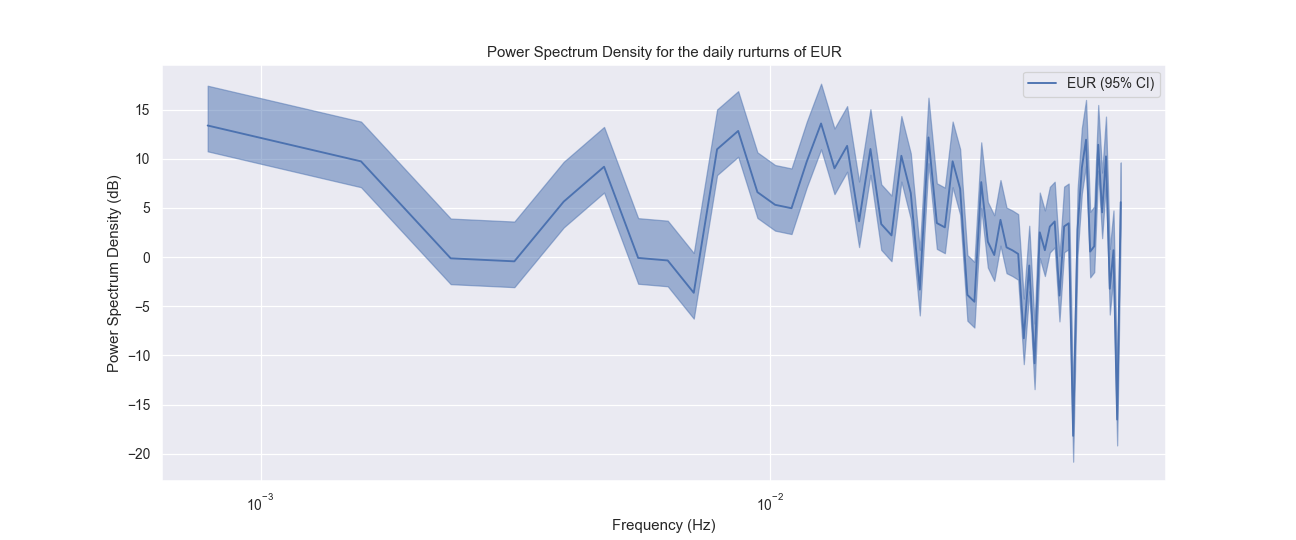}
  \end{subfigure}
\end{figure}

\begin{figure}[H]
  \centering
  \caption{The power spectrum density (PSD) of GOLD}
  \label{fig:Spec_GOLD}
  \begin{subfigure}[H]{1.0\textwidth}
  \centering
  \caption{The power spectrum density (PSD) of GOLD (before the M\"{u}ller-Watson filtering)}
  \includegraphics[width=1.0\linewidth]{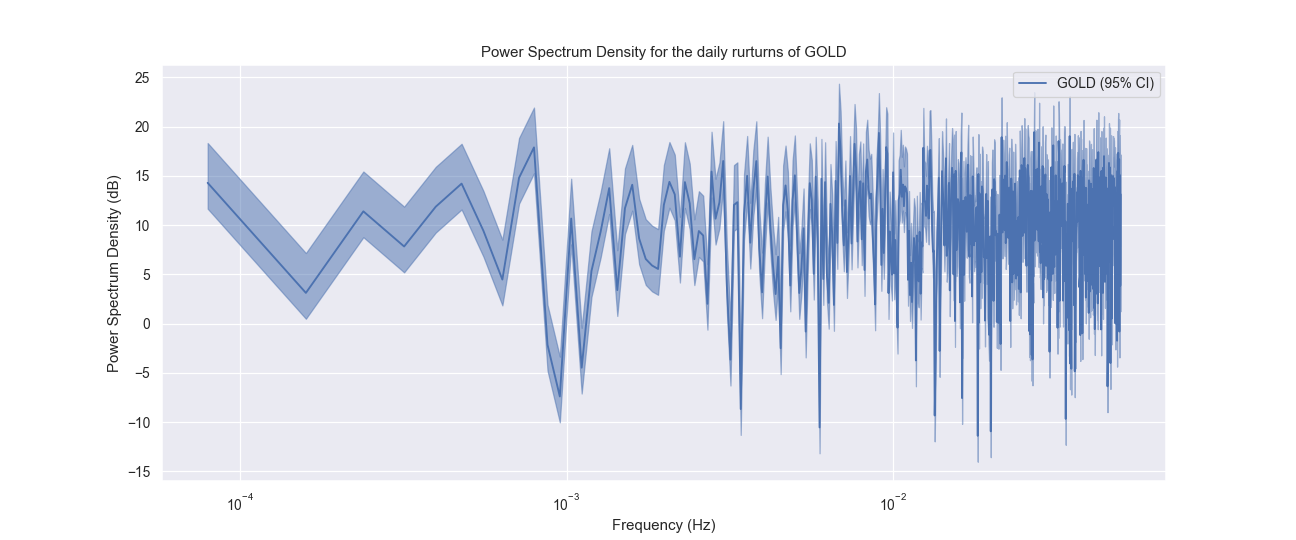}
  \end{subfigure}
  \hfill
  \begin{subfigure}[H]{1.0\textwidth}
  \centering
  \caption{The power spectrum density (PSD) of GOLD (after the M\"{u}ller-Watson filtering)}
  \includegraphics[width=1.0\linewidth]{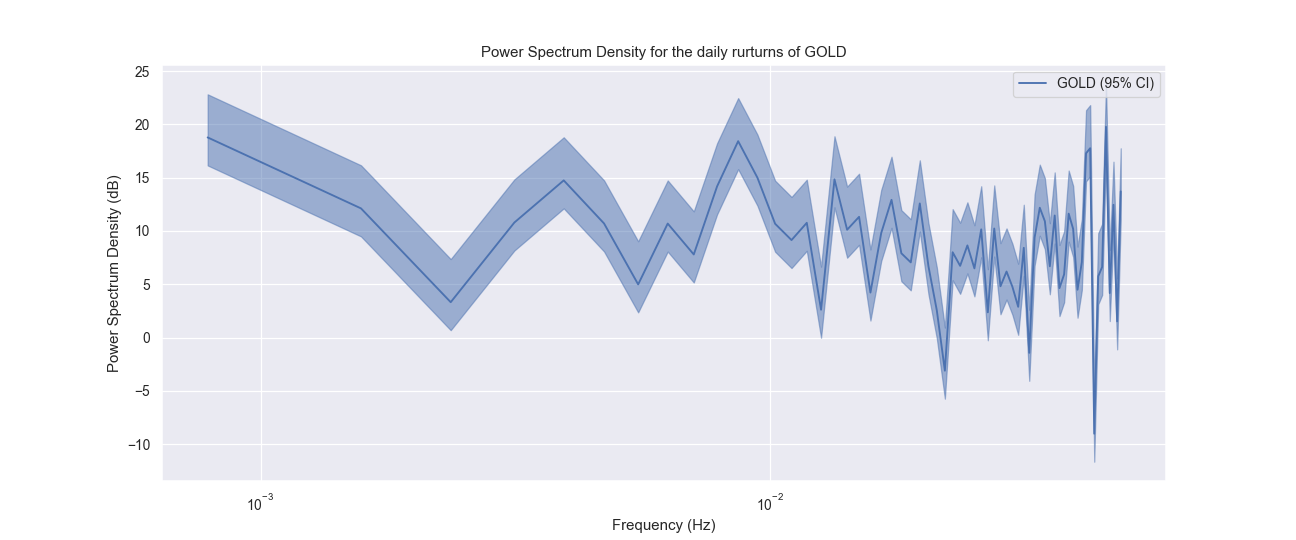}
  \end{subfigure}
\end{figure}

\begin{figure}[H]
  \centering
  \caption{The power spectrum density (PSD) of S\&P500}
  \label{fig:Spec_SP500}
  \begin{subfigure}[H]{1.0\textwidth}
  \centering
  \caption{The power spectrum density (PSD) of S\&P500 (before the M\"{u}ller-Watson filtering)}
  \includegraphics[width=1.0\linewidth]{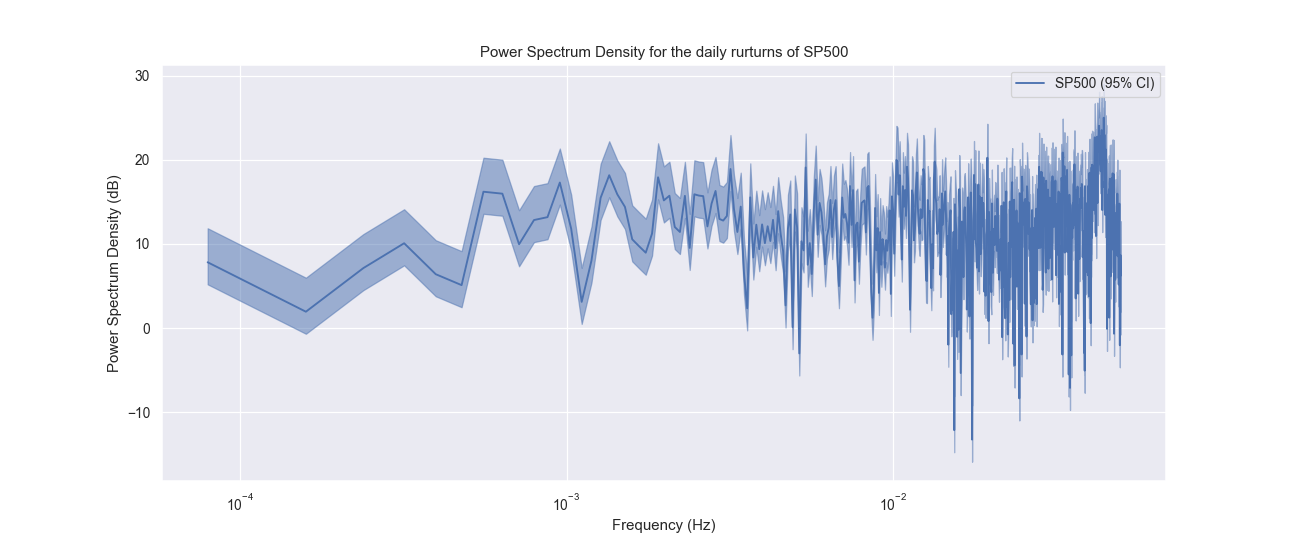}
  \end{subfigure}
  \hfill
  \begin{subfigure}[H]{1.0\textwidth}
  \centering
  \caption{The power spectrum density (PSD) of S\&P500 (after the M\"{u}ller-Watson filtering)}
  \includegraphics[width=1.0\linewidth]{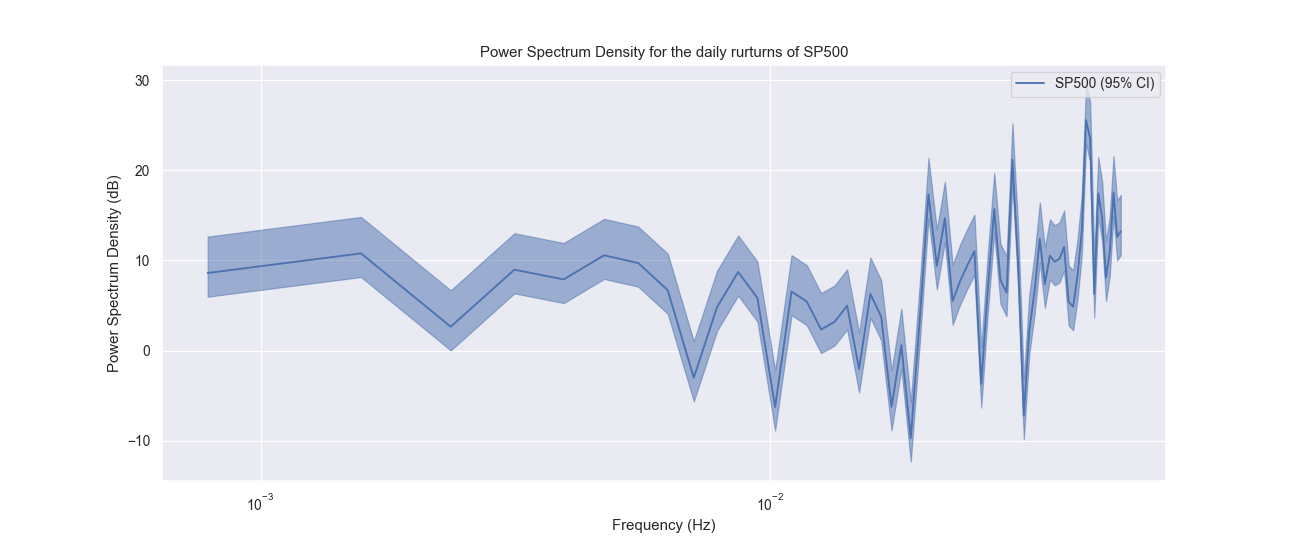}
  \end{subfigure}
\end{figure}

\begin{figure}[H]
  \centering
  \caption{The power spectrum density (PSD) of MSCI}
  \label{fig:Spec_MSCI}
  \begin{subfigure}[H]{1.0\textwidth}
  \centering
  \caption{The power spectrum density (PSD) of MSCI (before the M\"{u}ller-Watson filtering)}
  \includegraphics[width=1.0\linewidth]{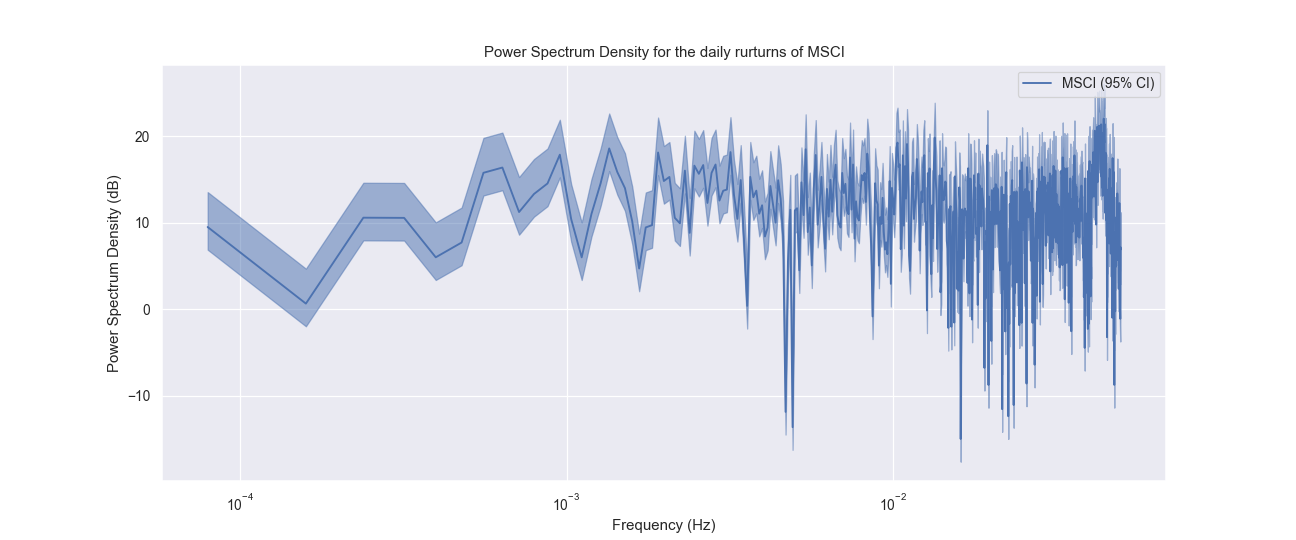}
  \end{subfigure}
  \hfill
  \begin{subfigure}[H]{1.0\textwidth}
  \centering
  \caption{The power spectrum density (PSD) of MSCI (after the M\"{u}ller-Watson filtering)}
  \includegraphics[width=1.0\linewidth]{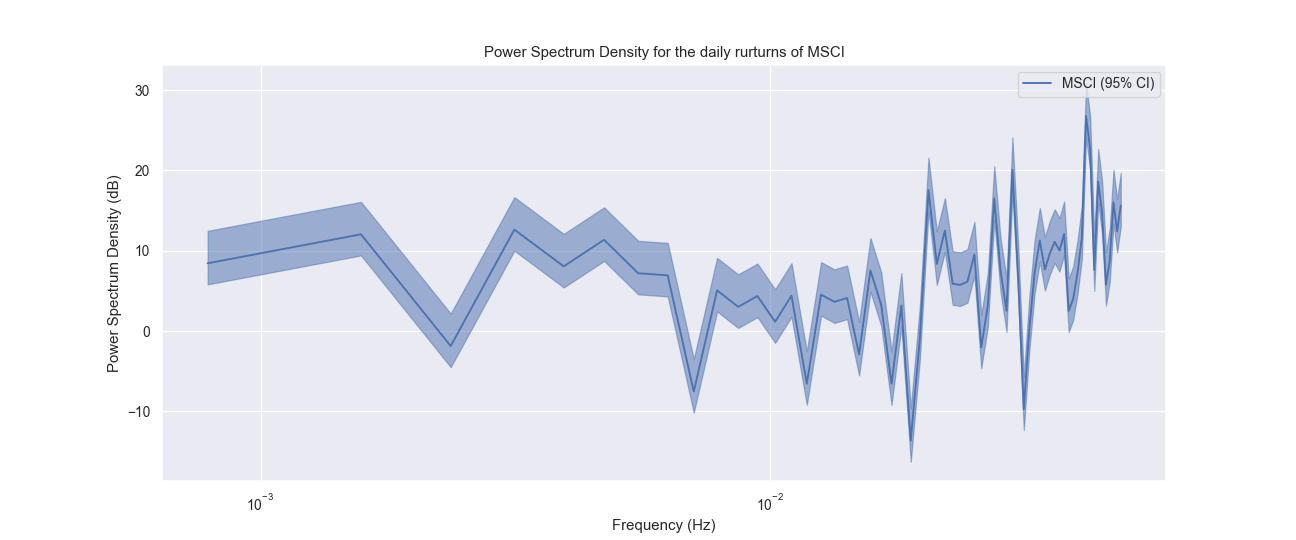}
  \end{subfigure}
\end{figure}

\begin{figure}[H]
  \centering
  \caption{DTW alignment between BTC and GOLD in 2020}
  \label{fig:FigBTC}
  \includegraphics[width=1.0\linewidth]{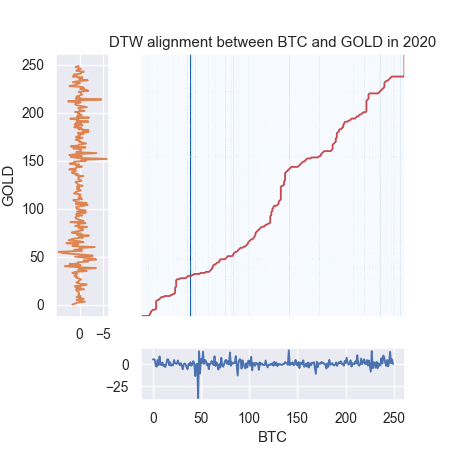}
\end{figure}

Second, frequency filtering does make some differences. For example, Table \ref{fig:Fig12} shows quite a different movement only in XRP in 2017 compared with the other assets. This result also seems to be supported in Figure \ref{fig:Fig1}. In particular, it is shown in Figure \ref{fig:Fig1} that the returns of XRP dramatically rises in 2017. However, after the MW filtering, the returns of XRP behave in a similar way, as shown in Table \ref{fig:Fig12q}. The following year, 2018, also show almost the same case for the returns of XRP before the MW filtering (see Table \ref{fig:Fig13}) and after the frequency filtering (see Table \ref{fig:Fig13q}). This is interesting as there was the well-known 2018 cryptocurrency crash (also known as the great crypto crash). Note, however, that even after the MW filtering, XRP seems to move somewhat differently compared to other assets in 2020.

\section{Stability from the Viewpoint of the Black-Scholes Model}\label{sec8}
Under a simple, efficient market hypothesis, can we assume the price stability of cryptocurrencies? Since the traditional financial asset returns can be considered as more or less following the Brownian motion, we test whether the cryptocurrency returns also follows a simple form of the Black-Scholes formula under stable relationships with key financial asset returns.

Following \citet{Elliott}, we consider the tests of the null hypothesis of a stable coefficient $\bar{\beta}$ in
\be
 z_t = {\bf X}^T_t \bar{\beta} + {\bf Z}^T_t \bar{\gamma} + \bar{\epsilon}_t
\ee
against the alternative of variable coefficient $\beta_t$ in
\be
 z_t = {\bf X}^T_t \beta_t + {\bf Z}^T_t \gamma + \epsilon_t \; .
\ee

The null hypothesis of stable $\bar{\beta}$ is rejected, if the cumulative sum (CUSUM) of residual with constant $\bar{\beta}$ exceed the limit that is consistent with the Brownian bridge (see, Appendix \ref{app}). 

First, we consider the structural stability based on the OLS-CUSUM test by using all the data. Here, $\bar{\beta}$ is obtained once by OLS. As is clearly explained in \citet{KPS}, the sum of the OLS residual always starts from $0$ and ends with $0$. In between, it has some bounds.

Second, we also employ the recursive CUSUM (Rec-CUSUM) test, which can be useful to monitor price stability every day. Here, the residuals are obtained as one-step ahead forecast errors. In other words, $\bar{\beta}$ is obtained using historical data at each date. The null hypothesis of the structural stability depends on the choice of the sampling periods. We divide our data into four sets of data samples as follows: (i) all the data; (ii) a sample data from 2017 to 2020; (iii) a sample data from 2018 to 2020; and (iv) a sample data from 2019 to 2020. We then proceed with the Rec-CUSUM test for each sample set.

We consider two types of boundaries for the CUSUM tests\footnote{The crucial difference between the Rec-CUSUM test and the OLS-CUSUM test is in the limiting process. In the former case, the limiting process is the Wiener process (Brownian motion), but in the latter case, the limiting process is the Brownian bridge.}. We use the following two types of boundaries for the OLS-CUSUM test. One is a constant in time and another is proportional to the standard deviation function of the corresponding theoretical process, as proposed in \citet{Zeileis2},
\beq
 b(t) &=& \nu   \\
 b(t) &=& 2 \nu \,  \sqrt{t (1-t)} 
\eeq
for the OLS-CUSUM path. We take $\nu = 1.358$ at the 95\% confidence interval. 

When we apply the Rec-CUSUM test, we consider the following two types of boundaries as proposed in \citet{Brown} and \citet{Zeileis2},
\beq
 b(t) &=& \lambda \, (2 \, t -1)  \\
 b(t) &=& 2 \nu \,  \sqrt{t} 
\eeq
with $\lambda$ takes the value $\lambda = 0.948$ at the 95\% confidence interval\footnote{There are several constructions of the boundaries of the stochastic process. In the monitoring context, a nearly linear boundary was considered in \citet{CSW} and \citet{LHK}, which is written as follows:
\be
 b(t) = \left[ t (t-1) \left( a^2 + \ln \left( \frac{t}{t-1} \right) \right) \right]^{1/2} \;, \nonumber
\ee
where $a$ only depends on the confidence level $\alpha$.}.

Note that there is no need to change conditions in the test of the structural stability even if we apply frequency filtering. This is known as pattern recovery in \citet{MC} and \citet{MC2}. Similar studies can be found, for example, in \citet{KKBG} and \citet{PJ}.

We perform the test of this structural stability, based on the following regression for the daily returns of each cryptocurrency or key financial asset on the returns of other financial assets. For example, for Bitcoin,
\beq
\tt{ BTC}_{t} &=& \bar{\beta}_0 + \tt{ BTC}_{t-1} ~ \bar{\beta}_1 + \tt{ JPY}_{t-1} ~\bar{\beta}_2 + 
\tt{ EUR}_{t-1} ~\bar{\beta}_3 + \tt{ GOLD}_{t-1} ~\bar{\beta}_4 +  \tt{ S\&P500}_{t-1} ~\bar{\beta}_5  \nonumber \\
&+& \tt{ MSCI}_{t-1} ~\bar{\beta}_7 + \bar{\epsilon}_t ~ ,
\eeq
where the symbolic notations stand for the daily returns, for example, $\tt{ BTC}_{t}$ for the BTC daily return at time $t$. 

\begin{figure}[H]
  \centering
  \caption{The OLS-CUSUM test of BTC (before the M\"{u}ller-Watson filtering)}
  \label{fig:CUSUM1}
  \includegraphics[width=1.0\linewidth]{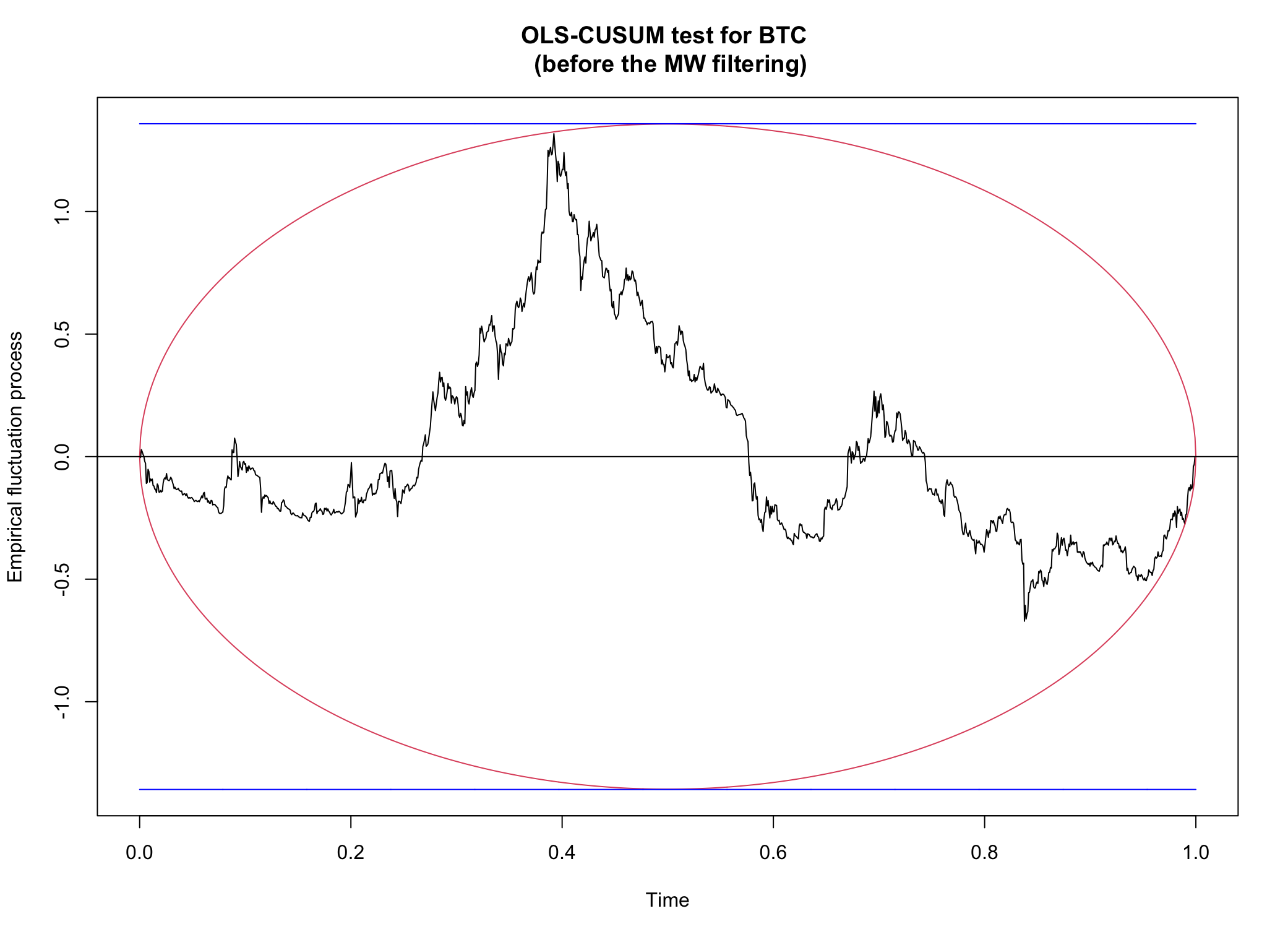}
\end{figure}

\begin{figure}[H]
  \centering
  \caption{The OLS-CUSUM test of ETH (before the M\"{u}ller-Watson filtering)}
  \label{fig:CUSUM2}
  \includegraphics[width=1.0\linewidth]{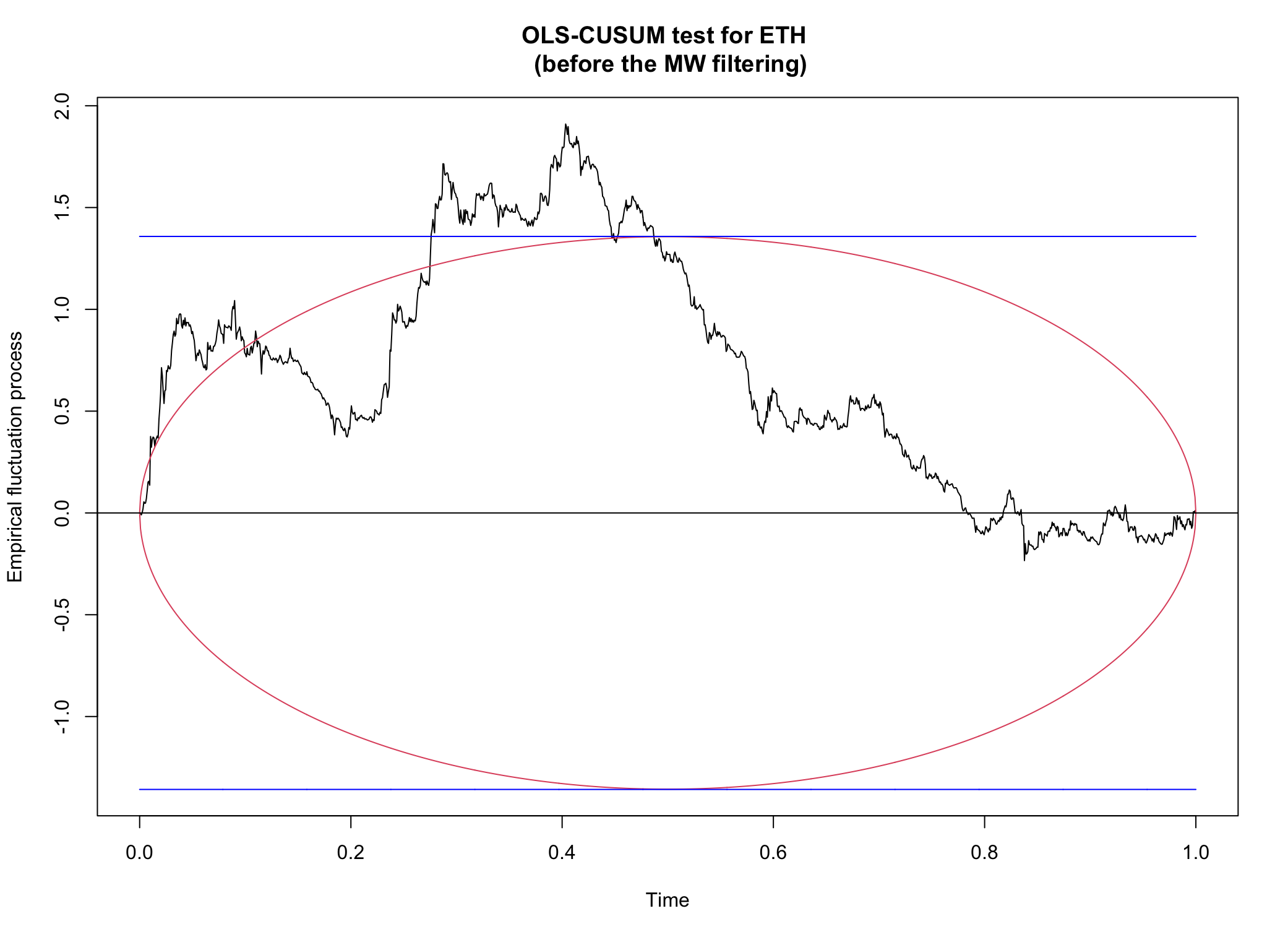}
\end{figure}

\begin{figure}[H]
  \centering
  \caption{The OLS-CUSUM test of XRP (before the M\"{u}ller-Watson filtering)}
  \label{fig:CUSUM3}
  \includegraphics[width=1.0\linewidth]{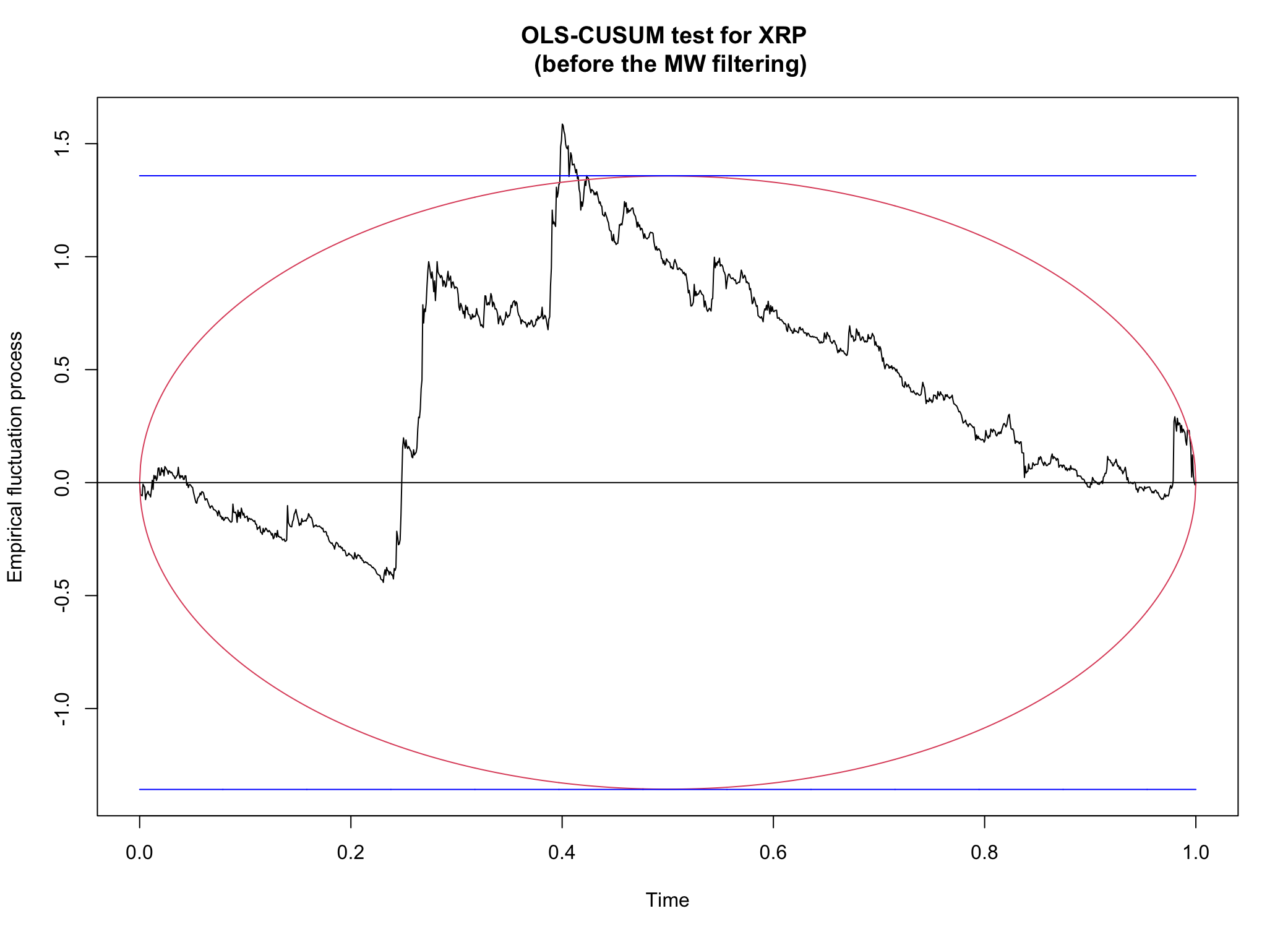}
\end{figure}

\begin{figure}[H]
  \centering
  \caption{The OLS-CUSUM test of JPY (before the M\"{u}ller-Watson filtering)}
  \label{fig:CUSUM4}
  \includegraphics[width=1.0\linewidth]{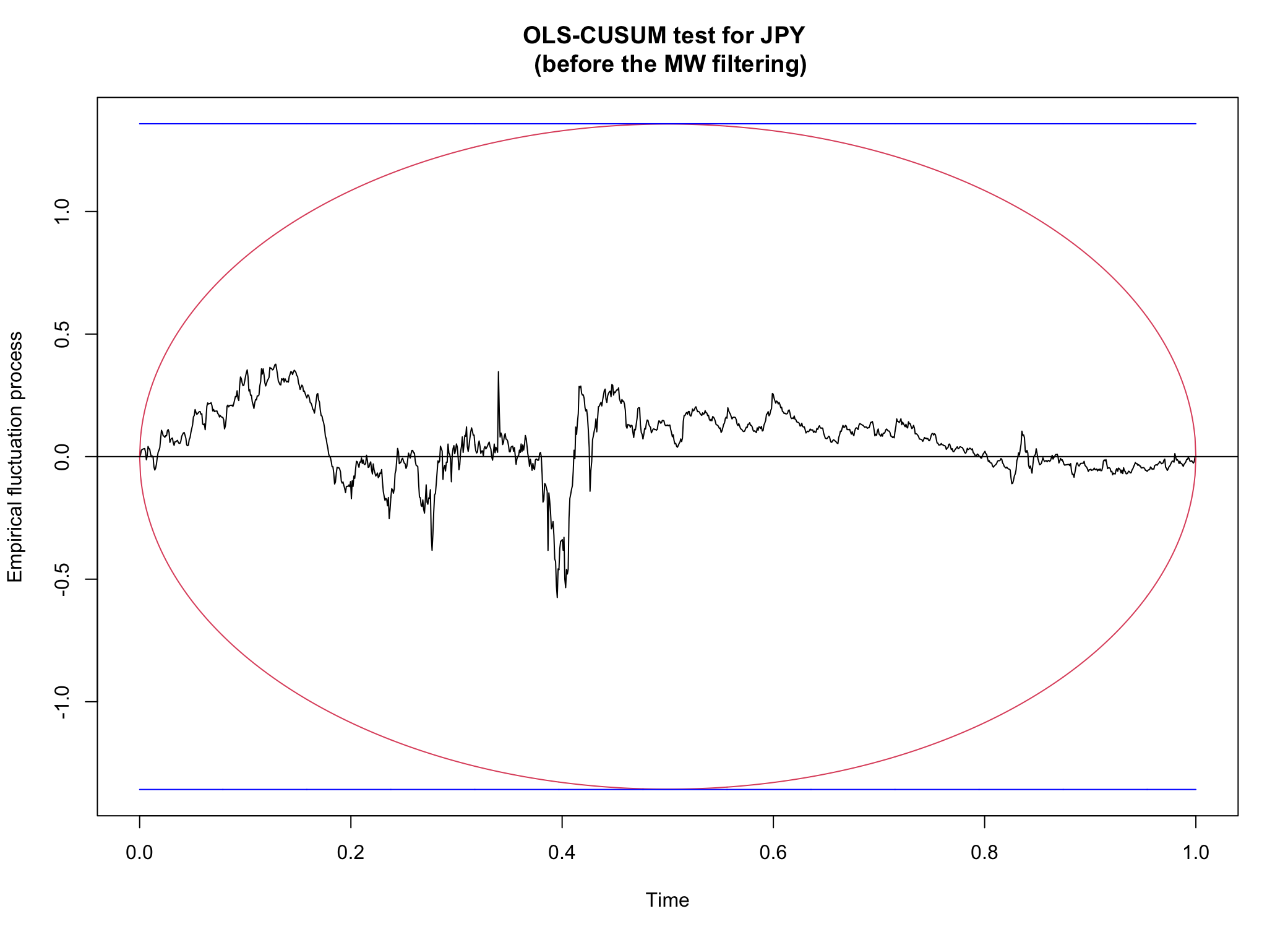}
\end{figure}

\begin{figure}[H]
  \centering
  \caption{The OLS-CUSUM test of EUR (before the M\"{u}ller-Watson filtering)}
  \label{fig:CUSUM5}
  \includegraphics[width=1.0\linewidth]{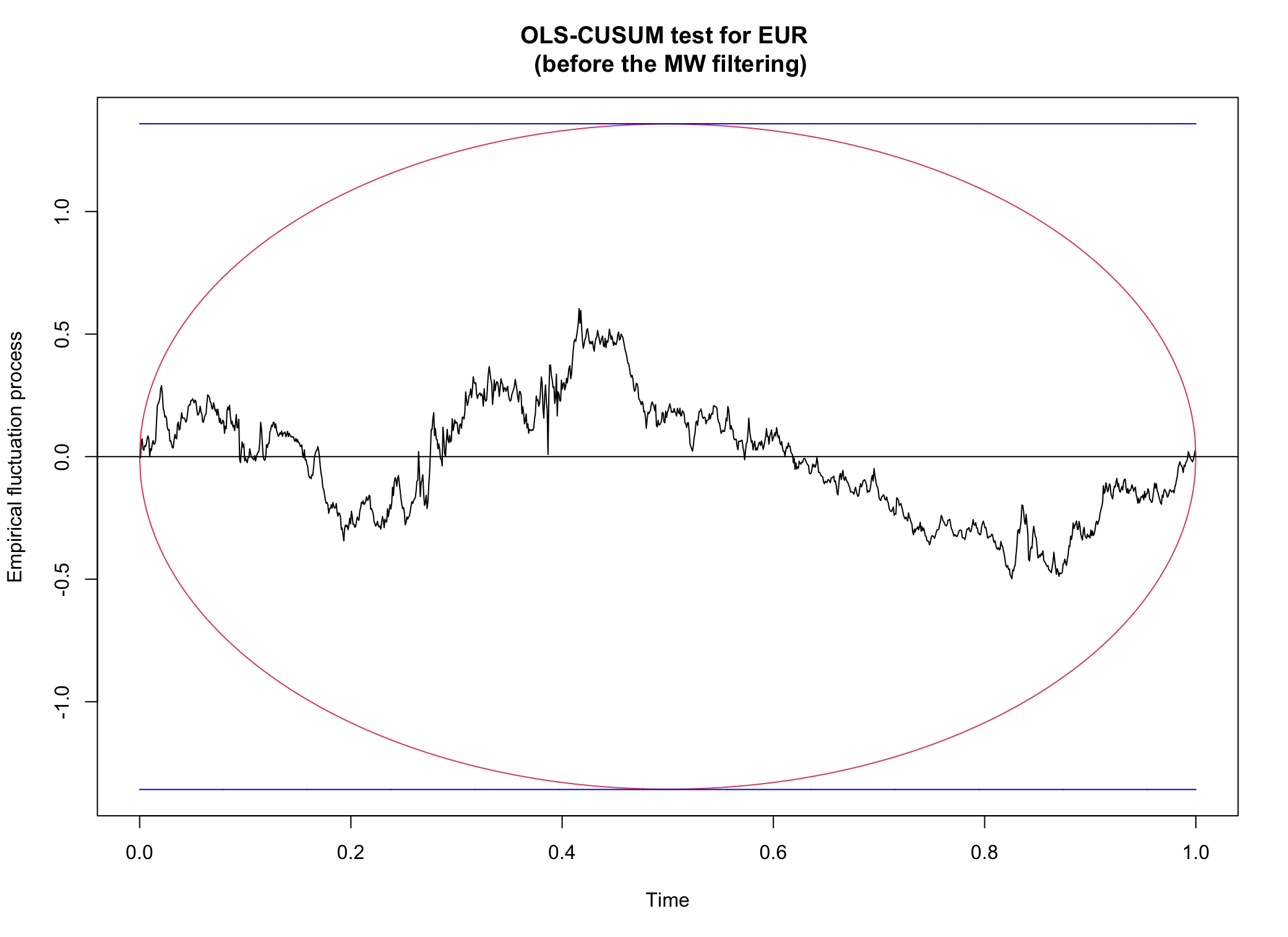}
\end{figure}

\begin{figure}[H]
  \centering
  \caption{The OLS-CUSUM test of GOLD (before the M\"{u}ller-Watson filtering)}
  \label{fig:CUSUM6}
  \includegraphics[width=1.0\linewidth]{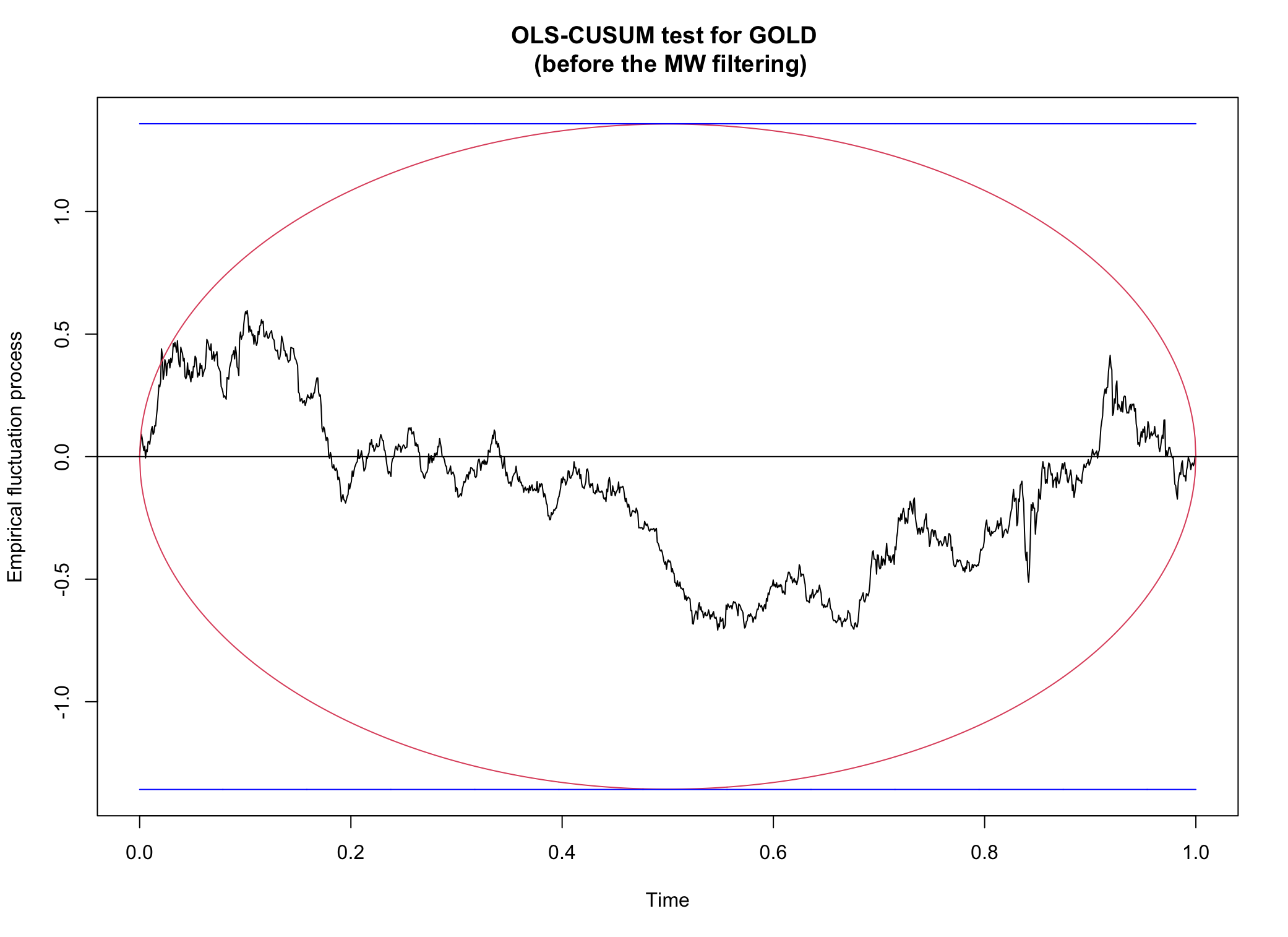}
\end{figure}

\begin{figure}[H]
  \centering
  \caption{The OLS-CUSUM test of S\&P500 (before the M\"{u}ller-Watson filtering)}
  \label{fig:CUSUM7}
  \includegraphics[width=1.0\linewidth]{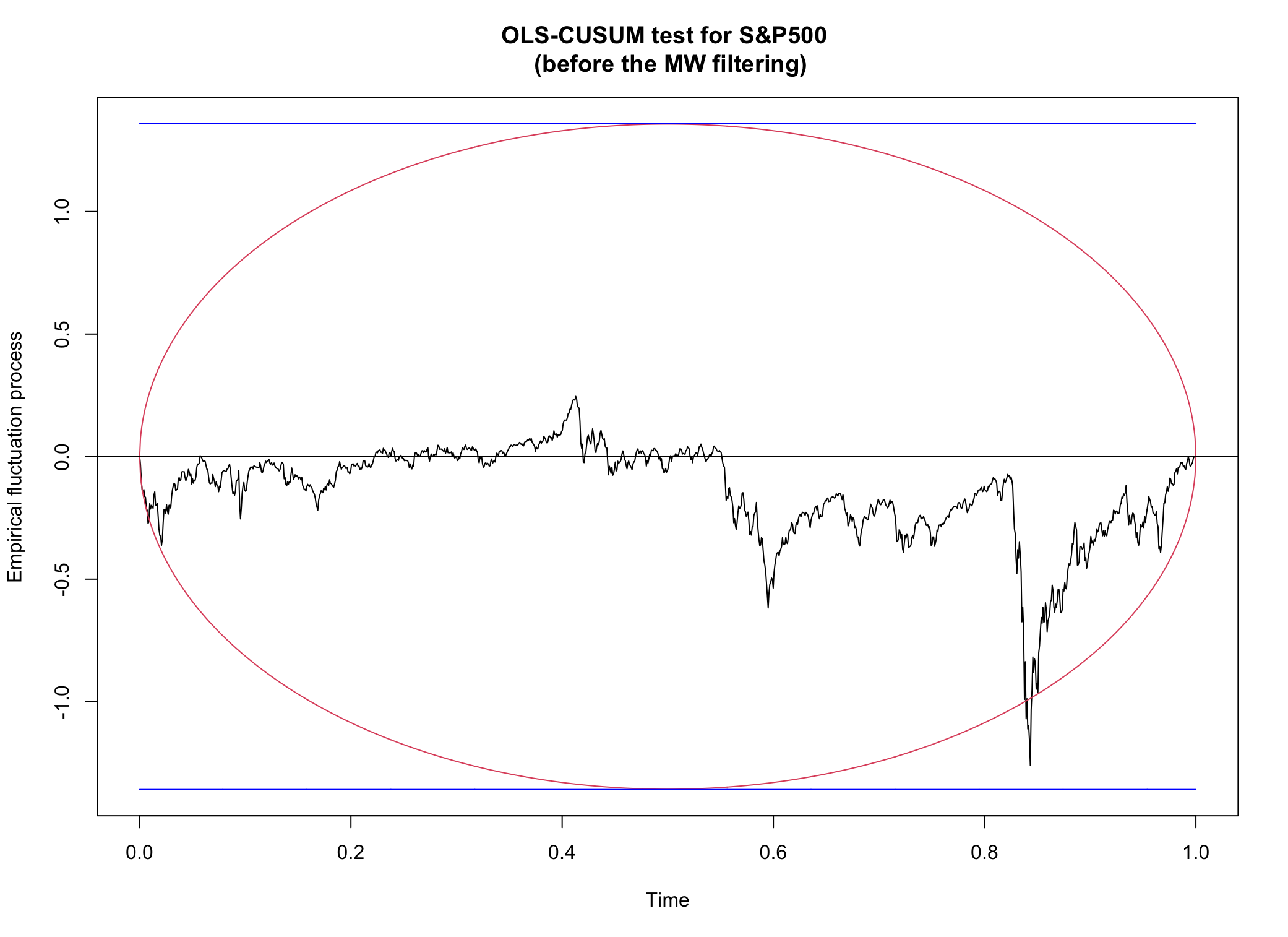}
\end{figure}

\begin{figure}[H]
  \centering
  \caption{The OLS-CUSUM test of MSCI (before the M\"{u}ller-Watson filtering)}
  \label{fig:CUSUM8}
  \includegraphics[width=1.0\linewidth]{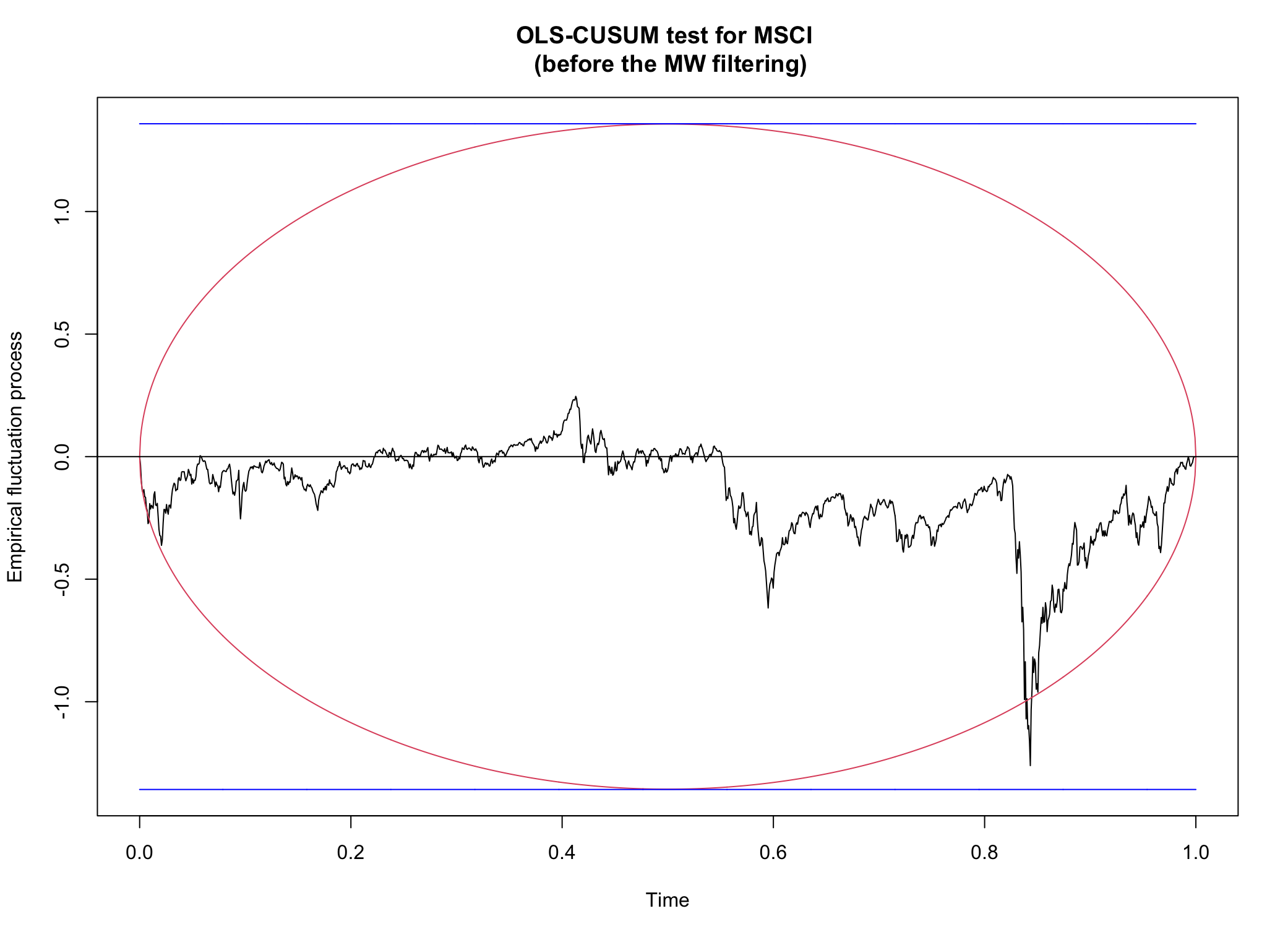}
\end{figure}

First, we show the results of the OLS-CUSUM test for both before and after the MW filtering is applied. Figures \ref{fig:CUSUM1}, \ref{fig:CUSUM2}, \ref{fig:CUSUM3}, \ref{fig:CUSUM4}, \ref{fig:CUSUM5}, \ref{fig:CUSUM6}, \ref{fig:CUSUM7} and \ref{fig:CUSUM8} illustrate the results, that is, whether the daily returns exceed the boundary consistent with the Black-Scholes formula under the assumption of a stable relationship with other financial assets before the M\"{u}ller-Watson filtering.

Based on the OLS-CUSUM test, there are two findings from the structural stability tests. We find that before the MW filtering, only ETH and XRP apparently exceed the boundaries of the Brownian bridge. Moreover, BTC almost reaches the boundaries of the Brownian bridge as well. 

\begin{figure}[H]
  \centering
  \caption{The OLS-CUSUM test of BTC (after the M\"{u}ller-Watson filtering)}
  \label{fig:CUSUM1q}
  \includegraphics[width=1.0\linewidth]{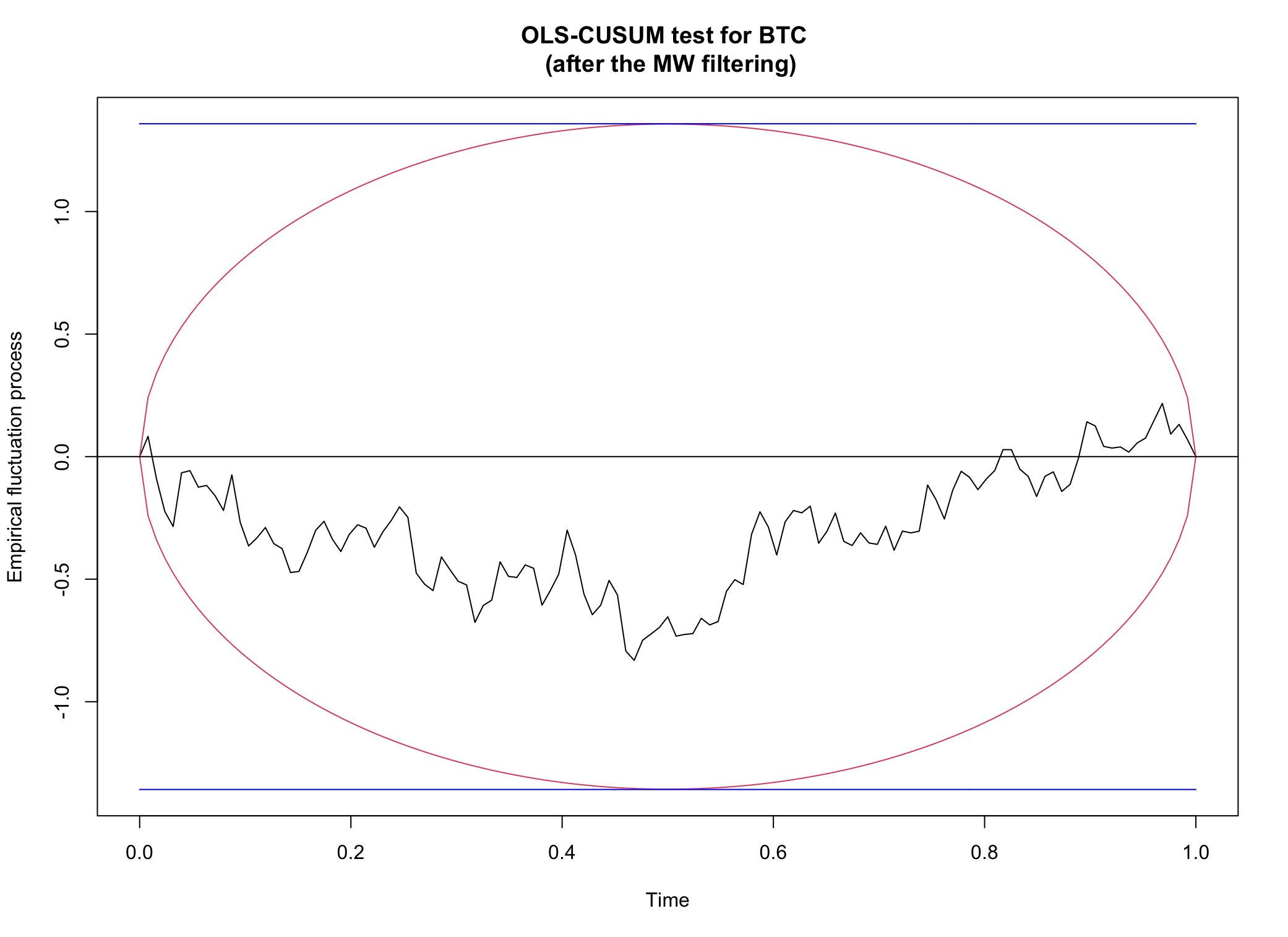}
\end{figure}

\begin{figure}[H]
  \centering
  \caption{The OLS-CUSUM test of ETH (after the M\"{u}ller-Watson filtering)}
  \label{fig:CUSUM2q}
  \includegraphics[width=1.0\linewidth]{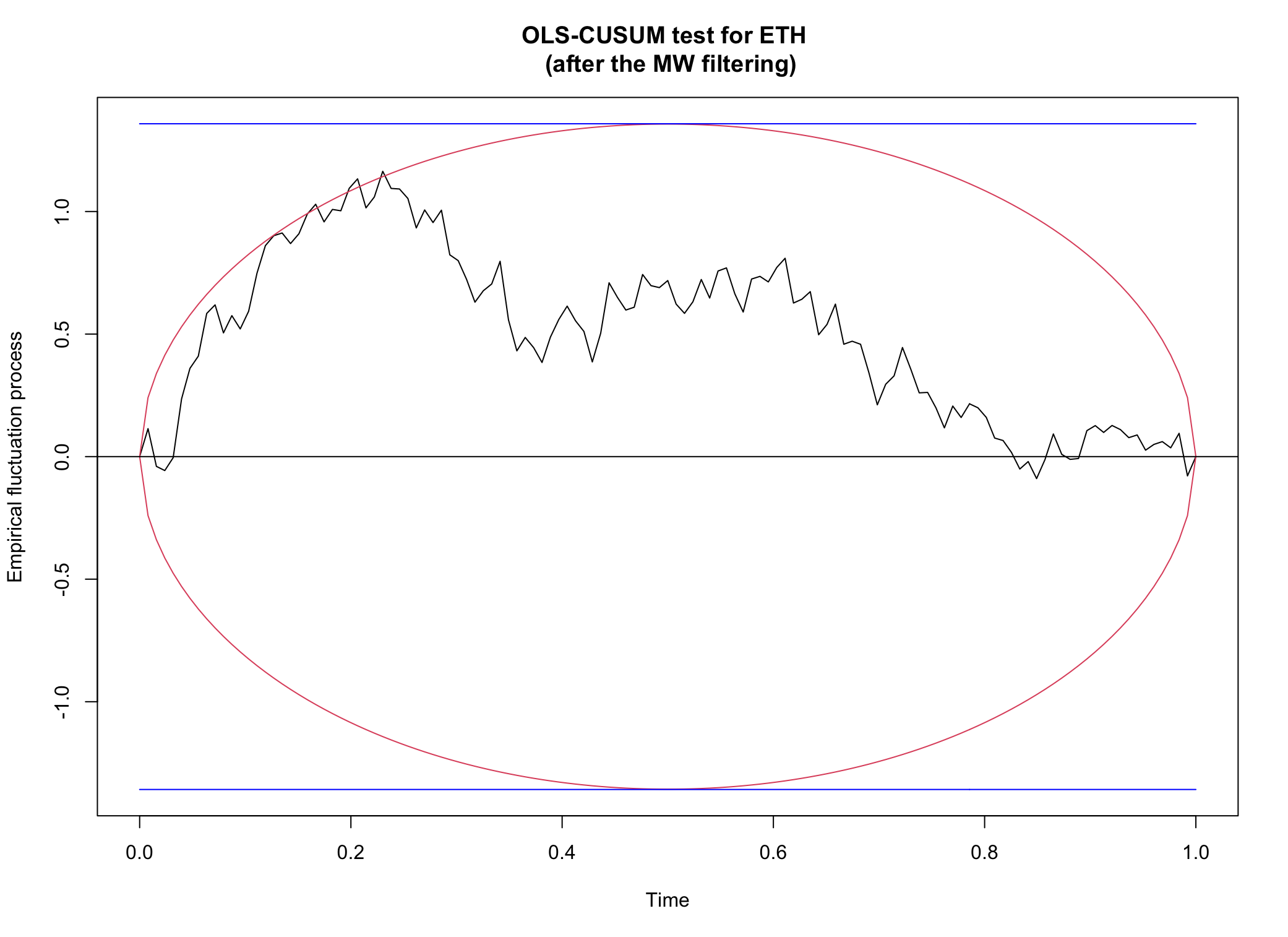}
\end{figure}

\begin{figure}[H]
  \centering
  \caption{The OLS-CUSUM test of XRP (after the M\"{u}ller-Watson filtering)}
  \label{fig:CUSUM3q}
  \includegraphics[width=1.0\linewidth]{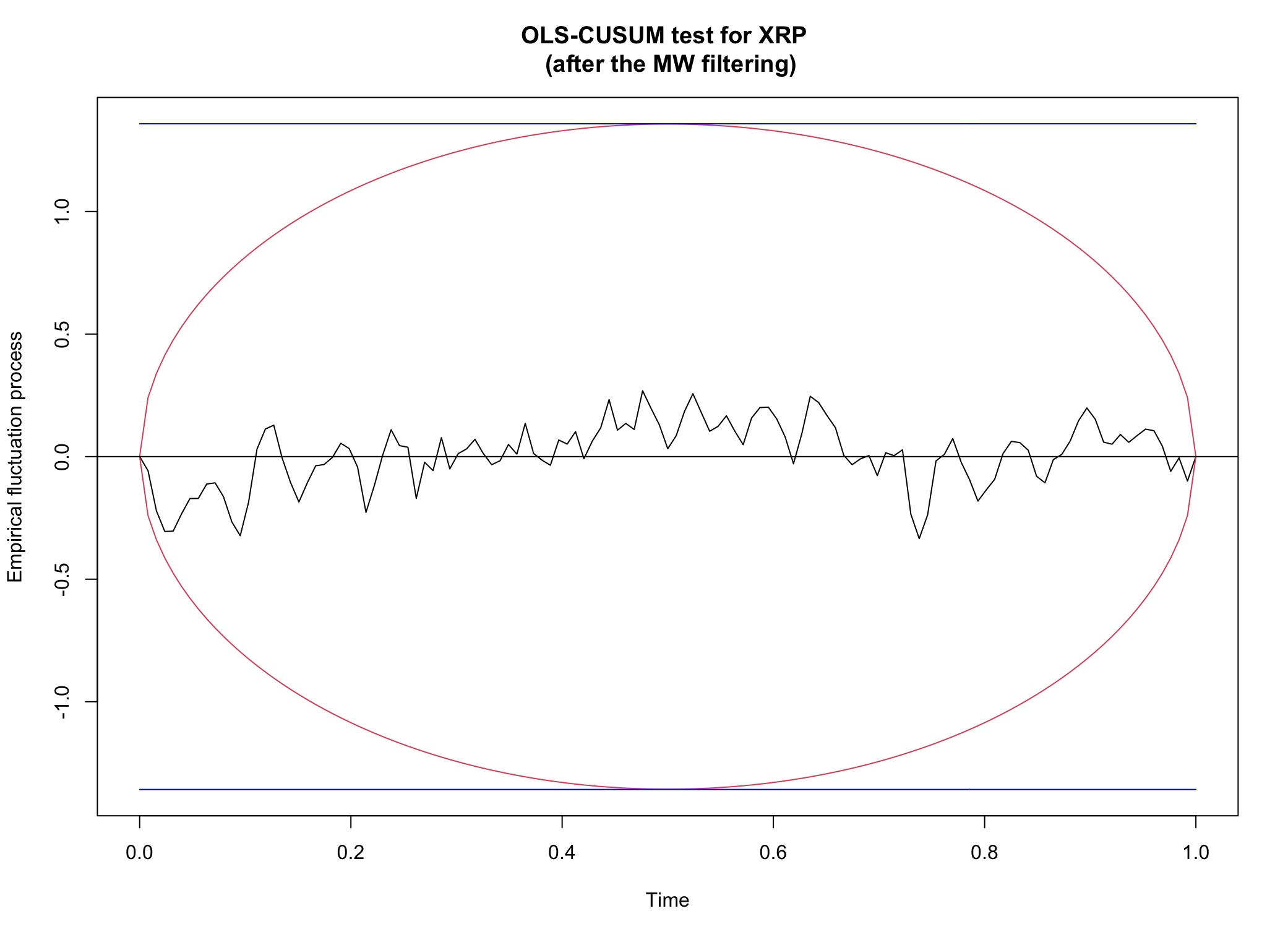}
\end{figure}

\begin{figure}[H]
  \centering
  \caption{The OLS-CUSUM test of JPY (after the M\"{u}ller-Watson filtering)}
  \label{fig:CUSUM4q}
  \includegraphics[width=1.0\linewidth]{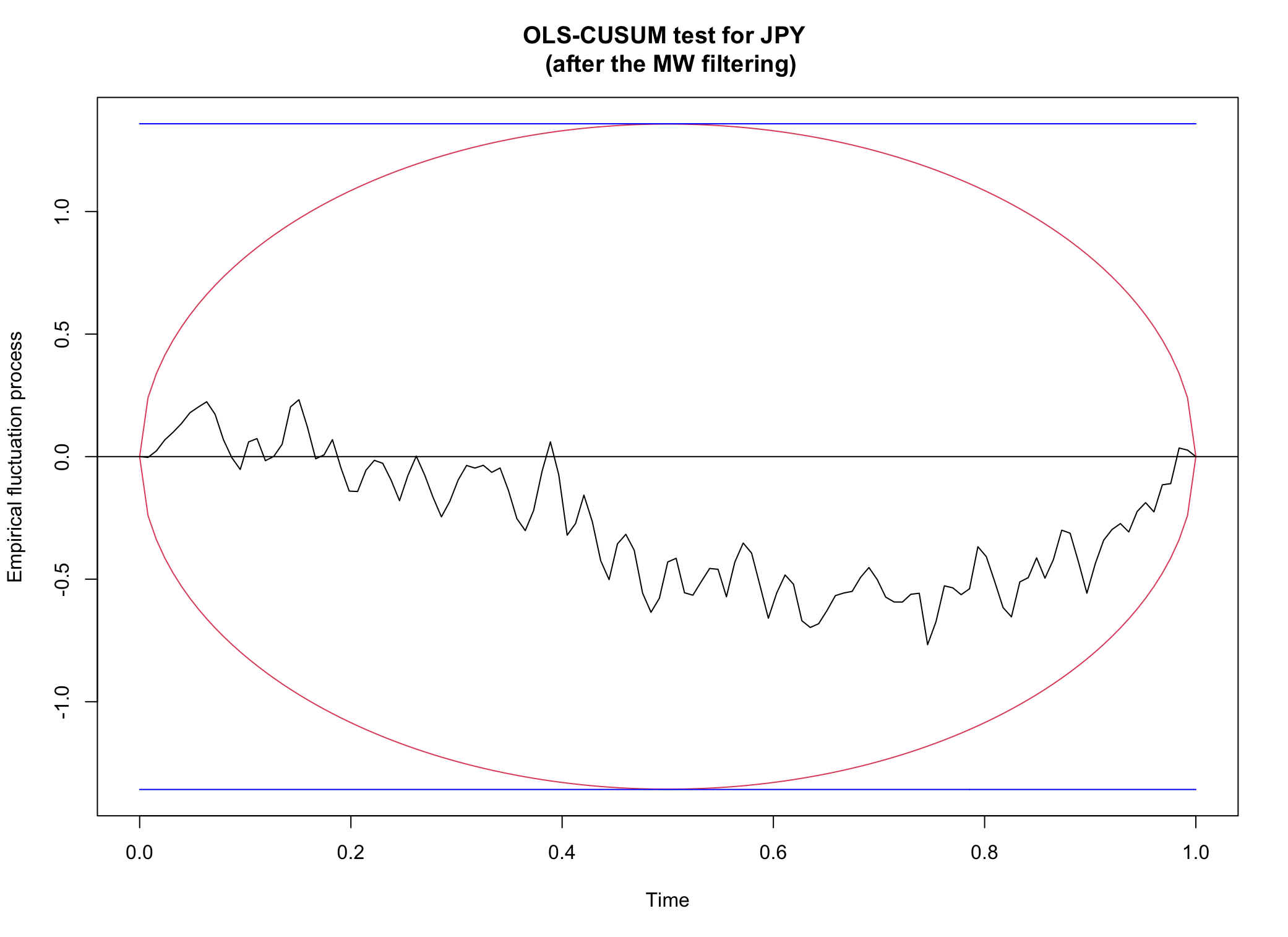}
\end{figure}

\begin{figure}[H]
  \centering
  \caption{The OLS-CUSUM test of EUR (after the M\"{u}ller-Watson filtering)}
  \label{fig:CUSUM5q}
  \includegraphics[width=1.0\linewidth]{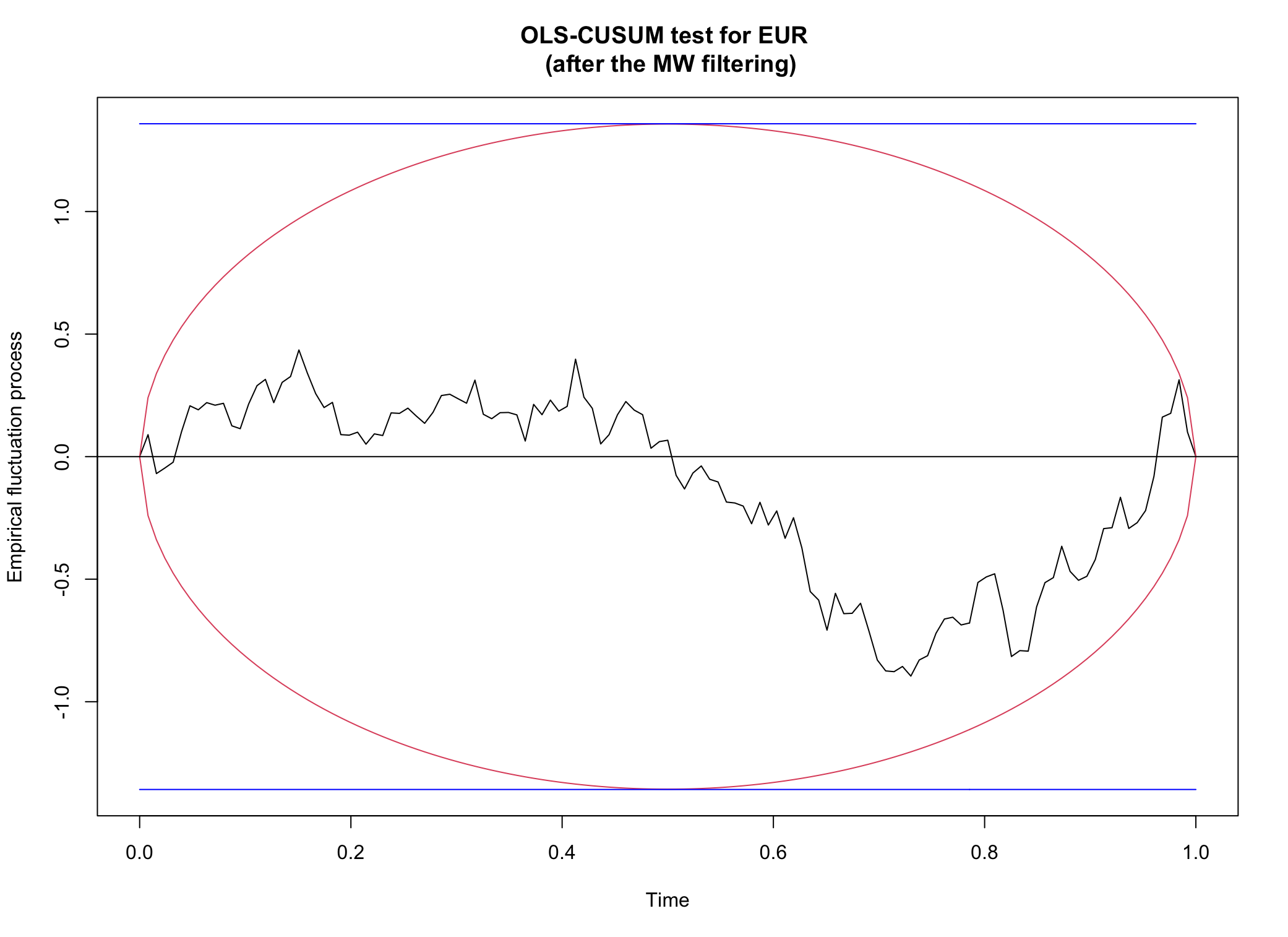}
\end{figure}

\begin{figure}[H]
  \centering
  \caption{The OLS-CUSUM test of GOLD (after the M\"{u}ller-Watson filtering)}
  \label{fig:CUSUM6q}
  \includegraphics[width=1.0\linewidth]{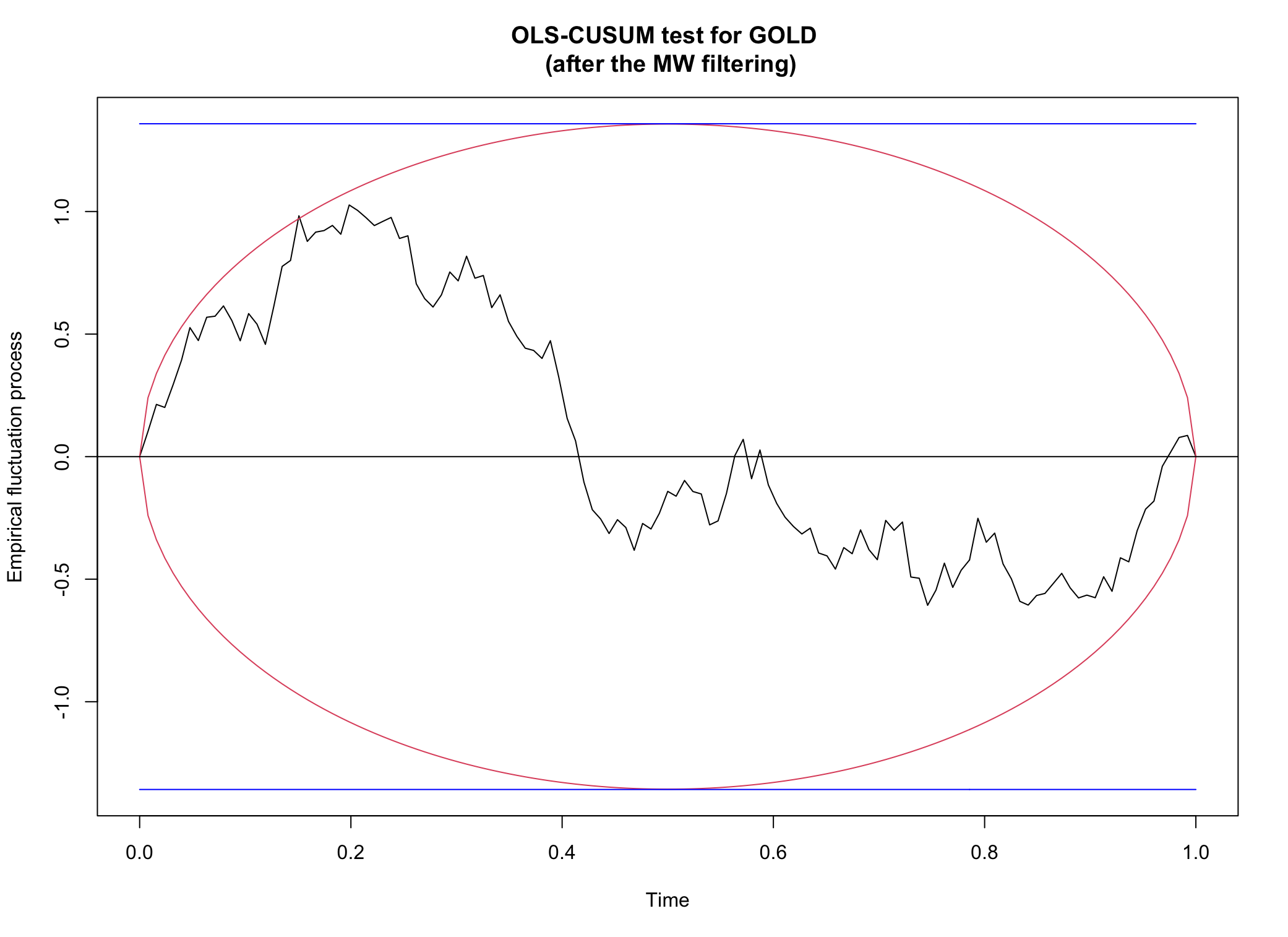}
\end{figure}

\begin{figure}[H]
  \centering
  \caption{The OLS-CUSUM test of S\&P500 (after the M\"{u}ller-Watson filtering)}
  \label{fig:CUSUM7q}
  \includegraphics[width=1.0\linewidth]{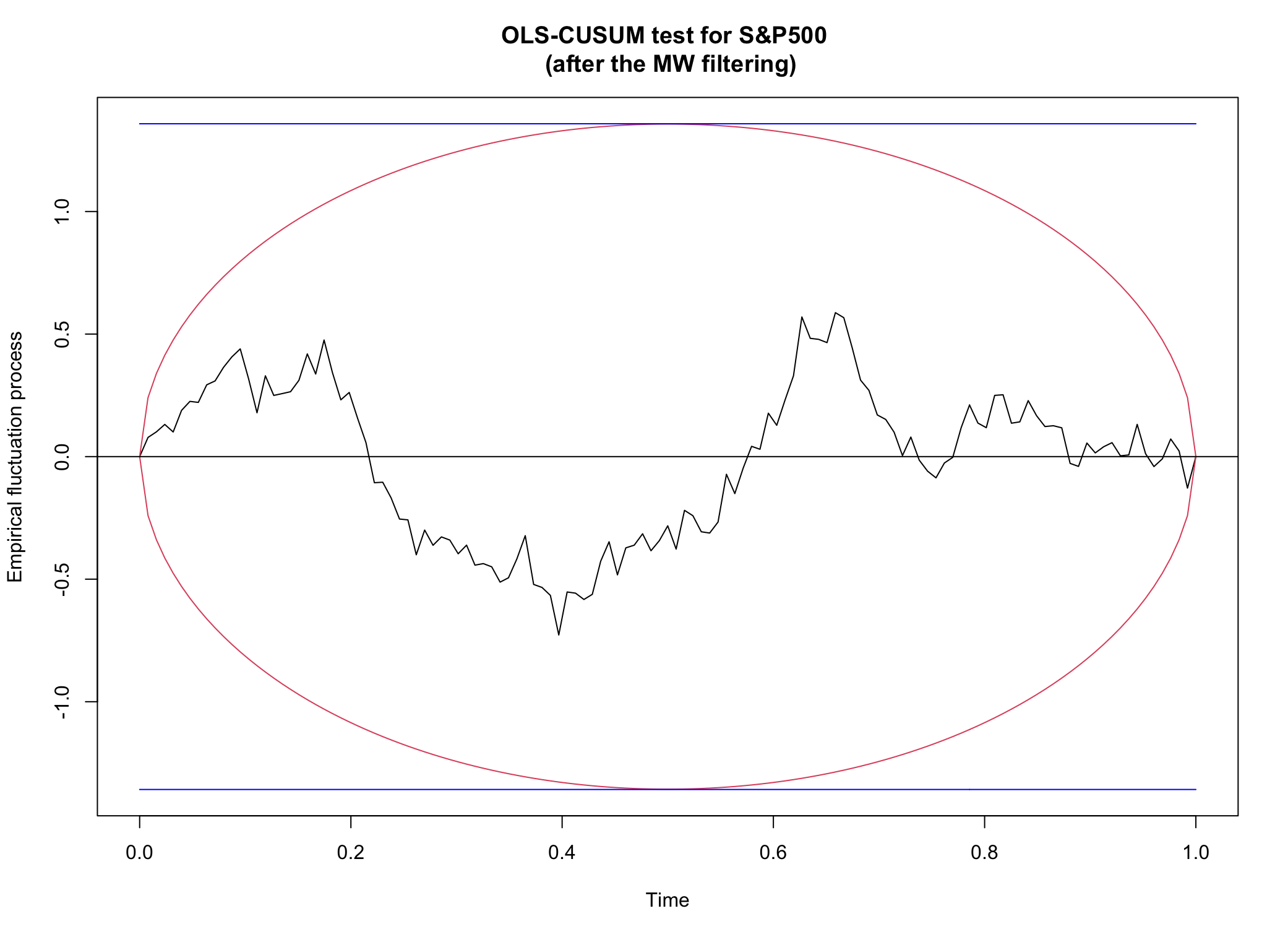}
\end{figure}

\begin{figure}[H]
  \centering
  \caption{The OLS-CUSUM test of MSCI (after the M\"{u}ller-Watson filtering)}
  \label{fig:CUSUM8q}
  \includegraphics[width=1.0\linewidth]{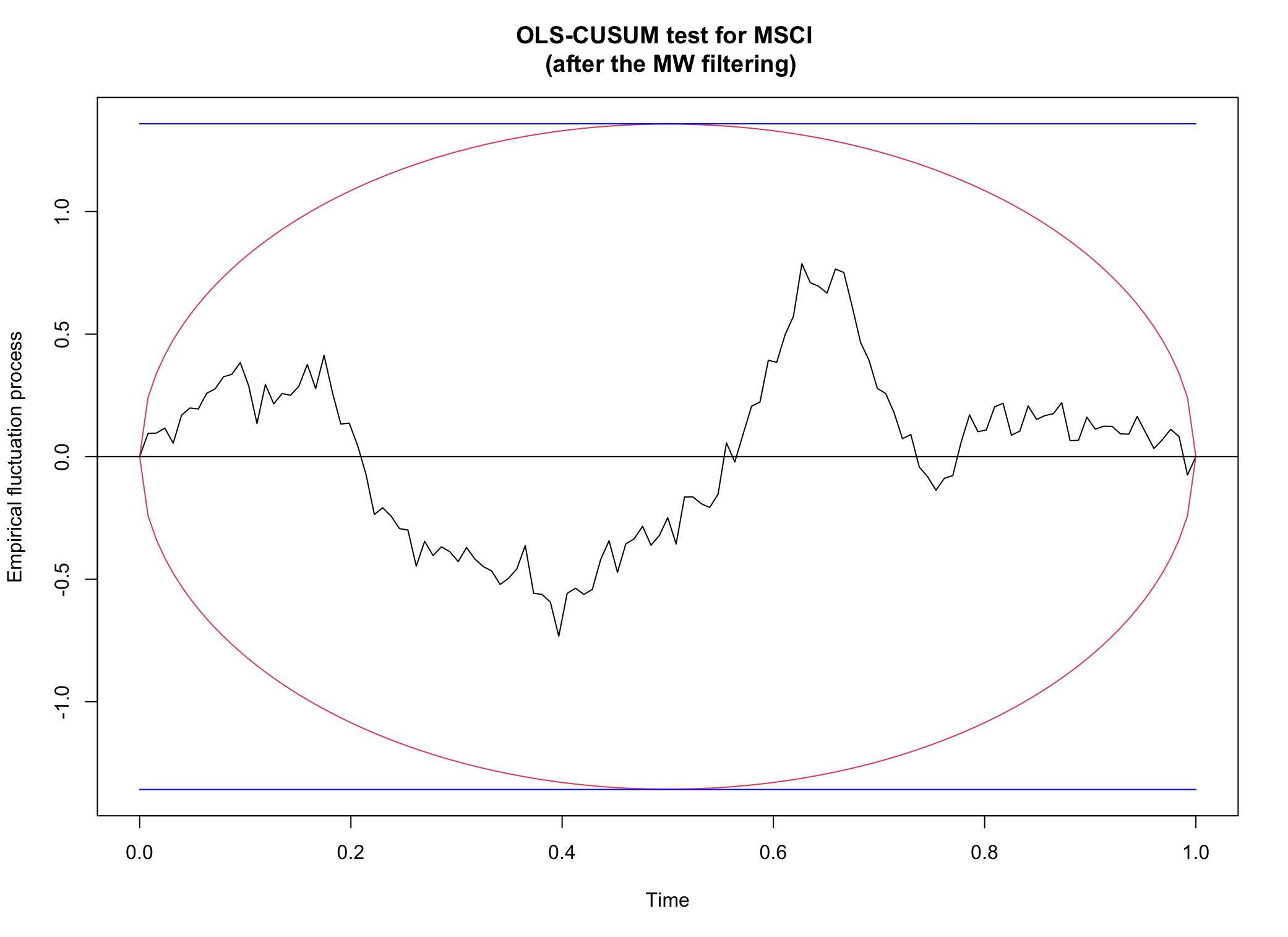}
\end{figure}

However, this is not the case after the MW filtering. Also, all the other assets do not exceed the boundaries either, before or after the MW filtering (see Figures \ref{fig:CUSUM1q}, \ref{fig:CUSUM2q}, \ref{fig:CUSUM3q}, \ref{fig:CUSUM4q}, \ref{fig:CUSUM5q}, \ref{fig:CUSUM6q}, \ref{fig:CUSUM7q} and \ref{fig:CUSUM8q}). In other words, cryptocurrencies are stable enough within the boundaries of the Brownian bridge after the MW filtering is applied, as is the case for the key financial assets.

\begin{figure}[H]
  \centering
  \caption{The Rec-CUSUM test of BTC [2016-2020] (before the M\"{u}ller-Watson filtering)}
  \label{fig:RE1}
  \includegraphics[width=1.0\linewidth]{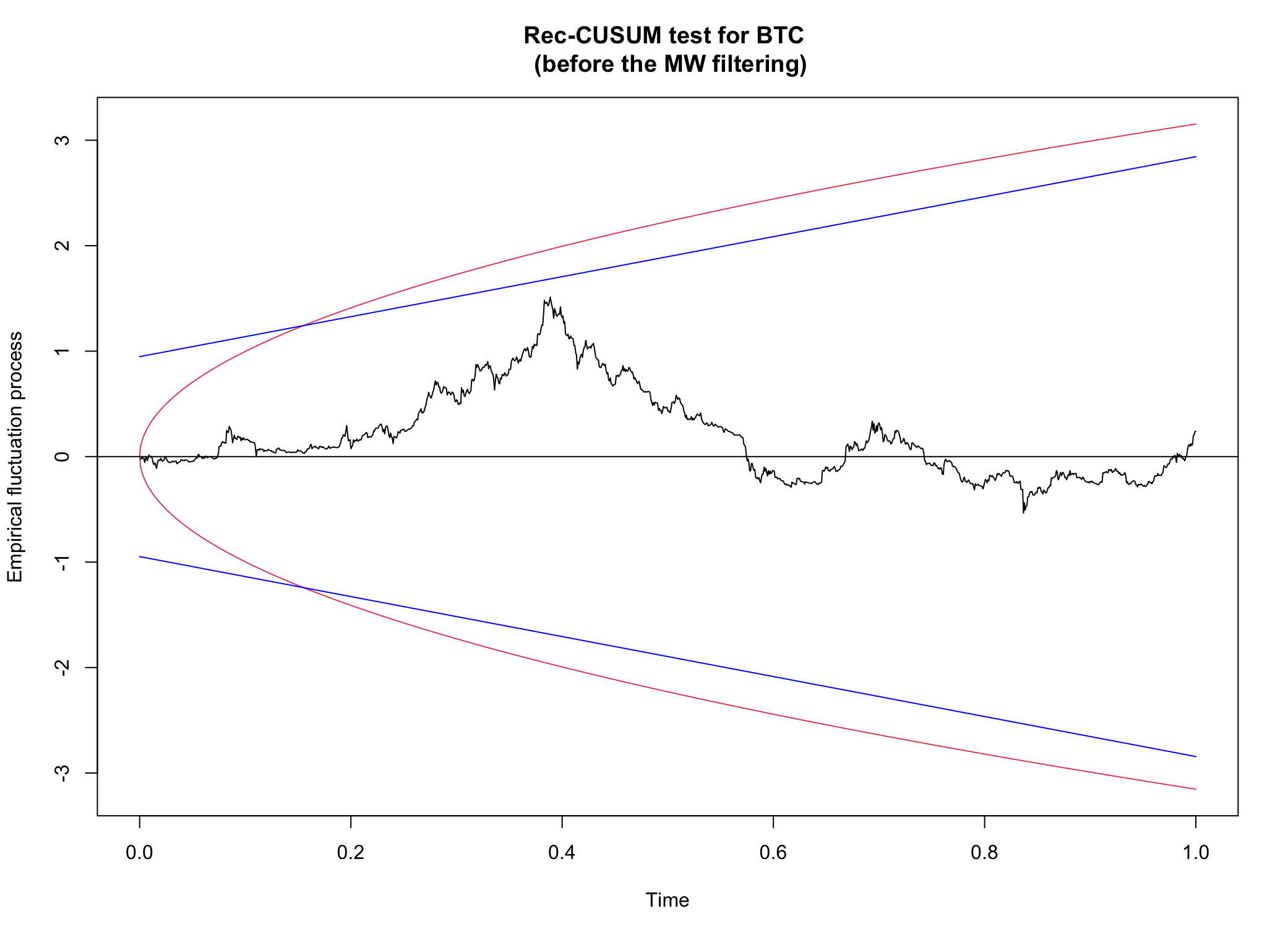}
\end{figure}

Second, we conduct the Rec-CUSUM test. Figures \ref{fig:RE1}, \ref{fig:RE2}, \ref{fig:RE3}, \ref{fig:RE4}, \ref{fig:RE5}, \ref{fig:RE6}, \ref{fig:RE7} and \ref{fig:RE8} illustrate the results using all the data, that is, whether the daily returns exceed the boundary consistent with the Black-Scholes formula before the MW filtering. 

Similar to the OLS-CUSUM test, we find that before the MW filtering, only ETH (and perhaps GOLD) exceeds the boundaries of the Brownian motion in the Rec-CUSUM test when we use all the data from 2016 to 2020.

However, once again, this is not the case after the MW filtering. All assets (perhaps except for GOLD) do not exceed the boundaries of the Brownian motion after the MW filtering (see Figures \ref{fig:RE1q}, \ref{fig:RE2q}, \ref{fig:RE3q}, \ref{fig:RE4q}, \ref{fig:RE5q}, \ref{fig:RE6q}, \ref{fig:RE7q} and \ref{fig:RE8q}). Once again, these results show that cryptocurrencies are stable enough within the boundaries of the Brownian motion after the MW filtering is applied, as is the case for the key financial assets.

\begin{figure}[H]
  \centering
  \caption{The Rec-CUSUM test of BTC [2017-2020] (before the M\"{u}ller-Watson filtering)}
  \label{fig:RE1a}
  \includegraphics[width=1.0\linewidth]{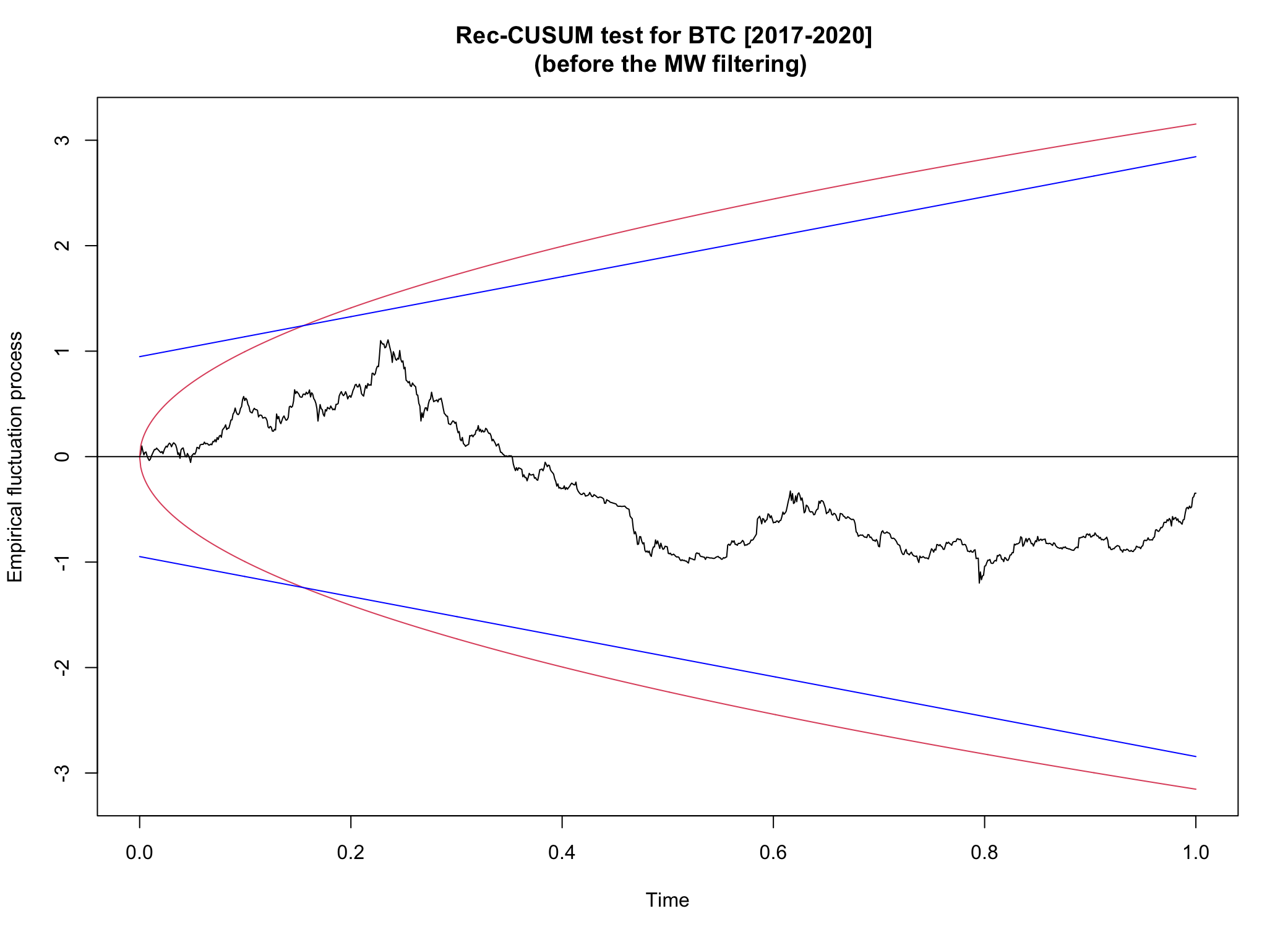}
\end{figure}

\begin{figure}[H]
  \centering
  \caption{The Rec-CUSUM test of BTC [2018-2020] (before the M\"{u}ller-Watson filtering)}
  \label{fig:RE1b}
  \includegraphics[width=1.0\linewidth]{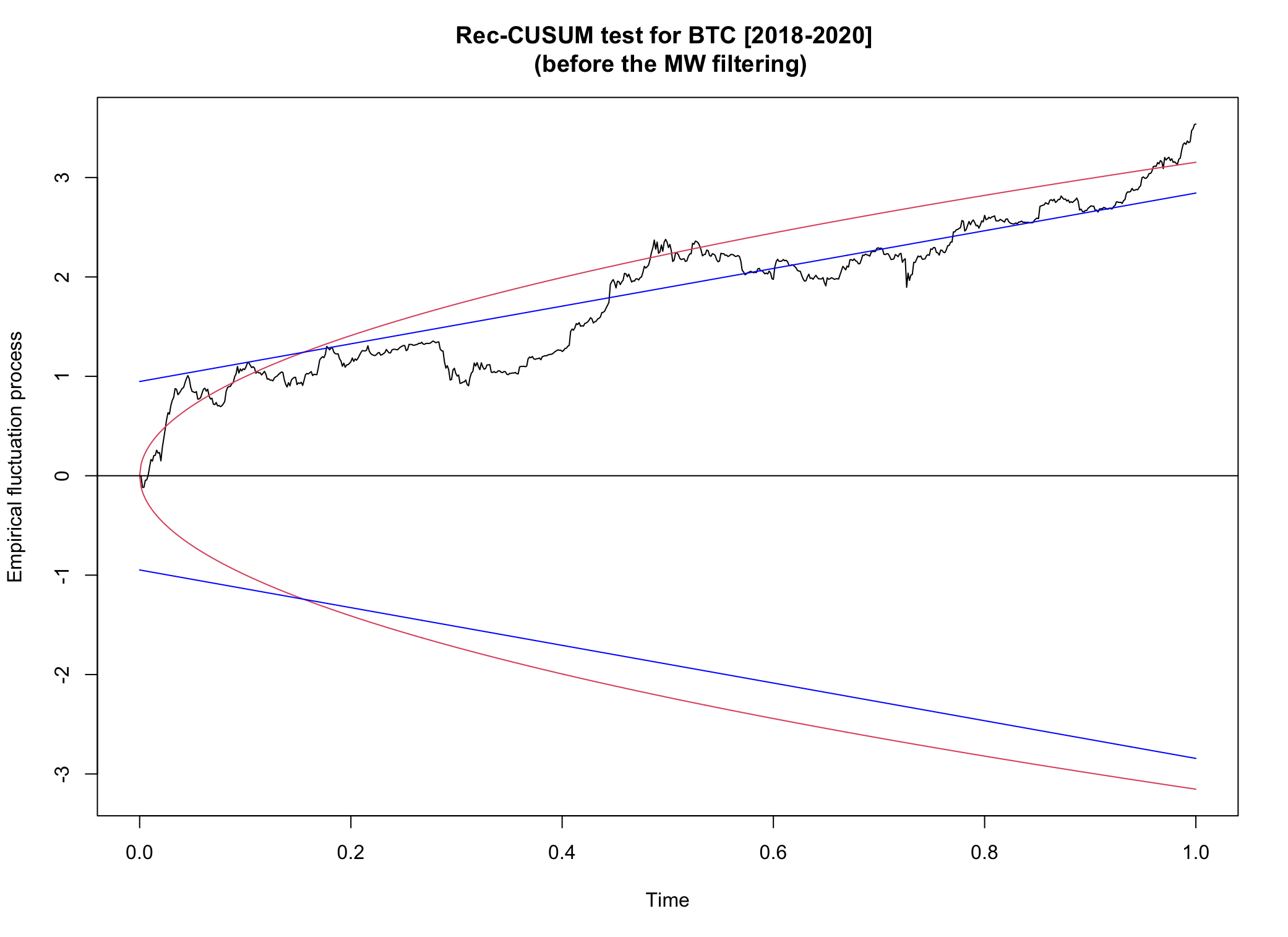}
\end{figure}

\begin{figure}[H]
  \centering
  \caption{The Rec-CUSUM test of BTC [2019-2020] (before the M\"{u}ller-Watson filtering)}
  \label{fig:RE1c}
  \includegraphics[width=1.0\linewidth]{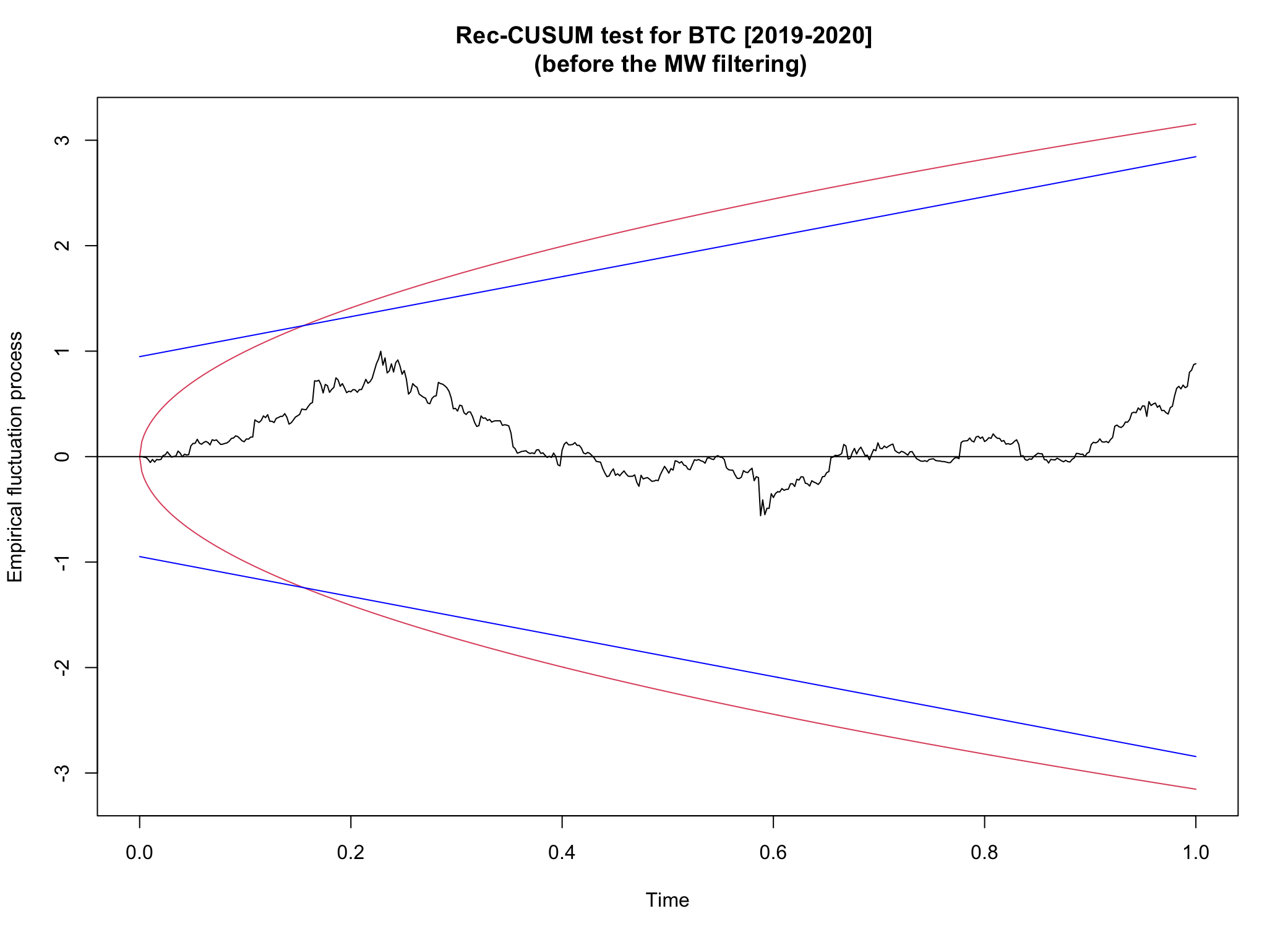}
\end{figure}

\begin{figure}[H]
  \centering
  \caption{The Rec-CUSUM test of ETH [2016-2020] (before the M\"{u}ller-Watson filtering)}
  \label{fig:RE2}
  \includegraphics[width=1.0\linewidth]{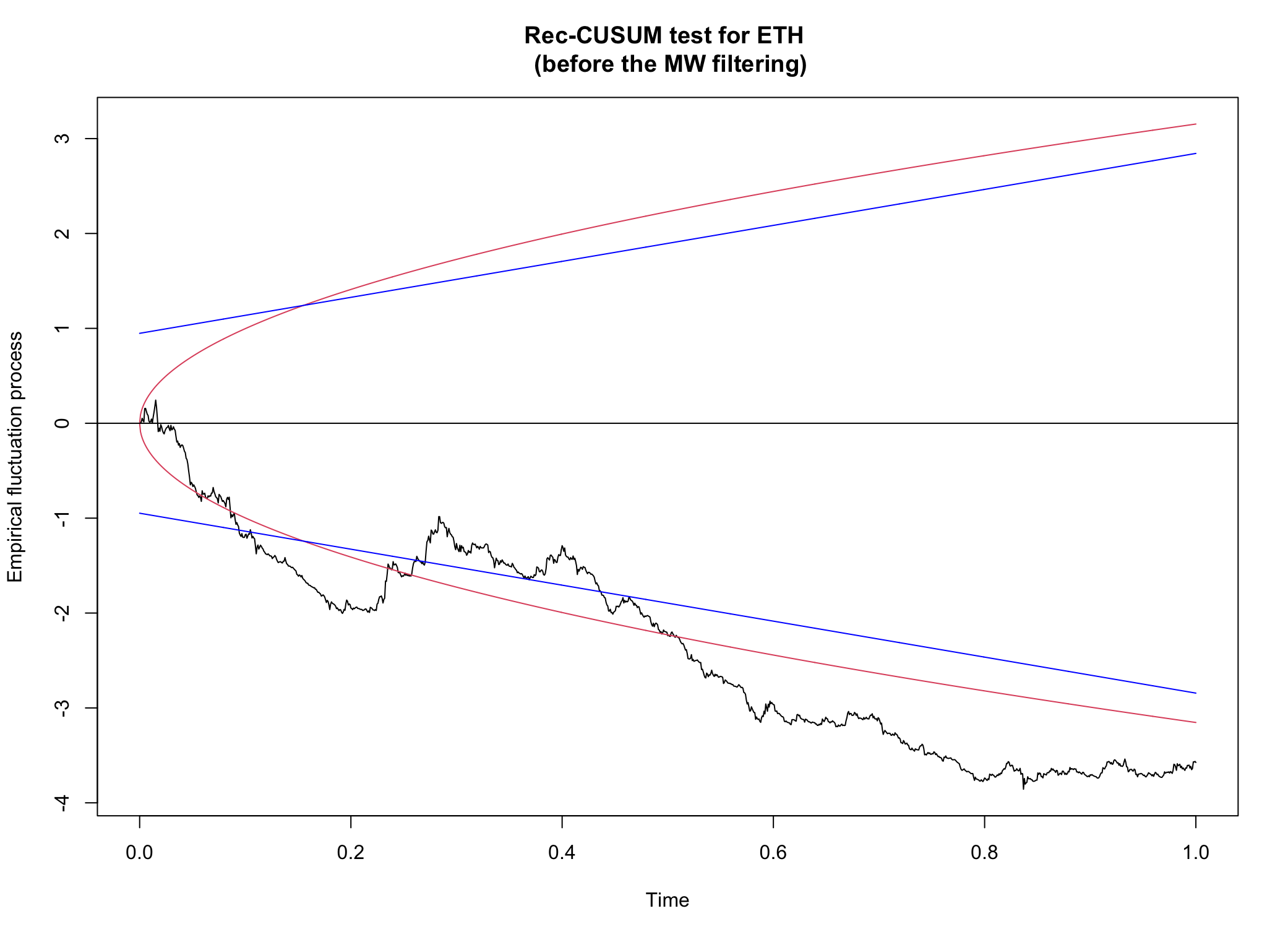}
\end{figure}

\begin{figure}[H]
  \centering
  \caption{The Rec-CUSUM test of ETH [2017-2020] (before the M\"{u}ller-Watson filtering)}
  \label{fig:RE2a}
  \includegraphics[width=1.0\linewidth]{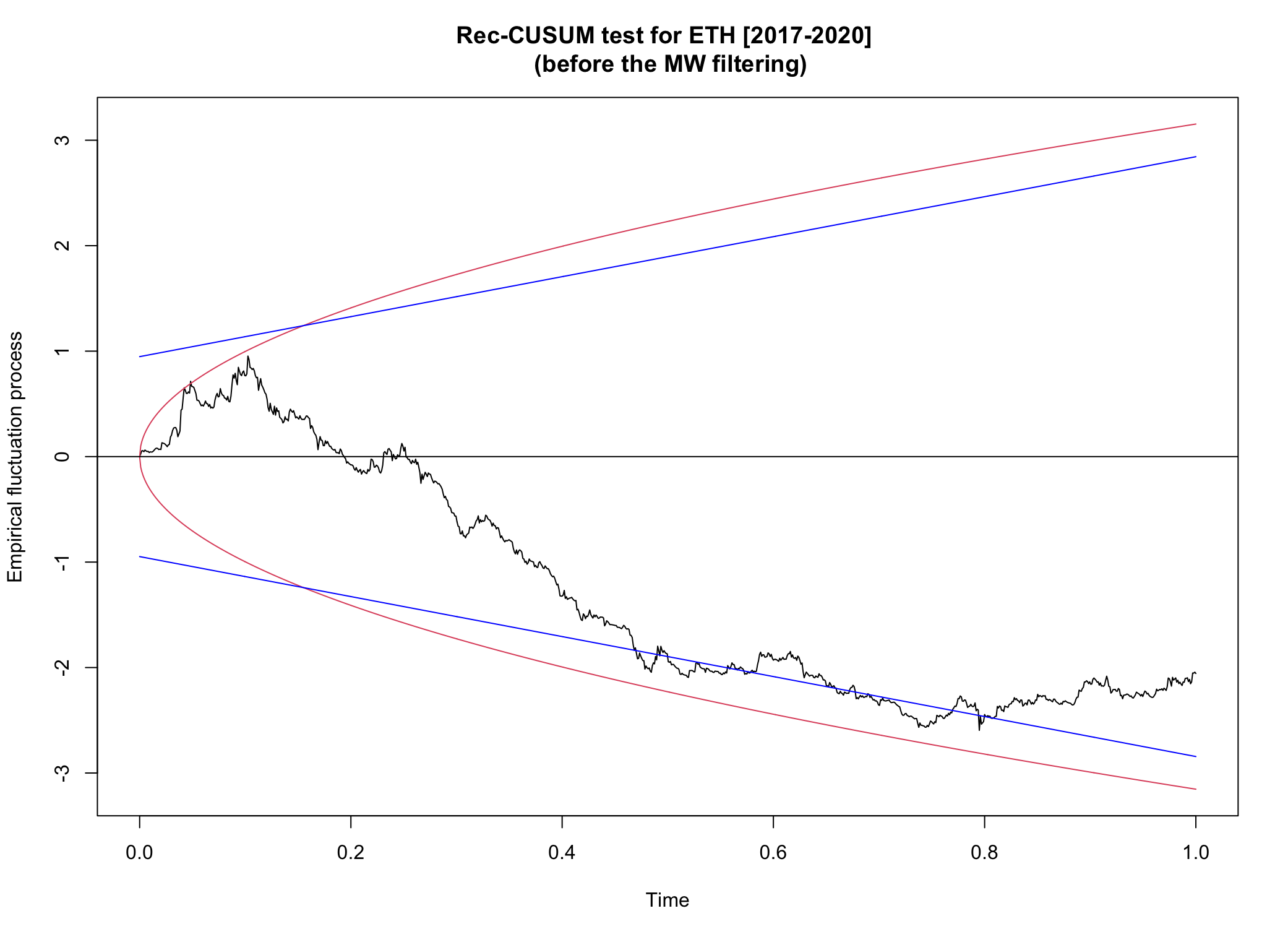}
\end{figure}

\begin{figure}[H]
  \centering
  \caption{The Rec-CUSUM test of ETH [2018-2020] (before the M\"{u}ller-Watson filtering)}
  \label{fig:RE2b}
  \includegraphics[width=1.0\linewidth]{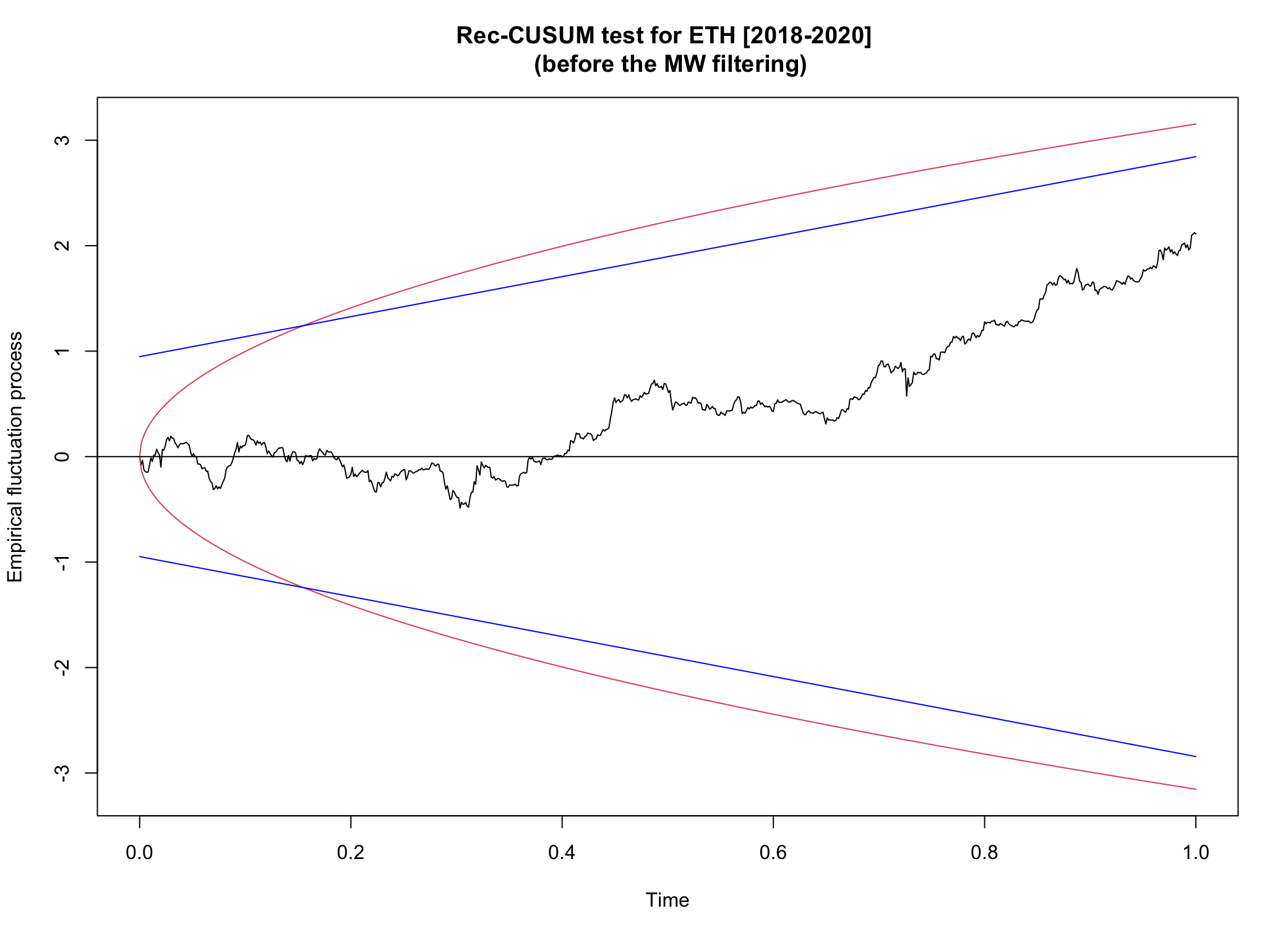}
\end{figure}

\begin{figure}[H]
  \centering
  \caption{The Rec-CUSUM test of ETH [2019-2020] (before the M\"{u}ller-Watson filtering)}
  \label{fig:RE2c}
  \includegraphics[width=1.0\linewidth]{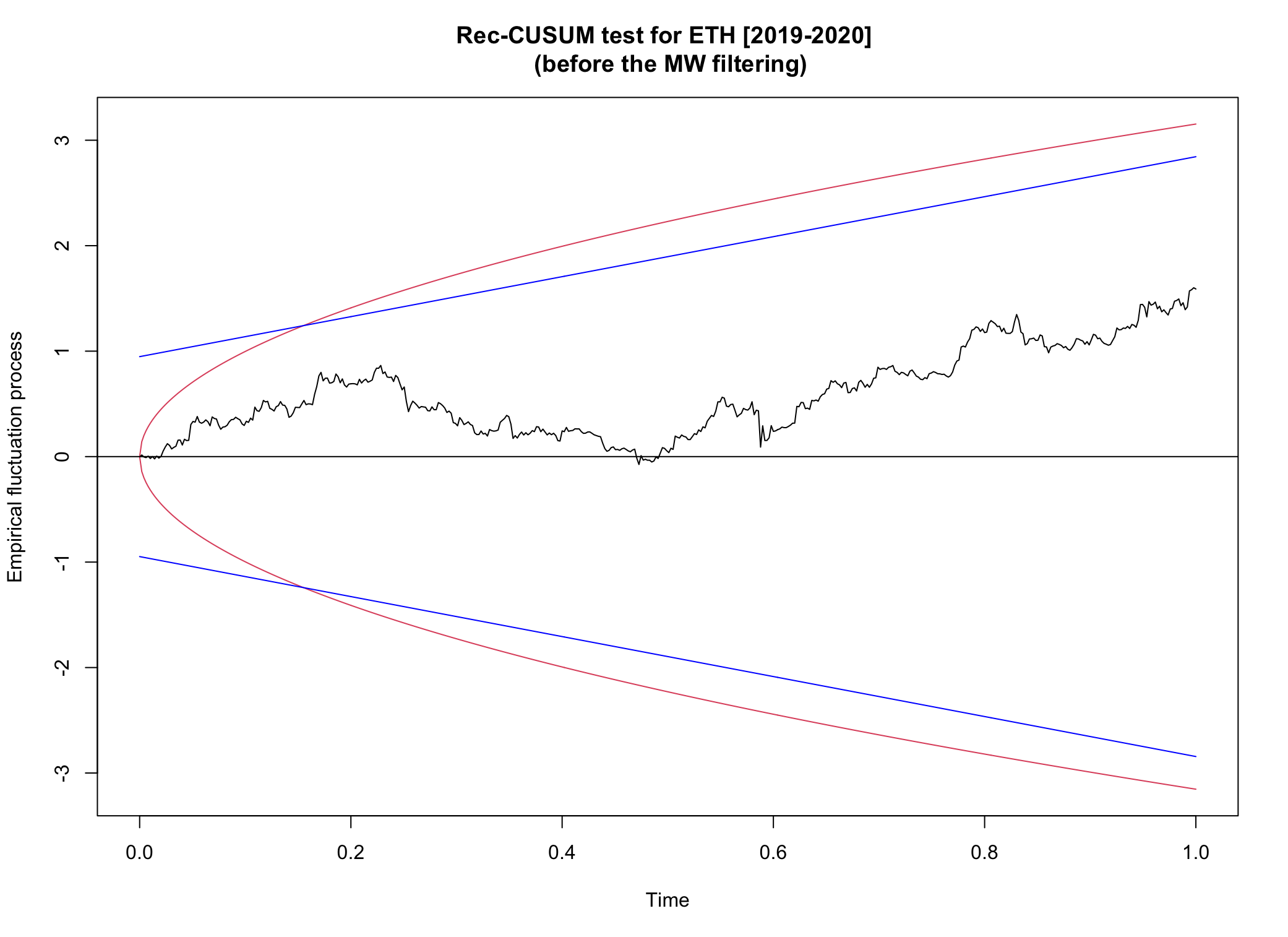}
\end{figure}

\begin{figure}[H]
  \centering
  \caption{The Rec-CUSUM test of XRP [2016-2020] (before the M\"{u}ller-Watson filtering)}
  \label{fig:RE3}
  \includegraphics[width=1.0\linewidth]{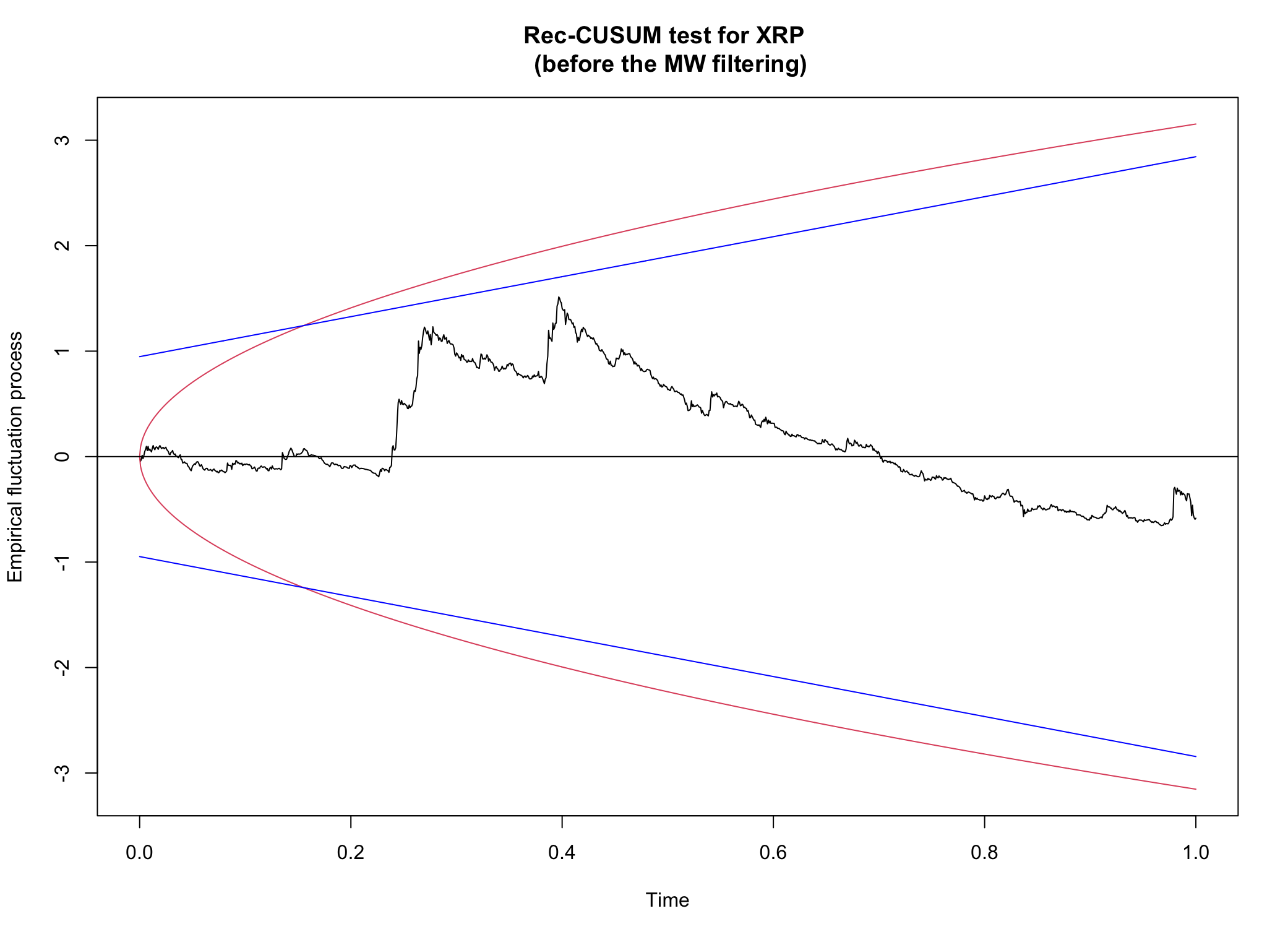}
\end{figure}

\begin{figure}[H]
  \centering
  \caption{The Rec-CUSUM test of XRP [2017-2020] (before the M\"{u}ller-Watson filtering)}
  \label{fig:RE3a}
  \includegraphics[width=1.0\linewidth]{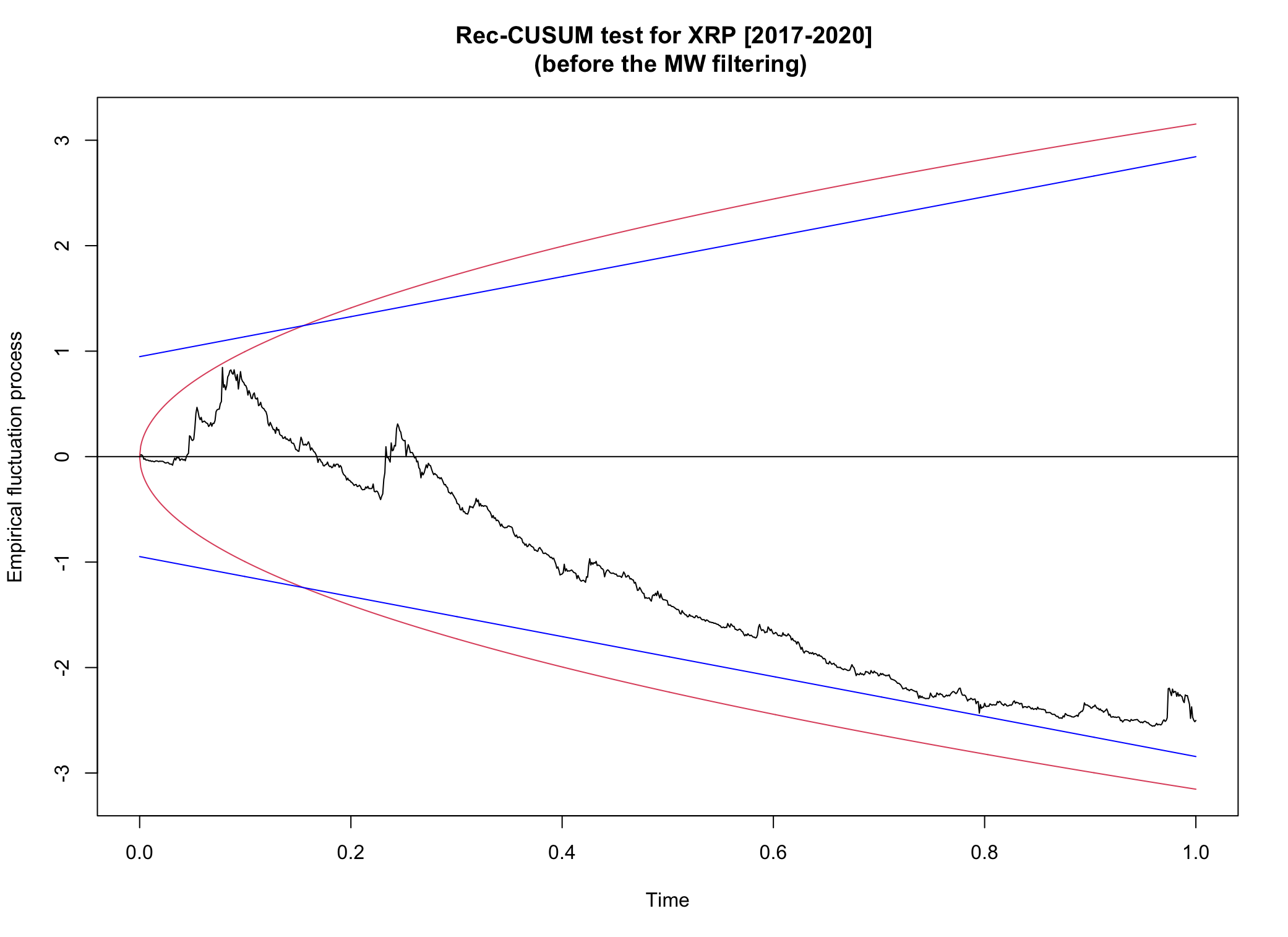}
\end{figure}

\begin{figure}[H]
  \centering
  \caption{The Rec-CUSUM test of XRP [2018-2020] (before the M\"{u}ller-Watson filtering)}
  \label{fig:RE3b}
  \includegraphics[width=1.0\linewidth]{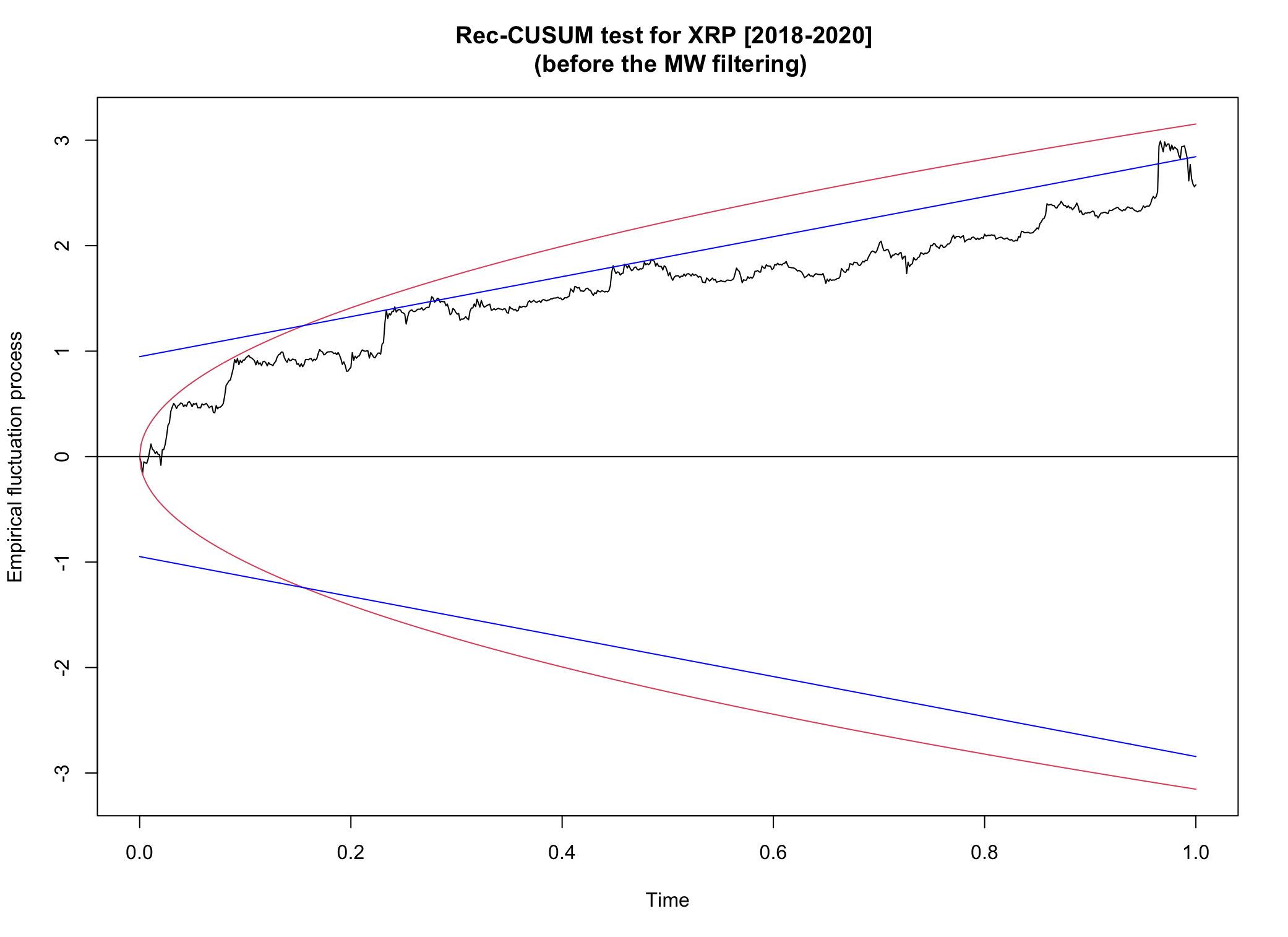}
\end{figure}

\begin{figure}[H]
  \centering
  \caption{The Rec-CUSUM test of XRP [2019-2020] (before the M\"{u}ller-Watson filtering)}
  \label{fig:RE3c}
  \includegraphics[width=1.0\linewidth]{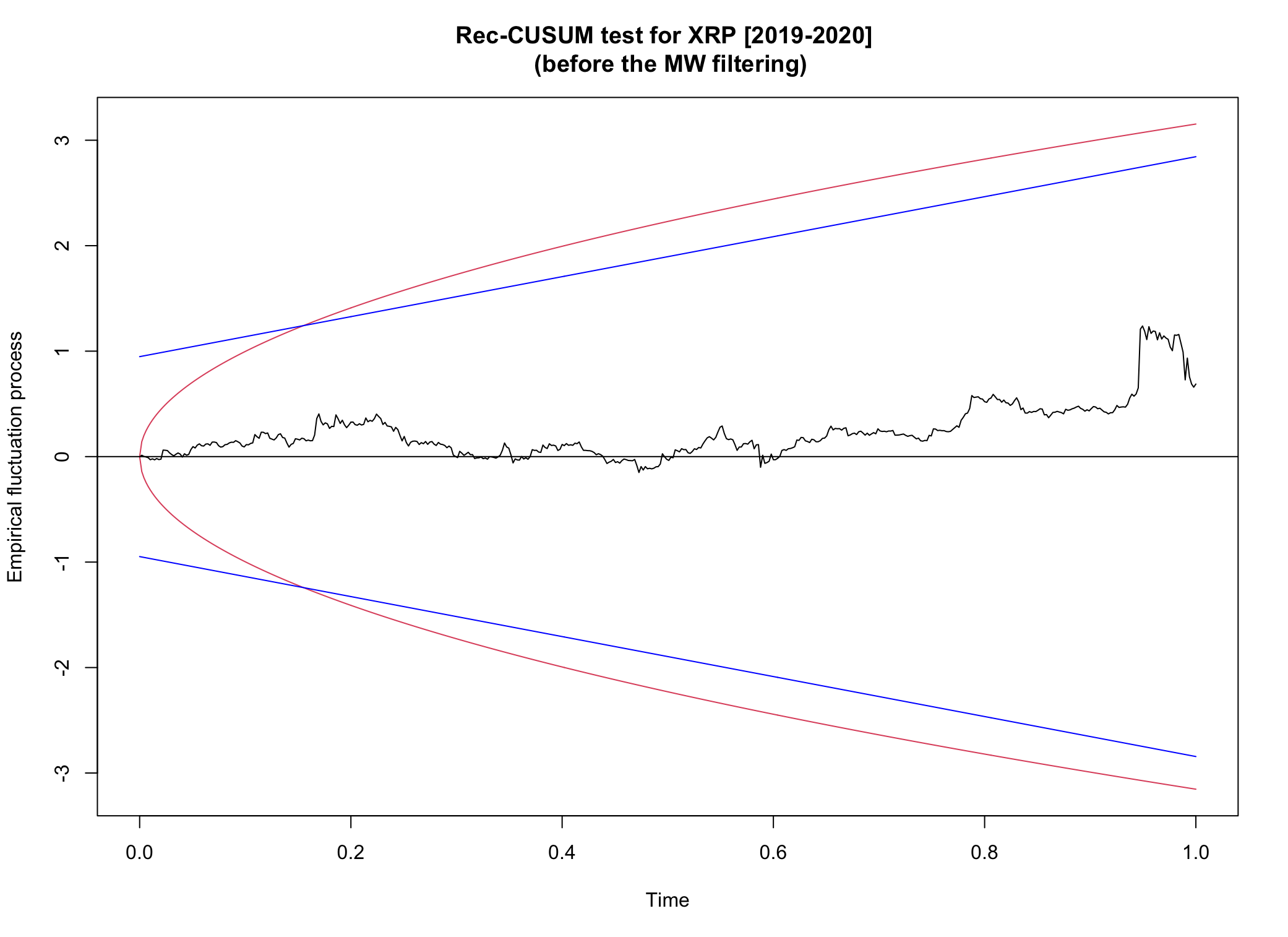}
\end{figure}

\begin{figure}[H]
  \centering
  \caption{The Rec-CUSUM test of JPY [2016-2020] (before the M\"{u}ller-Watson filtering)}
  \label{fig:RE4}
  \includegraphics[width=1.0\linewidth]{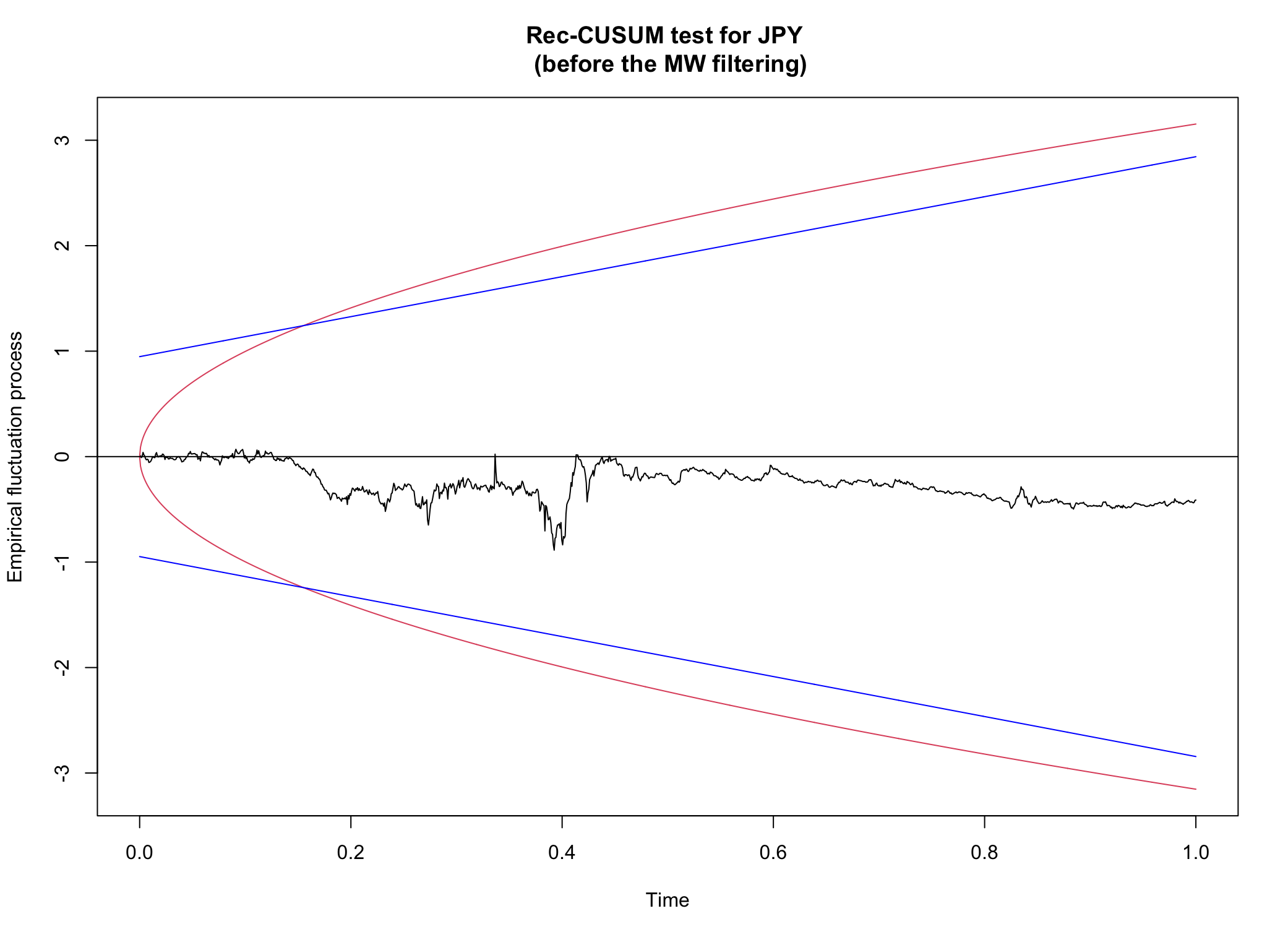}
\end{figure}

\begin{figure}[H]
  \centering
  \caption{The Rec-CUSUM test of JPY [2017-2020] (before the M\"{u}ller-Watson filtering)}
  \label{fig:RE4a}
  \includegraphics[width=1.0\linewidth]{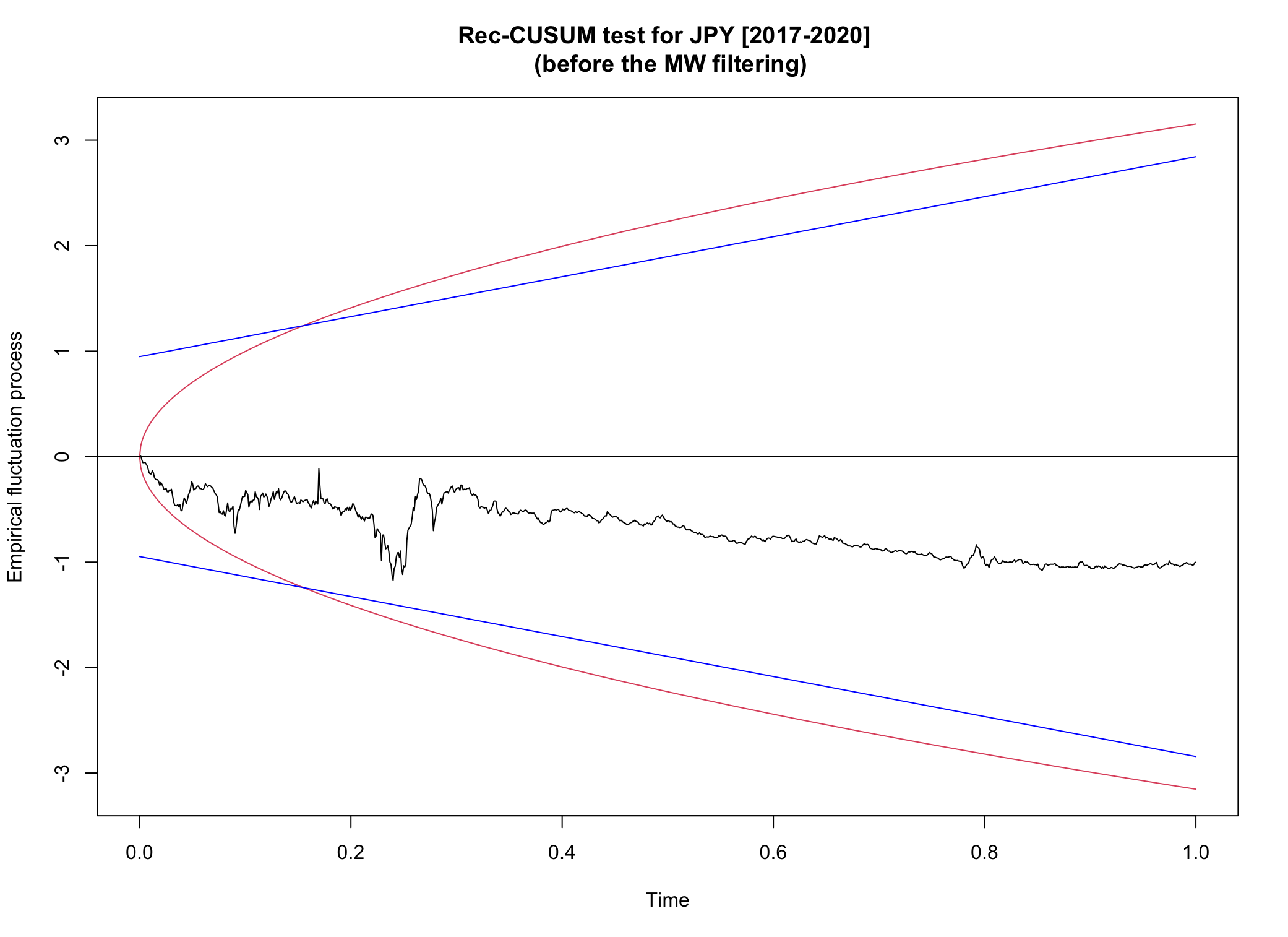}
\end{figure}

\begin{figure}[H]
  \centering
  \caption{The Rec-CUSUM test of JPY [2018-2020] (before the M\"{u}ller-Watson filtering)}
  \label{fig:RE4b}
  \includegraphics[width=1.0\linewidth]{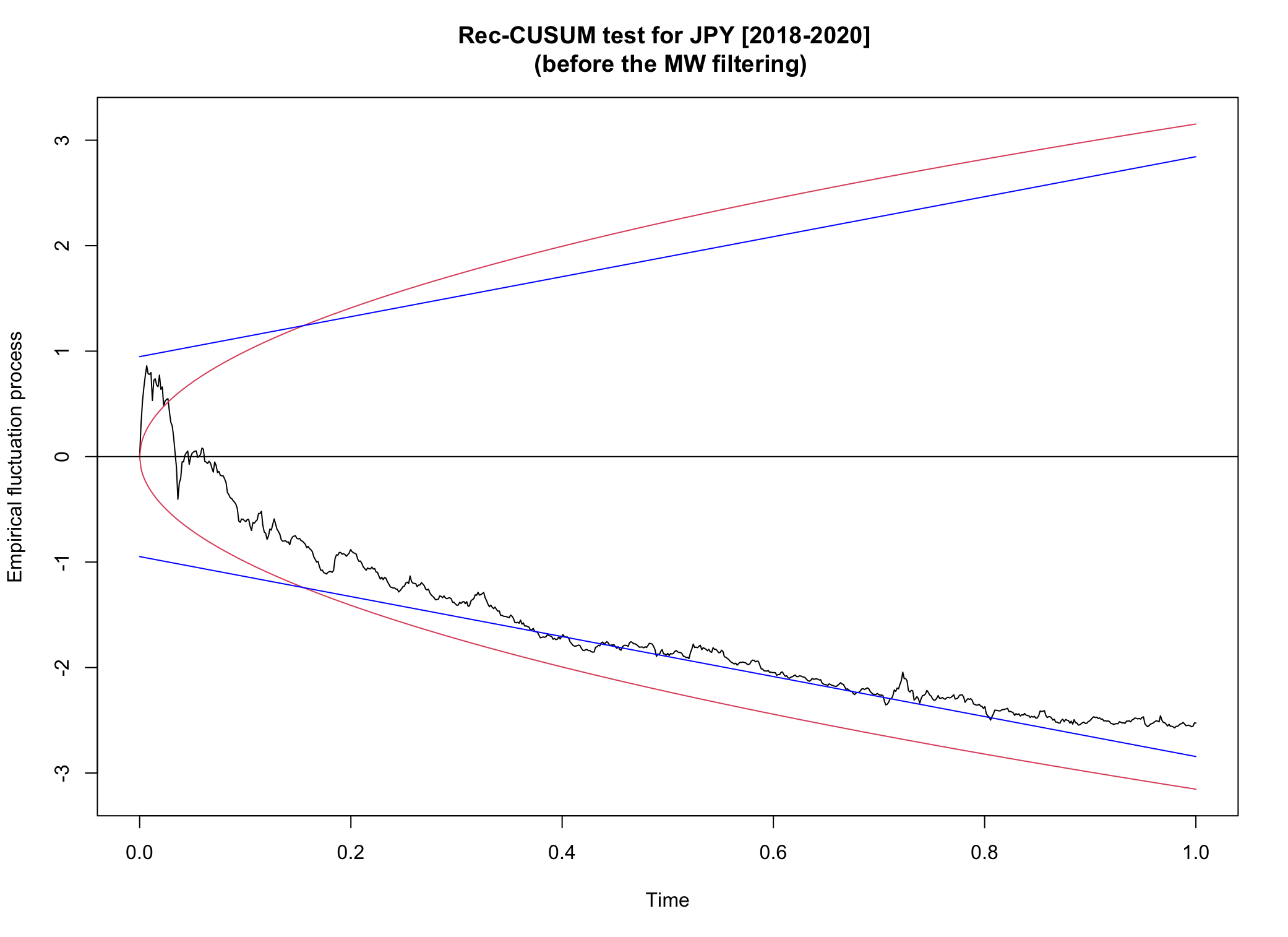}
\end{figure}

\begin{figure}[H]
  \centering
  \caption{The Rec-CUSUM test of JPY [2019-2020] (before the M\"{u}ller-Watson filtering)}
  \label{fig:RE4c}
  \includegraphics[width=1.0\linewidth]{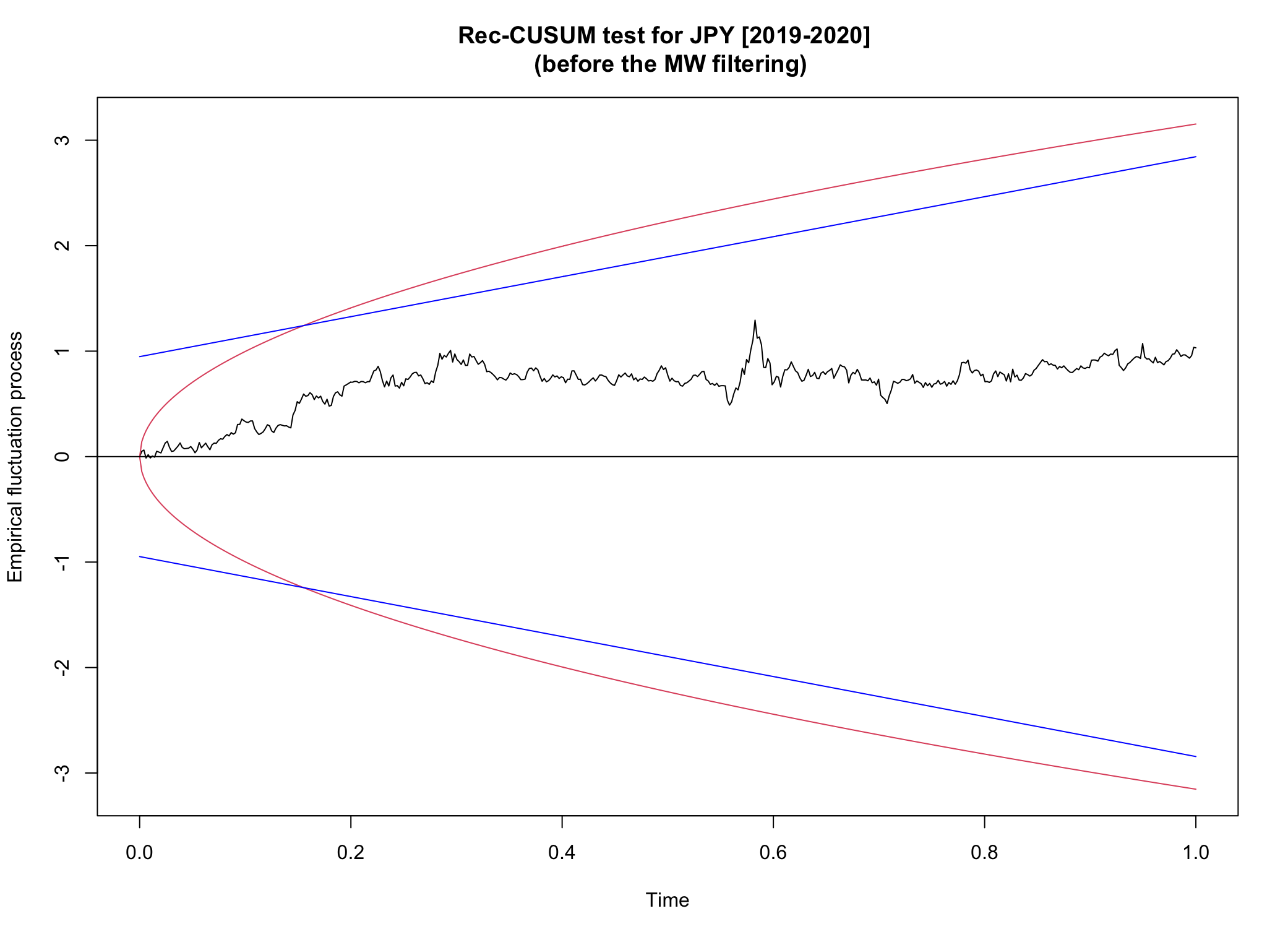}
\end{figure}

\begin{figure}[H]
  \centering
  \caption{The Rec-CUSUM test of EUR [2016-2020] (before the M\"{u}ller-Watson filtering)}
  \label{fig:RE5}
  \includegraphics[width=1.0\linewidth]{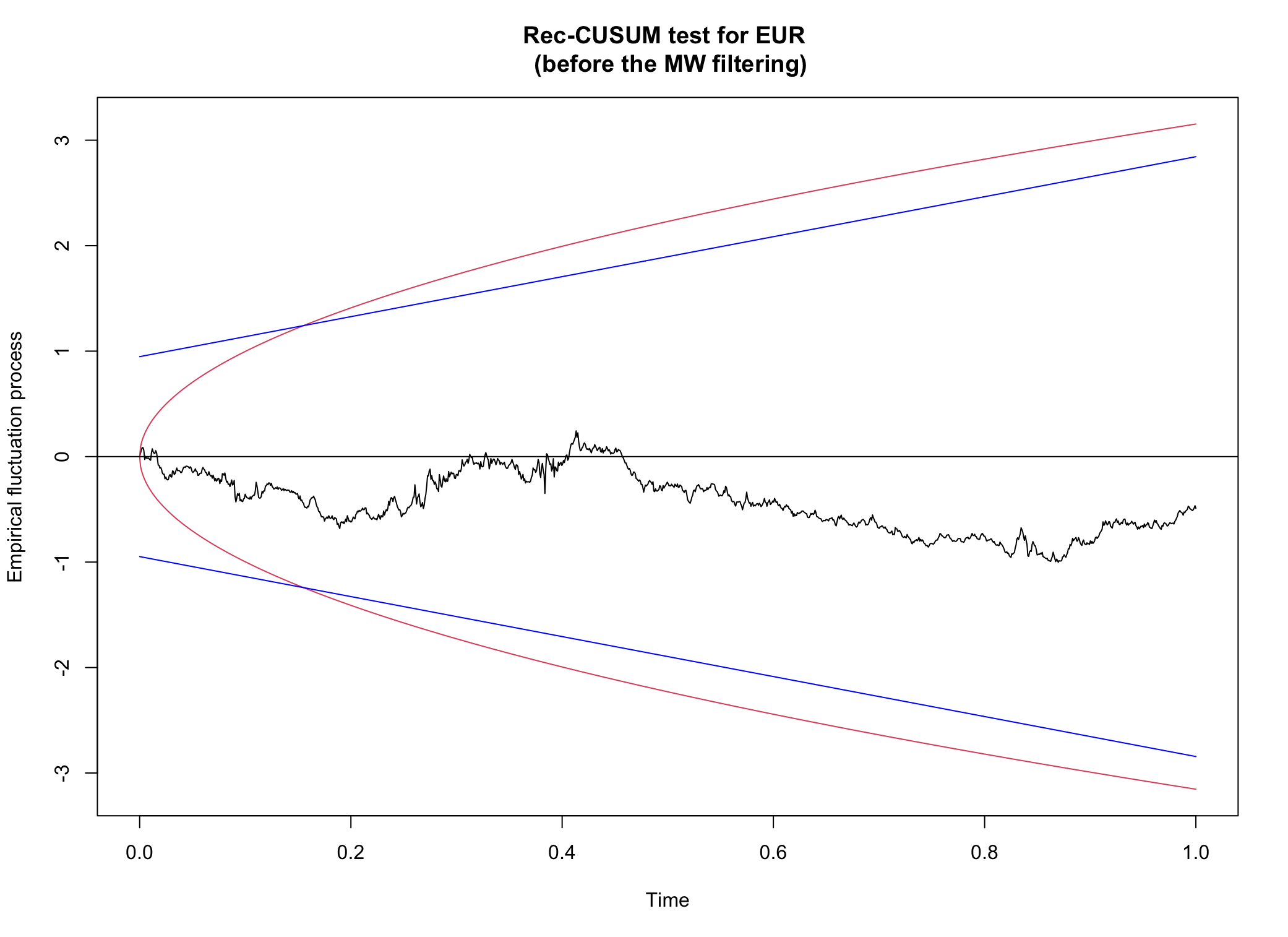}
\end{figure}

\begin{figure}[H]
  \centering
  \caption{The Rec-CUSUM test of EUR [2017-2020] (before the M\"{u}ller-Watson filtering)}
  \label{fig:RE5a}
  \includegraphics[width=1.0\linewidth]{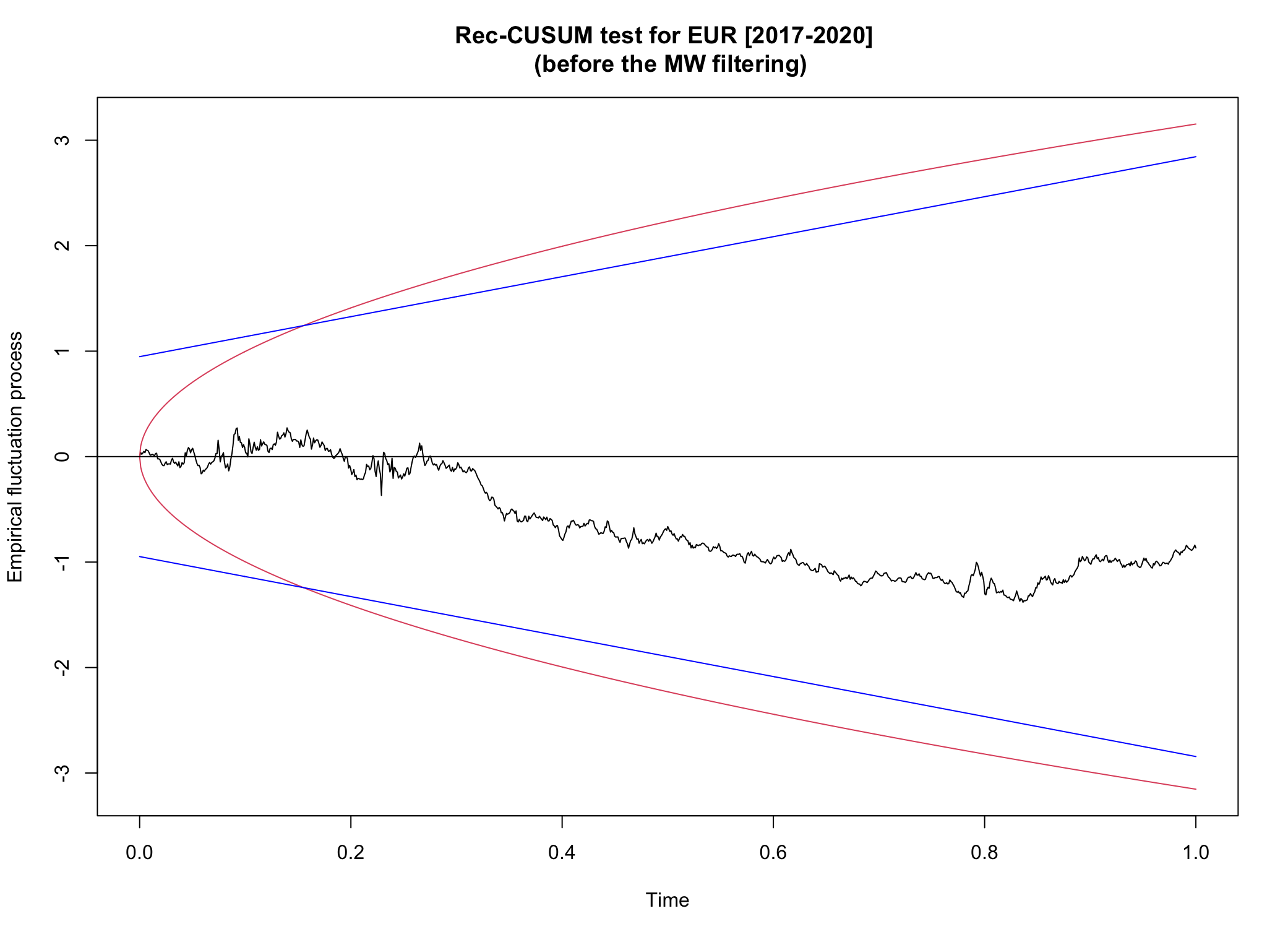}
\end{figure}

\begin{figure}[H]
  \centering
  \caption{The Rec-CUSUM test of EUR [2018-2020] (before the M\"{u}ller-Watson filtering)}
  \label{fig:RE5b}
  \includegraphics[width=1.0\linewidth]{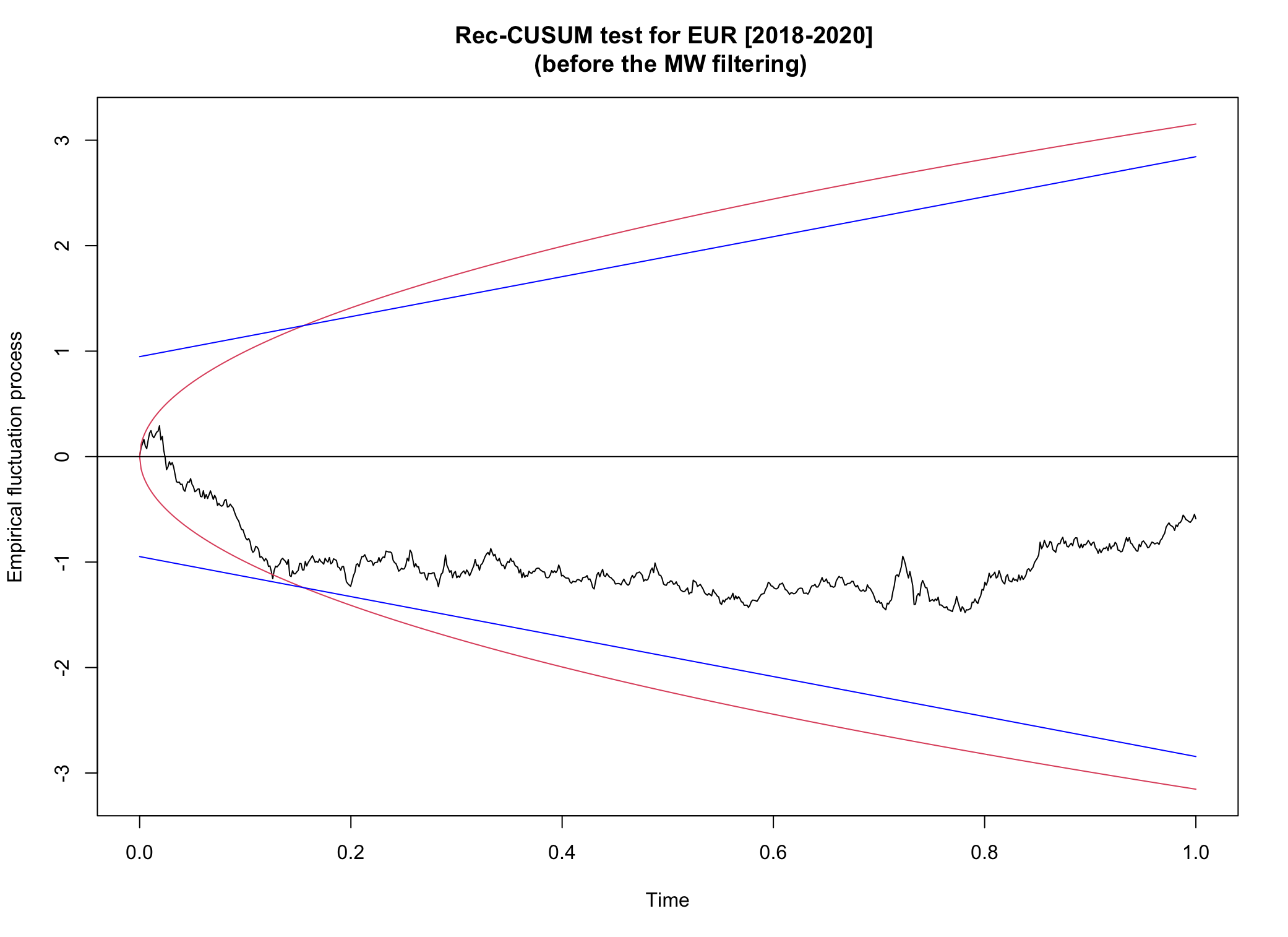}
\end{figure}

\begin{figure}[H]
  \centering
  \caption{The Rec-CUSUM test of EUR [2019-2020] (before the M\"{u}ller-Watson filtering)}
  \label{fig:RE5c}
  \includegraphics[width=1.0\linewidth]{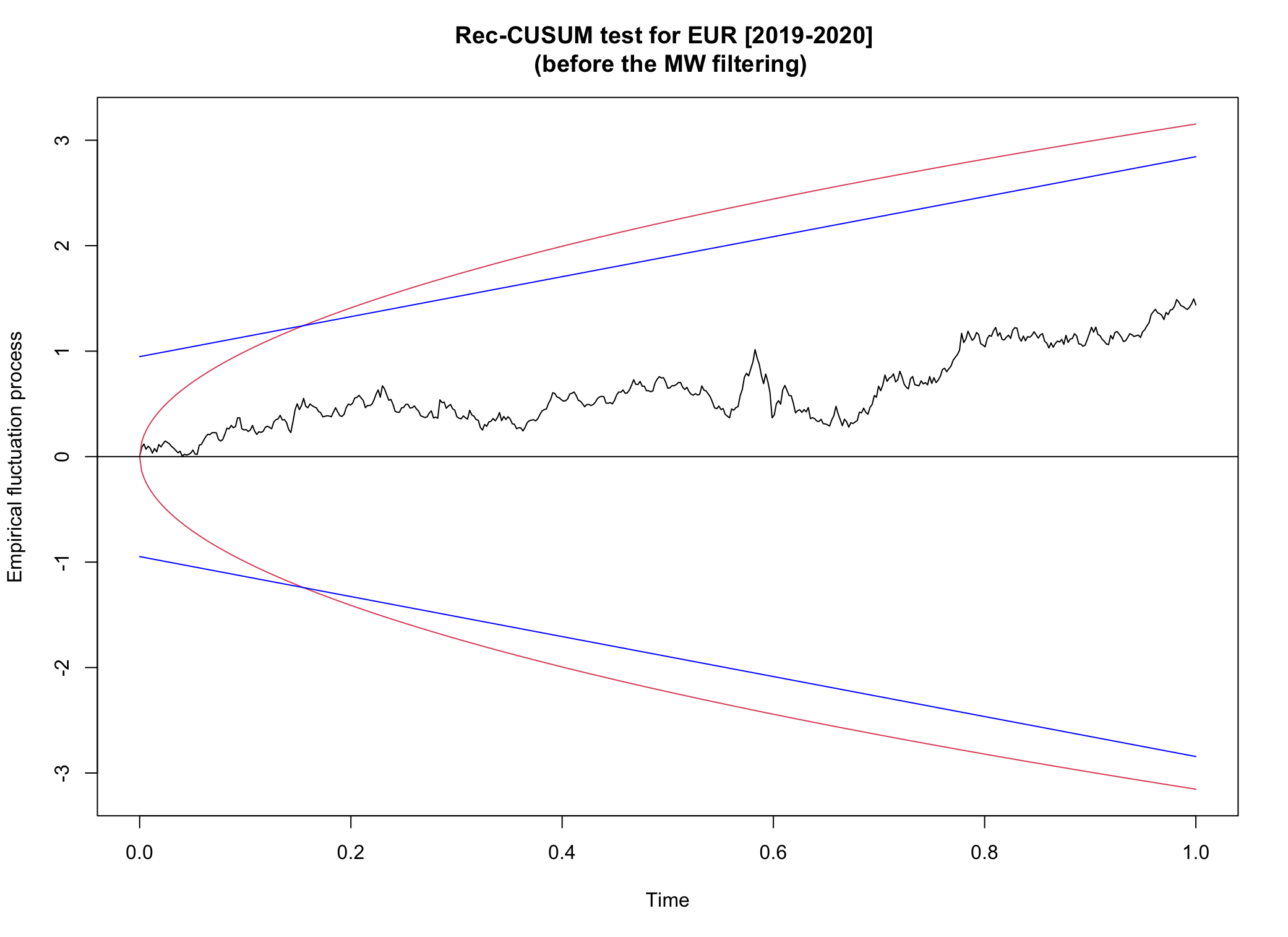}
\end{figure}

\begin{figure}[H]
  \centering
  \caption{The Rec-CUSUM test of GOLD [2016-2020] (before the M\"{u}ller-Watson filtering)}
  \label{fig:RE6}
  \includegraphics[width=1.0\linewidth]{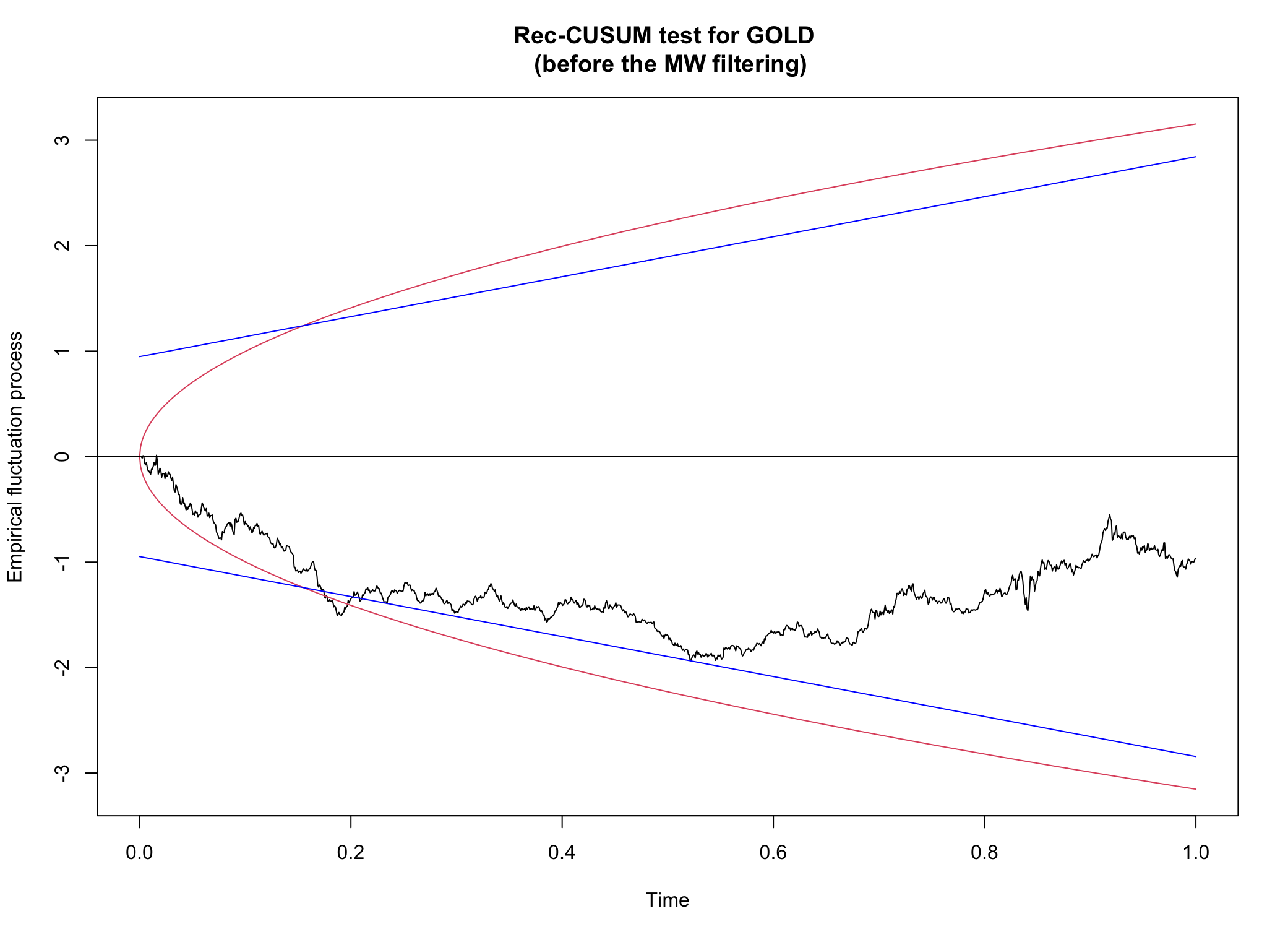}
\end{figure}

\begin{figure}[H]
  \centering
  \caption{The Rec-CUSUM test of GOLD [2017-2020] (before the M\"{u}ller-Watson filtering)}
  \label{fig:RE6a}
  \includegraphics[width=1.0\linewidth]{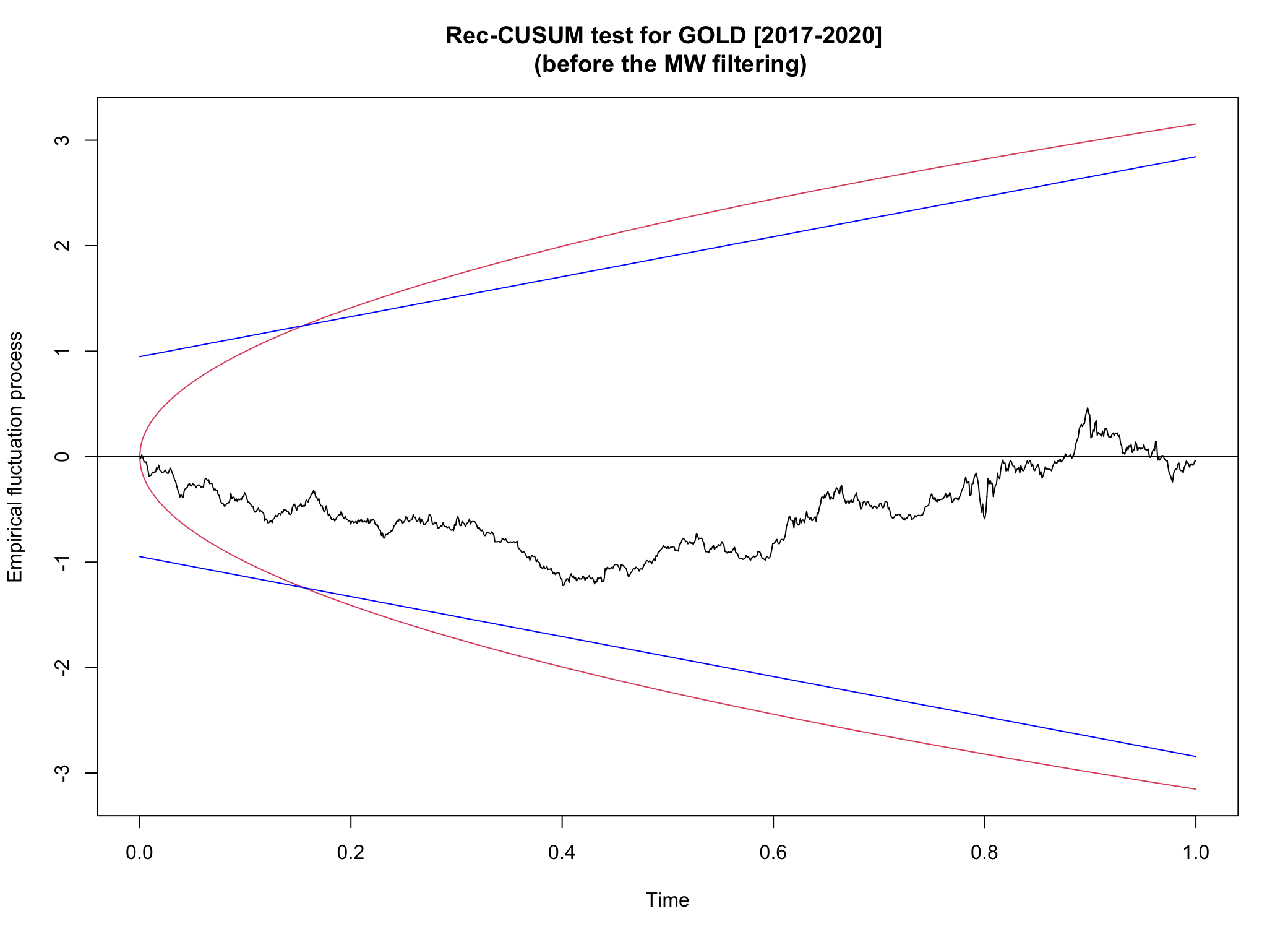}
\end{figure}

\begin{figure}[H]
  \centering
  \caption{The Rec-CUSUM test of GOLD [2018-2020] (before the M\"{u}ller-Watson filtering)}
  \label{fig:RE6b}
  \includegraphics[width=1.0\linewidth]{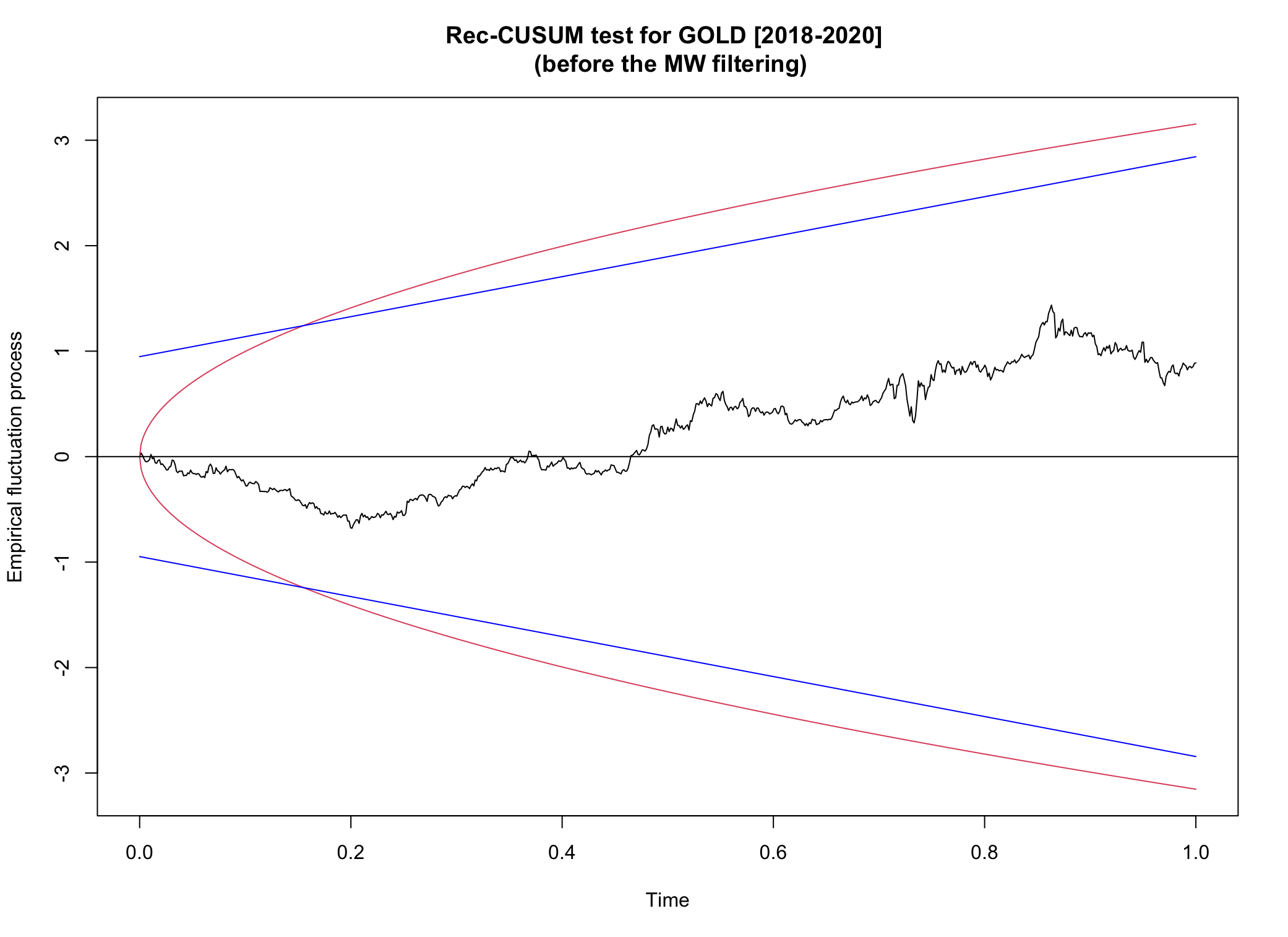}
\end{figure}

\begin{figure}[H]
  \centering
  \caption{The Rec-CUSUM test of GOLD [2019-2020] (before the M\"{u}ller-Watson filtering)}
  \label{fig:RE6c}
  \includegraphics[width=1.0\linewidth]{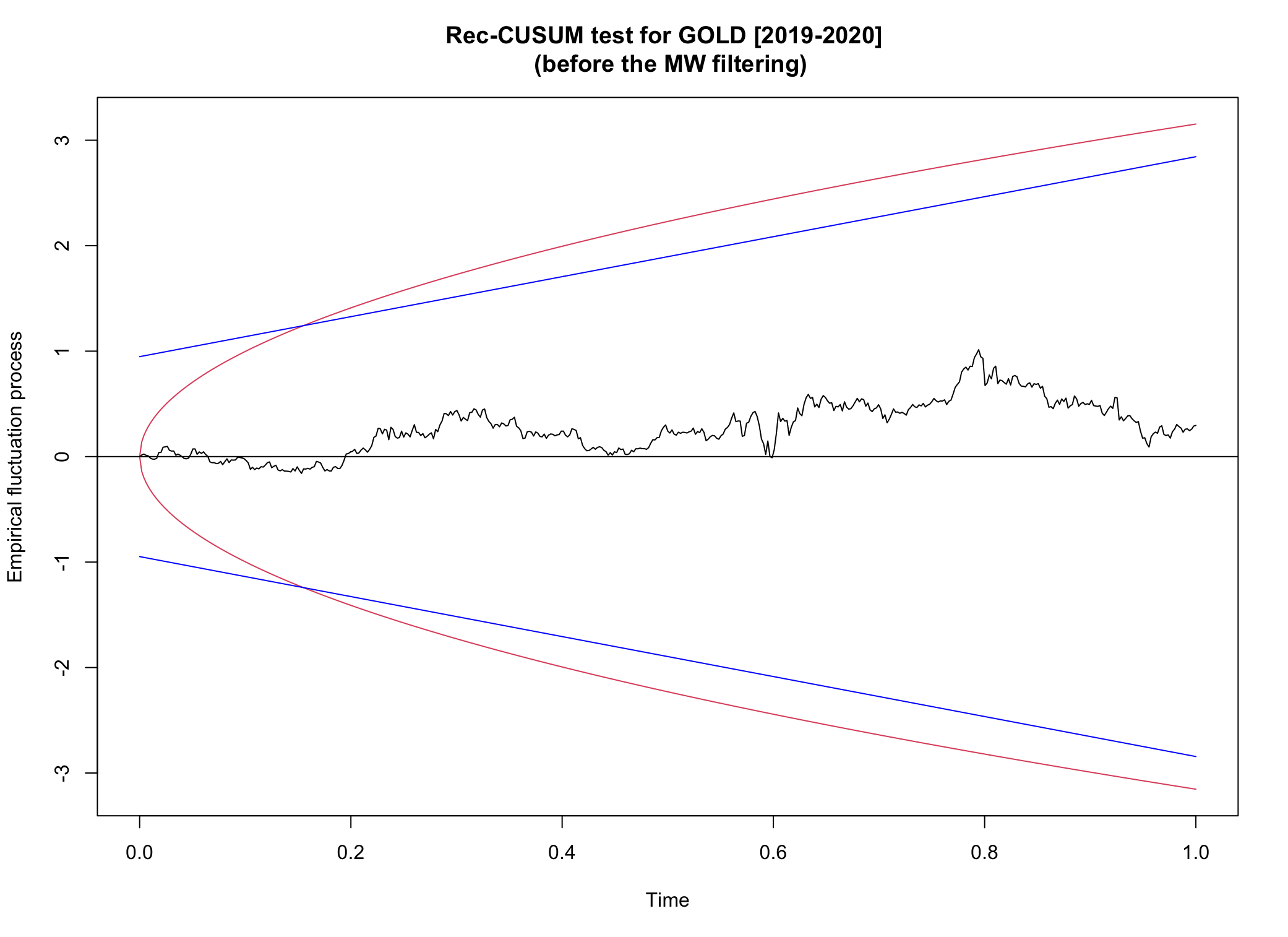}
\end{figure}

\begin{figure}[H]
  \centering
  \caption{The Rec-CUSUM test of S\&P500 [2016-2020] (before the M\"{u}ller-Watson filtering)}
  \label{fig:RE7}
  \includegraphics[width=1.0\linewidth]{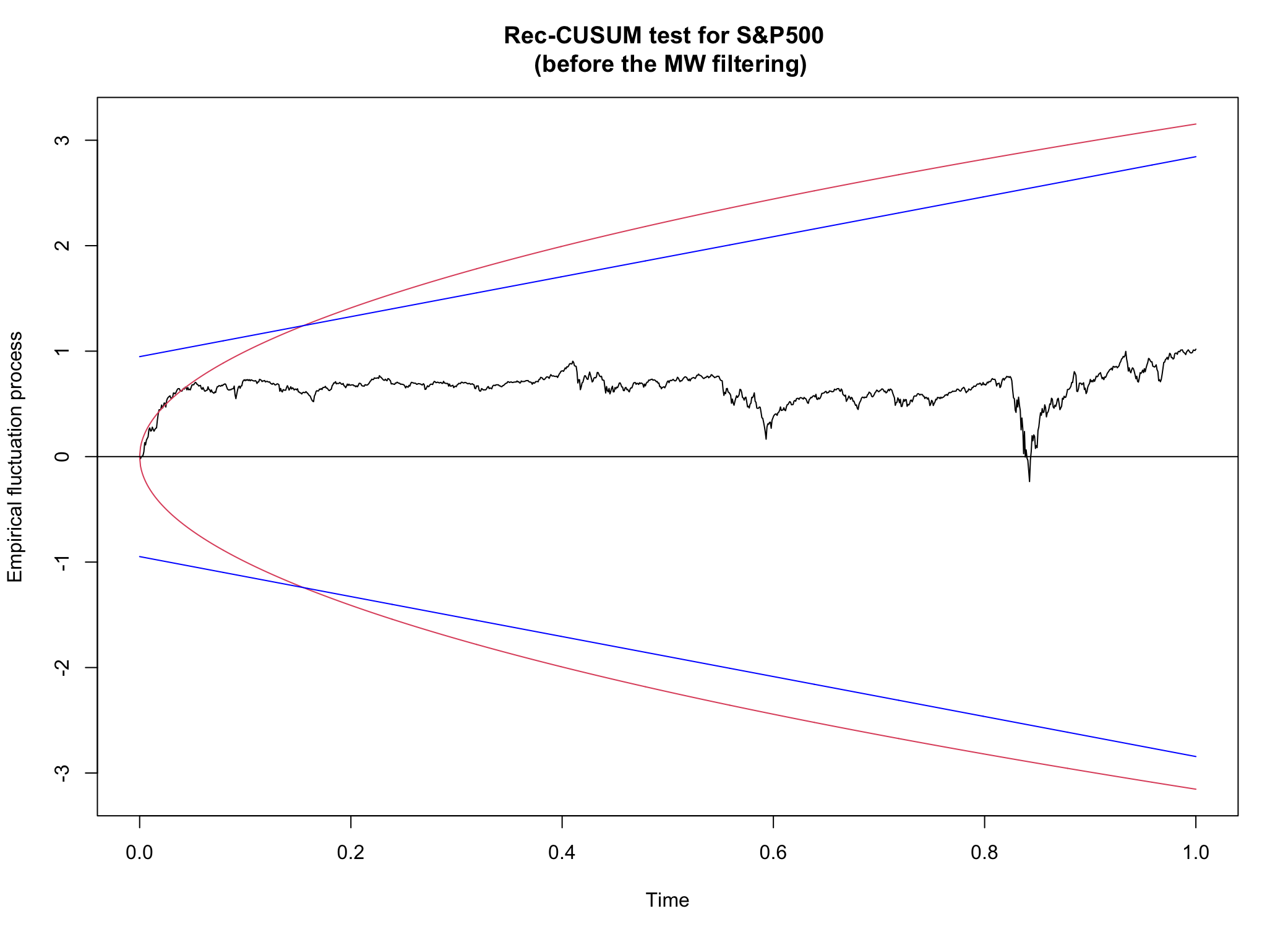}
\end{figure}

\begin{figure}[H]
  \centering
  \caption{The Rec-CUSUM test of S\&P500 [2017-2020] (before the M\"{u}ller-Watson filtering)}
  \label{fig:RE7a}
  \includegraphics[width=1.0\linewidth]{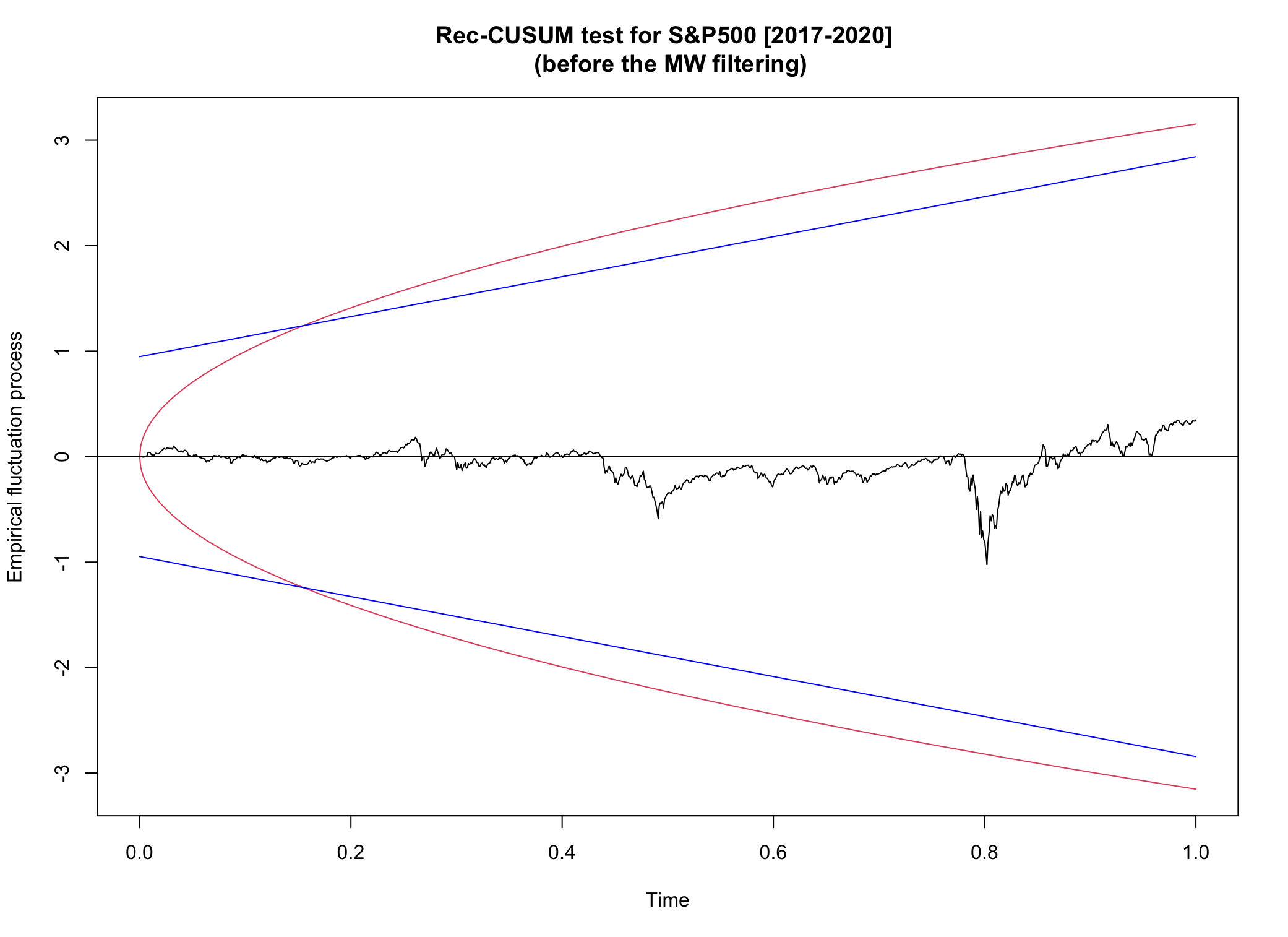}
\end{figure}

\begin{figure}[H]
  \centering
  \caption{The Rec-CUSUM test of S\&P500 [2018-2020] (before the M\"{u}ller-Watson filtering)}
  \label{fig:RE7b}
  \includegraphics[width=1.0\linewidth]{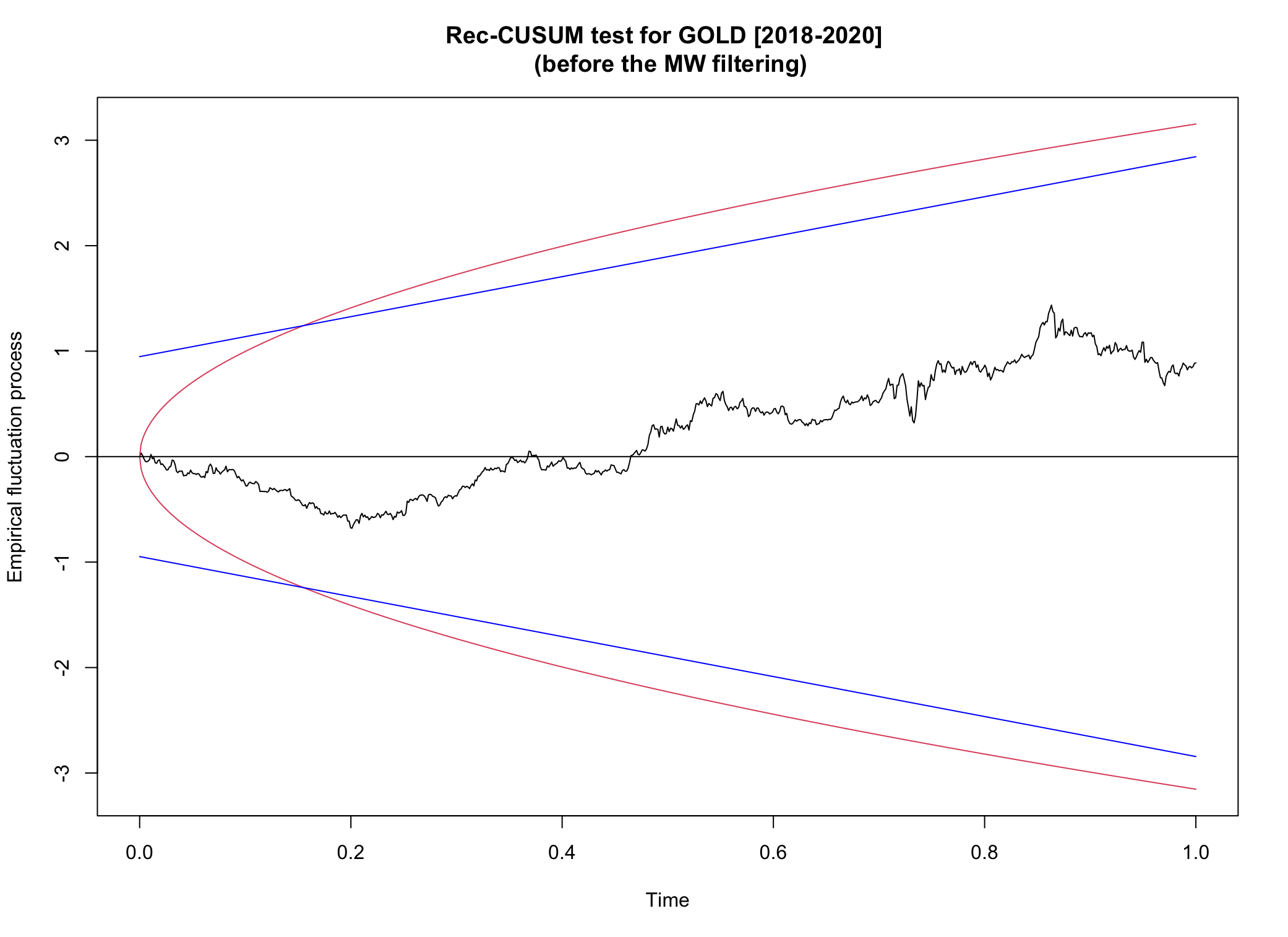}
\end{figure}

\begin{figure}[H]
  \centering
  \caption{The Rec-CUSUM test of S\&P500 [2019-2020] (before the M\"{u}ller-Watson filtering)}
  \label{fig:RE7c}
  \includegraphics[width=1.0\linewidth]{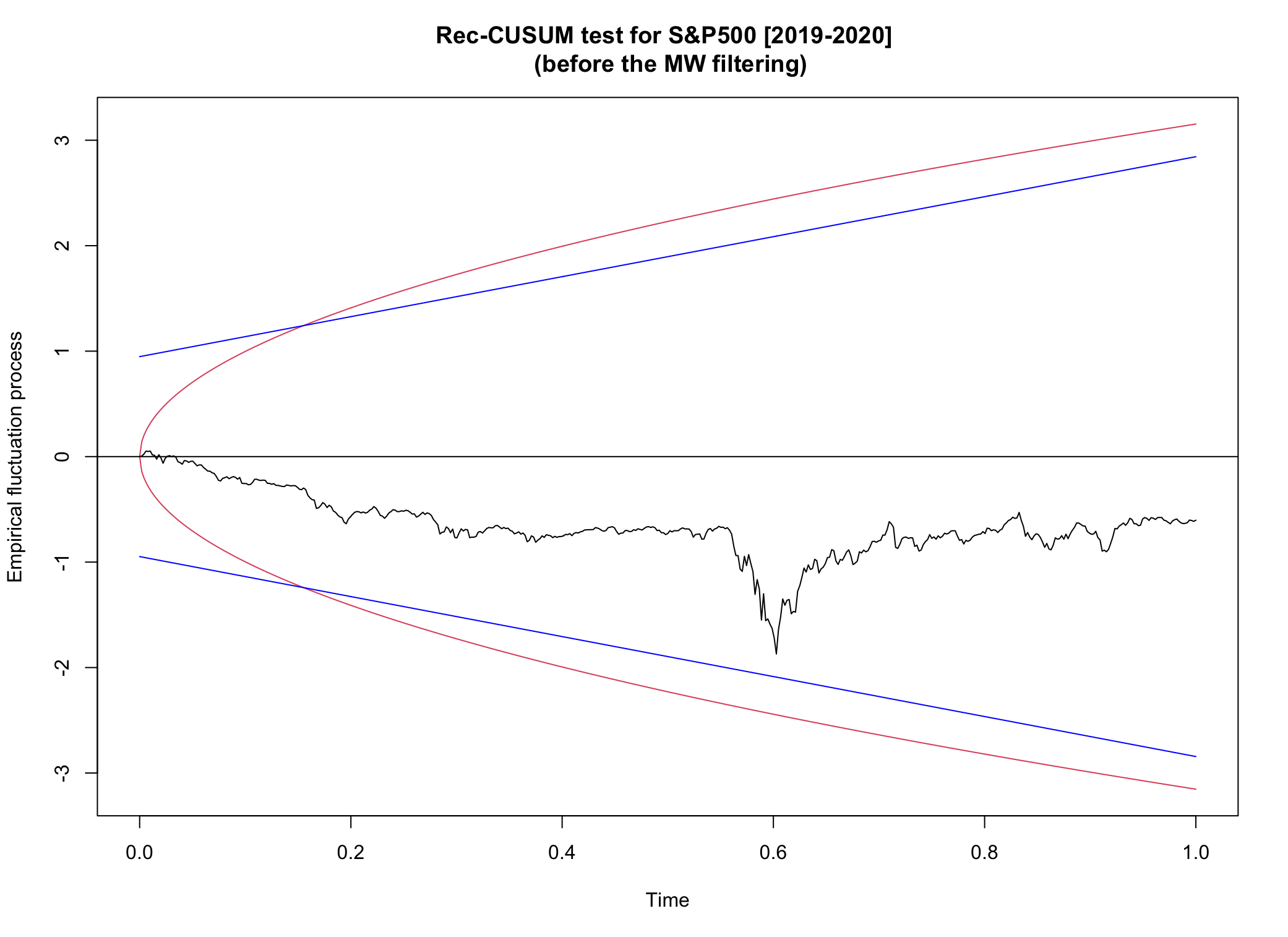}
\end{figure}

\begin{figure}[H]
  \centering
  \caption{The Rec-CUSUM test of MSCI [2016-2020] (before the M\"{u}ller-Watson filtering)}
  \label{fig:RE8}
  \includegraphics[width=1.0\linewidth]{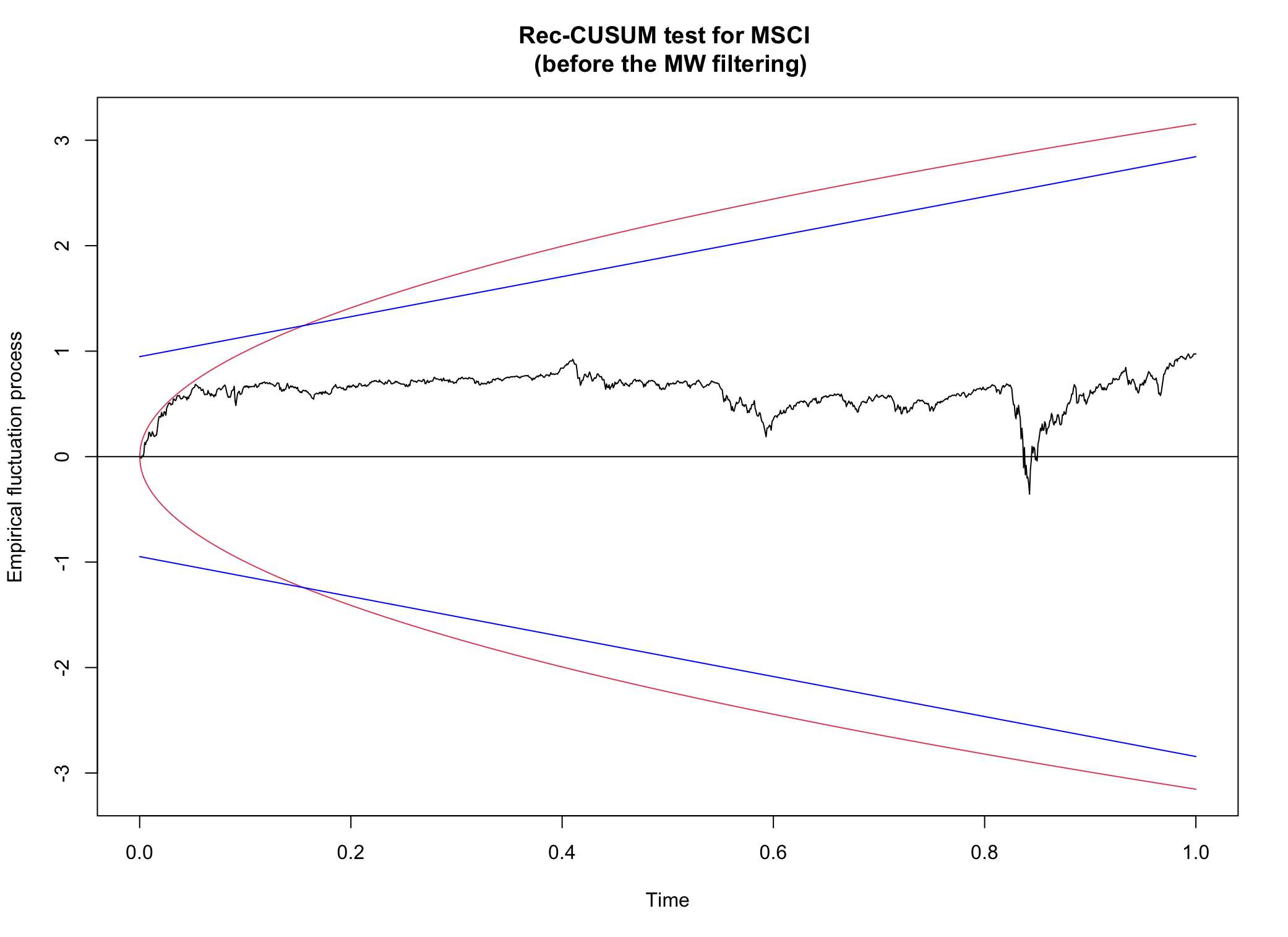}
\end{figure}

\begin{figure}[H]
  \centering
  \caption{The Rec-CUSUM test of MSCI [2017-2020] (before the M\"{u}ller-Watson filtering)}
  \label{fig:RE8a}
  \includegraphics[width=1.0\linewidth]{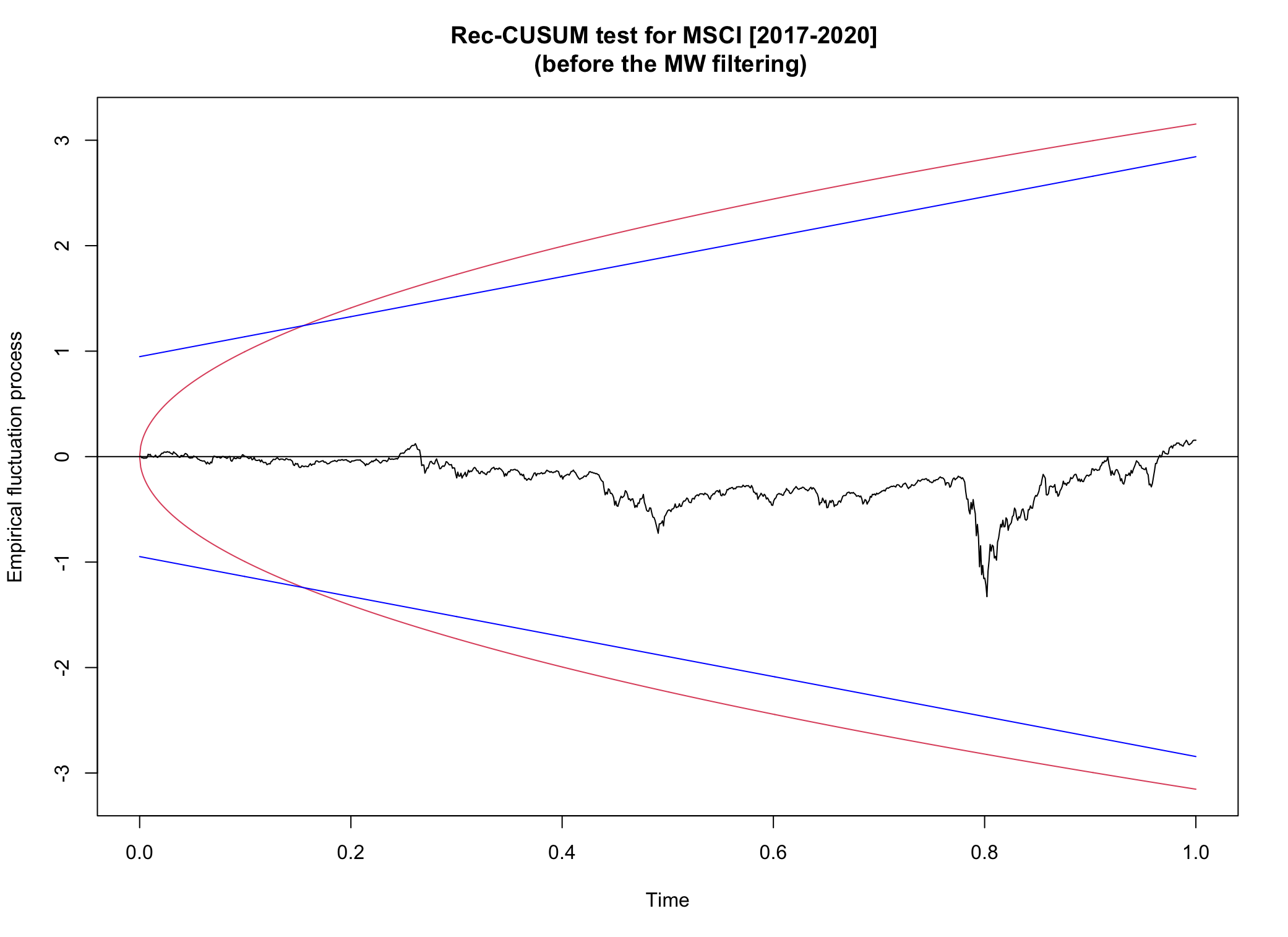}
\end{figure}

\begin{figure}[H]
  \centering
  \caption{The Rec-CUSUM test of MSCI [2018-2020] (before the M\"{u}ller-Watson filtering)}
  \label{fig:RE8b}
  \includegraphics[width=1.0\linewidth]{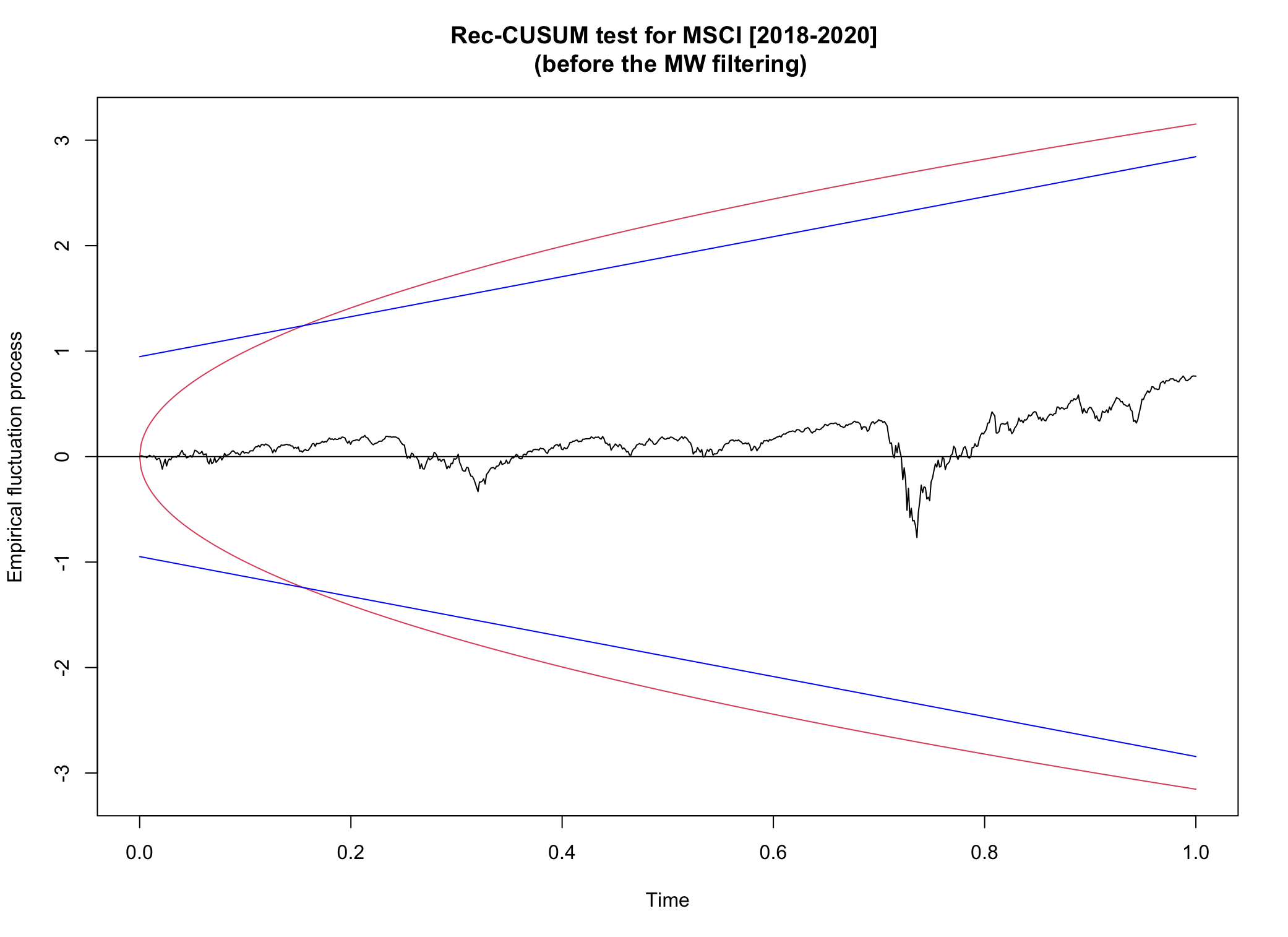}
\end{figure}

\begin{figure}[H]
  \centering
  \caption{The Rec-CUSUM test of MSCI [2019-2020] (before the M\"{u}ller-Watson filtering)}
  \label{fig:RE8c}
  \includegraphics[width=1.0\linewidth]{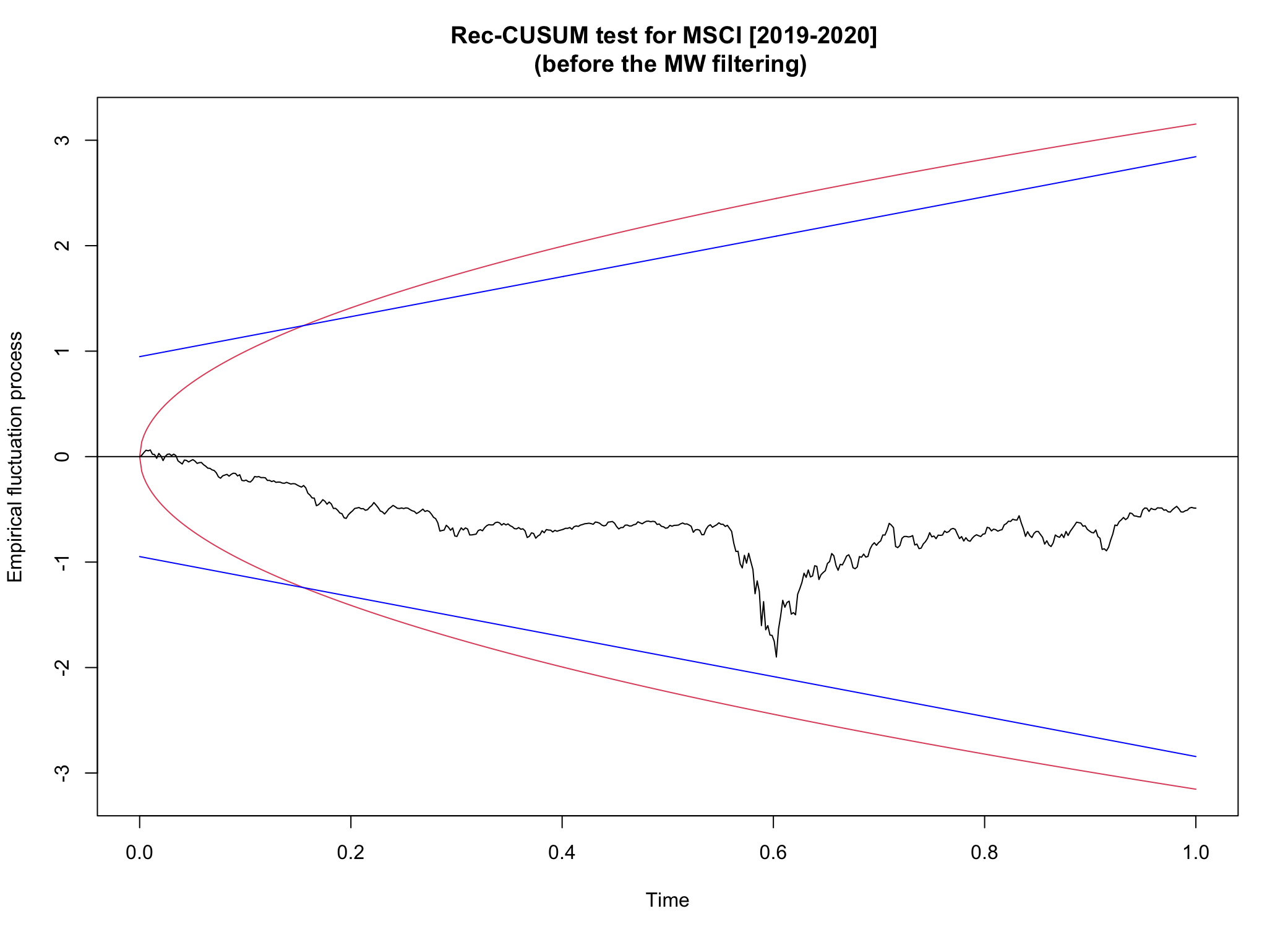}
\end{figure}

\begin{figure}[H]
  \centering
  \caption{The Rec-CUSUM test of BTC [2016-2020] (after the M\"{u}ller-Watson filtering)}
  \label{fig:RE1q}
  \includegraphics[width=1.0\linewidth]{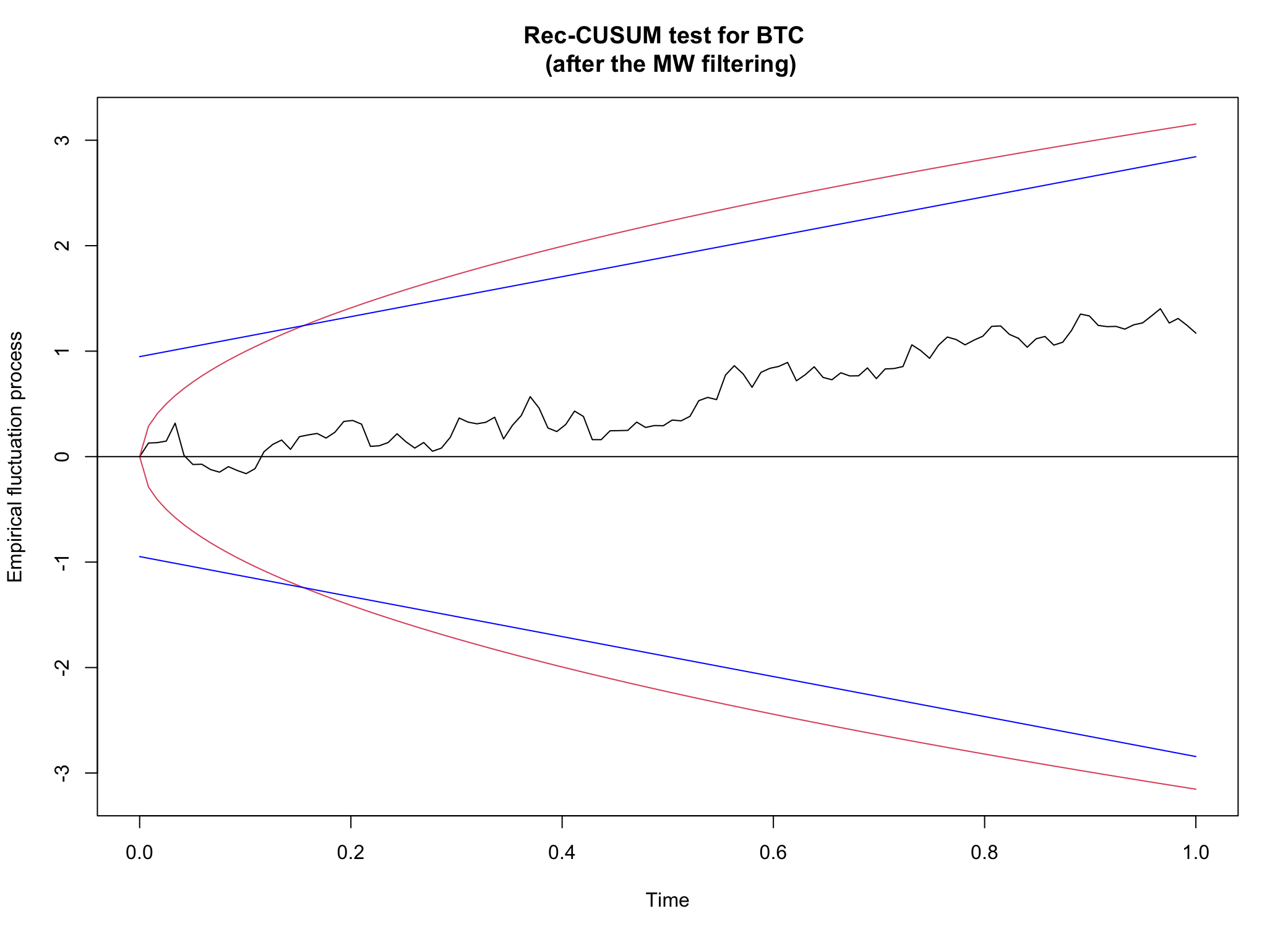}
\end{figure}

In addition to that, we also conduct the Rec-CUSUM tests using the latter parts of data before the MW filtering, that is, (i) a sample data from 2017 to 2020; (ii) a sample data from 2018 to 2020; and (iii) a sample data from 2019 to 2020. The results before the MW filtering can be seen in Figures \ref{fig:RE1a}, \ref{fig:RE1b}, and \ref{fig:RE1c} for BTC, Figures \ref{fig:RE2a}, \ref{fig:RE2b}, and \ref{fig:RE2c} for ETH, Figures \ref{fig:RE3a}, \ref{fig:RE3b}, and \ref{fig:RE3c} for XRP, Figures \ref{fig:RE4a}, \ref{fig:RE4b}, and \ref{fig:RE4c} for JPY, Figures \ref{fig:RE5a}, \ref{fig:RE5b}, and \ref{fig:RE5c} for EUR, Figures \ref{fig:RE6a}, \ref{fig:RE6b}, and \ref{fig:RE6c} for GOLD, Figures \ref{fig:RE7a}, \ref{fig:RE7b}, and \ref{fig:RE7c} for S\&P500, and Figures \ref{fig:RE8a}, \ref{fig:RE8b}, and \ref{fig:RE8c} for MSCI. 

For the sample data from 2018 to 2020, BTC and JPY before the MW filtering exceed the boundaries of the Brownian motion. For other subsample years, all the return movements are within the boundaries.

\begin{figure}[H]
  \centering
  \caption{The Rec-CUSUM test of BTC [2017-2020] (after the M\"{u}ller-Watson filtering)}
  \label{fig:RE1aq}
  \includegraphics[width=1.0\linewidth]{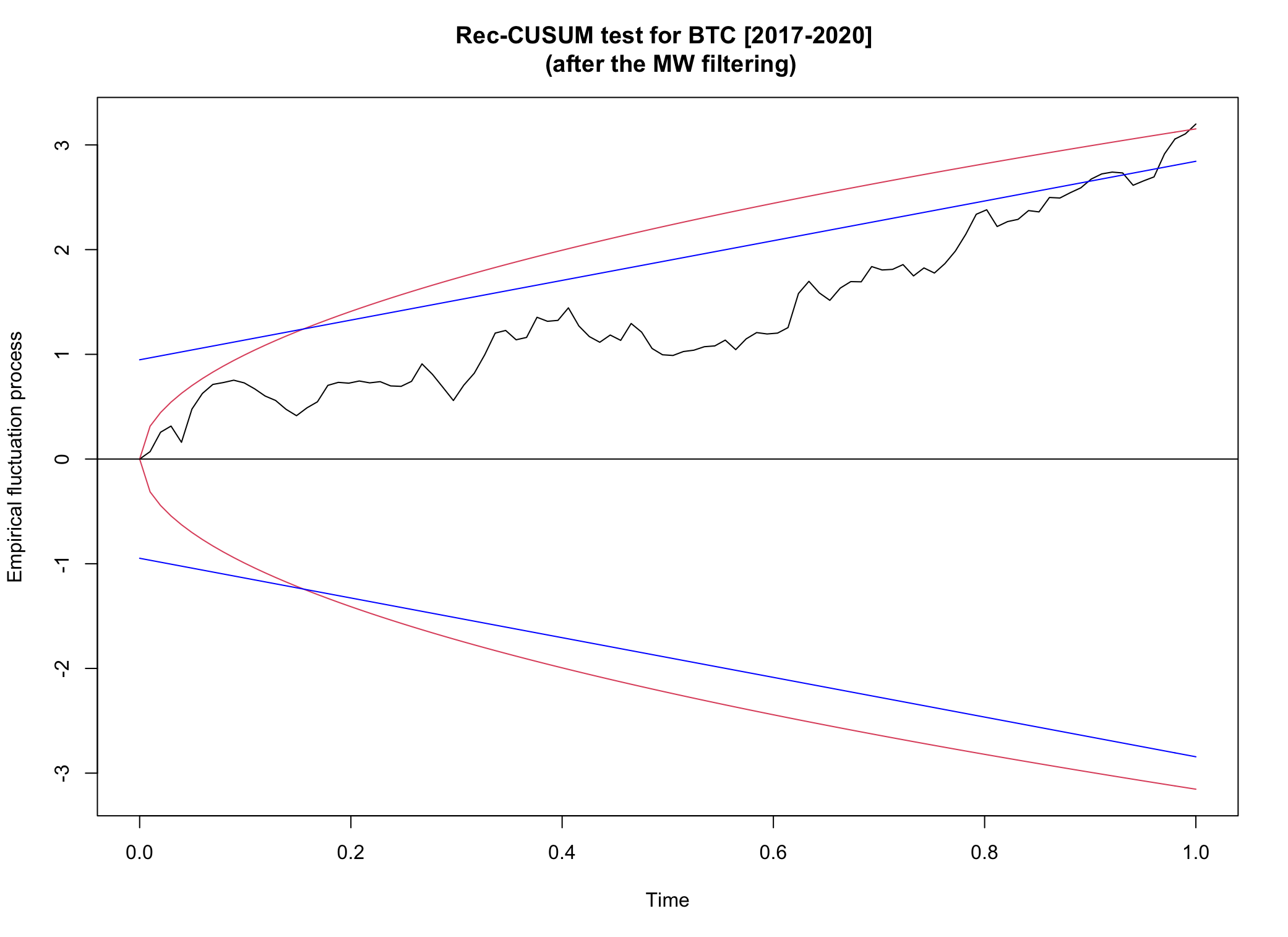}
\end{figure}

\begin{figure}[H]
  \centering
  \caption{The Rec-CUSUM test of BTC [2018-2020] (after the M\"{u}ller-Watson filtering)}
  \label{fig:RE1bq}
  \includegraphics[width=1.0\linewidth]{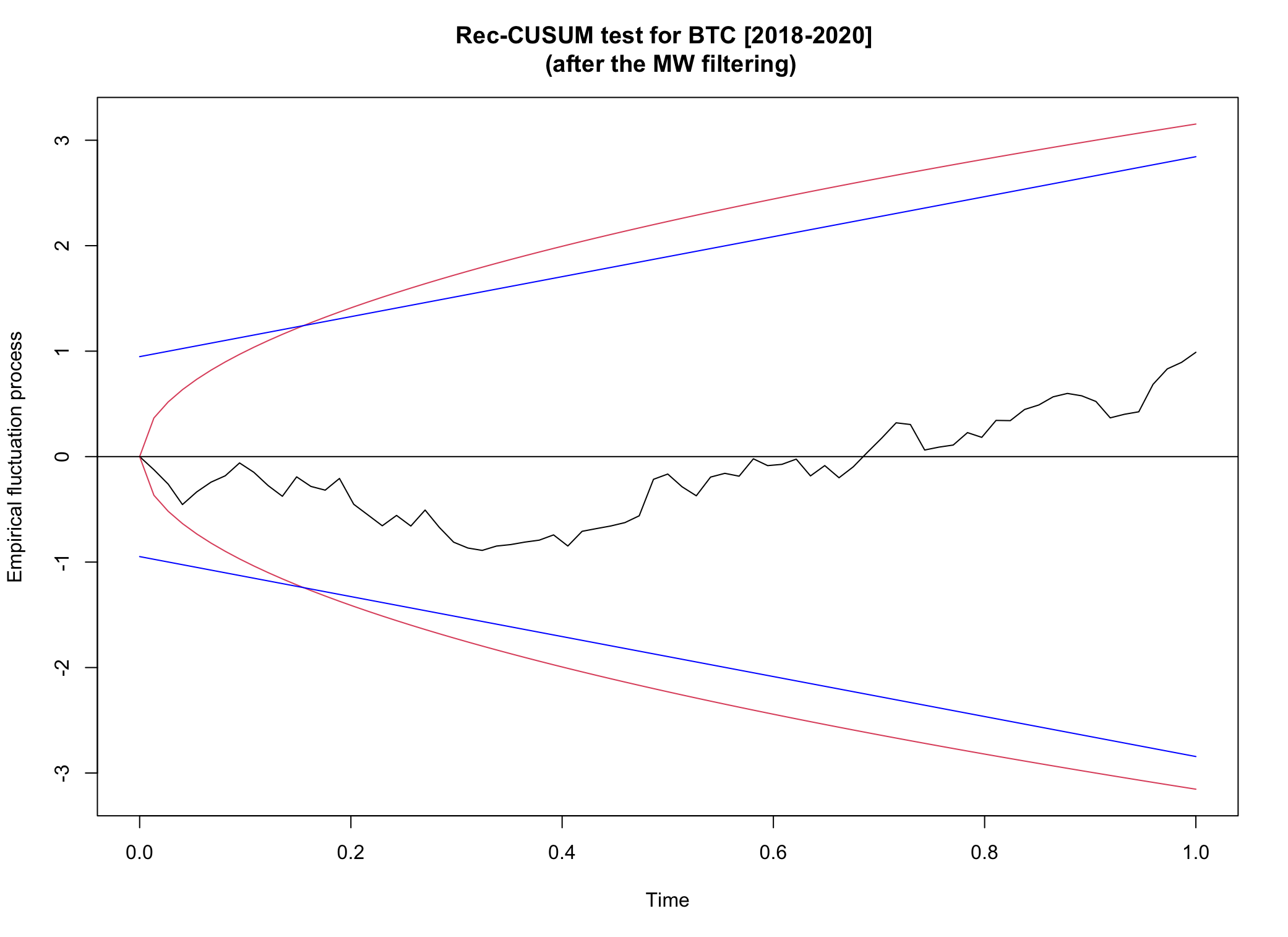}
\end{figure}

\begin{figure}[H]
  \centering
  \caption{The Rec-CUSUM test of BTC [2019-2020] (after the M\"{u}ller-Watson filtering)}
  \label{fig:RE1cq}
  \includegraphics[width=1.0\linewidth]{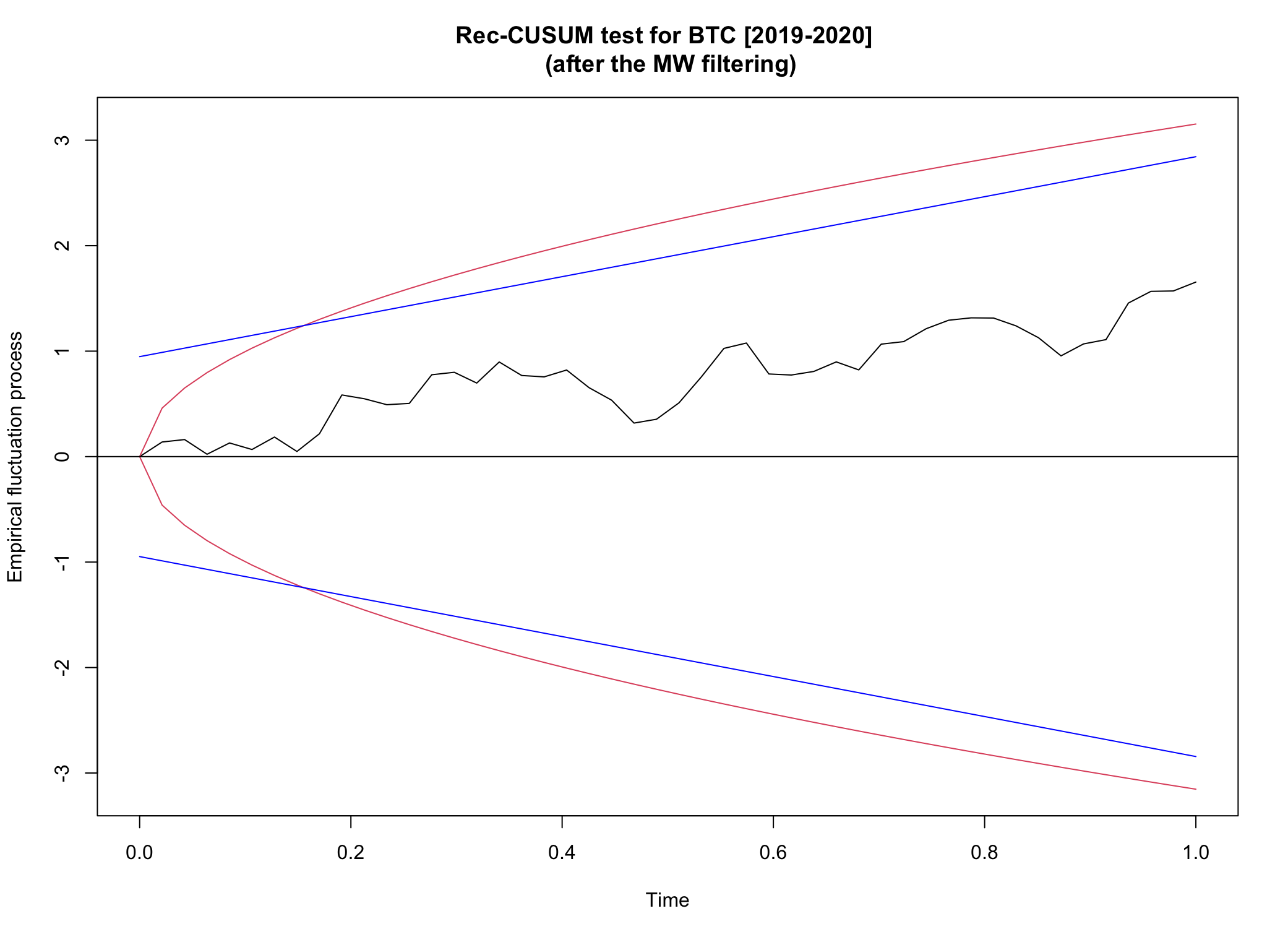}
\end{figure}

\begin{figure}[H]
  \centering
  \caption{The Rec-CUSUM test of ETH [2016-2020] (after the M\"{u}ller-Watson filtering)}
  \label{fig:RE2q}
  \includegraphics[width=1.0\linewidth]{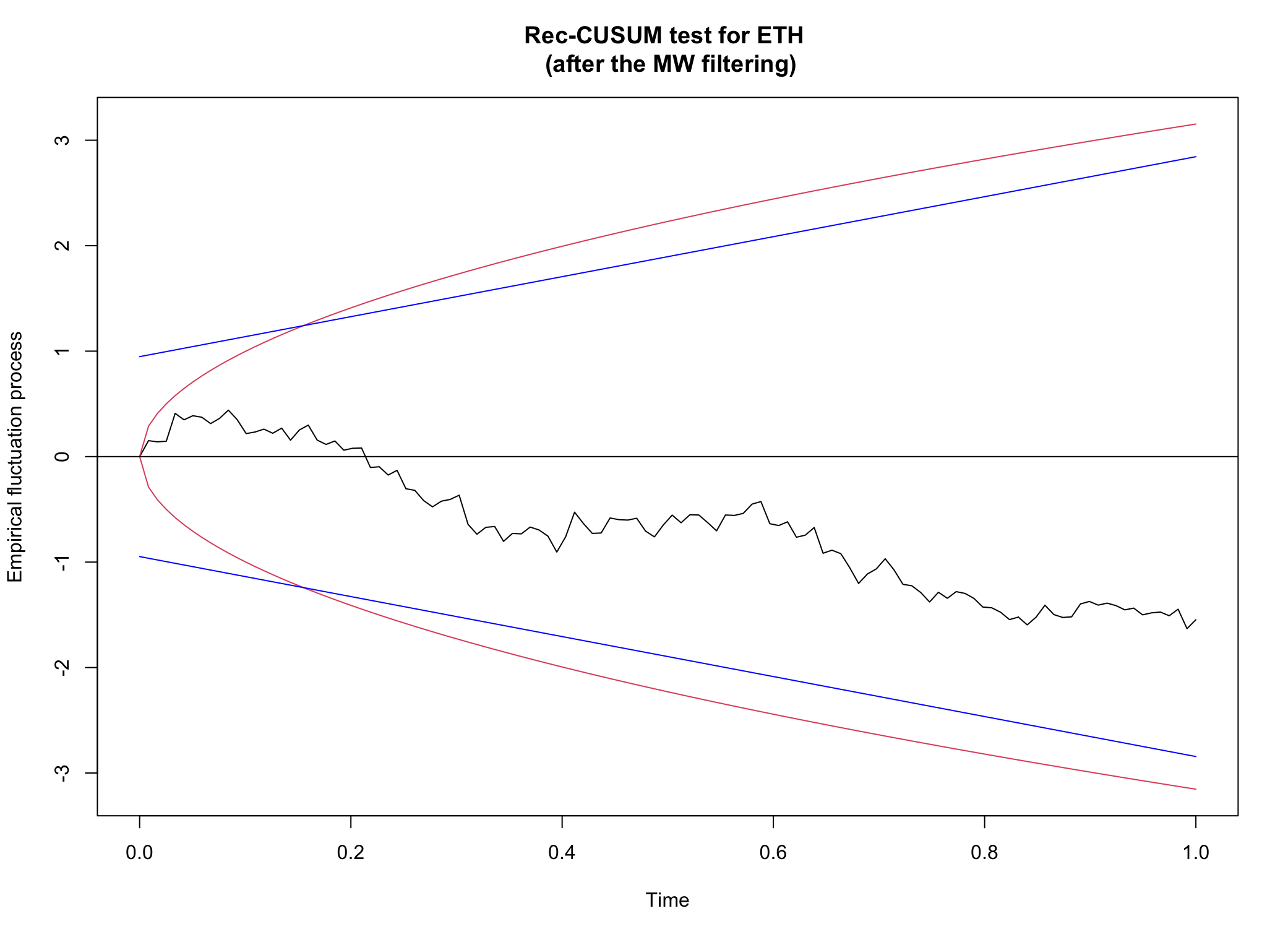}
\end{figure}

\begin{figure}[H]
  \centering
  \caption{The Rec-CUSUM test of ETH [2017-2020] (after the M\"{u}ller-Watson filtering)}
  \label{fig:RE2aq}
  \includegraphics[width=1.0\linewidth]{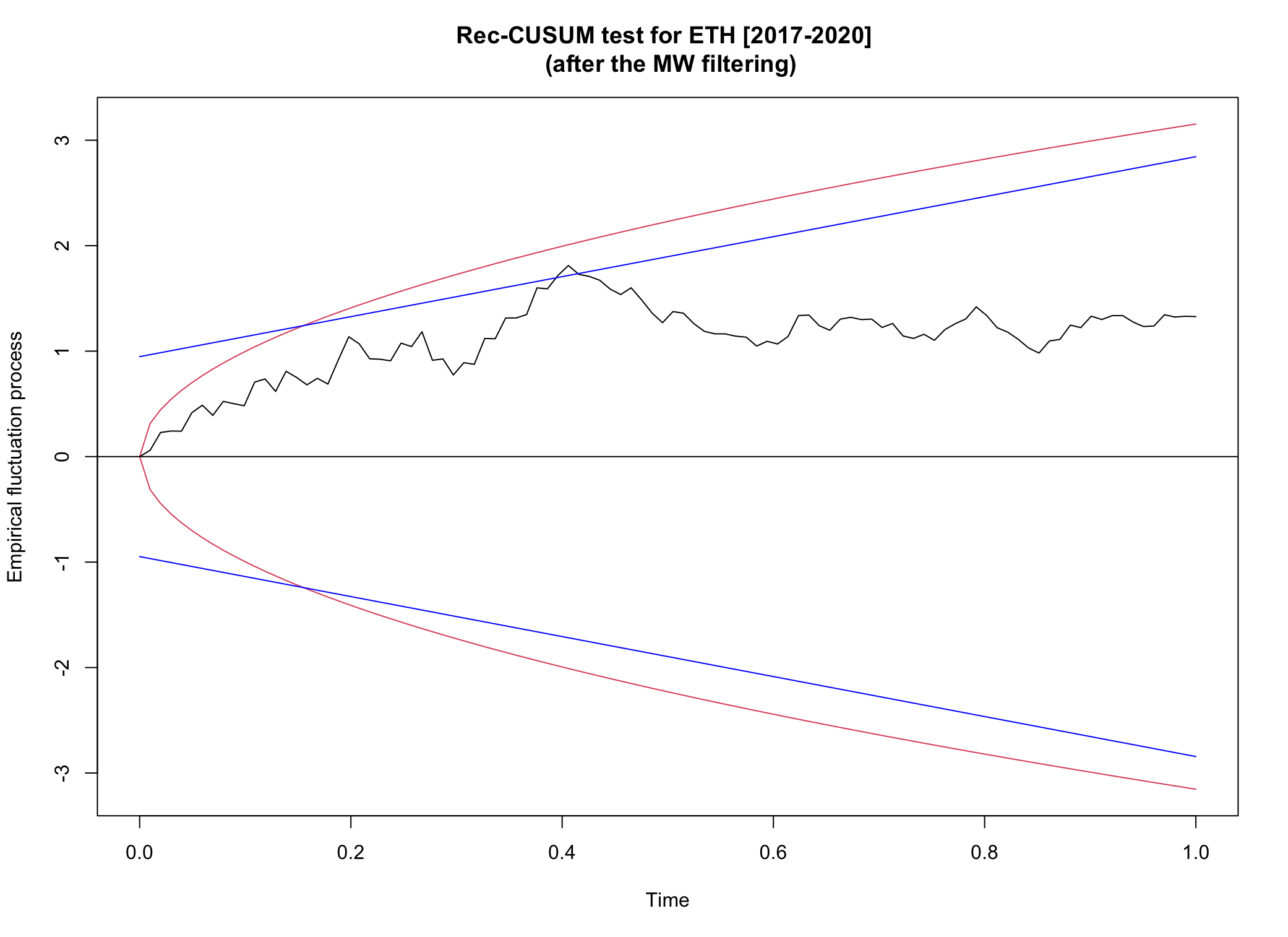}
\end{figure}

\begin{figure}[H]
  \centering
  \caption{The Rec-CUSUM test of ETH [2018-2020] (after the M\"{u}ller-Watson filtering)}
  \label{fig:RE2bq}
  \includegraphics[width=1.0\linewidth]{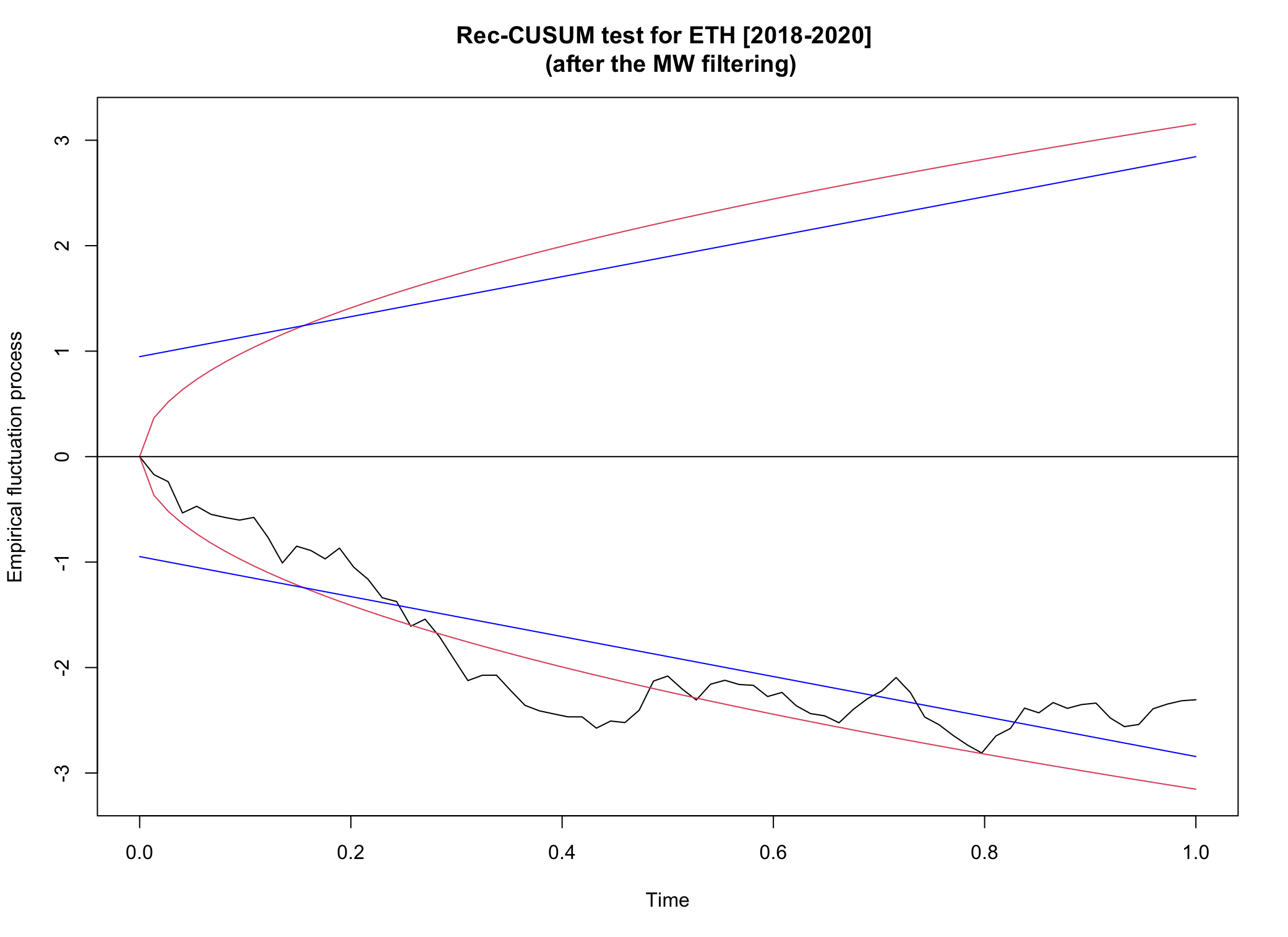}
\end{figure}

\begin{figure}[H]
  \centering
  \caption{The Rec-CUSUM test of ETH [2019-2020] (after the M\"{u}ller-Watson filtering)}
  \label{fig:RE2cq}
  \includegraphics[width=1.0\linewidth]{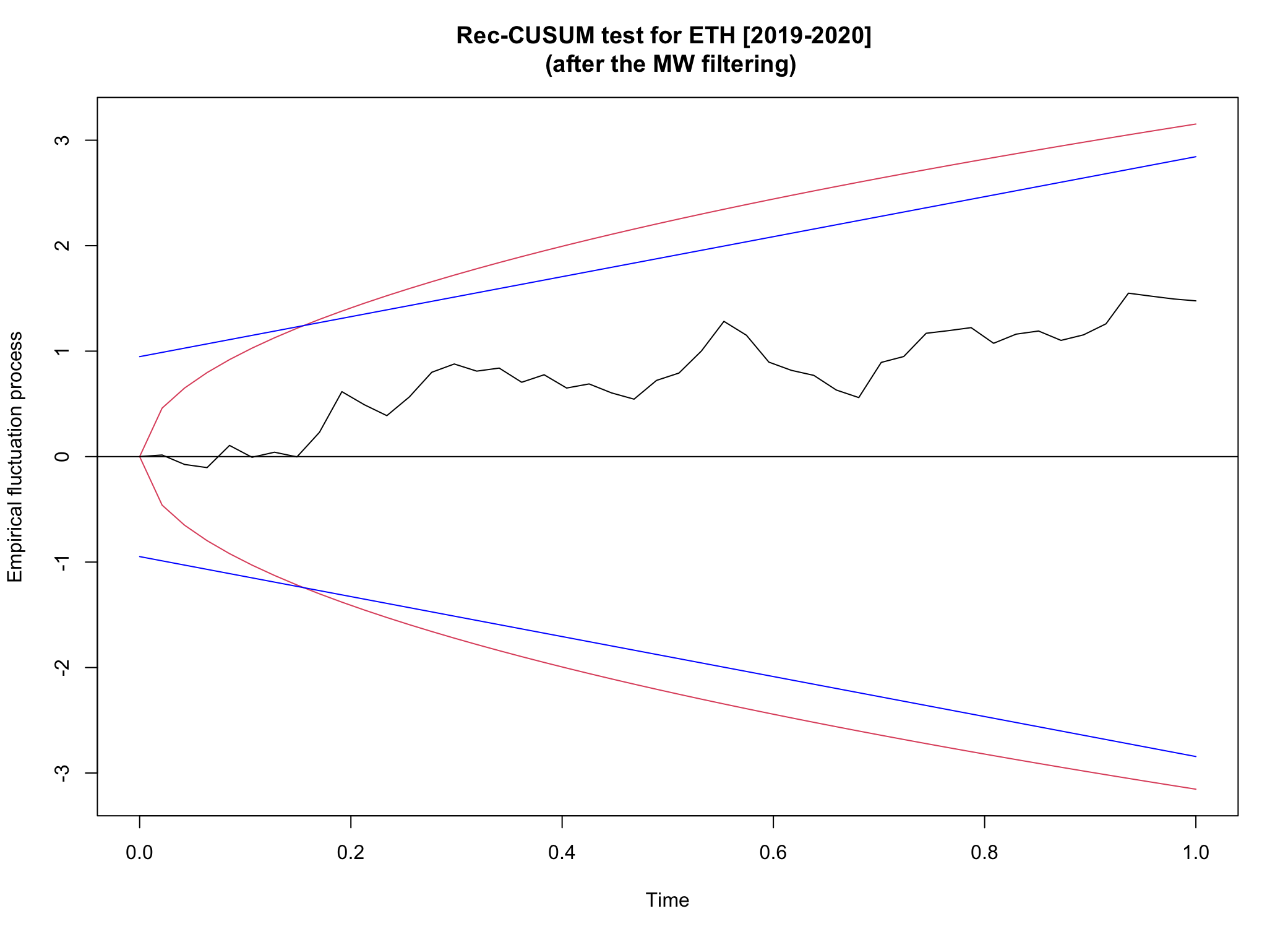}
\end{figure}

\begin{figure}[H]
  \centering
  \caption{The Rec-CUSUM test of XRP [2016-2020] (after the M\"{u}ller-Watson filtering)}
  \label{fig:RE3q}
  \includegraphics[width=1.0\linewidth]{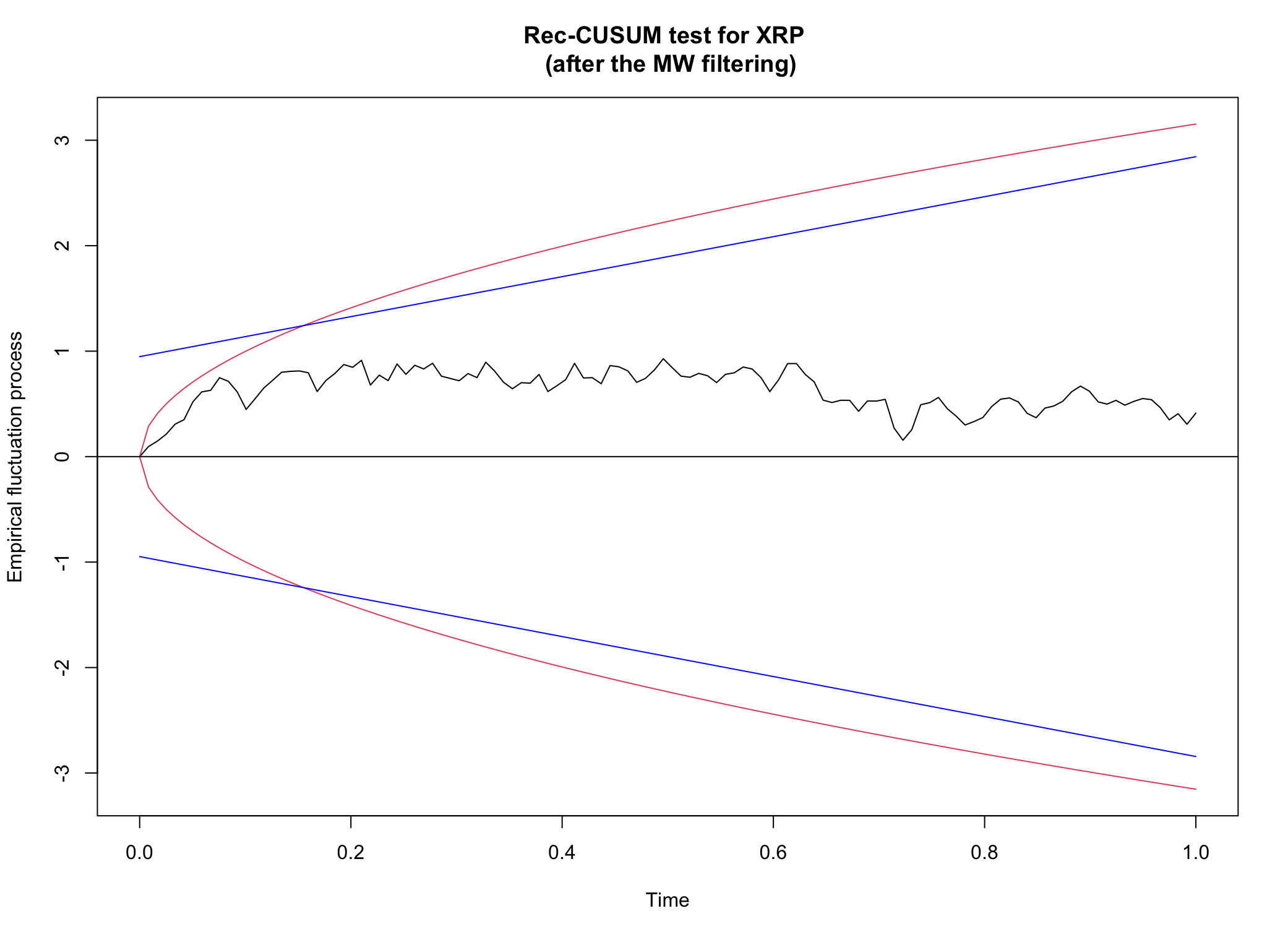}
\end{figure}

\begin{figure}[H]
  \centering
  \caption{The Rec-CUSUM test of XRP [2017-2020] (after the M\"{u}ller-Watson filtering)}
  \label{fig:RE3aq}
  \includegraphics[width=1.0\linewidth]{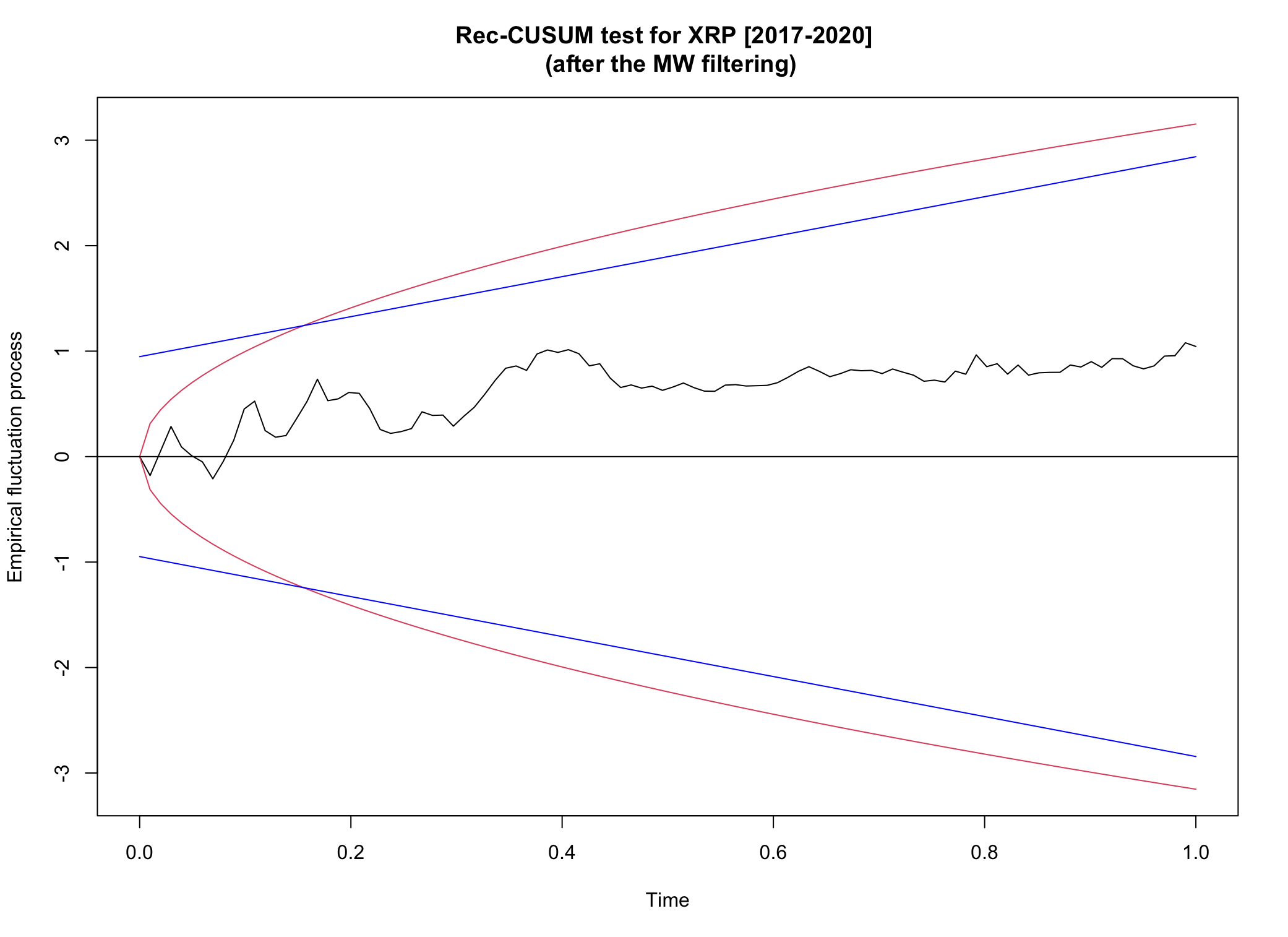}
\end{figure}

\begin{figure}[H]
  \centering
  \caption{The Rec-CUSUM test of XRP [2018-2020] (after the M\"{u}ller-Watson filtering)}
  \label{fig:RE3bq}
  \includegraphics[width=1.0\linewidth]{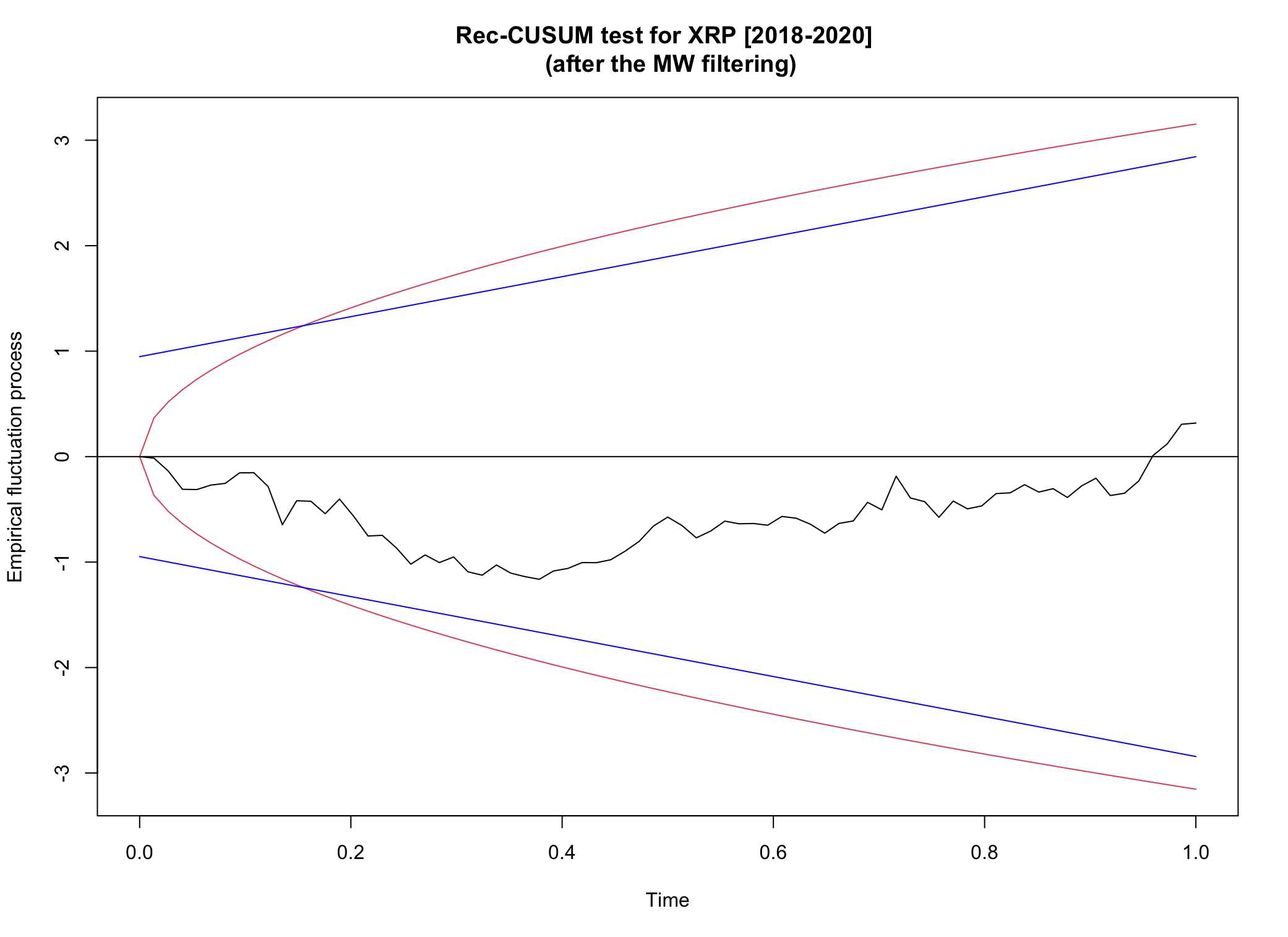}
\end{figure}

\begin{figure}[H]
  \centering
  \caption{The Rec-CUSUM test of XRP [2019-2020] (after the M\"{u}ller-Watson filtering)}
  \label{fig:RE3cq}
  \includegraphics[width=1.0\linewidth]{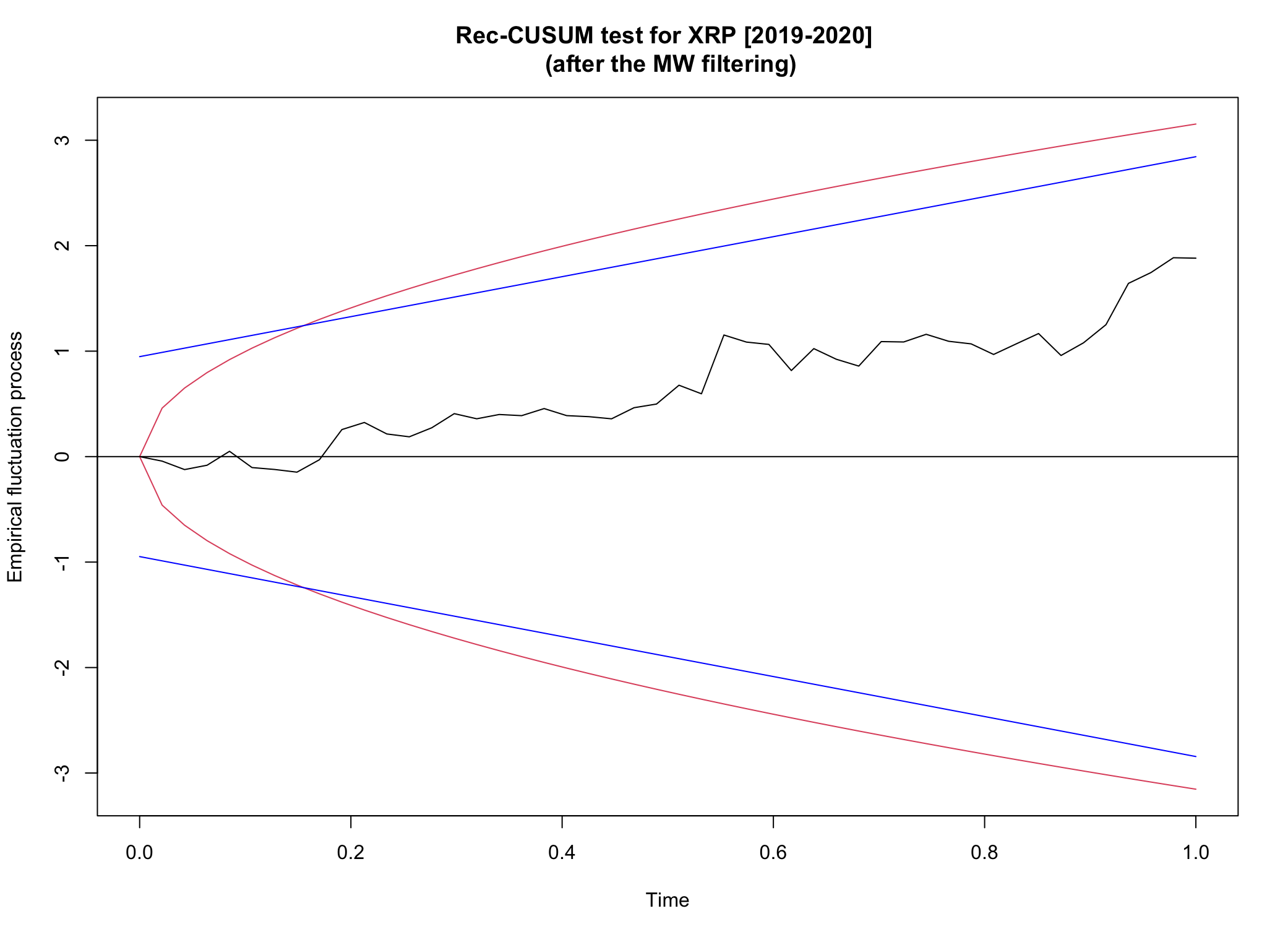}
\end{figure}

\begin{figure}[H]
  \centering
  \caption{The Rec-CUSUM test of JPY [2016-2020] (after the M\"{u}ller-Watson filtering)}
  \label{fig:RE4q}
  \includegraphics[width=1.0\linewidth]{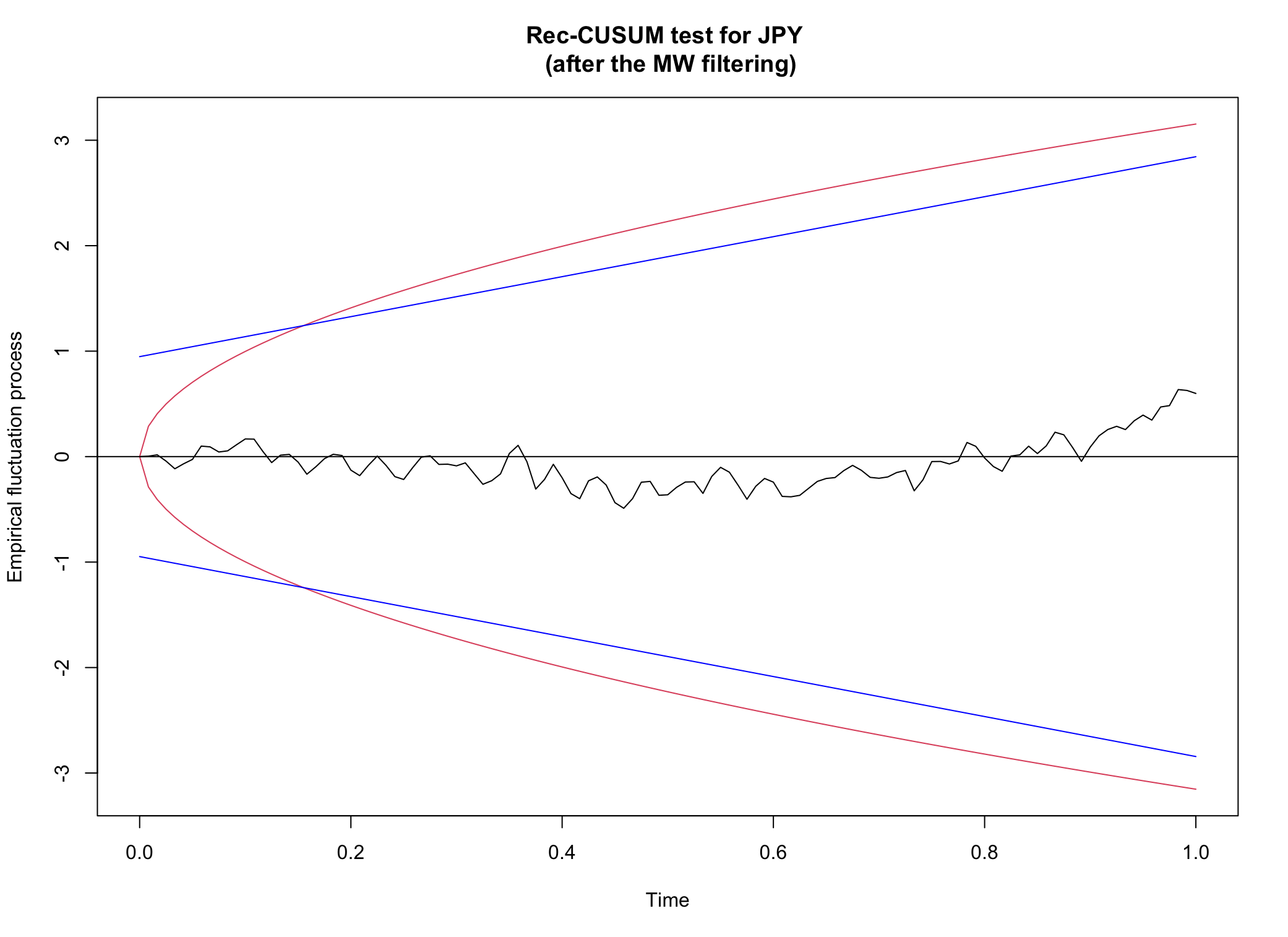}
\end{figure}

\begin{figure}[H]
  \centering
  \caption{The Rec-CUSUM test of JPY [2017-2020] (after the M\"{u}ller-Watson filtering)}
  \label{fig:RE4aq}
  \includegraphics[width=1.0\linewidth]{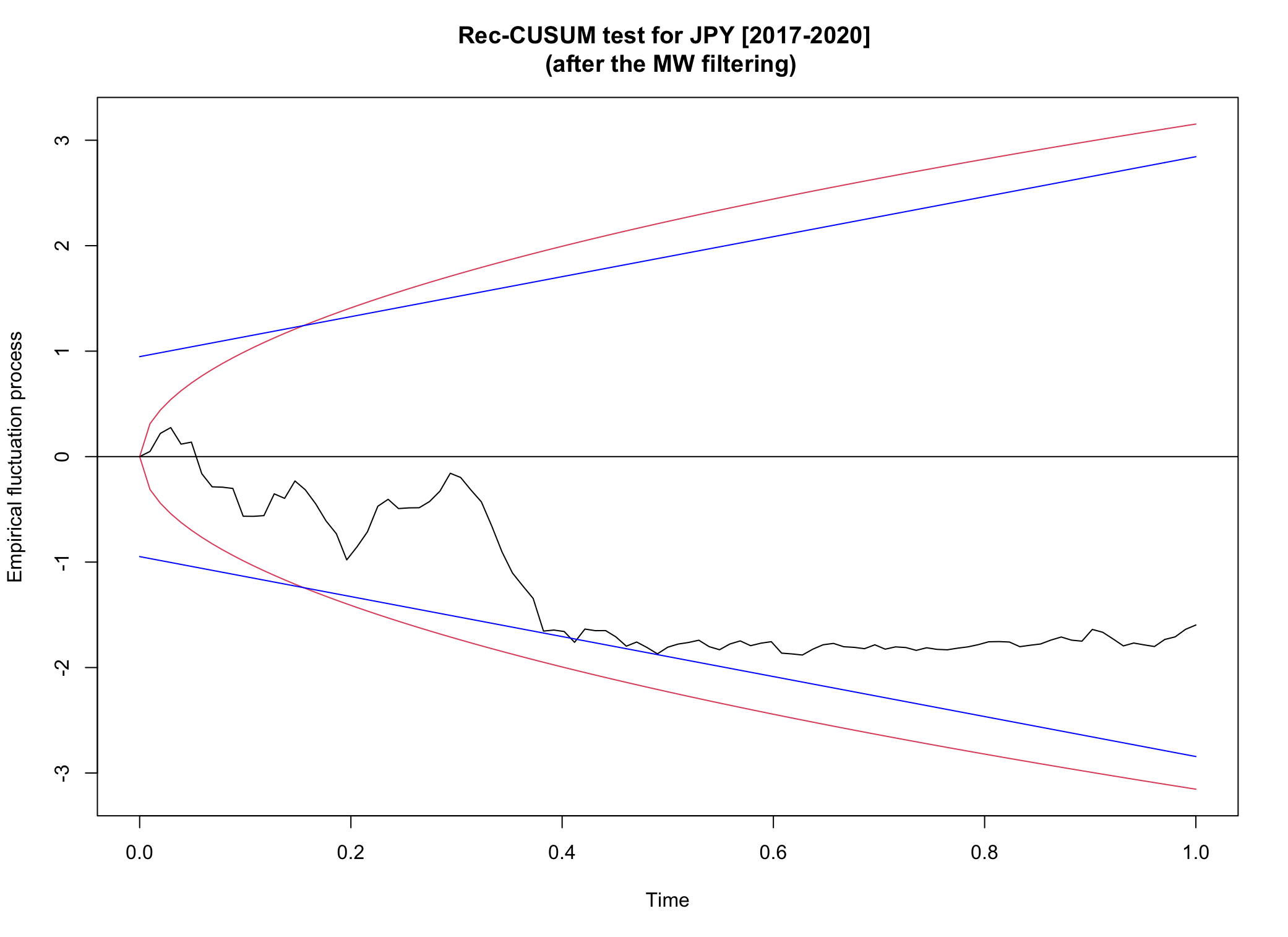}
\end{figure}

\begin{figure}[H]
  \centering
  \caption{The Rec-CUSUM test of JPY [2018-2020] (after the M\"{u}ller-Watson filtering)}
  \label{fig:RE4bq}
  \includegraphics[width=1.0\linewidth]{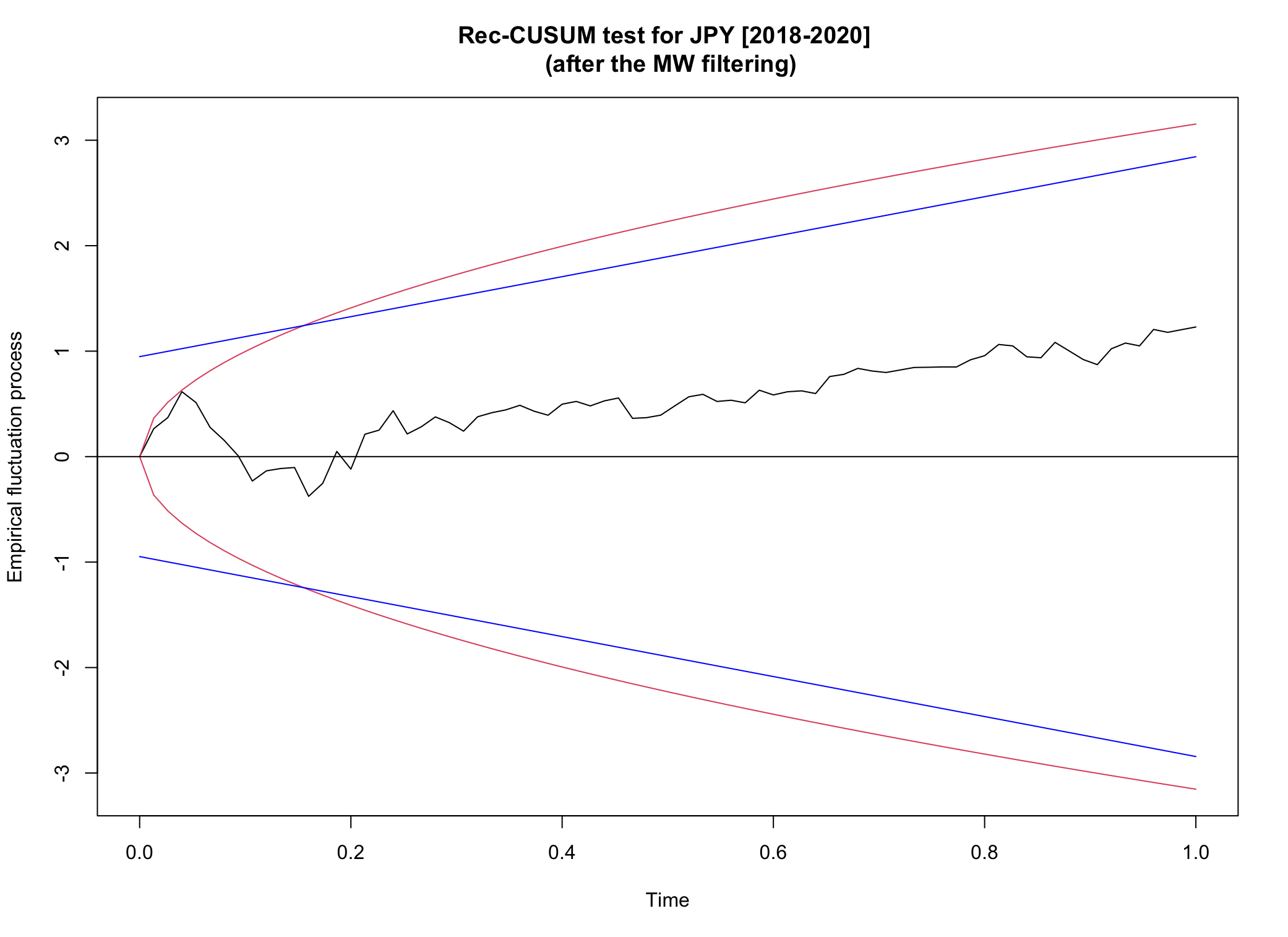}
\end{figure}

\begin{figure}[H]
  \centering
  \caption{The Rec-CUSUM test of JPY [2019-2020] (after the M\"{u}ller-Watson filtering)}
  \label{fig:RE4cq}
  \includegraphics[width=1.0\linewidth]{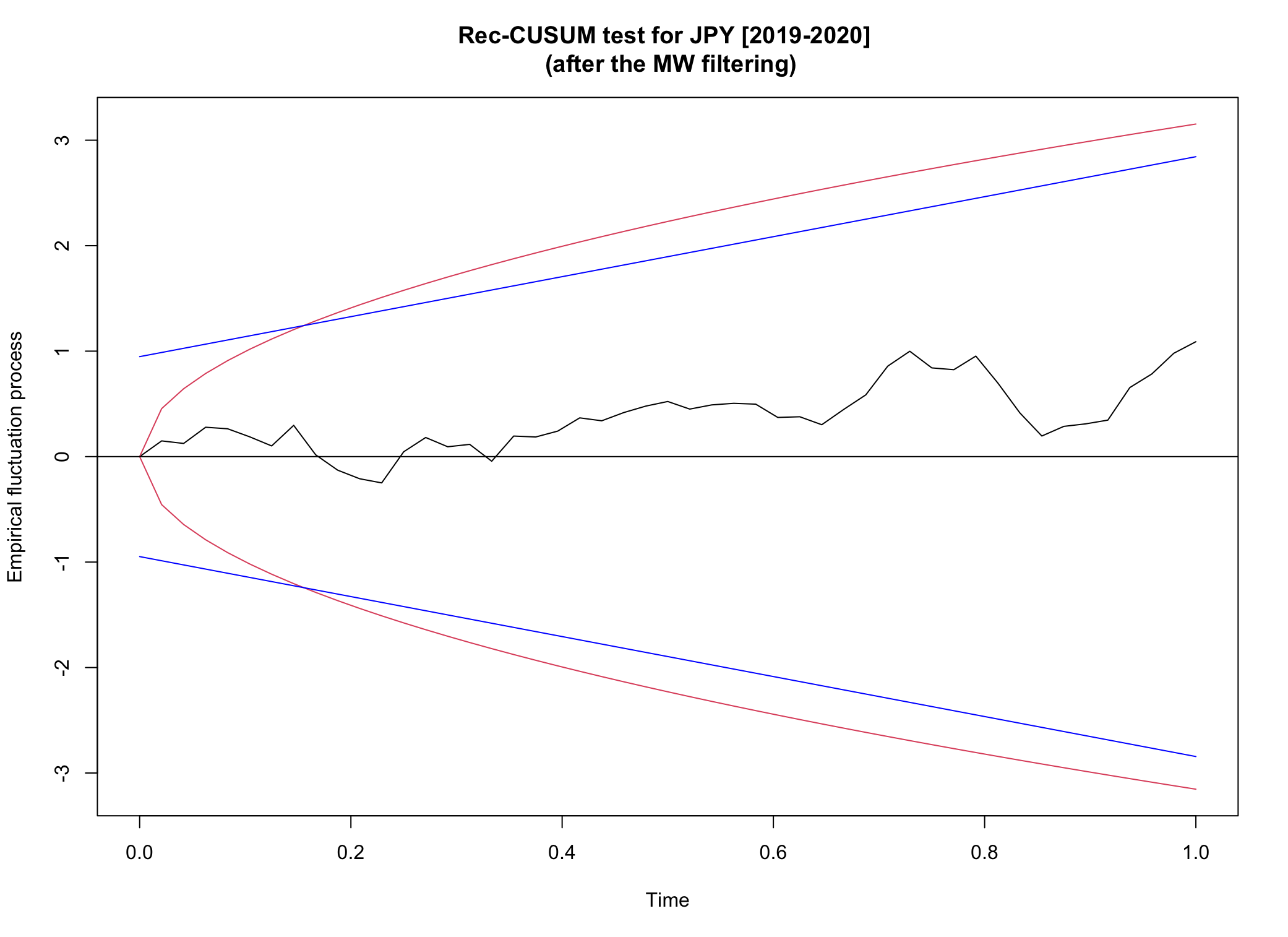}
\end{figure}

\begin{figure}[H]
  \centering
  \caption{The Rec-CUSUM test of EUR [2016-2020] (after the M\"{u}ller-Watson filtering)}
  \label{fig:RE5q}
  \includegraphics[width=1.0\linewidth]{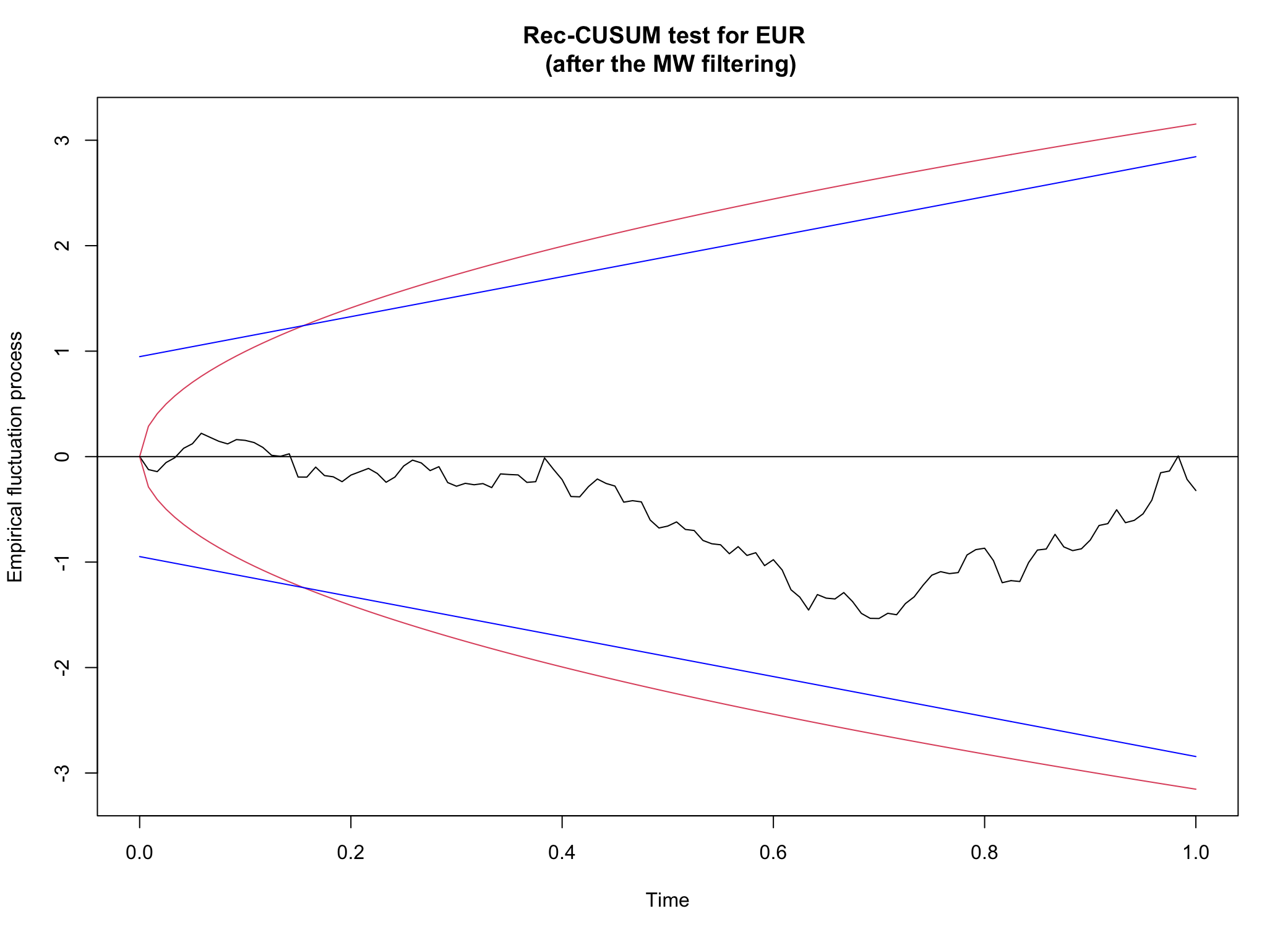}
\end{figure}

\begin{figure}[H]
  \centering
  \caption{The Rec-CUSUM test of EUR [2017-2020] (after the M\"{u}ller-Watson filtering)}
  \label{fig:RE5aq}
  \includegraphics[width=1.0\linewidth]{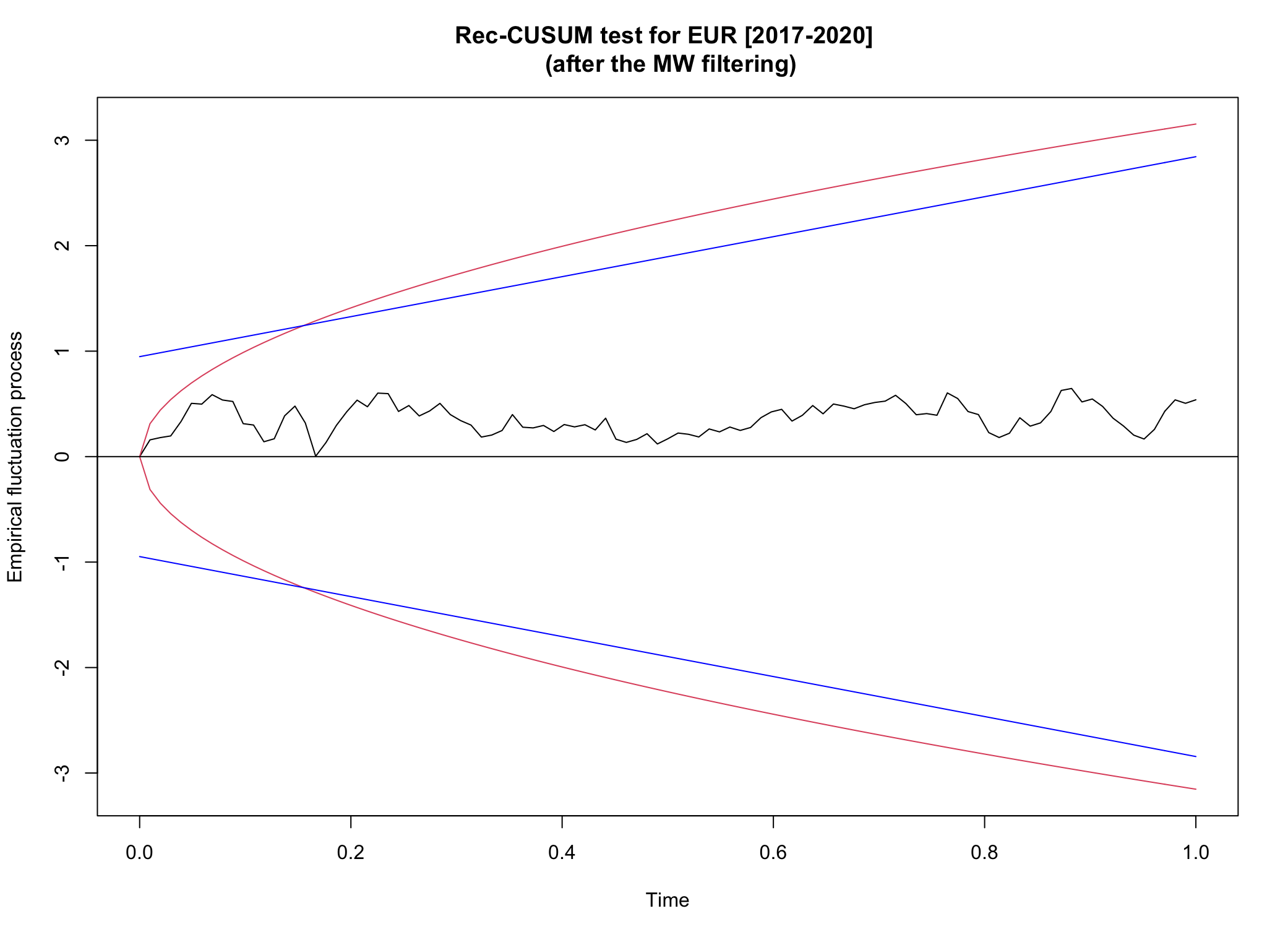}
\end{figure}

\begin{figure}[H]
  \centering
  \caption{The Rec-CUSUM test of EUR [2018-2020] (after the M\"{u}ller-Watson filtering)}
  \label{fig:RE5bq}
  \includegraphics[width=1.0\linewidth]{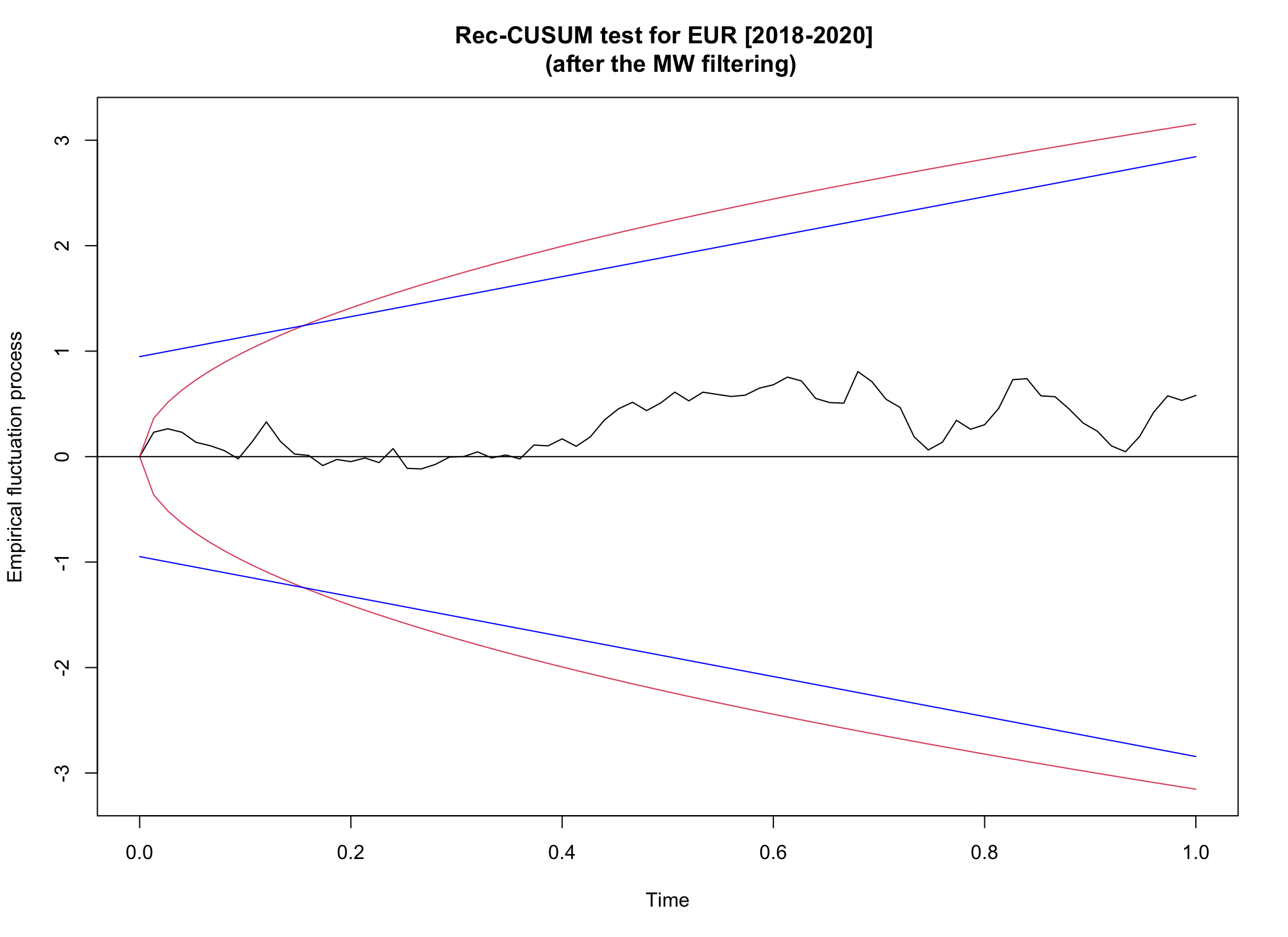}
\end{figure}

\begin{figure}[H]
  \centering
  \caption{The Rec-CUSUM test of EUR [2019-2020] (after the M\"{u}ller-Watson filtering)}
  \label{fig:RE5cq}
  \includegraphics[width=1.0\linewidth]{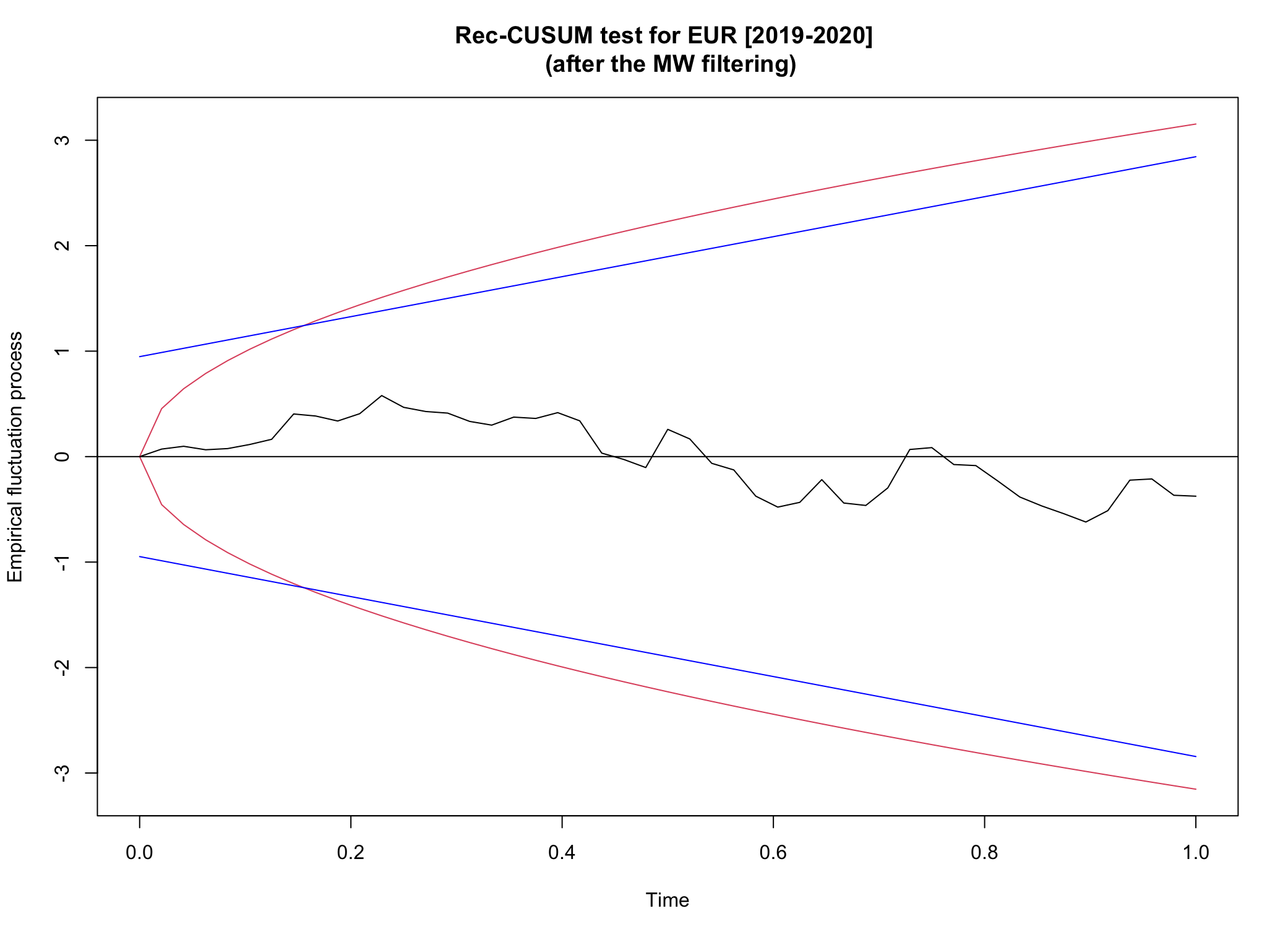}
\end{figure}

\begin{figure}[H]
  \centering
  \caption{The Rec-CUSUM test of GOLD [2016-2020] (after the M\"{u}ller-Watson filtering)}
  \label{fig:RE6q}
  \includegraphics[width=1.0\linewidth]{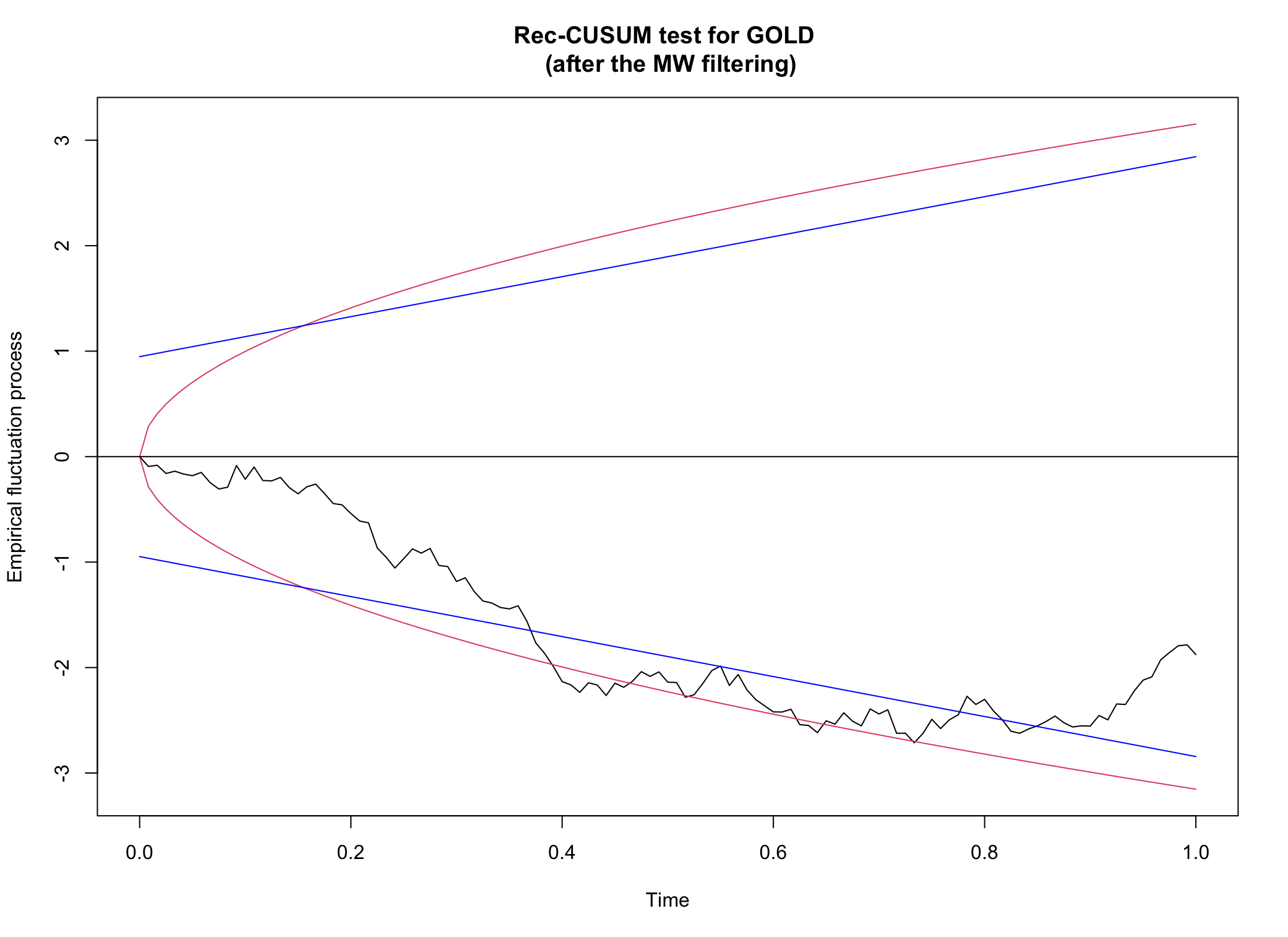}
\end{figure}

\begin{figure}[H]
  \centering
  \caption{The Rec-CUSUM test of GOLD [2017-2020] (after the M\"{u}ller-Watson filtering)}
  \label{fig:RE6aq}
  \includegraphics[width=1.0\linewidth]{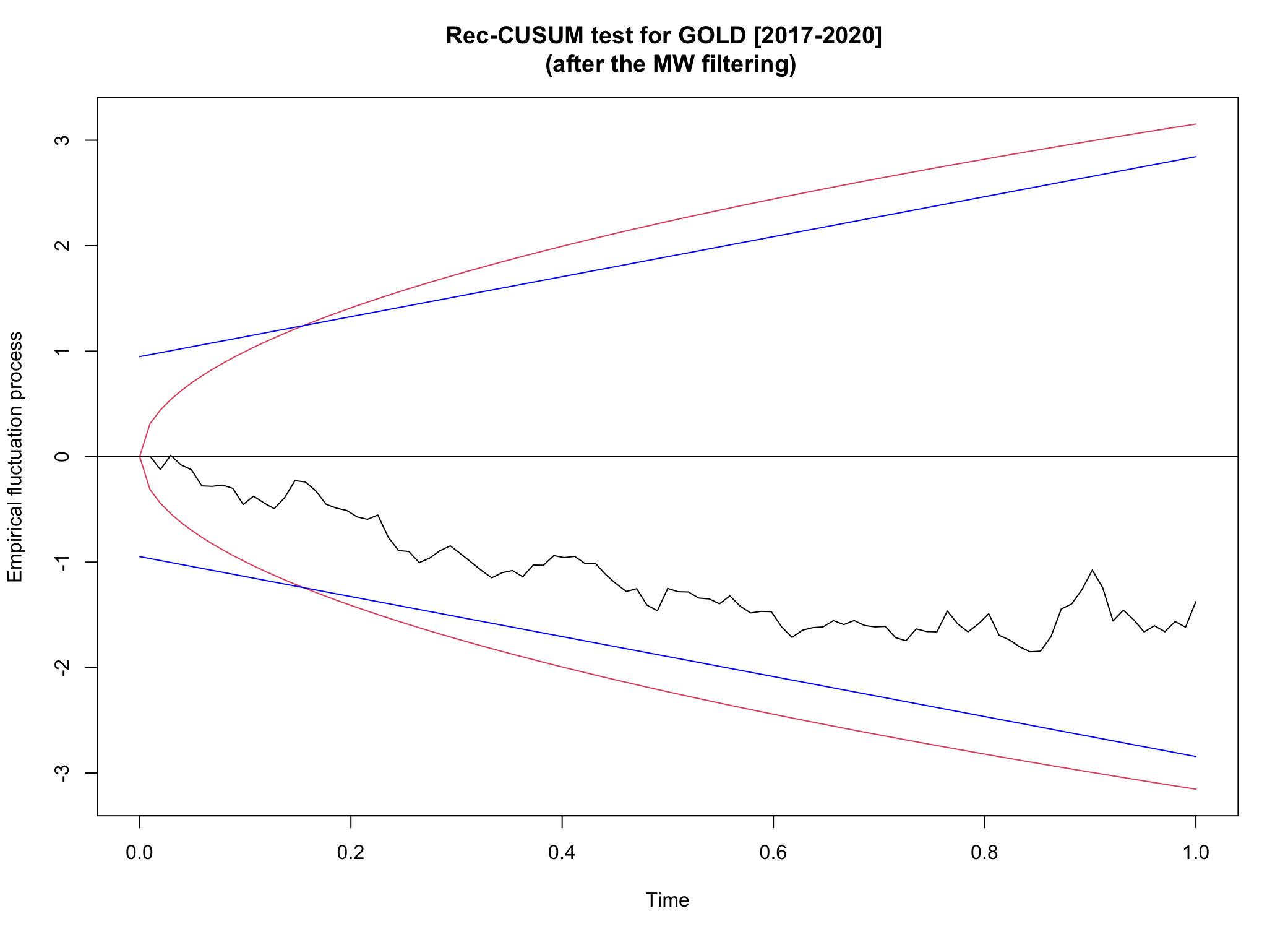}
\end{figure}

\begin{figure}[H]
  \centering
  \caption{The Rec-CUSUM test of GOLD [2018-2020] (after the M\"{u}ller-Watson filtering)}
  \label{fig:RE6bq}
  \includegraphics[width=1.0\linewidth]{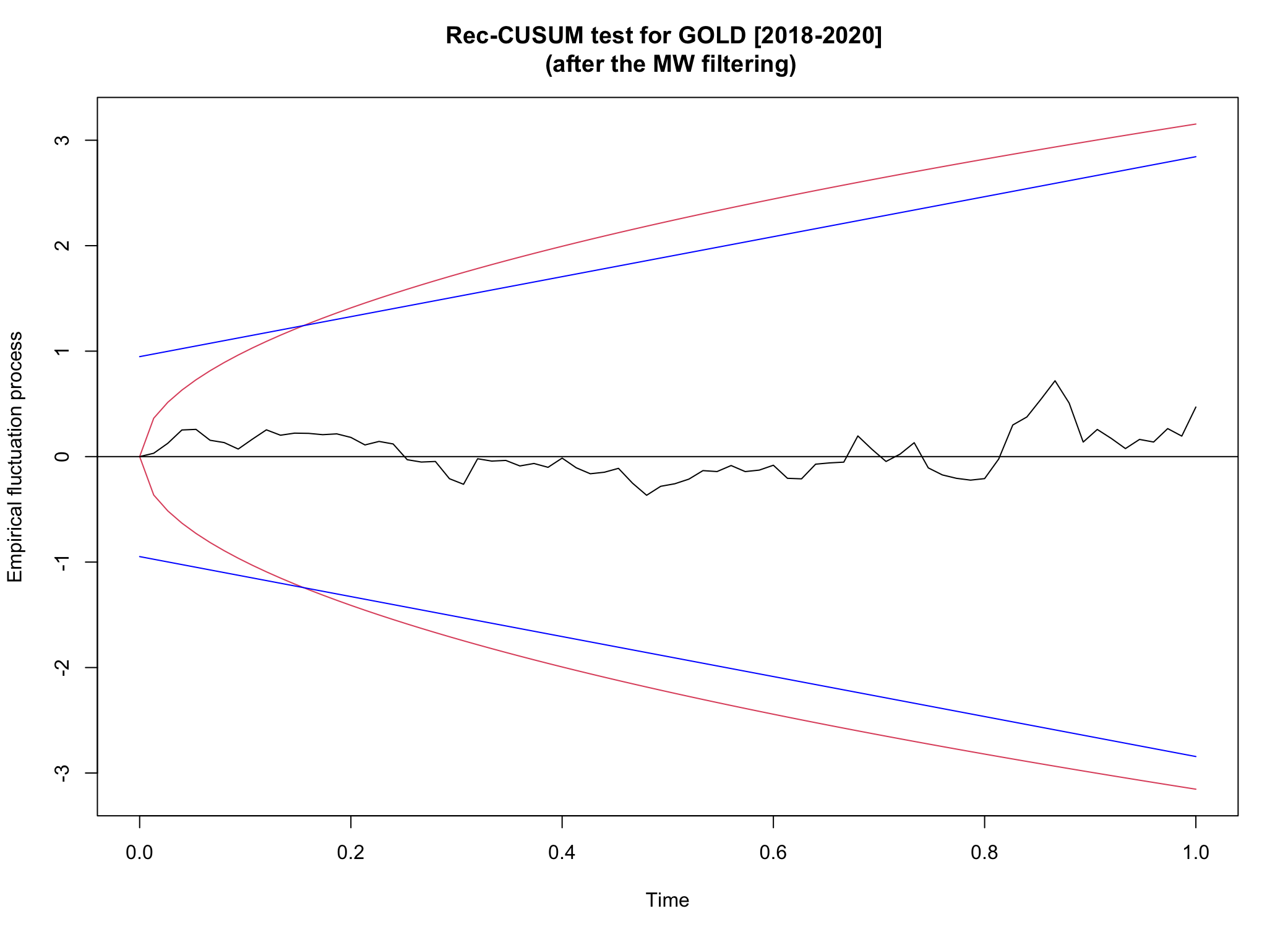}
\end{figure}

\begin{figure}[H]
  \centering
  \caption{The Rec-CUSUM test of GOLD [2019-2020] (after the M\"{u}ller-Watson filtering)}
  \label{fig:RE6cq}
  \includegraphics[width=1.0\linewidth]{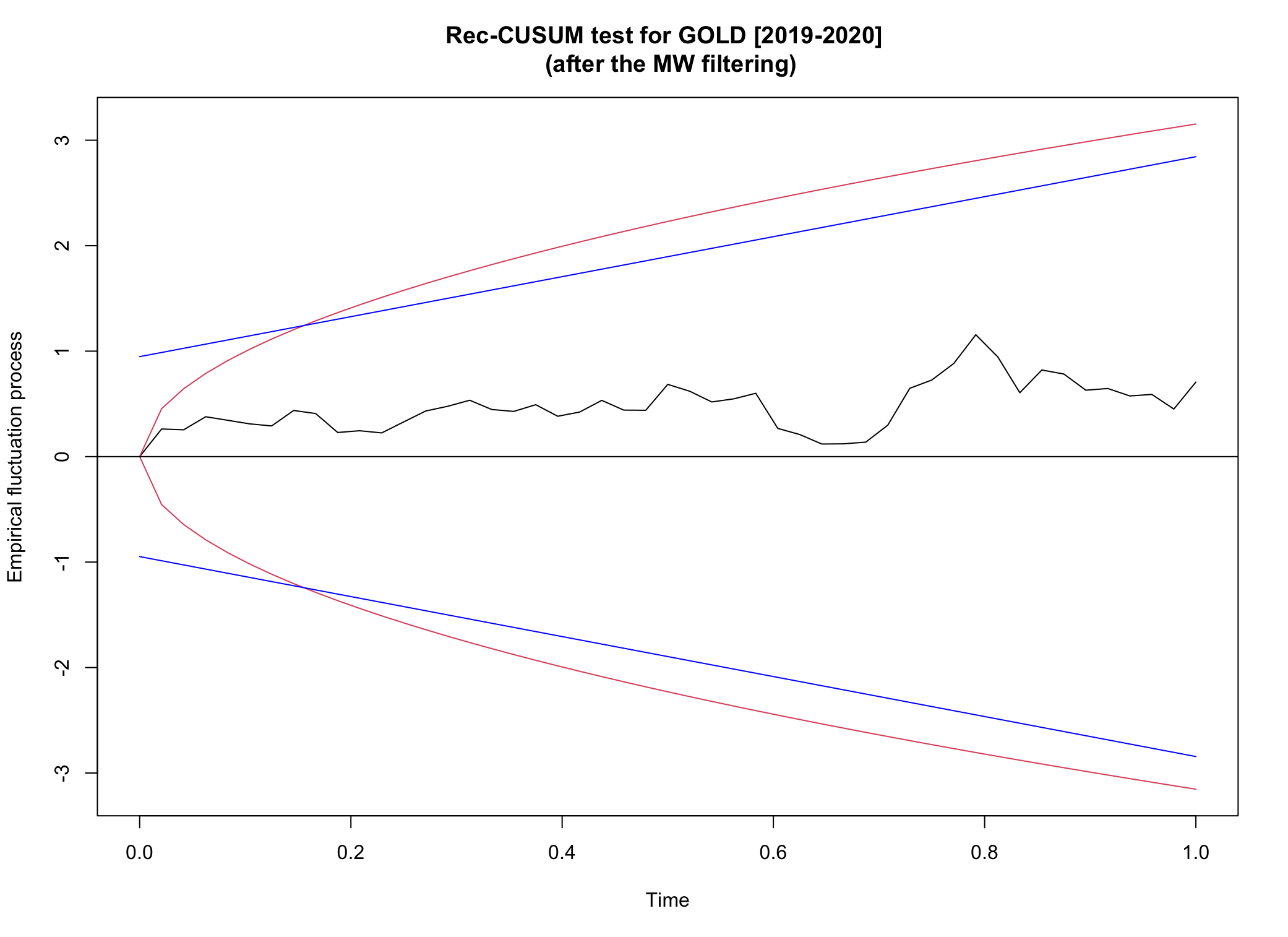}
\end{figure}

\begin{figure}[H]
  \centering
  \caption{The Rec-CUSUM test of S\&P500 [2016-2020] (after the M\"{u}ller-Watson filtering)}
  \label{fig:RE7q}
  \includegraphics[width=1.0\linewidth]{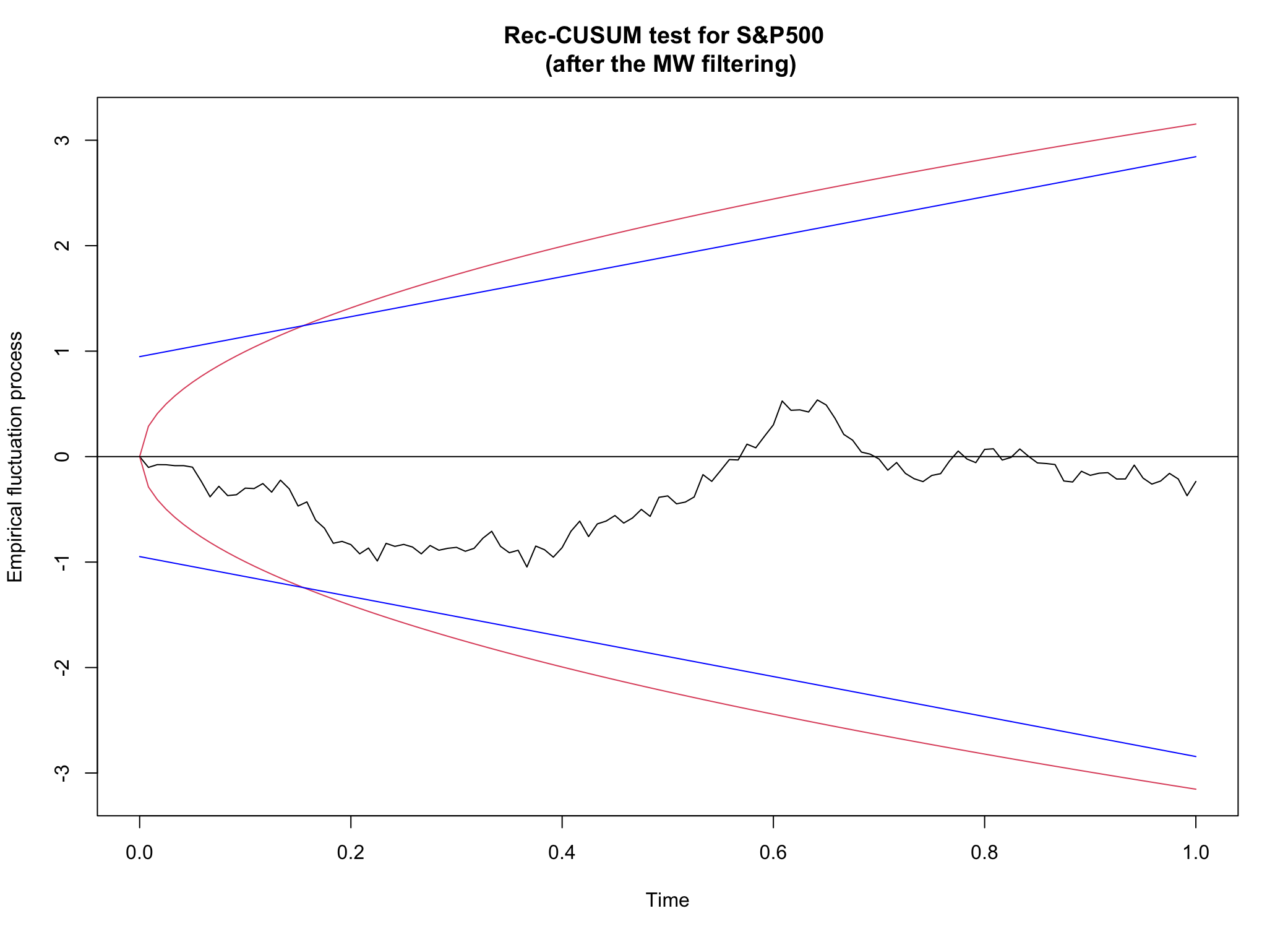}
\end{figure}

\begin{figure}[H]
  \centering
  \caption{The Rec-CUSUM test of S\&P500 [2017-2020] (after the M\"{u}ller-Watson filtering)}
  \label{fig:RE7aq}
  \includegraphics[width=1.0\linewidth]{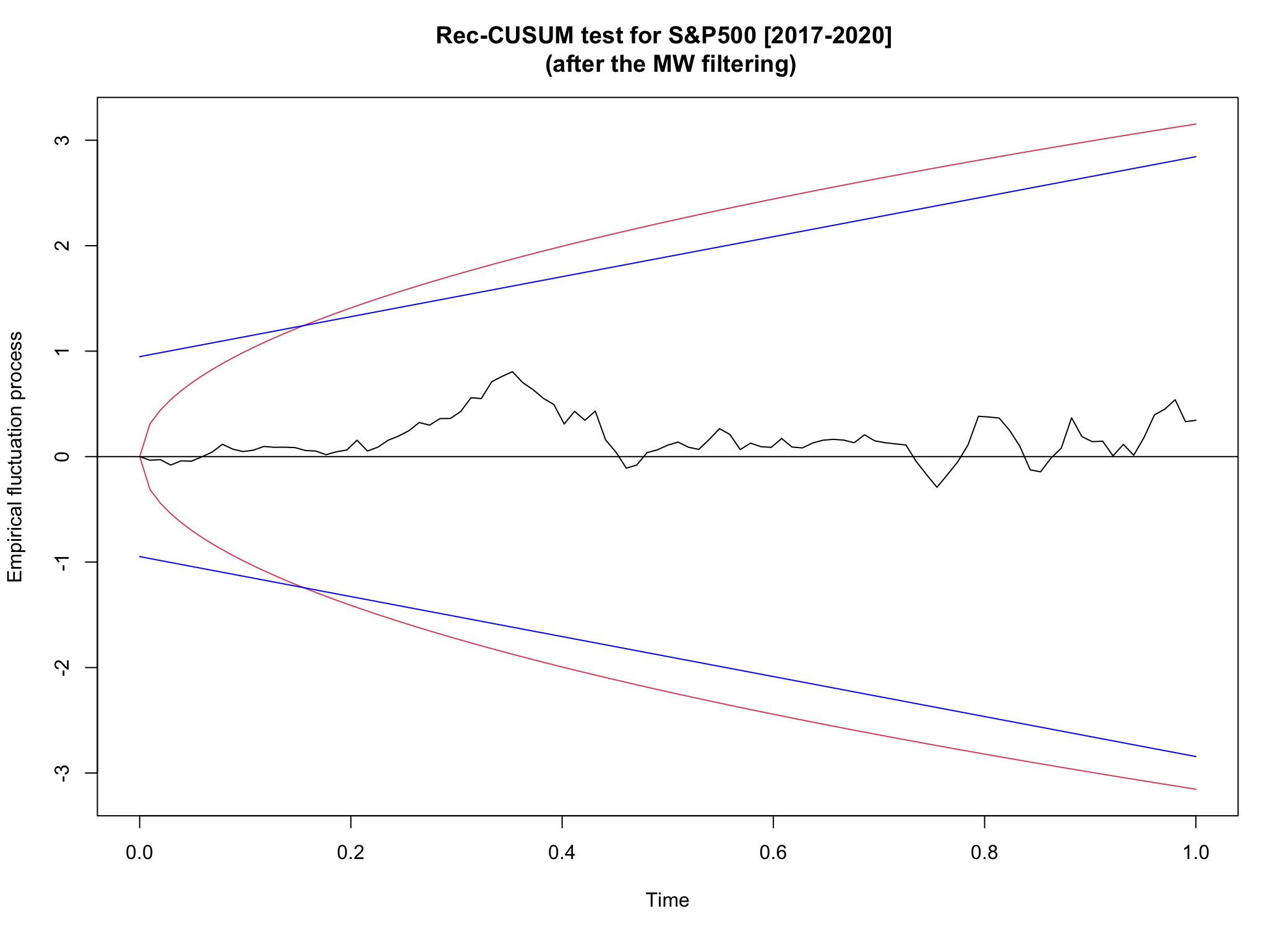}
\end{figure}

\begin{figure}[H]
  \centering
  \caption{The Rec-CUSUM test of S\&P500 [2018-2020] (after the M\"{u}ller-Watson filtering)}
  \label{fig:RE7bq}
  \includegraphics[width=1.0\linewidth]{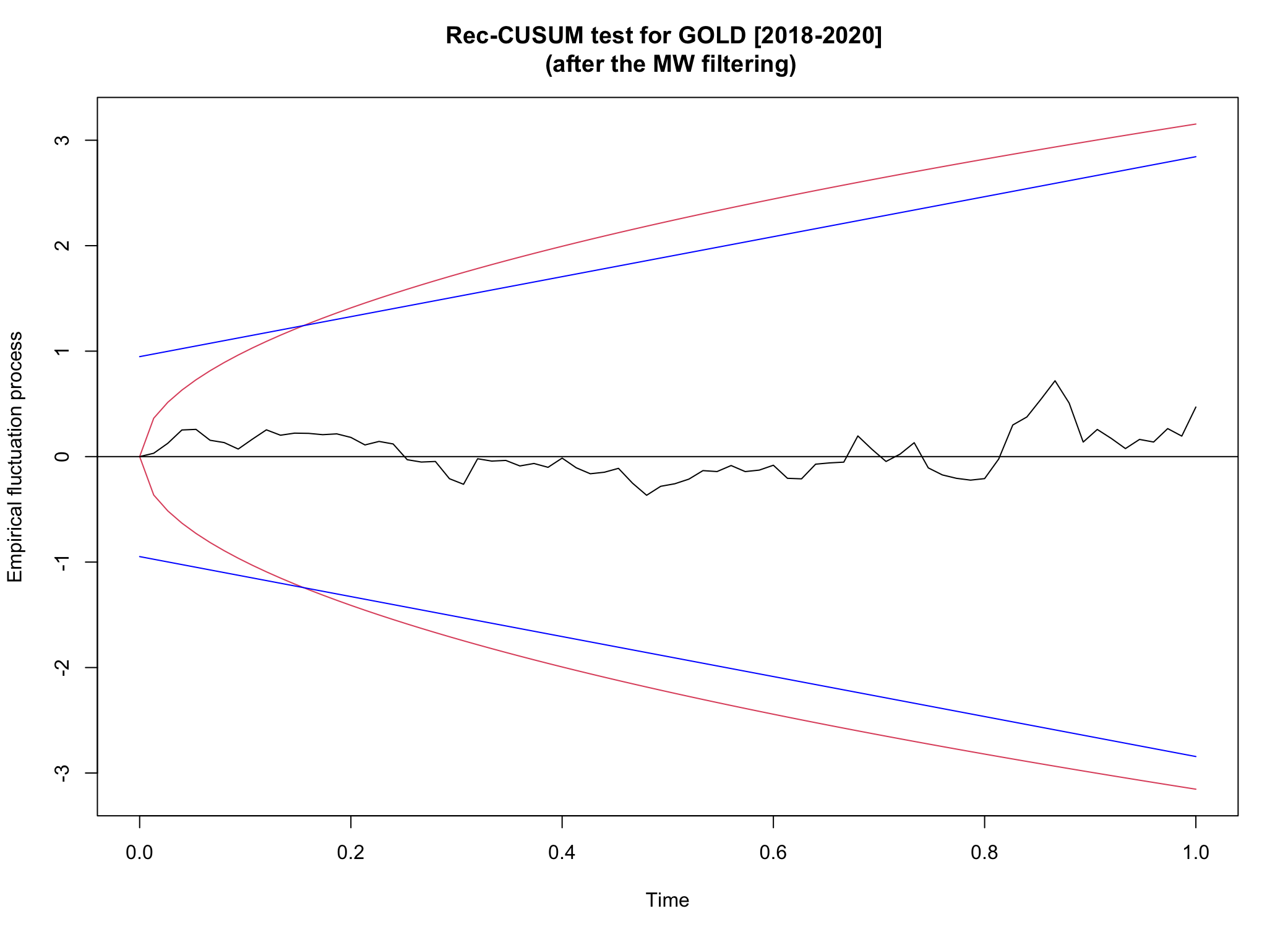}
\end{figure}

\begin{figure}[H]
  \centering
  \caption{The Rec-CUSUM test of S\&P500 [2019-2020] (after the M\"{u}ller-Watson filtering)}
  \label{fig:RE7cq}
  \includegraphics[width=1.0\linewidth]{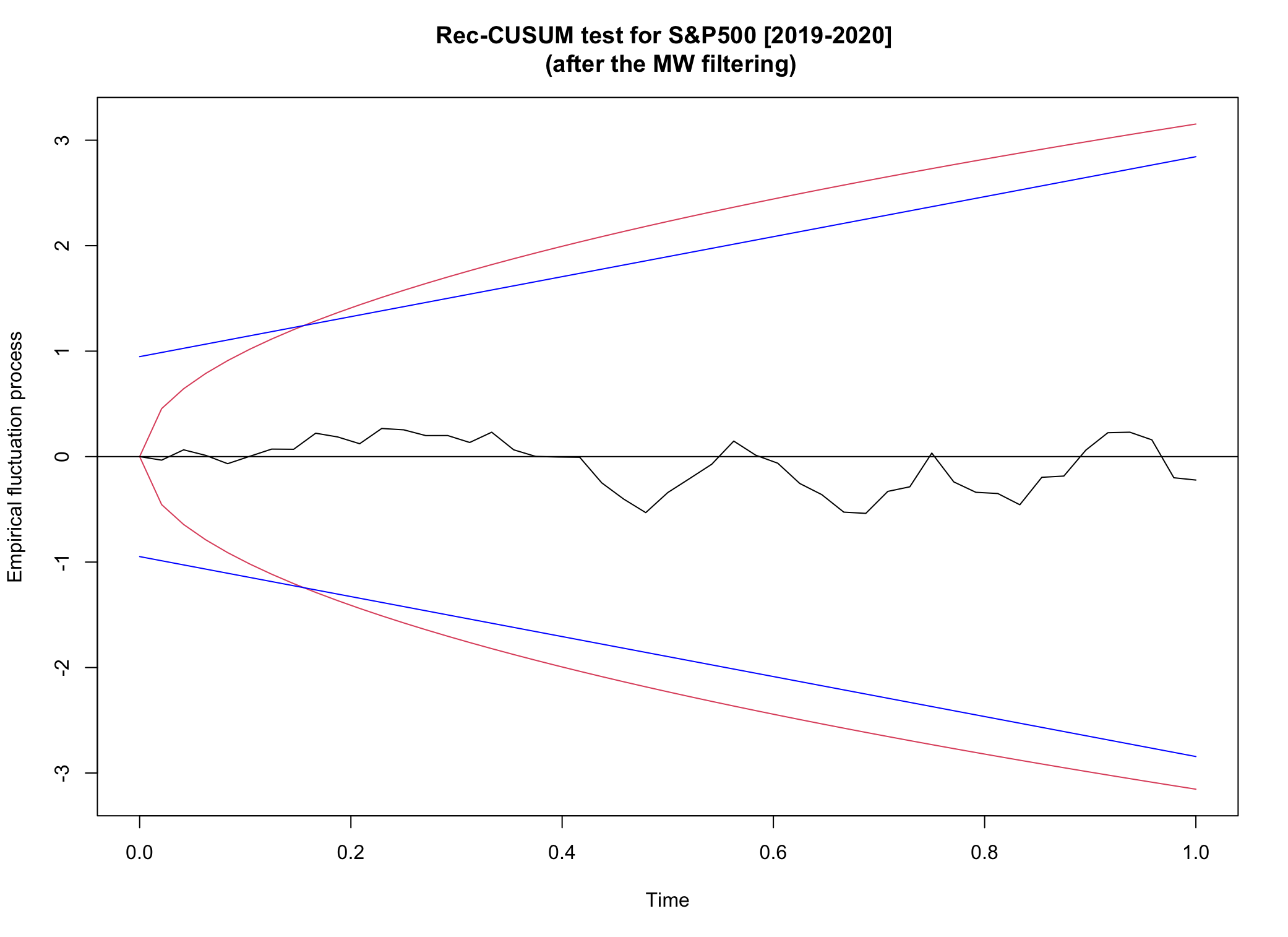}
\end{figure}

\begin{figure}[H]
  \centering
  \caption{The Rec-CUSUM test of MSCI [2016-2020] (after the M\"{u}ller-Watson filtering)}
  \label{fig:RE8q}
  \includegraphics[width=1.0\linewidth]{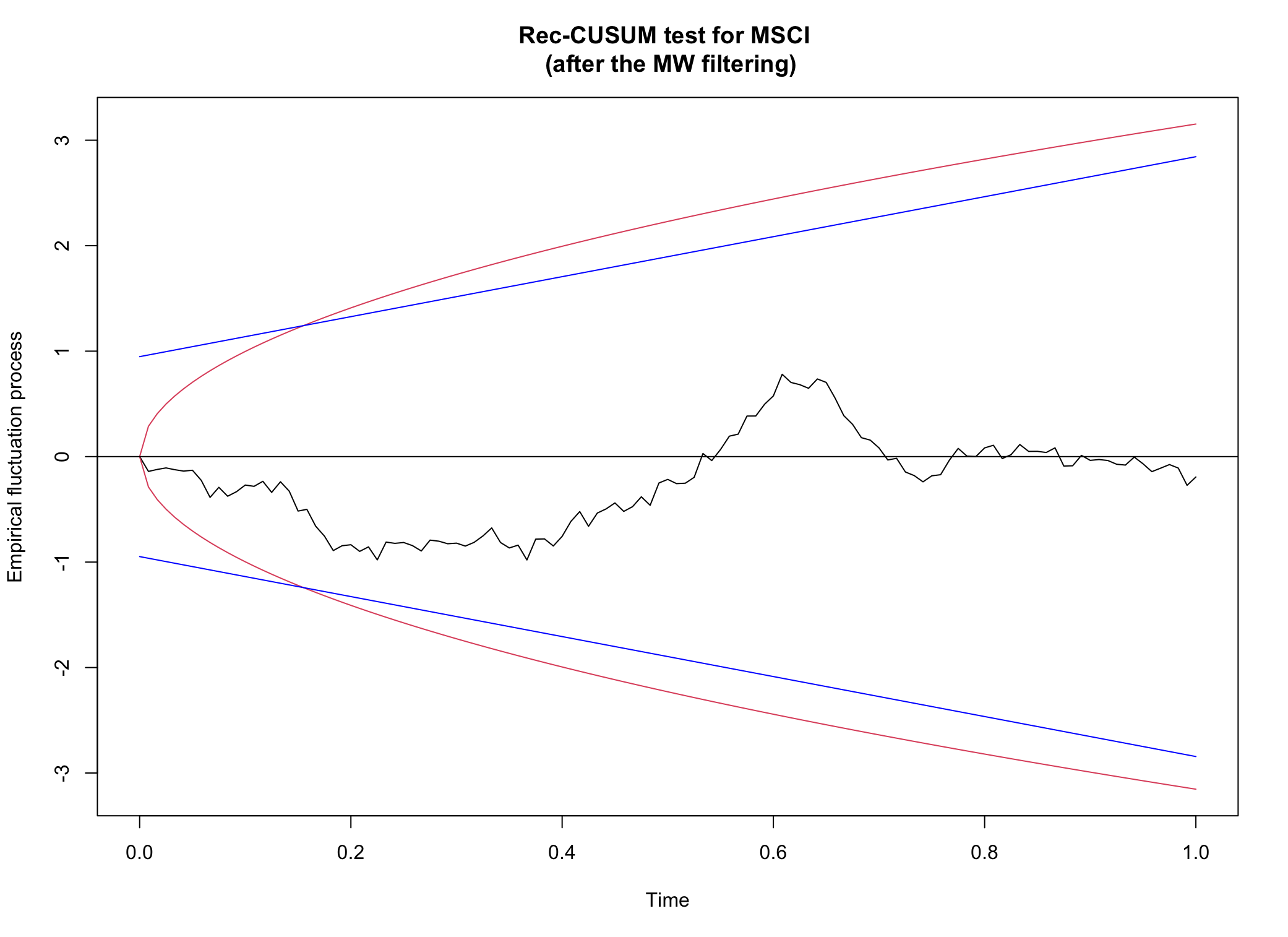}
\end{figure}

\begin{figure}[H]
  \centering
  \caption{The Rec-CUSUM test of MSCI [2017-2020] (after the M\"{u}ller-Watson filtering)}
  \label{fig:RE8aq}
  \includegraphics[width=1.0\linewidth]{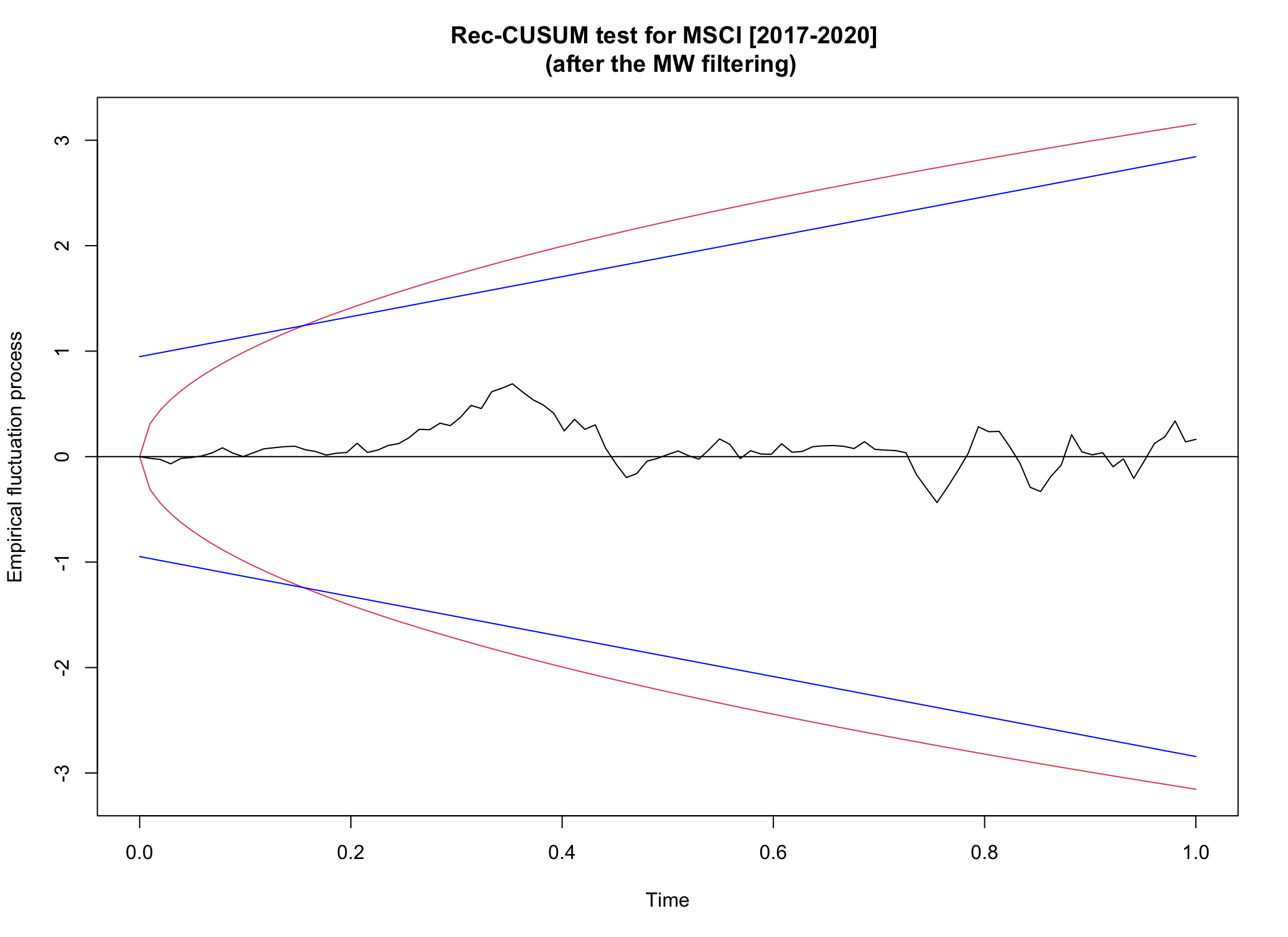}
\end{figure}

\begin{figure}[H]
  \centering
  \caption{The Rec-CUSUM test of MSCI [2018-2020] (after the M\"{u}ller-Watson filtering)}
  \label{fig:RE8bq}
  \includegraphics[width=1.0\linewidth]{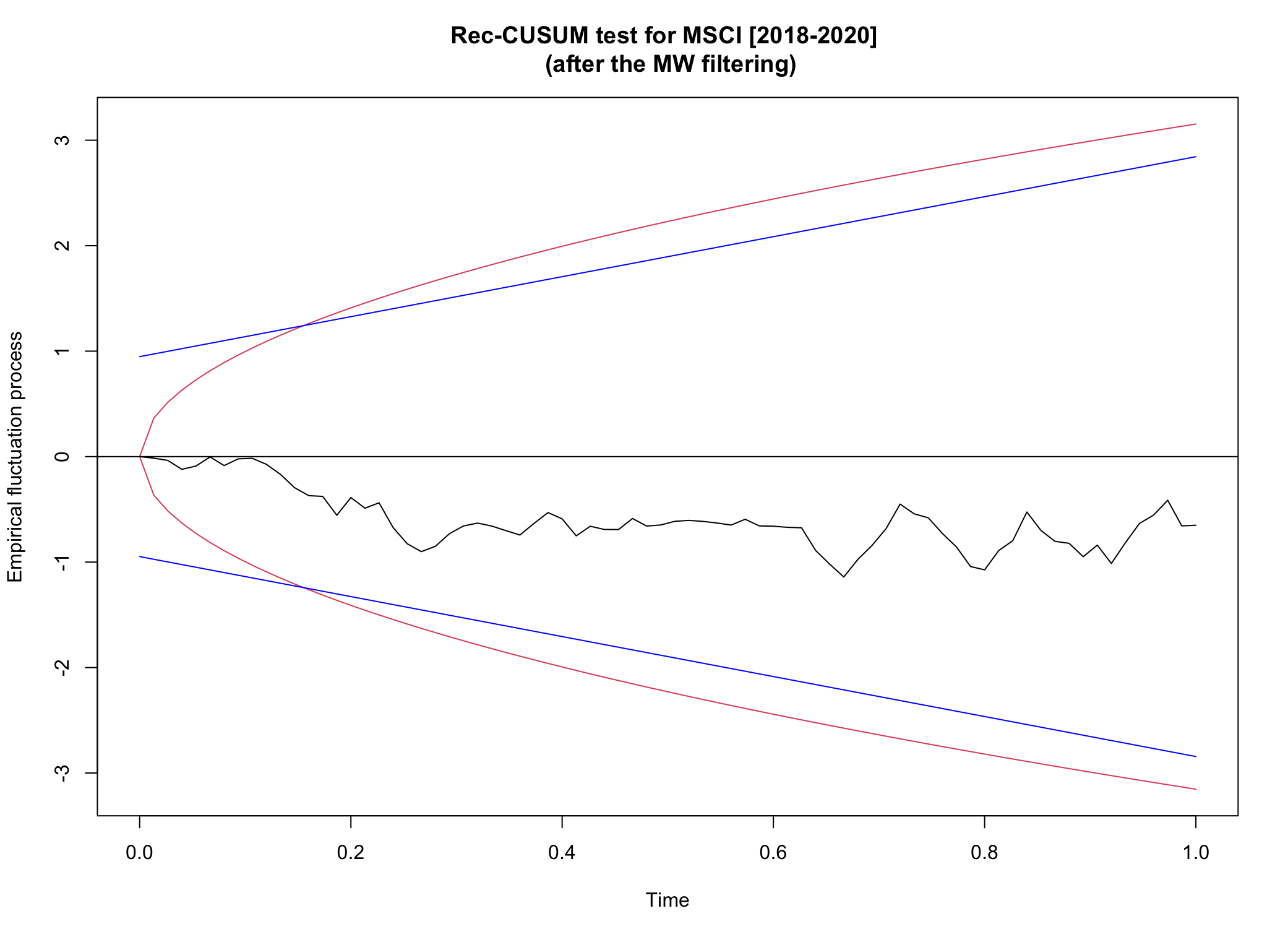}
\end{figure}

\begin{figure}[H]
  \centering
  \caption{The Rec-CUSUM test of MSCI [2019-2020] (after the M\"{u}ller-Watson filtering)}
  \label{fig:RE8cq}
  \includegraphics[width=1.0\linewidth]{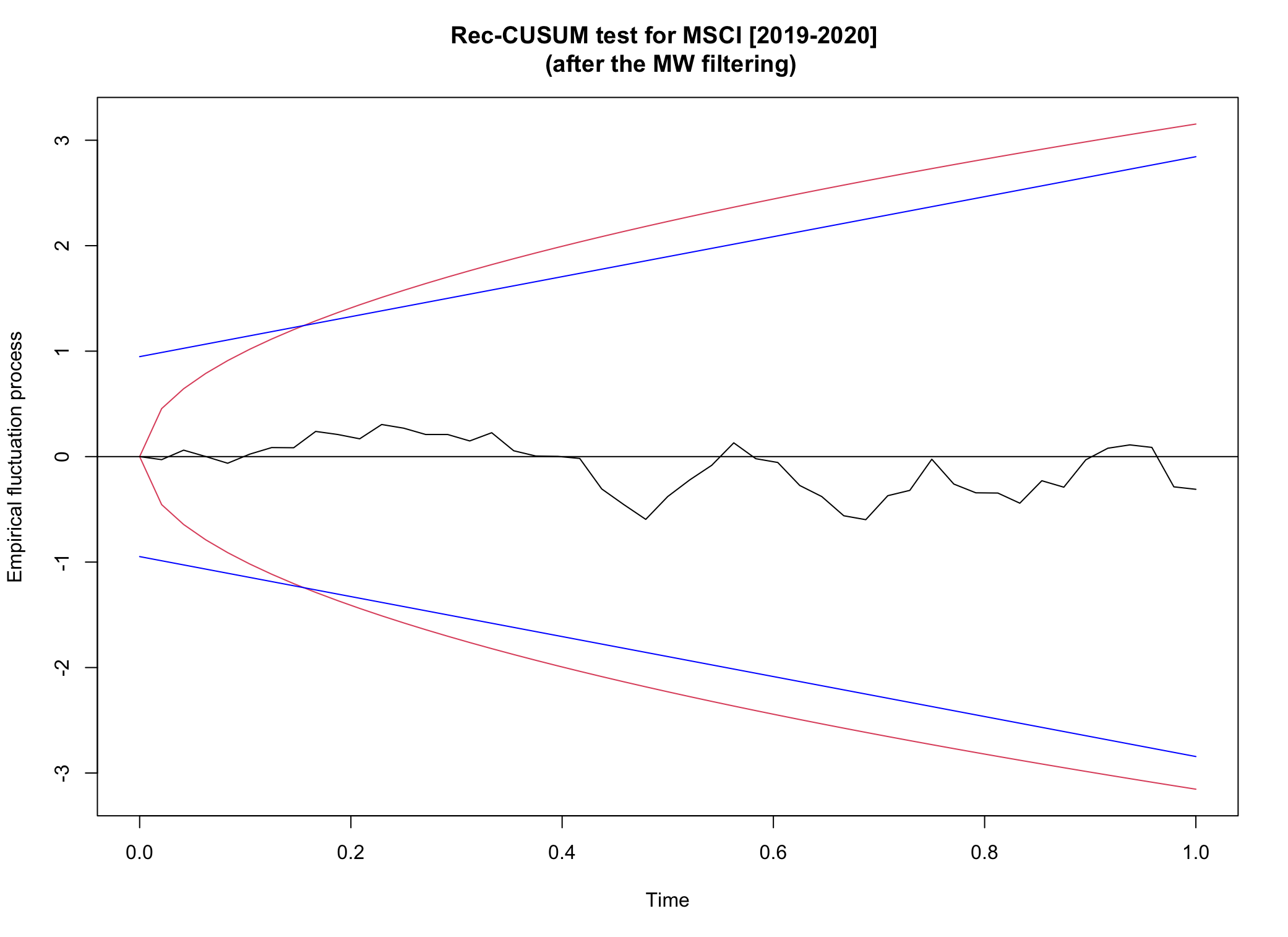}
\end{figure}

However, the results change in the Rec-CUSUM tests after the MW filtering (see Figures \ref{fig:RE1aq}, \ref{fig:RE1bq}, and \ref{fig:RE1cq} for BTC, Figures \ref{fig:RE2aq}, \ref{fig:RE2bq}, and \ref{fig:RE2cq} for ETH, Figures \ref{fig:RE3aq}, \ref{fig:RE3bq}, and \ref{fig:RE3cq} for XRP, Figures \ref{fig:RE4aq}, \ref{fig:RE4bq}, and \ref{fig:RE4cq} for JPY, Figures \ref{fig:RE5aq}, \ref{fig:RE5bq}, and \ref{fig:RE5cq} for EUR, Figures \ref{fig:RE6aq}, \ref{fig:RE6bq}, and \ref{fig:RE6cq} for GOLD, Figures \ref{fig:RE7aq}, \ref{fig:RE7bq}, and \ref{fig:RE7cq} for S\&P500, and Figures \ref{fig:RE8aq}, \ref{fig:RE8bq}, and \ref{fig:RE8cq} for MSCI). Even for the sample data from 2018 to 2020, all the assets, except for ETH, after the MW filtering do not exceed the boundaries of the Brownian motion.

In summary, we confirm that the returns on cryptocurrencies after the MW filtering are essentially stable enough within the boundaries of the Brownian bridge or the Brownian motion, as is the case for the key financial assets.

\section{Conclusion}\label{sec9}
A key question on cryptocurrencies is whether they can be used as money, a medium of exchange. Many argue they cannot. A major reason is that their price movements are too volatile to use as money. 

We have presented positive evidences of price stability of cryptocurrencies. We focus on the daily returns after filtering out the high frequency components, which are contaminated by technical forces. Also, for the daily transaction uses, people do not seem to care about the high frequency movements of money, say, the Euro or Japanese yen against the US dollar. 

Specifically, we apply the filter developed by \citet{MW} to the daily return data of major cryptocurrencies ({\it i.e.}, Bitcoin (BTC), Ethereum (ETH), and Ripple (XRP)) as well as their comparators ({\it i.e.}, major legal tenders, the Euro (EUR) and Japanese yen (JPY), and major stock indexes, the S\&P 500 and the MSCI World Index (MSCI)). By doing so, we essentially get rid of the less-than-one-month cycles of their price movements. 

We then investigate the stability of the filtered daily returns using three different measures. First, we find that the Pearson correlations in the daily returns between cryptocurrencies and other assets increased in latter years in our sample from the beginning of 2016 to the end of 2020. Second, however, based on the DTW method that allow lags and leads, we find that the similarity in the daily returns of cryptocurrencies with other assets has been present even since 2016, the beginning of our sample, and not much changed throughout our sample period. Third, we test the stability of the relationships between daily returns of cryptocurrencies and those of their comparators by checking if the cumulative sum of errors, under the assumption of stable coefficients on comparators' daily returns, does not exceeds the statistical bounds. This CUSUM test is based on the efficient market hypothesis and the results assure the market efficiency and structural price stability of cryptocurrencies, as is also the case for other major financial assets.

In summary, interestingly, for the years from 2016 to 2020, the prices of major cryptocurrencies are found to be stable. This conclusion is not well shown without filtering out the high frequency movements or without conducting deeper investigations than simple correlations. Still, apparently, an empirical question remains if such stability can be continued for the foreseeable future.

\section*{Acknowledgement}
The views expressed in this paper are those of the authors and should not be attributed to the Tokio Marine \& Nichido Fire Insurance Co., Ltd., or any institutions that the authors have been affiliated with. This work is supported by the Digital Economy Project at the University of Tokyo, funded by the Silicon Valley Community Foundation. We would like to thank these organizations. We are grateful for the helpful comments from participants of The University Blockchain Research Initiative (UBRI) Conference, hosted (online) by the University of London, the Digital Currency and Finance Workshop, hosted (online) by the University of Tokyo, and the Blockchain in Kyoto 2021 Conference, hosted (online) by the Kyoto University.

\newpage
\appendix
\section*{Appendix}
\section{CUSUM Test}\label{app}
Let $\{ z_t \}$, for $t=1,\cdots,n$, be the observed time series and we write it as.
\be
 z_t = \mu_t + \epsilon_t \;,
\ee
where $\mu_t = {\bf X}^T_t {\bf \beta}_t$ is a trend component in the linear regression model and $\epsilon_t$ is the {\it i.i.d.} disturbances, which are assumed to be stationary and ergodic with the following condition:
\be
 \mathbb{E}\left[ \epsilon_t \right] = 0 ~ \text{and}~ \mathbb{V}\left[ \epsilon_t \right] = \sigma^2 \;.
\ee
In the standard linear regression model, the coefficients $\beta_t$ is estimated as $\hat{\beta}_t$ by using the ordinary least squares (OLS) method. Then, the OLS residual $\hat{\epsilon}_t$, can be also estimated with
\be
 \mathbb{E}[ \hat{\epsilon}_t ] = 0~\text{and}~ \mathbb{V}[ \hat{\epsilon}_t ] = \hat{\sigma}^2 \;.
\ee
The key is to evaluate the fluctuations of the cumulative sum (CUSUM) of residuals. The null hypothesis is that $\hat{\sigma}^2$ is not explosive. See discussion by \citet{Brown}. 

It is useful to define a continuous time stochastic process (sometimes known as the empirical fluctuation process) of the sum of residuals as
\be
 W_n(\tau) 
= \frac{1}{\hat{\sigma} \sqrt{n}} \sum_{i=1}^{n} {\bf 1}_{\{ \tau \leq t \}} \hat{\epsilon}_i \;.
\ee

In the recursive CUSUM (Rec-CUSUM) test, the residual $\hat{\epsilon}_t$ is essentially the one-step-ahead forecast error and given by
\be
 \hat{\epsilon}_t = 
 \frac{z_t - {\bf X}^T_t \hat{\beta}_{t-1}}{\sqrt{1 + {\bf X}^T_t \left(\sum_{i=1}^{t-1} {\bf X}_i {\bf X}^T_i \right)^{-1}{\bf X}_t}} \;.
\ee
Here, $\hat{\beta}_{t-1}$ is estimated by the OLS using the data up to the previous period $t-1$. In this case, the limiting process becomes the Brownian motion ({\it i.e.}, Wiener process).

In fact, according to the functional central limit theorem or the Donsker's theorem, $W_n(\tau)$ converges to the Wiener process $W(\tau)$:
\be
 W(\tau) = \lim_{n \to \infty} \| W_n(\tau) \| \;, ~\text{for}~0 \leq \tau \leq 1 \;,
\ee
where the convergence means the weak convergence of the associated probability measures. The Wiener process ({\it i.e.}, the Brownian motion) has the following properties:
\be
\mathbb{E} \left[ W(\tau) \right] = 0 ~\text{and}~ \mathbb{V} \left[ W(\tau) \right] = \tau \;.
\ee
Based on this property, the Rec-CUSUM test bounds are obtained.

In the case of the OLS-CUSUM tests, $\hat{\beta}$ based on all samples is used instead of $\hat{\beta}_{t-1}$. As the OLS residuals are correlated to each other and their sum is zero by definition, the limiting process of the OLS-CUSUM is no longer a Brownian motion. Instead, the limiting process is known to become a Brownian bridge:
\be
 B(\tau) = W(\tau) - \tau \, W(1) \;, ~\text{for}~0 \leq \tau \leq 1 \;.
\ee
An alternative representation of the Brownian bridge is given by the stochastic differential equation,
\be
 dB(\tau) = dW(\tau) - \frac{B(\tau)}{1 - \tau} d\tau \;, ~\text{for}~0 \leq \tau \leq 1 \;,
\ee
whose solution is given by
\be
 B(\tau) = \int_{0}^{\tau} \frac{1-\tau}{1-t} \; dW(t) \;, ~\text{for}~0 \leq \tau \leq 1 \;.
\ee
The Brownian bridge has the following properties:
\be
\mathbb{E} \left[ B(\tau) \right] = 0 ~\text{and}~ \mathbb{V} \left[ B(\tau) \right] = \tau (1-\tau) \;.
\ee
Here, the OLS-CUSUM test uses the following quantity:
\be
 M_n(\tau) = \max_{0 \leq s \leq \tau} \|  W_n(s) \| \;.
\ee
The distribution $M_n(\tau)$ converges to the Kolmogorov distribution (or the Markov process) as follows:
\be
 M(\tau) = \lim_{n \to \infty} \| M_n(\tau) \| \;, 
\ee
then, it satisfies
\be
 M(\tau) = \sup_{0 \leq \tau \leq 1} \|  B(\tau) \| \;. 
\ee
In general, the cumulative distribution function of the Kolmogorov distribution $M(\tau)$ is given by
\be
 \mathbb{P} \left( M(\tau) \leq x \right) = 1 - 2 \sum_{k=1}^{\infty} (-1)^{k-1} \exp(-2 k^2 x^2) \;. 
\ee

The goodness-of-fit test or the Kolmogorov–Smirnov test can be constructed by using the critical values of the Kolmogorov distribution, 
\be
 \mathbb{P} \left( \sup_{0 \leq \tau \leq 1} \|  B(\tau) \| \leq b(\tau) \right) = 1 - \alpha \;, 
\ee
where $1-\alpha$ represents the confidence interval.

\newpage
\thispagestyle{empty}

\begin{table}[H]
  \centering
  \caption{Descriptive statistics of daily returns (annualized, \%) for the whole sample period}
  \begin{tabular}{ ccccccccccc }
    \toprule
    Statistics        & BTC     & ETH     & XRP    & JPY    & EUR   & GOLD  & S\&P500 & MSCI \\
    \midrule
    ${\bf N}$         & 1258    & 1258    & 1258    & 1249    & 1249   & 1249   & 1249    & 1258 \\
    ${\bf Mean~(\%)}$ & 0.446   & 0.804   & 0.648   & 0.022   & 0.012  & 0.049  & 0.057   & 0.045 \\
    ${\bf SD~(\%)}$   & 4.695   & 7.567   & 9.159   & 1.400   & 0.647  & 0.872  & 1.211   & 1.023 \\
    ${\bf Min~(\%)}$  & -38.118 & -43.420 & -43.069 & -12.702 & -4.160 & -5.723 & -11.984 & -9.915 \\
    ${\bf Max~(\%)}$  & 25.561  & 65.995  & 109.760 & 15.346  & 6.071  & 4.805  & 9.383   & 8.770 \\
    ${\bf Skewness}$  & -0.059  & 1.281   & 3.750   & 0.823   & 0.278  & -0.200 & -0.731  & -1.234 \\
    ${\bf Kurtosis}$  & 7.096   & 9.353   & 33.954  & 36.265  & 10.922 & 4.652  & 20.552  & 22.681 \\
    \bottomrule
  \end{tabular}
  \label{tab:Table1}
\end{table}

\begin{table}[H]
  \centering
  \caption{Correlation matrix in 2016 (before the M\"{u}ller-Watson filtering)}
  \includegraphics[width=1.0\linewidth]{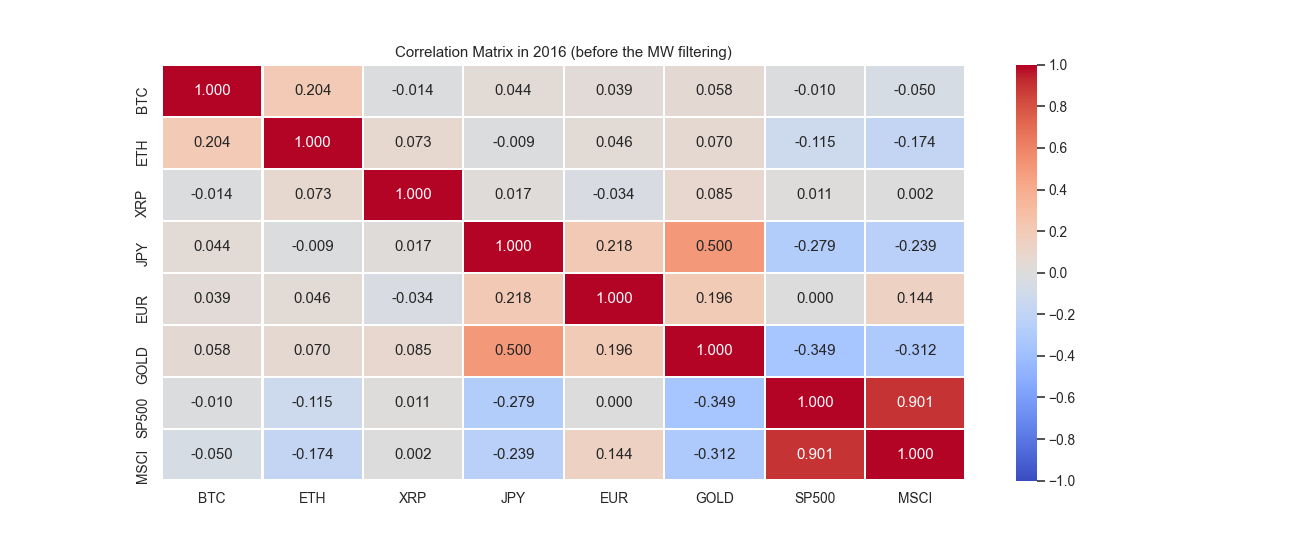}
  \label{fig:Fig1b}
\end{table}

\begin{table}[H]
  \centering
  \caption{Correlation matrix in 2017 (before the M\"{u}ller-Watson filtering)}
  \includegraphics[width=1.0\linewidth]{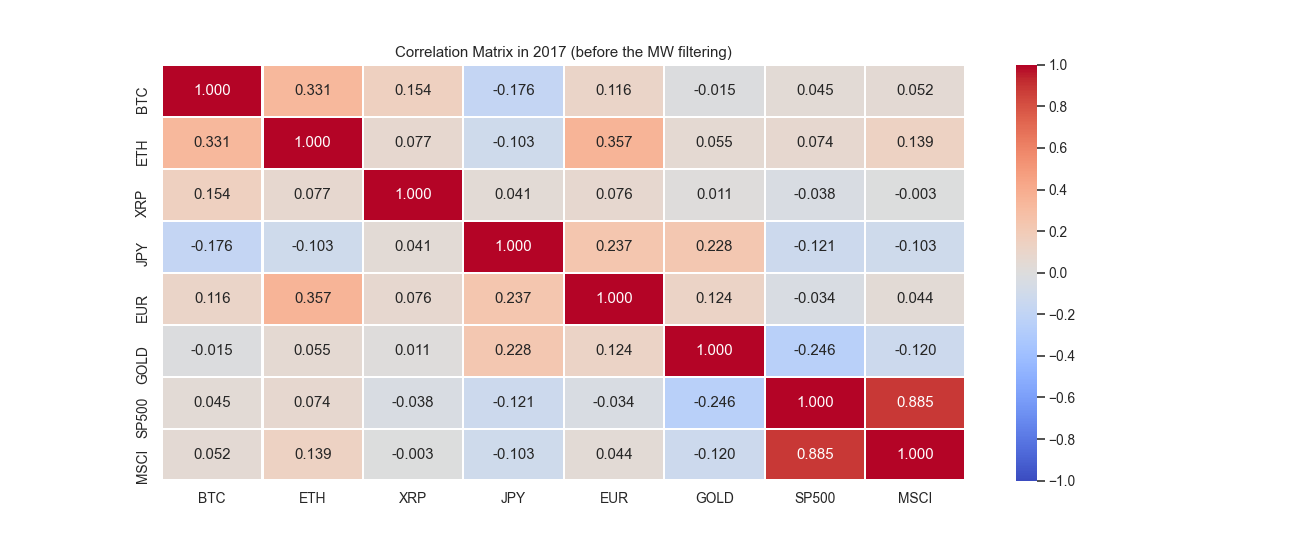}
  \label{fig:Fig2b}
\end{table}

\begin{table}[H]
  \centering
  \caption{Correlation matrix in 2018 (before the M\"{u}ller-Watson filtering)}
  \includegraphics[width=1.0\linewidth]{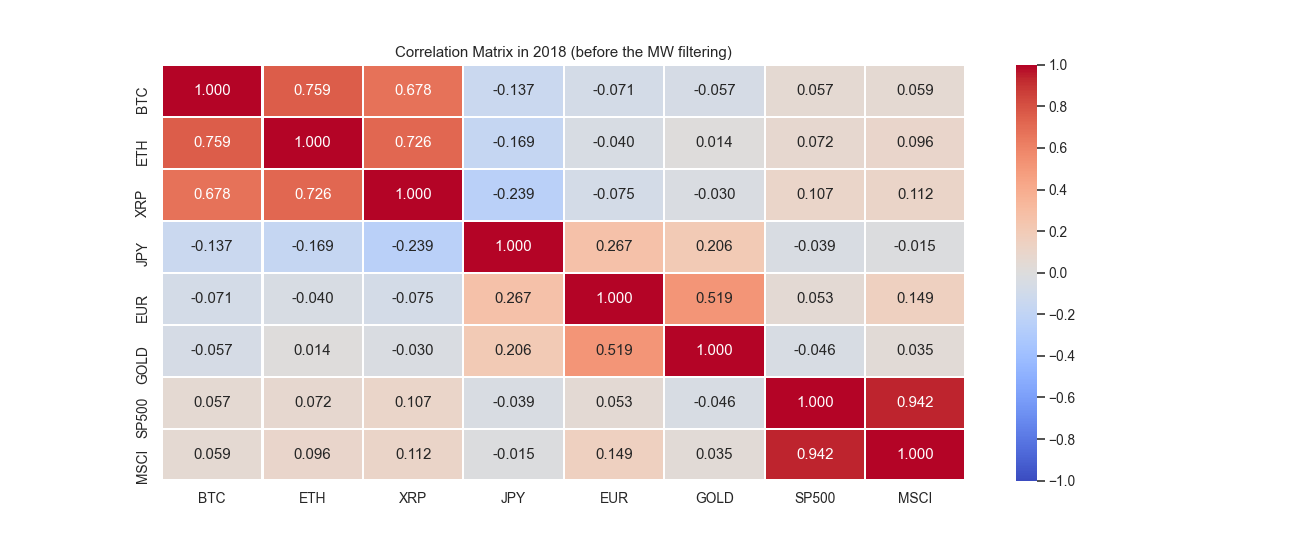}
  \label{fig:Fig3b}
\end{table}

\begin{table}[H]
  \centering
  \caption{Correlation matrix in 2019 (before the M\"{u}ller-Watson filtering)}
  \includegraphics[width=1.0\linewidth]{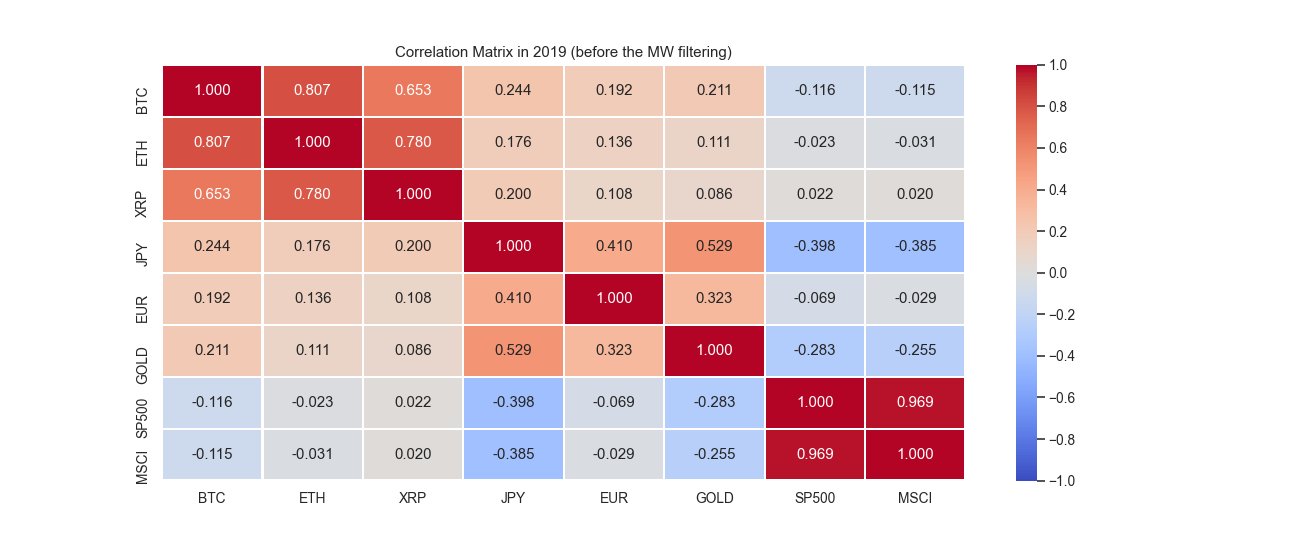}
  \label{fig:Fig4b}
\end{table}

\begin{table}[H]
  \centering
  \caption{Correlation matrix in 2020 (before the M\"{u}ller-Watson filtering)}
  \includegraphics[width=1.0\linewidth]{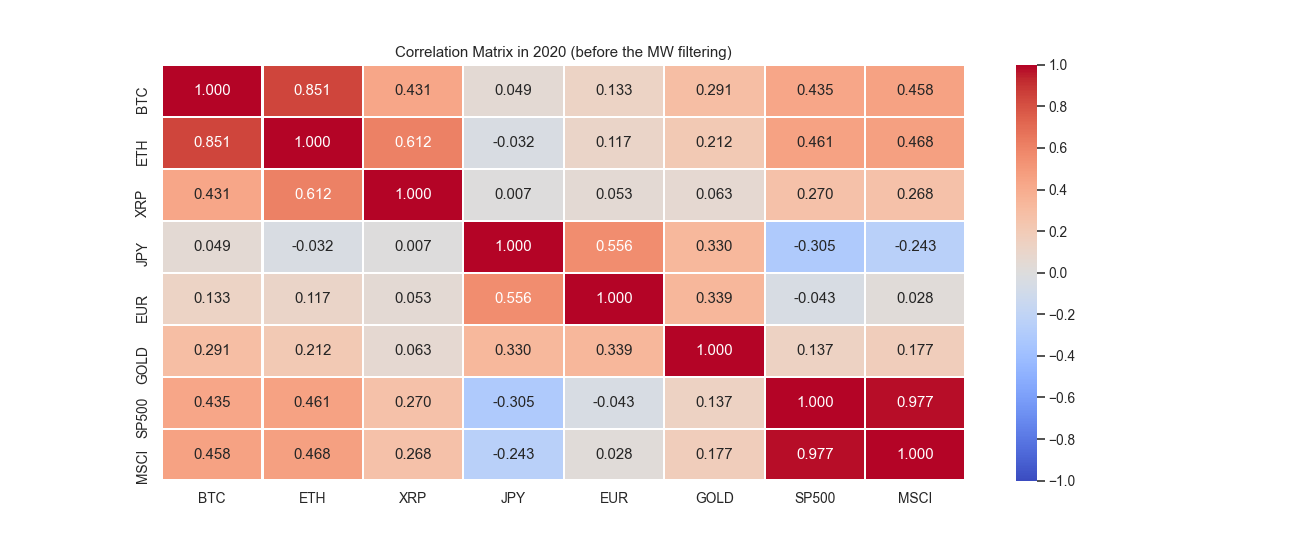}
  \label{fig:Fig5b}
\end{table}

\begin{table}[H]
  \centering
  \caption{Correlation matrix in 2016 (after the M\"{u}ller-Watson filtering)}
  \includegraphics[width=1.0\linewidth]{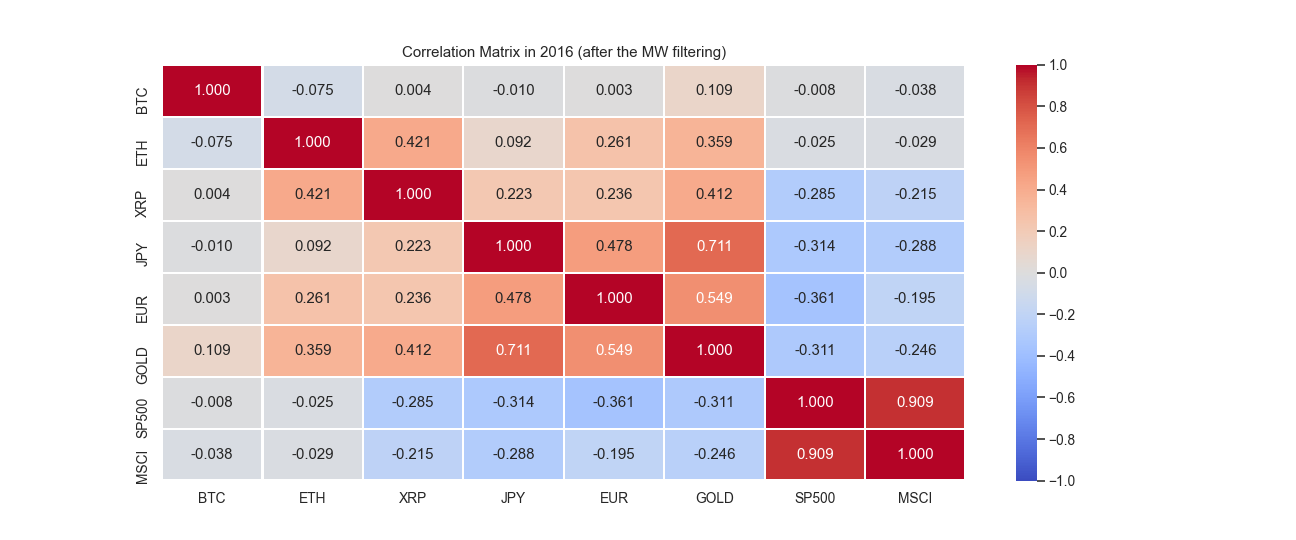}
  \label{fig:Fig1a}
\end{table}

\begin{table}[H]
  \centering
  \caption{Correlation matrix in 2017 (after the M\"{u}ller-Watson filtering)}
  \includegraphics[width=1.0\linewidth]{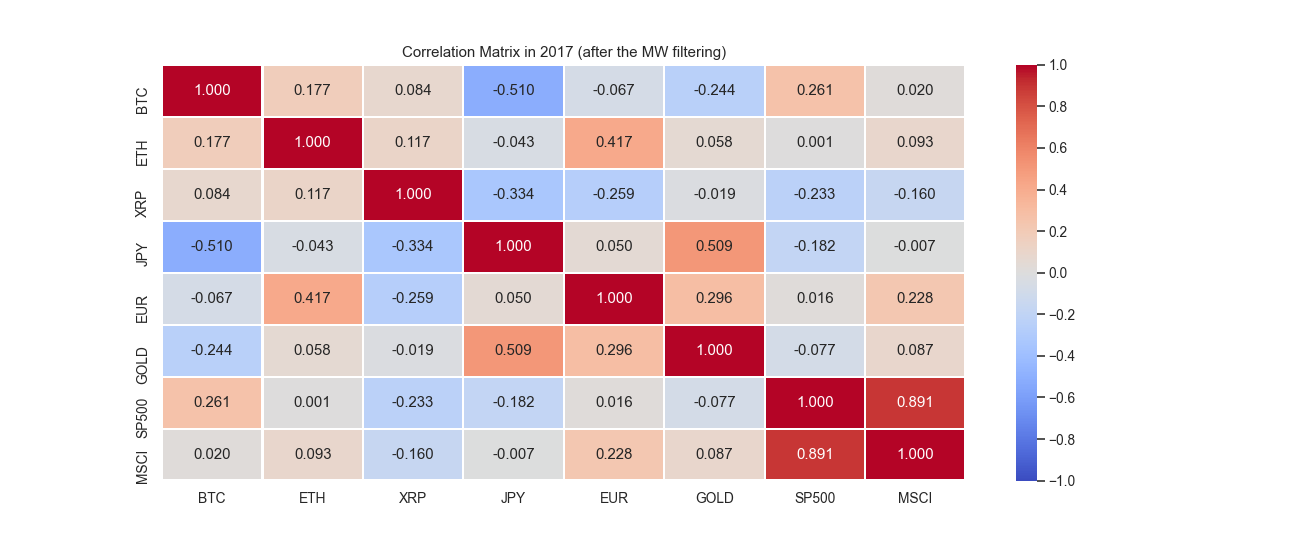}
  \label{fig:Fig2a}
\end{table}

\begin{table}[H]
  \centering
  \caption{Correlation matrix in 2018 (after the M\"{u}ller-Watson filtering)}
  \includegraphics[width=1.0\linewidth]{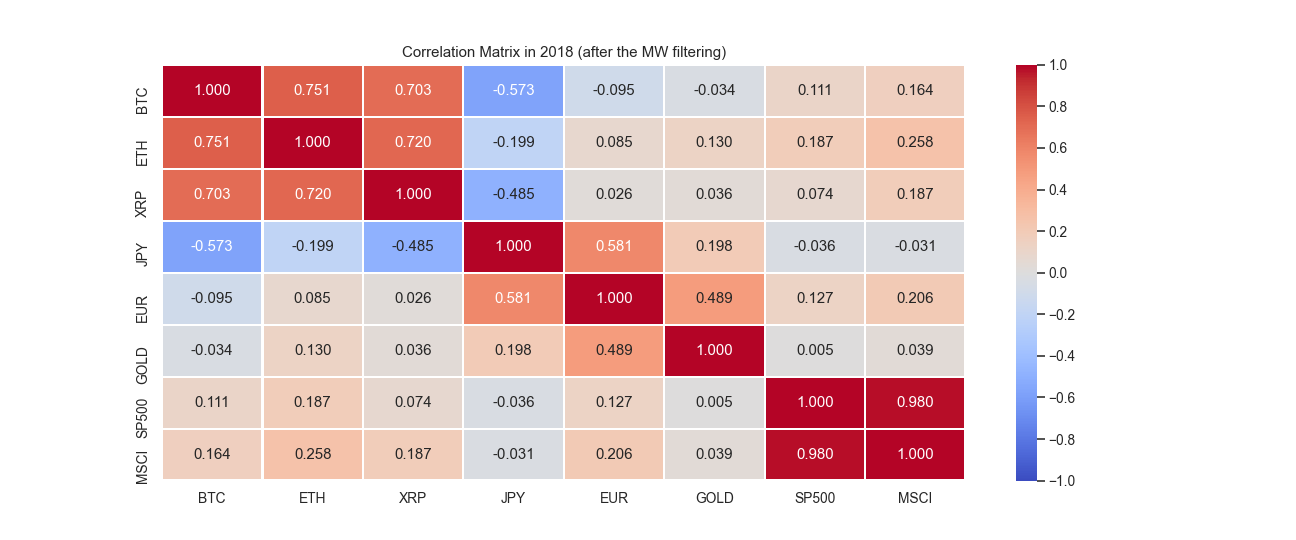}
  \label{fig:Fig3a}
\end{table}

\begin{table}[H]
  \centering
  \caption{Correlation matrix in 2019 (after the M\"{u}ller-Watson filtering)}
  \includegraphics[width=1.0\linewidth]{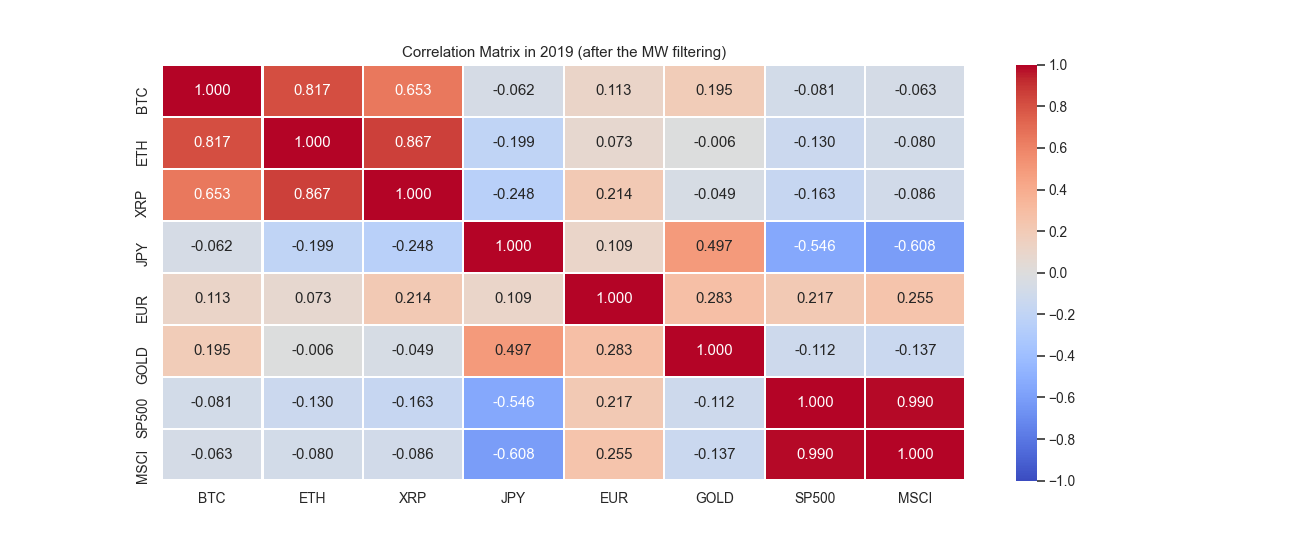}
  \label{fig:Fig4a}
\end{table}

\begin{table}[H]
  \centering
  \caption{Correlation matrix in 2020 (after the M\"{u}ller-Watson filtering)}
  \includegraphics[width=1.0\linewidth]{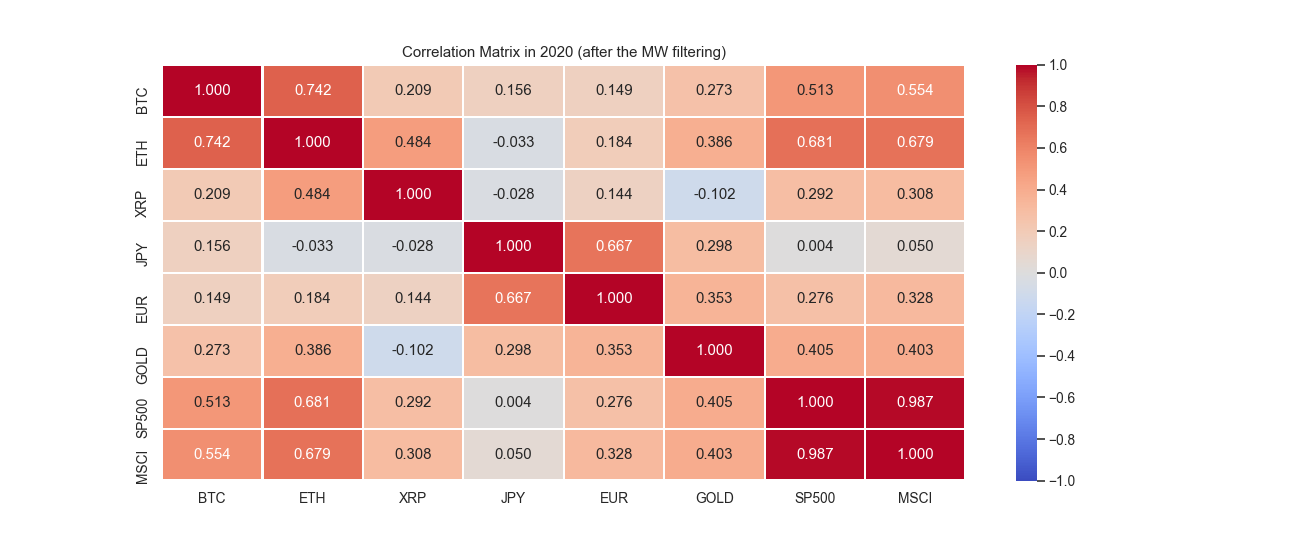}
  \label{fig:Fig5a}
\end{table}

\begin{table}[H]
  \centering
  \caption{DTW similarity matrix in 2016 (before the M\"{u}ller-Watson filtering)}
  \includegraphics[width=1.0\linewidth]{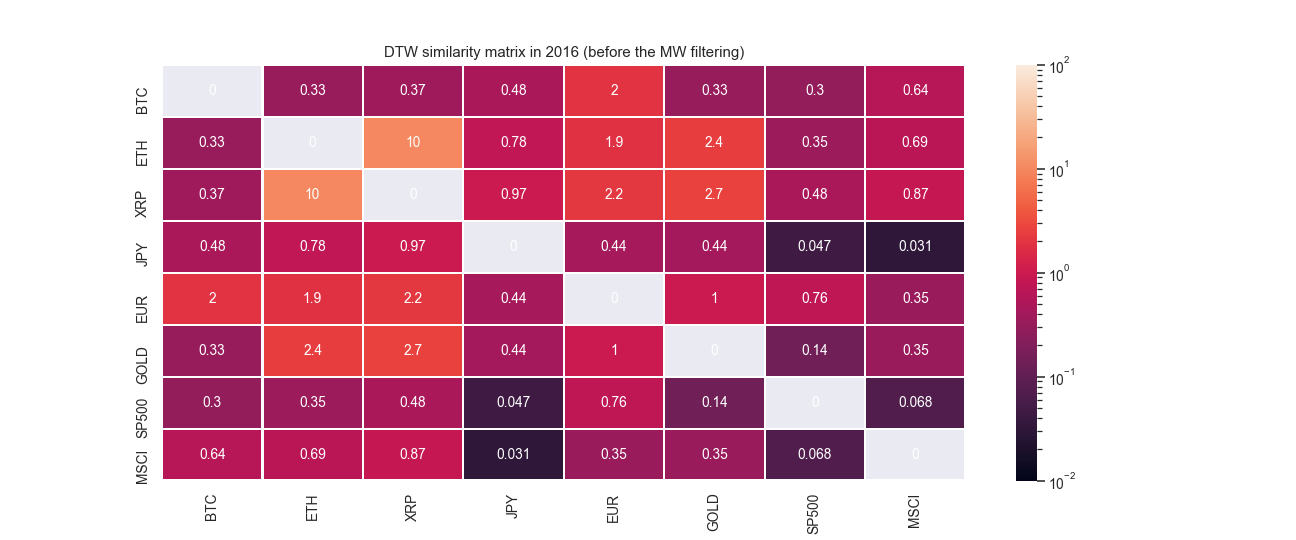}
  \label{fig:Fig11}
\end{table}

\begin{table}[H]
  \centering
  \caption{DTW similarity matrix in 2017 (before the M\"{u}ller-Watson filtering)}
  \includegraphics[width=1.0\linewidth]{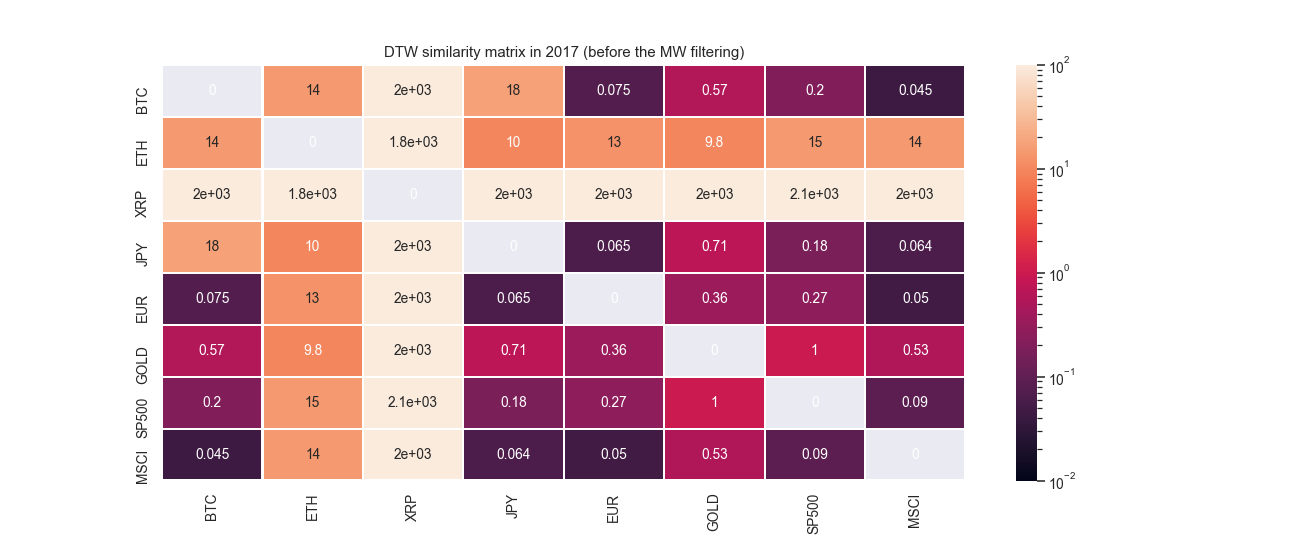}
  \label{fig:Fig12}
\end{table}

\begin{table}[H]
  \centering
  \caption{DTW similarity matrix in 2018 (before the M\"{u}ller-Watson filtering)}
  \includegraphics[width=1.0\linewidth]{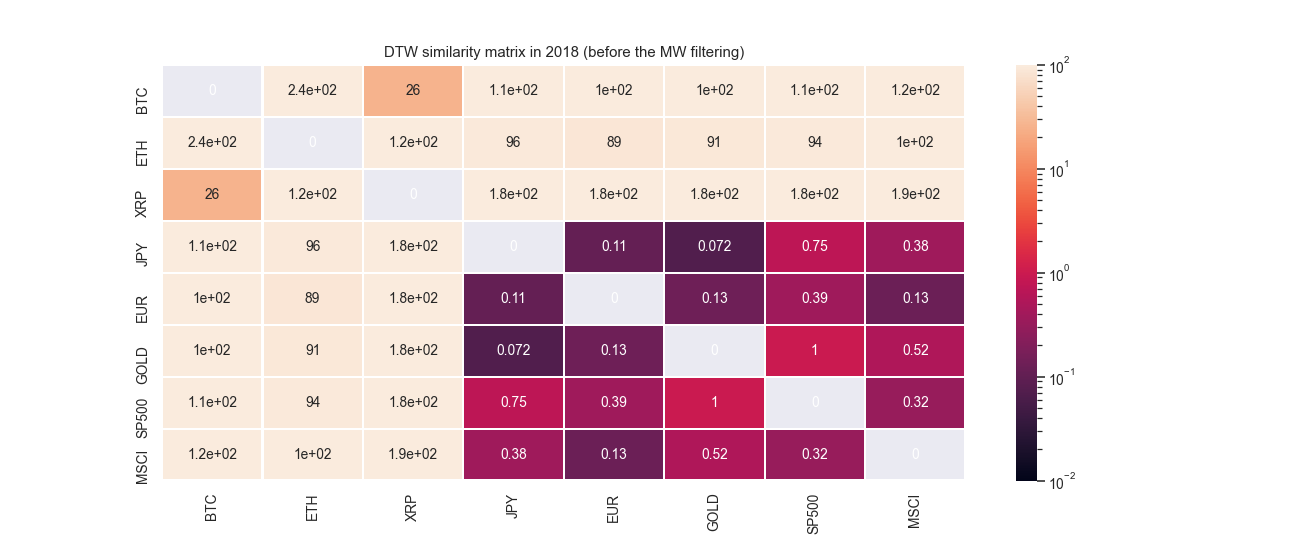}
  \label{fig:Fig13}
\end{table}

\begin{table}[H]
  \centering
  \caption{DTW similarity matrix in 2019 (before the M\"{u}ller-Watson filtering)}
  \includegraphics[width=1.0\linewidth]{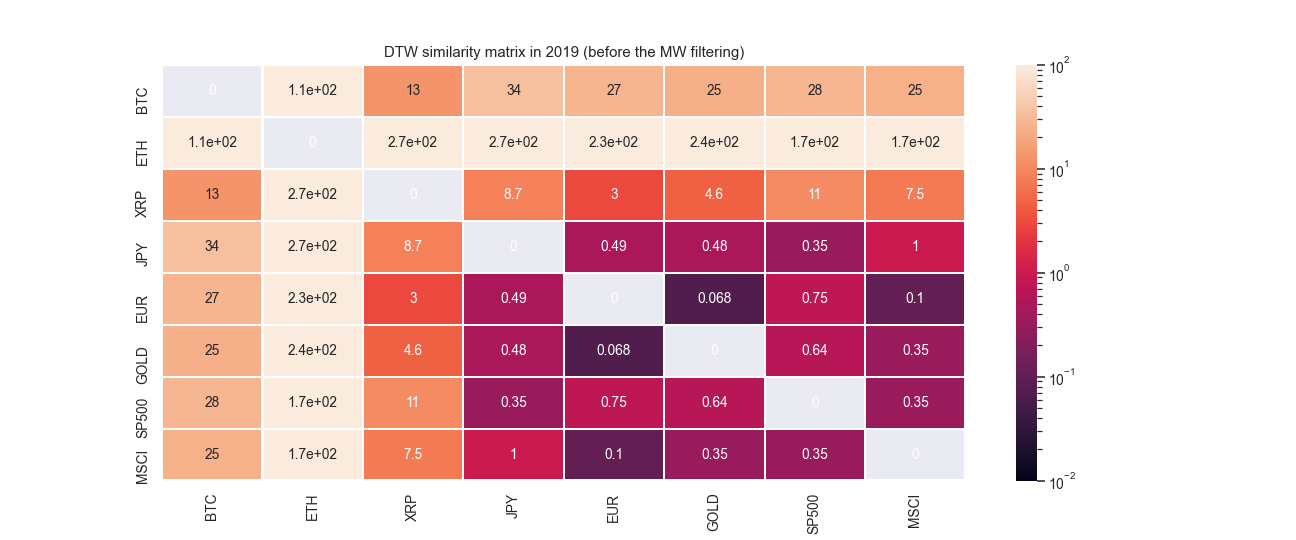}
  \label{fig:Fig14}
\end{table}

\begin{table}[H]
  \centering
  \caption{DTW similarity matrix in 2020 (before the M\"{u}ller-Watson filtering)}
  \includegraphics[width=1.0\linewidth]{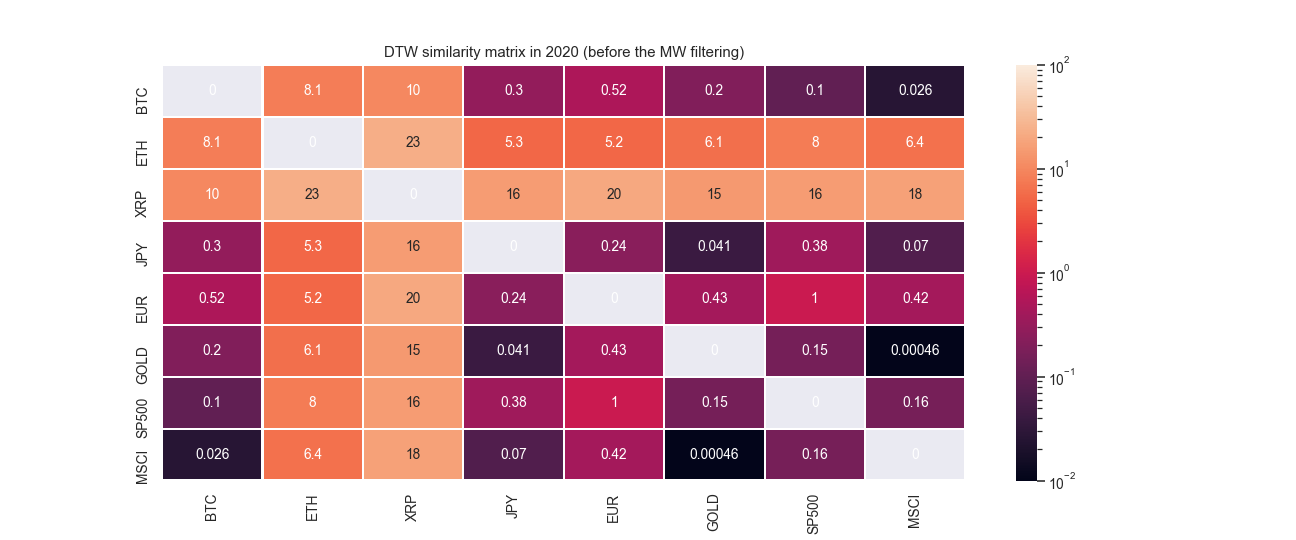}
  \label{fig:Fig15}
\end{table}

\begin{table}[H]
  \centering
  \caption{DTW similarity matrix in 2016 (after the M\"{u}ller-Watson filtering)}
  \includegraphics[width=1.0\linewidth]{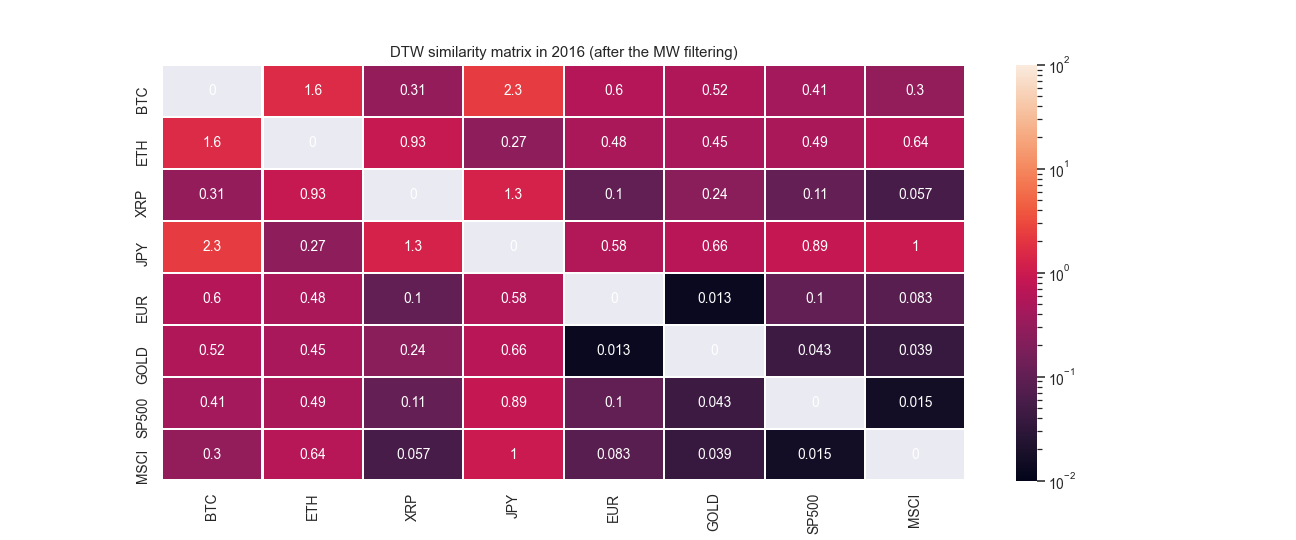}
  \label{fig:Fig11q}
\end{table}

\begin{table}[H]
  \centering
  \caption{DTW similarity matrix in 2017 (after the M\"{u}ller-Watson filtering)}
  \includegraphics[width=1.0\linewidth]{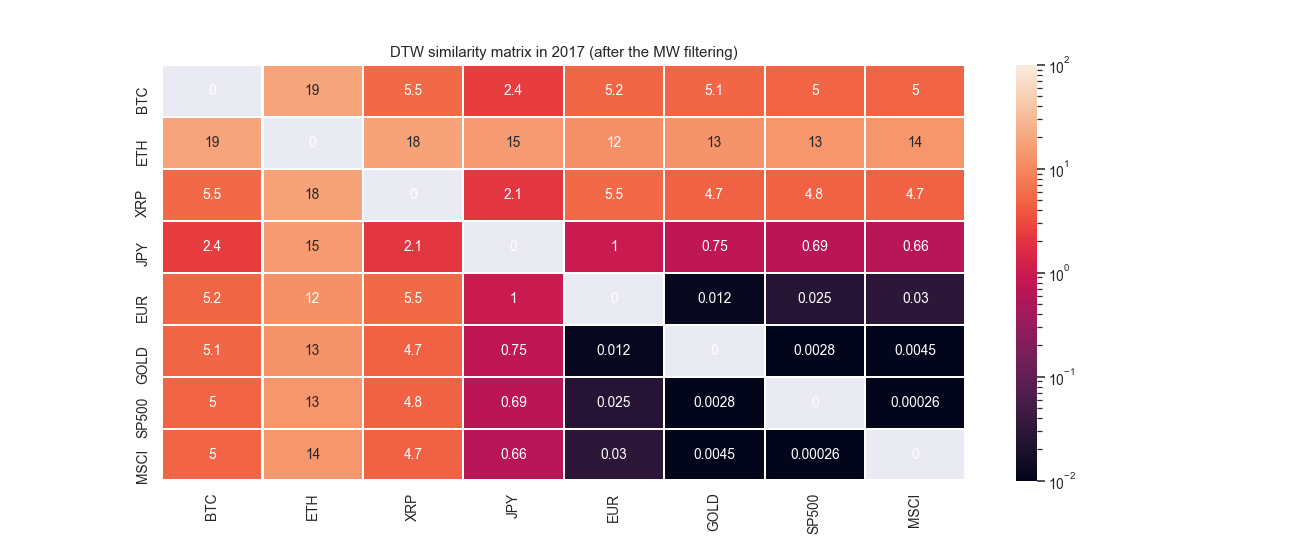}
  \label{fig:Fig12q}
\end{table}

\begin{table}[H]
  \centering
  \caption{DTW similarity matrix in 2018 (after the M\"{u}ller-Watson filtering)}
  \includegraphics[width=1.0\linewidth]{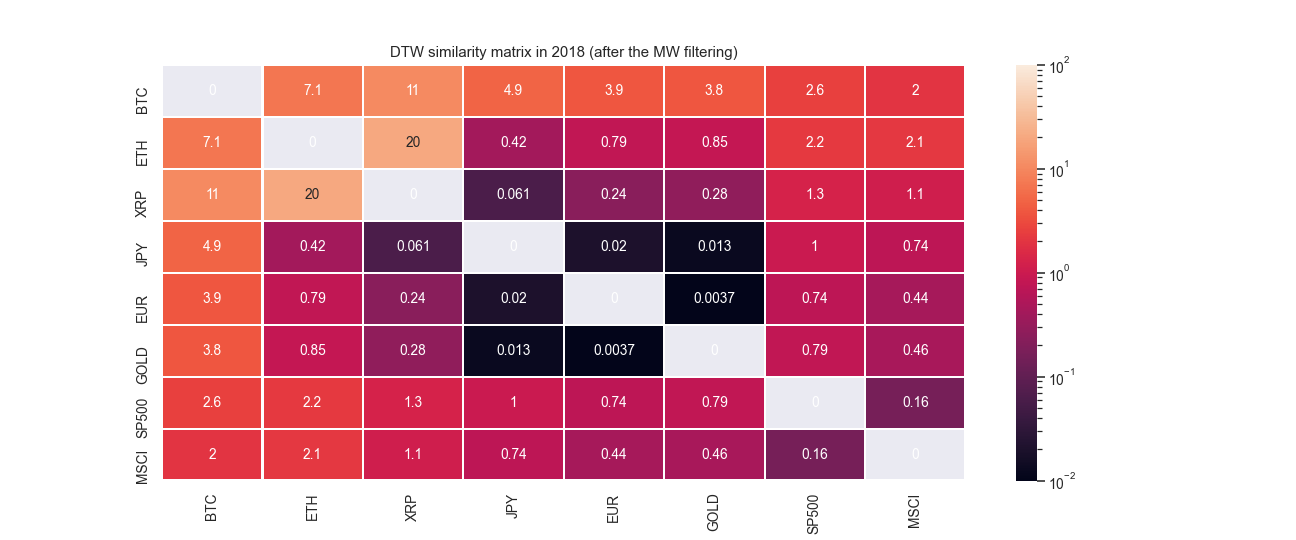}
  \label{fig:Fig13q}
\end{table}

\begin{table}[H]
  \centering
  \caption{DTW similarity matrix in 2019 (after the M\"{u}ller-Watson filtering)}
  \includegraphics[width=1.0\linewidth]{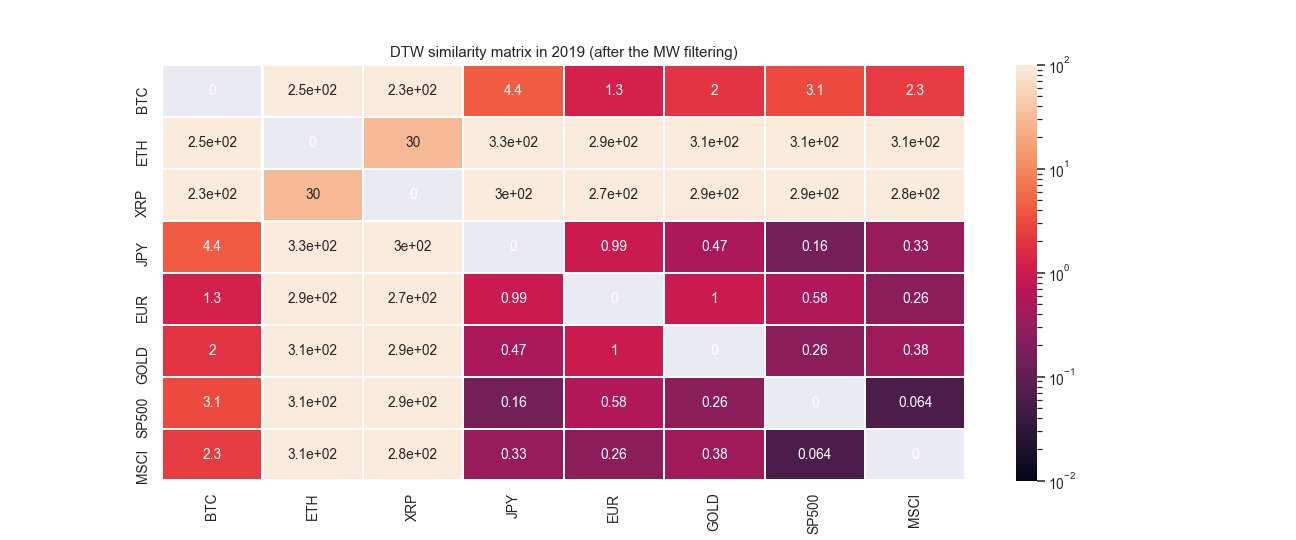}
  \label{fig:Fig14q}
\end{table}

\begin{table}[H]
  \centering
  \caption{DTW similarity matrix in 2020 (after the M\"{u}ller-Watson filtering)}
  \includegraphics[width=1.0\linewidth]{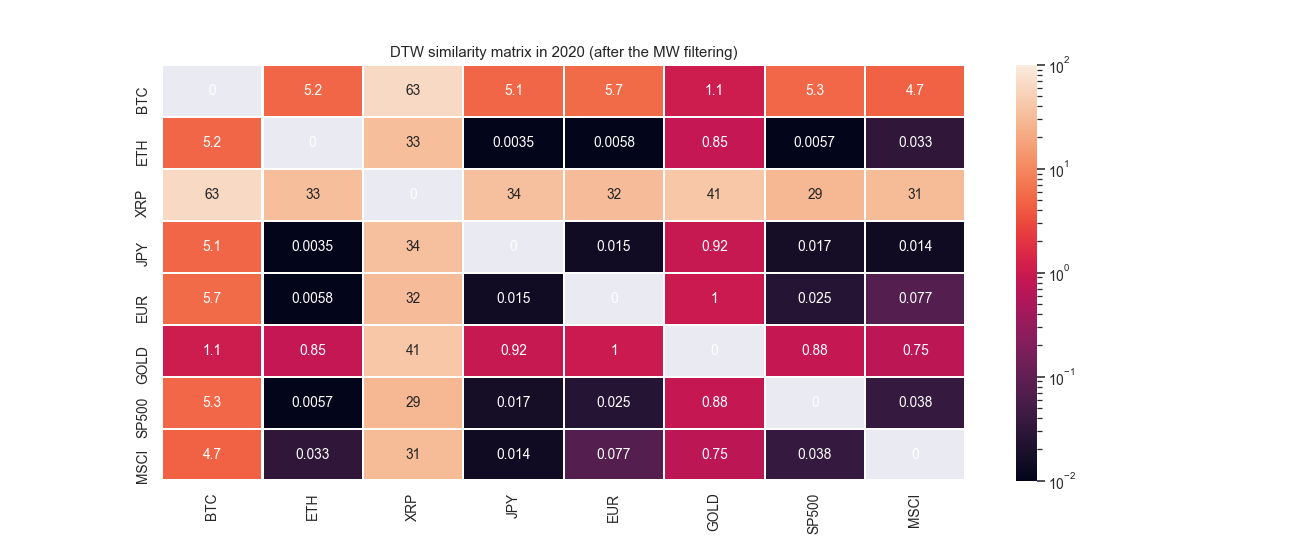}
  \label{fig:Fig15q}
\end{table}
\end{document}